\documentclass[11pt,a4paper]{article}
\usepackage{jheppub,xcolor}
\pdfoutput=1
\usepackage[utf8]{inputenc}
\graphicspath{{./figs/}}
\usepackage[a4paper,top=4cm,bottom=0cm,left=4.0cm,right=-0.8cm,bindingoffset=0mm]{geometry}

\usepackage{amssymb}
\usepackage{dcolumn}	
\usepackage{bm}			
\usepackage{bbm} 		
\usepackage{multirow}
\usepackage{slashed}
\usepackage{blkarray}
\usepackage{booktabs}
\usepackage{makecell}
\usepackage{longtable}
\usepackage{placeins}
\usepackage{listings}
\usepackage{amsmath}
\usepackage{xcolor}

\def\gsim{\mathrel{\rlap{\lower4pt\hbox{\hskip1pt$\sim$}}
    \raise1pt\hbox{$>$}}}         
\def\lsim{\mathrel{\rlap{\lower4pt\hbox{\hskip1pt$\sim$}}
    \raise1pt\hbox{$<$}}}         

\definecolor{codegreen}{rgb}{0,0.6,0}
\definecolor{codegray}{rgb}{0.5,0.5,0.5}
\definecolor{codepurple}{rgb}{0.58,0,0.82}
\definecolor{backcolour}{rgb}{0.95,0.95,0.92}

\lstdefinestyle{mystyle}{
    backgroundcolor=\color{backcolour},   
    commentstyle=\color{codegreen},
    keywordstyle=\color{magenta},
    numberstyle=\tiny\color{codegray},
    stringstyle=\color{codepurple},
    basicstyle=\ttfamily\footnotesize,
    breakatwhitespace=false,         
    breaklines=true,                 
    captionpos=b,                    
    keepspaces=true,                 
    numbers=left,                    
    numbersep=5pt,                  
    showspaces=false,                
    showstringspaces=false,
    showtabs=false,                  
    tabsize=2
}
\definecolor{Red}{rgb}{1.,0.,0.}

\newcommand{\mcO}{\mathcal{O}}

\newcommand{\OO}{\ensuremath{\mathcal{O}}}
\newcommand{\Op}[1]{\OO_{\sss #1}}
\newcommand{\sss}{\scriptscriptstyle}
\newcommand{\pdp}{\ensuremath{\varphi^\dagger\varphi}}
\newcommand{\ccc}[3]{c_{#2}^{#1 (#3)}}
\newcommand{\qq}[3]{\ensuremath{\mathcal{O}_{#2}^{#1 (#3)}}}

\newcommand{\lp}{\left(}
\newcommand{\rp}{\right)}
\newcommand{\lc}{\left[}
\newcommand{\rc}{\right]}

\newcommand{\bc}{\begin{center}}
\newcommand{\ec}{\end{center}}
\newcommand{\ba}{\begin{array}}
\newcommand{\ea}{\end{array}}

\newcommand{\be}{\begin{equation}}
\newcommand{\ee}{\end{equation}}
\newcommand{\bea}{\begin{align}}
\newcommand{\eea}{\end{align}}

 \newcolumntype{C}[1]{>{\centering\arraybackslash}p{#1}}

 \def\lra#1{\overset{\text{\scriptsize$\leftrightarrow$}}{#1}}

\numberwithin{equation}{section}
\numberwithin{figure}{section}
\numberwithin{table}{section}

\title{Mapping the SMEFT at High-Energy Colliders: \\[0.05cm] from LEP and the (HL-)LHC to the FCC-ee}

\author[a]{Eugenia~Celada,}
\author[b,c]{Tommaso~Giani,}
\author[b,c]{Jaco~ter~Hoeve,}
\author[d]{Luca~Mantani,}
\author[b,c]{Juan~Rojo,}
\author[a]{Alejo~N.~Rossia,}
\author[a]{Marion~O.~A.~Thomas,}
\author[a]{and Eleni~Vryonidou}
\affiliation[a]{Department of Physics and Astronomy, University of Manchester, Oxford Road, Manchester M13 9PL, United Kingdom}
\affiliation[b]{Nikhef Theory Group, Science Park 105, 1098 XG Amsterdam, The Netherlands}
\affiliation[c]{Department of Physics and Astronomy, Vrije Universiteit Amsterdam, NL-1081 HV Amsterdam, The Netherlands}
\affiliation[d]{DAMTP, University of Cambridge, Wilberforce Road, Cambridge, CB3 0WA, United Kingdom}

\emailAdd{eugenia.celada@manchester.ac.uk}
\emailAdd{tgiani@nikhef.nl}
\emailAdd{j.j.ter.hoeve@vu.nl}
\emailAdd{luca.mantani@maths.cam.ac.uk}
\emailAdd{j.rojo@vu.nl}
\emailAdd{alejo.rossia@manchester.ac.uk}
\emailAdd{marion.thomas@manchester.ac.uk}
\emailAdd{eleni.vryonidou@manchester.ac.uk}

\date{\today}
\abstract{
We present {\sc\small SMEFiT3.0}, an updated global SMEFT analysis of Higgs, top quark, and diboson production data from the LHC complemented by electroweak precision observables (EWPOs) from LEP and SLD.
We consider recent inclusive and differential measurements from the LHC Run II, alongside with a novel implementation of the EWPOs based on independent calculations of the relevant EFT contributions.
We estimate the impact of HL-LHC measurements on the SMEFT parameter space when added on top of {\sc\small SMEFiT3.0}, through dedicated projections extrapolating from Run II data.
We quantify the significant constraints that measurements from two proposed high-energy circular $e^+e^-$ colliders, the FCC-ee and the CEPC, would impose on both the SMEFT parameter space and on representative UV-complete models.
Our analysis considers projections for the FCC-ee and the CEPC based on the latest running scenarios and includes $Z$-pole EWPOs, fermion-pair, Higgs, diboson, and top quark production, using optimal observables for both the $W^+W^-$ and the $t\bar{t}$ channels.
The framework presented in this work may be extended to other future colliders and running scenarios, providing timely input to ongoing studies towards future high-energy particle physics facilities.  
}


\begin{document}
\begin{flushright}
\end{flushright}

\maketitle
\flushbottom

\newpage
\section{Introduction}
\label{sec:intro}

The vast amount of data collected by the LHC during Run II has significantly increased our knowledge of fundamental particle physics. 
A global interpretation of these measurements is required to understand how well our best current theory, the Standard Model (SM), describes nature at the TeV scale and how much room (and where) is left for its extensions.
The magnitude of this task, as well as our knowledge, keeps increasing thanks to the ongoing LHC Run III, which will provide exciting new insights.

Following Run III of the LHC, its High-Luminosity upgrade (HL-LHC)~\cite{Cepeda:2019klc,Azzi:2019yne} is scheduled to start in 2029 and operate for the next decade, accumulating a total integrated luminosity of up to 3 ab$^{-1}$ per experiment.
Beyond the HL-LHC program, several proposals for future particle colliders have been put forward and are being actively discussed by the global community.
These proposals include electron-positron colliders, either circular such as the FCC-ee~\cite{FCC:2018byv,FCC:2018evy} and the CEPC~\cite{CEPCPhysicsStudyGroup:2022uwl}, or linear such as the ILC~\cite{Behnke:2013xla,ILC:2013jhg}, the C3~\cite{Vernieri:2022fae}, and CLIC~\cite{Linssen:2012hp}, high-energy proton-proton colliders such as the FCC-hh~\cite{FCC:2018byv,FCC:2018vvp} and the SppC~\cite{Tang:2015qga}, muon colliders~\cite{Accettura:2023ked,Aime:2022flm}, and high-energy electron-proton/ion colliders such as the LHeC and the FCC-eh~\cite{LHeC:2020van,FCC:2018byv}. 
Furthermore, with a focus on QCD and hadronic physics but also with a rich program of electroweak measurements and BSM searches, the Electron Ion Collider (EIC)~\cite{AbdulKhalek:2021gbh} has already been approved and is expected to see its first collisions in the early 2030s.

These proposed facilities envisage a significant expansion of our knowledge of nature at the smallest accessible scales.
Novel insights would be provided via unprecedented sensitivity to subtle quantum effects, distorting the properties of known particles, and via the direct production of new particles, for instance, heavy (TeV-scale) particles or lighter ones but feebly interacting. 
While opening unique opportunities, the theoretical interpretation of future collider measurements also poses several challenges that must be tackled beforehand.
In the specific case of leptonic colliders, significant theoretical progress both within the SM and beyond would be required to match the expected precision of the experimental measurements. 
This is illustrated with the Tera-$Z$ running program~\cite{TLEPDesignStudyWorkingGroup:2013myl} of circular $e^+e^-$ colliders, which would lead to up to $10^{12}$ $Z$-bosons at the FCC-ee, with hence extremely small statistical uncertainties.

Making an informed decision about which of these future particle colliders should be built demands, in addition to feasibility and cost/effectiveness studies, the quantitative assessment of their scientific reach.
Several dedicated studies comparing the reach of future colliders have been presented in the last years, in particular in the context of the European Strategy for Particle Physics Update (ESPPU)~\cite{EuropeanStrategyforParticlePhysicsPreparatoryGroup:2019qin} and of the Snowmass US-based community process~\cite{Narain:2022qud}. 
The latter has recently culminated in the P5 report, one of whose main recommendations is endorsing an off-shore Higgs factory located
in either Europe or Japan.
As the European community ramps up its activities towards the next update of its long-term strategy, continuing, diversifying, and deepening these quantitative assessments of the reach of future colliders is more timely than ever.

The Standard Model Effective Field Theory (SMEFT) framework~\cite{Grzadkowski:2010es,Brivio:2017vri,Isidori:2023pyp} is particularly well suited to perform global interpretations of current data and study the physics potential of future colliders. 
This feature follows from its ability to parametrize simultaneously the effects of a vast family of models of New Physics (NP) and to correlate effects in several observables.
The models that are covered by the SMEFT are those where all new particles are heavier than the electroweak scale and that decouple from the SM 
~\cite{Falkowski:2019tft,Cohen:2020xca}.
Hence, projected global SMEFT fits allow us to compare the proposed future colliders in generic NP scenarios. 
The development of flexible and user-friendly frameworks for such global interpretations is thus a timely priority~\cite{DeBlas:2019ehy,Ellis:2020unq,Brivio:2021alv,Giani:2023gfq}.
Furthermore, the predictions computed within the SMEFT framework can then be reused to repeat the comparison within specific UV models by imposing the constraints dictated by the UV matching onto the SMEFT~\cite{deBlas:2017xtg,DasBakshi:2018vni,Ellis:2020unq,Brivio:2021alv,Carmona:2021xtq,Fuentes-Martin:2022jrf,terHoeve:2023pvs}.

The advantages of using SMEFT (and EFTs in general) for the assessment of future colliders have been leveraged by several groups in the past. 
First, in studies that considered a limited set of measurements and then in fits to a much larger data set, see e.g.~\cite{Ellis:2015sca,deBlas:2016nqo,Ellis:2017kfi,Durieux:2017rsg,Barklow:2017suo,Barklow:2017awn,DiVita:2017vrr,Chiu:2017yrx,Durieux:2018tev,deBlas:2019rxi,DeBlas:2019qco,LCCPhysicsWorkingGroup:2019fvj,Durieux:2019rbz,Jung:2020uzh,deBlas:2021jlt,MuonCollider:2022xlm,deBlas:2022ofj,Allwicher:2023shc}. 
The latter group of studies has highlighted the need for the inclusion of LEP and SLD data and HL-LHC projections to avoid overestimating the benefits of future colliders~\cite{deBlas:2019rxi,DeBlas:2019qco,Durieux:2019rbz,deBlas:2022ofj}. 
SMEFT studies at future lepton colliders could also probe some of the fundamental principles underlying Quantum Field Theory such as unitarity, locality and Lorentz invariance~\cite{Gu:2020thj,Gu:2020ldn}.

A key component of the LEP and SLD legacy are the electroweak precision observables (EWPOs)~\cite{ALEPH:2005ab} from electron-positron collisions at the $Z$-pole and beyond.
The high precision achieved by the measurements of EWPOs imposes stringent stress tests of the Standard Model (SM)~\cite{Altarelli:1991fk,Baak:2012kk} and indirectly constrains a wide range of models beyond
the Standard Model (BSM).
In the language of the SMEFT, EWPOs provide information on several directions in the EFT parameter space~\cite{Brivio:2017bnu}, some of them of great relevance when matching onto compelling UV-completions~\cite{terHoeve:2023pvs}.
Indeed, as compared to measurements at the LHC, these EWPOs provide complementary, and in many cases dominant, sensitivity to a wealth of BSM scenarios.
In this respect, the EWPOs supplement uniquely the direct information on the properties of the Higgs and electroweak sectors obtained at the LHC~\cite{Dawson:2018dcd,Butter:2016cvz,Azatov:2019xxn}.
Furthermore, they show an intriguing interplay with low-energy flavour observables~\cite{Cirigliano:2016nyn,Aoude:2020dwv,Crivellin:2020ebi,Bruggisser:2021duo,Bruggisser:2022rhb,Cirigliano:2022qdm,Bartocci:2023nvp,Cirigliano:2023nol,Allwicher:2023shc,Bellafronte:2023amz}.
The consistent implementation of the latter, which also have an interplay with top-quark measurements~\cite{Garosi:2023yxg}, requires considering the renormalisation group effects~\cite{Jenkins:2013zja,Jenkins:2013wua,Alonso:2013hga,Aoude:2022aro}.
Within a global SMEFT fit, the implementation of EWPOs  must deal with several theoretical subtleties, and ensure that the underlying settings such as the choice of electroweak input scheme, flavour assumptions, and operator basis are consistent with those adopted in the associated interpretation of LHC measurements.

The goal of this paper is two-fold.
First, we present {\sc\small SMEFiT3.0}, an updated global SMEFT analysis of Higgs, top quark, and diboson data from the LHC complemented by EWPOs from LEP and SLD.
This analysis,  carried out within the {\sc\small SMEFiT} framework~\cite{Hartland:2019bjb,Ethier:2021ydt,Ethier:2021bye,vanBeek:2019evb,Giani:2023gfq,terHoeve:2023pvs}, considers recent inclusive and differential measurements from the LHC Run II, in several cases based on its full integrated luminosity, alongside with a new implementation of the EWPOs.
The latter is based on independent calculations of the relevant EFT contributions, ensures consistent theory settings between the $e^+e^-$ and hadron collider processes, and is benchmarked with previous studies in the literature. 
Constraining 45~(50) independent directions in the parameter space within linear (quadratic) SMEFT fits, this analysis provides a state-of-the-art set of EFT bounds enabled by available LEP and LHC data.

Second, starting from this {\sc\small SMEFiT3.0} baseline, we quantify the constraints on the EFT operators of projected measurements, first from the HL-LHC and subsequently from two of the proposed high-energy electron-positron colliders, namely the circular variants FCC-ee and the CEPC. 
This result is achieved by extending the {\sc\small SMEFiT} framework with novel functionalities streamlining the inclusion of projections for future experimental facilities into the global fit.
Beyond its specific application to FCC-ee and CEPC projections, this proof-of-concept study illustrates the relevance of {\sc\small SMEFiT} for ongoing studies assessing the physics reach of new particle colliders. 
As such, we expect that it will represent a useful tool for the particle physics community in the ongoing discussions towards charting its long-term future.
Our framework is open-source, extensively documented, and user-friendly.
We provide all the inputs needed to reproduce our results, such that the community can easily reuse and expand them towards other future experiments.

Our analysis emphasises the profound interplay between measurements at leptonic and hadronic colliders to constrain complementary directions in the EFT parameter space. 
It illustrates the potential of future colliders, first the HL-LHC and then the FCC-ee and CEPC, to inform indirect BSM searches via high-precision measurements extending the sensitivity provided by existing data.
This unprecedented reach is quantified both at the level of Wilson coefficients as well as in terms of the parameters (masses and couplings) of representative UV-complete models, in the latter case benefiting from progress in the interfacing with automated matching tools~\cite{Carmona:2021xtq,Fuentes-Martin:2022jrf}, as presented in~\cite{terHoeve:2023pvs}.

The outline of this paper is as follows.
First, in Sect.~\ref{sec:ewpos} we describe the new implementation of EWPOs in {\sc\small SMEFiT} and benchmark our results with related studies in the literature.
Sect.~\ref{sec:updated_global_fit} presents the {\sc\small SMEFiT3.0} global analysis, including the LHC Run II measurements alongside the new EWPO implementation.
By extrapolating from Run II data, this fit is used as a baseline to estimate the constraints on the SMEFT coefficients which may be achieved at the HL-LHC.
In Sect.~\ref{sec:results}, projections for FCC-ee and CEPC measurements are added to the HL-LHC baseline to determine the ultimate sensitivity of future circular $e^+e^-$ colliders to the SMEFT parameter space.
Sect.~\ref{subsec:uvmodels} presents the results of our global analyses of LHC Run II, HL-LHC and FCC-ee measurements at the level of the couplings and masses of UV-complete models matched onto the SMEFT. 
Finally, we summarise and outline possible
developments in Sect.~\ref{sec:summary}.

Technical information is collected in the appendices.
App.~\ref{app-list-fits} collects the input settings adopted for the SMEFT fits presented in this work.
App.~\ref{app-operators} describes the EFT operator basis adopted, whilst App.~\ref{app-benchmarking}
describes the benchmarking and validation
of the EWPO implementation.
App.~\ref{app:hl_lhc_projections} describes the procedure to extrapolate Run II datasets to the HL-LHC data-taking period.
App.~\ref{app:observables} summarises the FCC-ee and CEPC observables considered in this study and compiles their projected uncertainties.
App.~\ref{app:optimal} reviews our treatment of optimal observables within the global SMEFT fit and their application to $W^+W^-$ and $t\bar{t}$ production from $e^+e^-$ collisions.
Finally, App.~\ref{subsec:app-widths} describes the approach adopted to include SMEFT effects in electroweak gauge boson decays.

\section{Electroweak precision observables in  {\sc\small SMEFiT}}
\label{sec:ewpos}

Here we describe the new implementation and validation
of EWPOs in the {\sc\small SMEFiT} analysis framework.
In the next section we compare our results with those obtained with the previous approximation.

\subsection{EWPOs in the SMEFT}
\label{subsec:EWPOs_recap}

For completeness and to set up the notation, we provide a concise overview of how SMEFT operators affect the EWPOs measured at electron-positron colliders operating at the $Z$-pole and beyond such as LEP and SLD.
We work in the $\{\hat{m}_W, \hat{m}_Z, \hat{G}_F\}$ input electroweak scheme.
In the following, physical quantities which are either measured or
derived from measurements
are indicated with a hat, while canonically normalized Lagrangian parameters are denoted with a bar.
Through this paper, Wilson coefficients follow the definitions and conventions of the Warsaw basis~\cite{Grzadkowski:2010es} and we use a U$(2)_q\times$U$(3)_d\times$U$(2)_u\times(\text{U}(1)_\ell\times\text{U}(1)_e)^3$ flavour assumption.
The operators are defined in App.~\ref{app-operators} following the same conventions as \cite{Ethier:2021bye}.

In the presence of dimension-six SMEFT operators and adapting ~\cite{Brivio:2017bnu} to our conventions, the SM values of Fermi's constant and the electroweak boson masses are shifted as follows:
\begin{align}
\delta G_F &= \frac{1}{2 \hat{G}_F} \Bigg( c_{\varphi \ell_1}^{(3)}+c_{\varphi \ell_2}^{(3)} -  c_{\ell\ell}  \Bigg) \, , \nonumber\\
\frac{\delta m_Z^2}{\hat{m}_Z^2} &= \frac{1}{2\sqrt{2}\hat{G}_F}c_{\varphi D} + \frac{\sqrt{2}}{\hat{G}_F}\frac{\hat{m}_W}{\hat{m}_Z}\sqrt{1-\frac{\hat{m}_W^2}{\hat{m}_Z^2}}c_{\varphi WB} \, , \label{eq:input_EW_shifts}\\
\frac{\delta m_W^2}{\hat{m}_W^2} &= 0 \, . \nonumber
\end{align}
In the following, we adopt a notation in which the new physics cutoff scale $\Lambda$ has been reabsorbed into the Wilson coefficients in Eq.~(\ref{eq:input_EW_shifts}), which therefore here and in the rest of the section should be understood to be dimensionful and with mass-energy units of $\lc c\rc = -2$.
We note that in this notation $\delta G_F$ is dimensionless, and hence indicates a relative shift.

These SMEFT-induced shifts in the electroweak input parameters defining the $\{\hat{m}_W, \hat{m}_Z, \hat{G}_F\}$ scheme modify the  interactions of the electroweak gauge bosons.
Specifically, the vector (V) and axial (A) couplings $g_{V,A}$ of the $Z$-boson are shifted in comparison
to the SM reference $\bar{g}_{V, A}$ (recall that the bar indicates renormalised Lagrangian parameters)
according to the following relation:
\begin{align}
\label{eq:VA-shifts}
g_{V, A}^x = \bar{g}_{V, A}^x + \delta g_{V, A}^x \, , \qquad x=\{\ell_i, u_i, d_i, \nu_i\} \, ,
\end{align}
where the superscript $x$ denotes the
fermion to which the $Z$-boson couples: either a charged (neutral) lepton $\ell_i$ ($\nu_i$), an up-type quark $u_i$ or a down-type quark $d_i$, respectively.
The flavour index $i=1,2,3$ runs over fermionic generations.
The SM couplings in Eq.~(\ref{eq:VA-shifts})
are given in the adopted notation by
\be
\bar{g}_V^x = T_3^x/2 - Q^xs_{\hat{\theta}}^2  \, ,\qquad \bar{g}_{A}^{x} = T_3^x/2 \, ,
\ee
\be
Q^x = \{-1, 2/3, -1/3, 0\} \, ,\qquad T_3^x = \{-1/2, 1/2, -1/2, 1/2\} \, ,\qquad s_{\hat{\theta}}^2=1-\hat{m}_W^2/\hat{m}_Z^2 \, ,
\ee
where $ s_{\hat{\theta}}\equiv \sin \hat{\theta}$.
This shift in the SM couplings of the $Z$-boson arising from the dimension-six operators
in Eq.~(\ref{eq:VA-shifts})
can be further decomposed as
\begin{equation}
\label{eq:Zcoupling_EFT_shifts}
\delta g_V^x = \delta\bar{g}_Z\,\bar{g}_V^x + Q^x\delta s_\theta^2 + \Delta_V^x, \qquad \delta g_A^x = \delta\bar{g}_Z\,\bar{g}_A + \Delta_A^x \, ,
\end{equation}
for the vector and axial couplings respectively.
In Eq.~(\ref{eq:Zcoupling_EFT_shifts}) we have defined the (dimensionless) shifts in terms of the Wilson coefficients in the Warsaw basis
\begin{align}
\nonumber \delta \bar{g}_Z &= -\frac{1}{\sqrt{2}}\delta G_F - \frac{1}{2}\frac{\delta m_Z^2}{\hat{m}_Z^2} + \frac{s_{\hat{\theta}}c_{\hat{\theta}}}{\sqrt{2}\hat{G}_F}c_{\varphi WB}\\
&= -\frac{1}{4\sqrt{2} \hat{G}_F}\left(c_{\varphi D} + 2 c_{\varphi \ell_1}^{(3)} + 2 c_{\varphi \ell_2}^{(3)} - 2c_{\ell\ell}\right) \, ,\\
\delta s_{\theta}^2 &= \frac{1}{2\sqrt{2}\hat{G}_F}\frac{\hat{m}_W^2}{\hat{m}_Z^2}c_{\varphi D} + \frac{1}{\sqrt{2}\hat{G}_F}\frac{\hat{m}_W}{\hat{m}_Z}\sqrt{1-\frac{\hat{m}_W^2}{\hat{m}_Z^2}}c_{\varphi WB} \, ,
\end{align}
where the cosine of the weak mixing angle is given by $ c_{\hat{\theta}}\equiv \cos \hat{\theta} = \hat{m}_W/\hat{m}_Z$.

In this notation, the contributions to the shifts $\delta g_V^x$ and $\delta g_A^x$ which are not proportional to either $\bar{g}_{V,A}^x$ or $Q^x$  are  denoted as $\Delta^{x}_{V,A}$ and are given by
\begin{align}
\nonumber\Delta^{\ell_i}_V &= -\frac{1}{4\sqrt 2 \hat{G}_F}\left(c_{\varphi\ell_i}+c_{\varphi\ell_i}^{(3)}+c_{\varphi e/\mu/\tau}\right) \, ,&
\Delta^{\ell_i}_A &= -\frac{1}{4\sqrt 2 \hat{G}_F}\left(c_{\varphi \ell_i}+c_{\varphi\ell_i}^{(3)}-c_{\varphi e/\mu/\tau}\right),\\
\nonumber\Delta^{\nu_i}_V &=-\frac{1}{4\sqrt 2 \hat{G}_F}\left(c_{\varphi\ell_i}-c_{\varphi\ell_i}^{(3)} \right) \, ,&
\Delta^{\nu_i}_A &= -\frac{1}{4\sqrt 2 \hat{G}_F}\left(c_{\varphi \ell_i}-c_{\varphi\ell_i}^{(3)}\right) \, ,\\
\nonumber\Delta^{u_j}_V &= -\frac{1}{4\sqrt2 \hat{G}_F}\left(c_{\varphi q}^{(1)}-c_{\varphi q}^{(3)}+c_{\varphi u}\right) \, ,&
\Delta^{u_j}_A &= -\frac{1}{4\sqrt2 \hat{G}_F}\left(c_{\varphi q}^{(1)}-c_{\varphi q}^{(3)}-c_{\varphi u}\right),\\
\Delta^{d_j}_V &= -\frac{1}{4\sqrt2 \hat{G}_F}\left(c_{\varphi q}^{(1)}+c_{\varphi q}^{(3)}+c_{\varphi d}\right) \, ,&
\Delta^{d_j}_A &= -\frac{1}{4\sqrt2 \hat{G}_F}\left(c_{\varphi q}^{(1)}+c_{\varphi q}^{(3)}-c_{\varphi d}\right) 
\end{align}
where $i=1,2,3$ for the leptonic generations and $j=1,2$ runs over the two light quark generations. 
Note that in the above equations there is some ambiguity in the definition of $\ell_i$: $\Delta^{\ell_i}$ refers to the shift for the charged leptons, while $\ell_i$ in the coefficient names refers to the left-handed lepton doublet. 
For the heavy third-generation quarks ($j=3$) we have instead:
\begin{align}
\nonumber\Delta^{t}_V &= -\frac{1}{4\sqrt2 \hat{G}_F}\left(c_{\varphi Q}^{(1)}-c_{\varphi Q}^{(3)}+c_{\varphi t}\right) \, ,&
\Delta^{t}_A &= -\frac{1}{4\sqrt2 \hat{G}_F}\left(c_{\varphi Q}^{(1)}-c_{\varphi Q}^{(3)}-c_{\varphi t}\right),\\
\Delta^{b}_V &= -\frac{1}{4\sqrt2 \hat{G}_F}\left(c_{\varphi Q}^{(1)}+c_{\varphi Q}^{(3)}+c_{\varphi d}\right) \, ,&
\Delta^{b}_A &= -\frac{1}{4\sqrt2 \hat{G}_F}\left(c_{\varphi Q}^{(1)}+c_{\varphi Q}^{(3)}-c_{\varphi d}\right).
\end{align}
Concerning the SMEFT-induced shifts to the $W$-boson couplings, these are as follows:
\begin{align}
\nonumber g_{V, A}^{W_\pm, \ell_i} &= \bar{g}_{V, A}^{W_\pm, \ell_i} + \delta\left(g_{V, A}^{W_\pm, \ell_i}\right) \, , \\
g_{V, A}^{W_\pm, q} &= \bar{g}_{V, A}^{W_\pm, q} + \delta\left(g_{V, A}^{W_\pm, q}\right),
\label{eq:VA-shifts-W}
\end{align}
where the SM values are given by $\bar{g}_{V, A}^{W_\pm, \ell_i} = \bar{g}_{V, A}^{W_\pm, q} = 1/2$ and $\ell_i$ refers again to the lepton doublet. 
The SMEFT-induced shifts are given by
\begin{align}
\delta\left(g_{V, A}^{W_\pm, \ell_i}\right) &= \frac{1}{2\sqrt{2}\hat{G}_F}c_{\varphi \ell_i}^{(3)} - \frac{\delta G_F}{2\sqrt{2}}\, ,\\
\delta\left(g_{V, A}^{W_\pm, q}\right) &= \frac{1}{2\sqrt{2}\hat{G}_F}c_{\varphi q}^{(3)} - \frac{\delta G_F}{2\sqrt{2}} \, .
\label{eq:wshift}
\end{align}
where Eq.~\eqref{eq:wshift} applies only to the first two quark generations.

The corrections derived in this section for the couplings of leptons and quarks to the  $Z$ ($g_{V,A}^x$) and to the $W$ $\lp g_{V, A}^{W_\pm, x}\rp$ can be constrained by measurements of $Z$-pole observables at LEP and SLD together with additional electroweak measurements, as discussed below. 

\subsection{Approximate implementation}
\label{subsec:approx-ewpo}

The previous implementation of the EWPOs in the {\sc SMEFiT} analysis as presented in \cite{Ethier:2021bye} relied on the assumption that measurements at LEP and SLD were precise enough (compared to LHC measurements), and in agreement with the SM, to constrain the SMEFT-induced shifts modifying the $W$- and $Z$-boson couplings to fermions to be exactly zero.

This assumption results in a series of linear combinations of EFT coefficients appearing in Eqns.~\eqref{eq:VA-shifts} and~\eqref{eq:VA-shifts-W} being set to zero, inducing a number of relations between the relevant coefficients. 
Accounting for the three leptonic generations, this corresponds to 14 constraints parameterised in terms of 16 independent Wilson coefficients such that $14$ of them can be expressed in terms of the remaining two.
 For instance, it was chosen in \cite{Ethier:2021bye} to include $c_{\varphi WB}$ and $c_{\varphi D}$ as the two independent fit parameters and then to parameterise the other 14 coefficients entering in the EWPOs in terms of them as follows:
\begin{flalign}
	\left(
 \def\arraystretch{1.3}
\begin{array}{c}
c_{\varphi \ell_i}^{(3)} \\
 c_{\varphi \ell_i} \\
 c_{\varphi e/\mu/\tau} \\
 c_{\varphi q}^{(-)} \\
 c_{\varphi q}^{(3)} \\
 c_{\varphi u} \\
 c_{\varphi d} \\
 c_{\ell\ell} \\
\end{array}
\right)
= 
\left(
\def\arraystretch{1.3}
\begin{array}{cc}
 -\frac{1}{t_{\hat{\theta}}} & -\frac{1}{4 t_{\hat{\theta}}^2} \\
 0 & -\frac{1}{4} \\
 0 & -\frac{1}{2} \\
 \frac{1}{t_{\hat{\theta}}} & \frac{1}{4 s_{\hat{\theta}}^2}-\frac{1}{6} \\
 -\frac{1}{t_{\hat{\theta}}} & -\frac{1}{4 t_{\hat{\theta}}^2} \\
 0 & \frac{1}{3} \\
 0 & -\frac{1}{6} \\
 0 & 0 \\
\end{array}
\right)
\left(
\begin{array}{c}
	c_{\varphi WB}\\ c_{\varphi D}
\end{array}
\right) \, ,
\label{eq:2independents}
\end{flalign}
where $i=1,2,3$, and $t_{\hat{\theta}}=s_{\hat{\theta}}/c_{\hat{\theta}}$ indicates the tangent of the weak mixing angle. 

We refer to the linear system of equations defined by Eq.~(\ref{eq:2independents}) as the ``approximate'' implementation of the EWPOs used in previous {\sc\small SMEFiT} analyses,
meaning that only $c_{\varphi WB}$ and $c_{\varphi D}$ enter as independent degrees of freedom in the fit, while all other Wilson coefficients in the LHS of Eq.~(\ref{eq:2independents}) are then determined from those two rather than being constrained separately from the data.
Likewise, whenever theory predictions depend on some of these 14 dependent coefficients, for example in LHC processes, they can be reparameterised in terms of  only $c_{\varphi WB}$ and $c_{\varphi D}$.

\subsection{Exact implementation}

The approximate implementation of EWPO constraints as described by Eq.~(\ref{eq:2independents}) encodes a two-fold assumption.
First, it assumes that EWPO measurements coincide with the SM expectations, which in general is not the case.
Second, it also implies that the precision of LEP and SLD measurements is infinite compared to the LHC measurements, which is not necessarily true as demonstrated by LHC diboson production~\cite{Grojean:2018dqj,Banerjee:2018bio}.

To  bypass these two assumptions, which also prevent a robust use of matching results between SMEFT and UV-complete models~\cite{terHoeve:2023pvs}, here we implement an exact treatment of the EWPOs and include the LEP and SLD measurements in the global fit alongside with the LHC observables.
That is, all 16 Wilson coefficients appearing in Eq.~\eqref{eq:2independents} become independent degrees of freedom, and are constrained by experimental data
from LEP/SLD and LHC sensitive to the shifts in the weak boson couplings given by Eqns.~\eqref{eq:VA-shifts}-(\ref{eq:VA-shifts-W}).
As a consequence, we had to also recompute the dependence of all the observables included in the global fit on these $16$ Wilson coefficients. 

Here we present an overview of the EWPOs included in the fit and discuss the computation of the corresponding theory predictions.
We consider the LEP and SLD legacy measurements specified in Table \ref{tab:ew-datasets}.
They consist of 19 $Z$-pole observables from LEP-1, 21 bins in $\cos(\theta)$ for various center of mass energies of Bhabha scattering ($e^+e^- \to e^+e^-$) from LEP-2, the weak coupling $\alpha_{\mathrm{EW}}$ as measured at $m_Z$, the three $W$ branching ratios to all generations of leptons, and 40 bins in $\cos(\theta)$ for four center of mass energies of four-fermion production mediated by $W$-pairs at LEP-2.
To facilitate comparison with previous results, we adopt the same bin choices as in Table 9 of~\cite{Berthier:2015gja} in the case of Bhabha scattering, which provides an independent constraint on $\alpha_{\mathrm{EW}}$ as this is not fixed by the inputs. 
The information provided by $\alpha_{\mathrm{EW}}$ and by Bhabha scattering is equivalent from the point of view of constraining the SMEFT parameter space, and here we include for completeness both datasets to increase the precision of the resulting fit. 

\begin{table}[t]
\small
\renewcommand{\arraystretch}{1.8}
\centering
\resizebox{\textwidth}{!}{
\begin{tabular}{c|c|c|c|c}
\toprule
Input & Observables & Central values & Covariance & SM predictions \\
\midrule
\multirow{4}{*}{$Z$-pole EWPOs}&$\Gamma_Z$, $\sigma_{\text{had}}^0$, $R_e^0$,
$R_\mu^0$, $R_\tau^0$, 
$A_{FB}^{0,e}$, $A_{FB}^{0,\mu}$, $A_{FB}^{0,\tau}$&  \cite{ALEPH:2005ab} (Table $2.13$) & \cite{ALEPH:2005ab} (Table $2.13$)&\multirow{4}{*}{\cite{Corbett:2021eux} (Table $1$), \cite{Awramik:2003rn,Freitas:2014hra}} \\
&$R_b^0$, $R_c^0$, $A_{FB}^{0,b}$, $A_{FB}^{0,c}$, $A_b$, $A_c$                            & \cite{ALEPH:2005ab} (Table $5.10$) & \cite{ALEPH:2005ab} (Table $5.11$)        \\
&$A_\tau$ ($\mathcal{P}_\tau$), $A_e$ ($\mathcal{P}_\tau$)                                                               & \cite{ALEPH:2005ab} (Table $4.3$)  & n/a          \\
&$A_e$ (SLD), $A_\mu$ (SLD), $A_\tau$ (SLD) & \cite{ALEPH:2005ab} (Table $3.6$)  & \cite{ALEPH:2005ab}  (Table $3.6$)   \\
 \midrule
 \multirow{2}{*}{Bhabha scattering}& $d\sigma/d\cos{\theta}$ ($n_{\rm dat}=21$)
 &\multirow{2}{*}{ \cite{LEP-2} (Tables 3.11-12)}&\multirow{2}{*}{\cite{LEP-2} (App. B.3)}& \multirow{2}{*}{ \cite{LEP-2} (Tables 3.11-12)}\\
 & $\sqrt{s}=189, 192, 196, 200, 202, 205, 207\:\mathrm{GeV}$&&\\
 \midrule
  \multirow{1}{*}{$\alpha_{\mathrm{EW}}$}&$\alpha^{-1}_{\mathrm{EW}}(m_Z)$&\cite{PDG}&\cite{PDG}& \cite{Awramik:2003rn,Corbett:2021eux,PDG} (See text)\\
  \midrule
   \multirow{3}{*}{$W$ branching ratios}&Br($W\rightarrow e \nu_e$)&\multirow{3}{*}{\cite{LEP-2} (Table 5.5) }&\multirow{3}{*}{\cite{LEP-2} (Table E.6) }&\multirow{3}{*}{\cite{Efrati:2015eaa} (Table $2$)}\\
   &Br($W\rightarrow \mu \nu_{\mu}$)&&\\
   &Br($W\rightarrow \tau \nu_{\tau}$)&&\\
    \midrule
 \multirow{2}{*}{$W^+W^-$ production}& $d\sigma/d\cos{\theta}$ ($n_{\rm dat}=40$)
 &\multirow{2}{*}{\cite{LEP-2}}&\multirow{2}{*}{n/a} &\multirow{2}{*}{\cite{LEP-2} (Figure $5.4$)}\\
  & $\sqrt{s}=182, 189, 198, 206\:\mathrm{GeV}$&&\\
 \bottomrule
\end{tabular}}
\caption{Overview of the EWPOs
from LEP and SLD considered in this
work.
For completeness, we also indicate the differential $WW$ measurements
from LEP entering the fit and which
were already included in~\cite{Ethier:2021bye}.
For each measurement, we indicate
the observables considered and
the corresponding 
references for the experimental
central values and covariance matrix. 
In the absence of a covariance matrix,
measurements are assumed to be uncorrelated.
}
\label{tab:ew-datasets}
\end{table}

\paragraph{Theoretical calculations.}
We discuss now the corresponding theory implementation of the observables reported in Table \ref{tab:ew-datasets}, i.e. $Z$-pole data, $W$ branching ratios, Bhabha scattering, $\alpha_{\mathrm{EW}}$, and $WW$ production. The EFT predictions of these observables are implemented at leading order.
As mentioned above, we adopt the $\{\hat{m}_W, \hat{m}_Z, \hat{G}_F\}$ input scheme, with the following numerical values of the input electroweak parameters:
\be
G_F = 1.1663787\cdot 10^{-5}\;\mathrm{GeV}^{-2} \,, \quad m_Z=91.1876\;\mathrm{GeV}\, , \quad m_W=80.387\;\mathrm{GeV} \, .
\label{eq:inputparam}
\ee
Concerning the flavour assumptions, we adopt the U(2)$_q\times$ U(2)$_u\times $U(3)$_d\times [$U(1)$_\ell\times $U(1)$_e]^3$ flavour symmetry.
Starting with the $Z$-pole observables, we adopt the following definitions:
\begin{align}
\label{eq:gammaZ}
    \Gamma_Z &= \sum_{i=1}^3 \Gamma_{\ell_i} + \Gamma_{\mathrm{
had}} + \Gamma_{\mathrm{inv}}, \qquad &\Gamma_{\mathrm{had}} &= \sum_{i=1}^2\Gamma_{u_i} + \sum_{i=1}^3\Gamma_{d_i}, \qquad &\Gamma_{\mathrm{inv}} &= \sum_{i=1}^3\Gamma_{\nu_i},\qquad \qquad\\
\sigma_{\mathrm{had}}^0 &= \frac{12 \pi}{\hat{m}_Z^2}\frac{\Gamma_e\Gamma_{\mathrm{had}}}{\Gamma_Z^2},\qquad 
&R_{\ell_i}^0 &= \frac{\Gamma_{\mathrm{had}}}{\Gamma_{\ell_i}}\,,\qquad &R^0_{b,c} &= \frac{\Gamma_{b,c}}{\Gamma_{\mathrm{had}}},
\\
A_f &= \frac{2g_V^{f}g_A^{f}}{\lp g_V^{f}\rp^2 + \lp g_A^{f}\rp^2},\qquad
&A_{FB}^{0, \ell_i} &= \frac{3}{4}A_e A_{\ell_i},\qquad
&A_{FB}^{0, b/c} &= \frac{3}{4}A_e A_{b/c}\,
\label{eq:Rbc}
\end{align}
where $\ell_i=\{e, \mu, \tau\}$ and where the partial decay widths of the $Z$ boson to (massless) quarks and leptons are expressed in terms of their electroweak couplings as
\be 
\Gamma_i = \frac{\sqrt{2}\hat{G}_F\hat{m}_Z^3 C}{3\pi}\lp \left|g_V^i\right|^2 + \left|g_A^i\right|^2 \rp
\ee 
where $C=3 \, (1)$ for quarks (leptons) is a colour normalisation factor. 
Substituting the SMEFT-induced shifts to the $Z$-boson couplings Eq.~(\ref{eq:VA-shifts}) into the  $Z$-pole observables Eqs.~(\ref{eq:gammaZ})-(\ref{eq:Rbc}) and expanding up to quadratic order in the EFT expansion, i.e. $\mathcal{O}\lp\Lambda^{-4}\rp$, one obtains the corresponding EFT theory predictions. 

One can proceed in the same manner concerning the $W$-boson branching ratios. 
The starting point is
\begin{align}
\label{eq:Wmass_BR}
\Gamma_W &= \sum_{i=1}^3 \Gamma_{W, \ell_i} + \Gamma_{W, u} + \Gamma_{W, c} \, , \\
\Gamma_{W, i} &= \frac{\sqrt{2}\hat{G}_F \hat{m}_W^3 C} {3\pi}|g_{V,A}^{W,i}|^2\, , 
\end{align}

We then expand $\Gamma_{W,i}/\Gamma_W$ up to quadratic order to end up with the EFT theory predictions for the $W$ branching ratios.
Note that no exotic decays of the $W$-boson are allowed.

The tree-level theoretical expressions for Bhabha scattering in the SMEFT were obtained analytically.
%
We generated all tree-level diagrams with up to one insertion of SMEFT operators for $e^+e^-\rightarrow e^+e^-$ using FeynArts~\cite{Hahn:2000kx} and then obtained the expressions for the cross-section $\sigma({e^+e^-\rightarrow e^+e^-})$ up to order $\mathcal{O}\lp\Lambda^{-4}\rp$ using Feyncalc~\cite{Mertig:1990an,Shtabovenko:2016sxi,Shtabovenko:2020gxv}. We cross-checked our SM expressions with Table $9$ of Ref.~\cite{Berthier:2015gja} and our SMEFT predictions with those obtained using the SMEFT@NLO~\cite{Degrande:2020evl} model in {\sc\small mg5\_aMC@NLO} \cite{Alwall:2014hca}, finding agreement in both cases.

Concerning the EW coupling constant $\alpha_{\mathrm{EW}}$, this is a derived quantity in the $\{\hat{m}_W, \hat{m}_Z, \hat{G}_F\}$ input scheme, which can be expressed in terms of the input parameters as follows
\be
\bar{\alpha}_{\mathrm{EW}} = \frac{\bar{e}^2}{4\pi} = \frac{\left(\hat{e}-\delta e\right)^2}{4\pi}
=\frac{\hat{e}^2}{4\pi}\left(1-2\frac{\delta e}{\hat{e}}+\left(\frac{\delta e}{\hat{e}}\right)^2+\dots\right),
\label{eq:alpha_ew}
\ee
where the ellipsis indicates higher-order corrections. 
In Eq.~(\ref{eq:alpha_ew}), the SMEFT-induced shift in the electric charge is given by \cite{Brivio:2017bnu}
\begin{equation}
\delta e = \hat{e}\left(-\frac{\delta G_F}{\sqrt{2}}+\frac{\delta m_Z^2}{2\hat{m}_Z^2}\frac{\hat{m}_W^2}{\hat{m}_W^2 - \hat{m}_Z^2}-\frac{\hat{m}_W s_{\hat{\theta}}}{\sqrt{2}\hat{G}_F\hat{m}_Z}c_{\varphi WB}\right) \, ,
\end{equation}
with the measured value of the electric charge given in this electroweak scheme by 
$
\hat{e} = 2^{5/4}\hat{m}_W\sqrt{\hat{G}_F}s_{\hat{\theta}}
$.
We expand Eq.~(\ref{eq:alpha_ew}) up to quadratic order to obtain the sought-for EFT theory predictions for $a_{\mathrm{EW}}$. 
The SM prediction is obtained by solving the on-shell expression for $m_W$ from \cite{Awramik:2003rn} for $\Delta \alpha$ and using
\be
    \alpha_{EW}(m_Z) = \frac{\alpha_{EW}}{1-(\Delta \alpha + 0.007127)} \, ,
\ee
 where $\alpha_{EW}$ is the fine-structure constant at zero energy, $0.007127$ represents the conversion factor between the on-shell and $\overline{\rm MS}$ renormalisation schemes \cite{PDG}, and we substitute in the input parameters from \eqref{eq:inputparam} along with those from \cite{Corbett:2021eux}.

Regarding the theory calculations for $WW$-production at LEP-2, we compute linear and quadratic SMEFT contributions to four-fermion production mediated by charged currents using the SMEFT@NLO model in {\sc\small mg5\_aMC@NLO}. 
Only semileptonic final states where a $W$-boson decays to either a $e\nu$ or $\mu\nu$ pair were considered. We computed the angular distribution in $\cos{\theta}$, where $\theta$ is the angle formed by the momentum of the $W^-$ and the incoming $e^-$, and we applied a kinematic cut on the charged lepton angle $\theta_{\ell}$, $|\cos(\theta_{\ell})|< 0.94$, corresponding to the detector acceptance of $20^{\circ}$ around the beam. The four distributions, corresponding to four luminosity-weighted values of center of mass energy $\sqrt s =182.66,\, 189.09,\, 198.38,\,$ and $ 205.92\:\mathrm{GeV}$, were divided into 10 bins. 
SMEFT corrections to the $W$-boson decays were added {\it a posteriori} as discussed in App.~\ref{subsec:app-widths}. 

In addition to the LEP and SLD datasets listed in Table ~\ref{tab:ew-datasets}, new theory predictions were also computed for LHC processes sensitive to operators entering in the EWPOs and hence in the fit as new independent degrees of freedom.
For this, we used {\sc\small mg5\_aMC@NLO} interfaced to SMEFT@NLO to evaluate linear and quadratic EFT corrections at NLO QCD whenever available. 
In these calculations, in order to avoid any possible overlap between datasets entering simultaneously PDF and EFT fits~\cite{Carrazza:2019sec,Kassabov:2023hbm,Greljo:2021kvv}, we used NNPDF4.0 NNLO no-top~\cite{NNPDF:2021njg} as input PDF set. We refer to Tables 3.1-3.7 in~\cite{Ethier:2021bye} for an overview of the datasets that we include on top of those already presented in Table \ref{tab:ew-datasets}. 
Furthermore, in comparison to the LHC datasets in~\cite{Ethier:2021bye}, we now include additional datasets from Run II, described in Sect.~\ref{subsec:new_datasets}. At the moment, we include theory uncertainties on the SM predictions, and make sure the Monte Carlo statistical uncertainty on the EFT theory predictions are below $1\%$. Theory uncertainties on the EFT predictions may be included following the prescription outlined in  \cite{Altmannshofer:2021qrr}, although we neglect its effect in this work as we already include NLO QCD corrections in the EFT on the LHC cross-sections.

\paragraph{Benchmarking and validation}
Our implementation of EWPOs described above has been cross-checked and validated with previous studies in the literature, in particular with the analysis of~\cite{Brivio:2017bnu}. 
%
 %
 First, we note that a complete interpretation of EWPOs depends on the 16 Wilson coefficients that enter in Eq.~(\ref{eq:2independents}) together with $c_{\varphi Q}^{(-)}$ and $c_{\varphi Q}^{(3)}$, thus giving 18 directions to probe in total.
 However, only 15 of these can be probed in an EWPOs-only fit such as that of~\cite{Brivio:2017bnu}, which leaves three directions unconstrained.
 Hence, a valuable cross-check is to make sure we reproduce the same flat directions as those found in~\cite{Brivio:2017bnu}.
 Secondly, one must be aware of different flavour assumptions while doing this comparison: the {\sc\small  SMEFiT} flavour assumption singles out the top quark, while \cite{Brivio:2017bnu} adopts a flavour universal scenario where all three generations are treated on the same footing and leading to a significantly smaller number of degrees of freedom.

In total, one expects to obtain three flat directions in an EWPO-only fit, two of them originating from the TGCs (as in the case of the flavour universal scenario) and a third from the left-handed $Zt\bar{t}$-coupling, as already alluded to in Sect.~\ref{subsec:approx-ewpo}.
Indeed, we find three unconstrained directions $\mathbf{v}_i$ in the SMEFT parameter space (or linear combinations thereof) in a fit to the data listed in Table~\ref{tab:ew-datasets}.
These three flat directions are given by:
\begin{align}
    \label{eq:fd-1}
    \mathbf{v}_1 &=  c_{\varphi Q}^{(1)} - c_{\varphi Q}^{(3)} \, ,\\ \label{eq:fd-2}
    \nonumber \mathbf{v}_2 &= \bigg(1.60\, c_{\varphi WB} + \frac{1}{2}\bigg( c_{\varphi\ell_1} + c_{\varphi\ell_2} + c_{\varphi\ell_3}\bigg) -1.24 \bigg( c_{\varphi \ell_1}^{(3)} + c_{\varphi \ell_2}^{(3)} + c_{\varphi \ell_3}^{(3)} + c_{\varphi q}^{(3)} + c_{\varphi Q}^{(3)}\bigg) \\
    &\hspace{4cm}- 2\, c_{\varphi D} + c_{\varphi e} + c_{\varphi \mu} + c_{\varphi \tau} -\frac{2}{3} c_{\varphi u} +\frac{1}{3} c_{\varphi d} - 0.167\bigg( c_{\varphi q}^{(1)}+ c_{\varphi Q}^{(1)}\bigg) \bigg) \, ,\\
    \nonumber \mathbf{v}_3 &= \bigg(-0.24\,c_{\varphi WB} + \frac{1}{2}\bigg( c_{\varphi\ell_1} + c_{\varphi\ell_2} + c_{\varphi\ell_3}\bigg) +2.20 \bigg( c_{\varphi \ell_1}^{(3)} + c_{\varphi \ell_2}^{(3)} + c_{\varphi \ell_3}^{(3)} + c_{\varphi q}^{(3)} + c_{\varphi Q}^{(3)}\bigg) \\
    &\hspace{4cm}- 2\, c_{\varphi D} + c_{\varphi e} + c_{\varphi \mu} + c_{\varphi \tau} -\frac{2}{3} c_{\varphi u} +\frac{1}{3} c_{\varphi d} - 0.167\bigg( c_{\varphi q}^{(1)}+ c_{\varphi Q}^{(1)}\bigg) \bigg) \, .
    \label{eq:fd-3}
\end{align}
In the flavour universal scenario, one can verify that Eqs.~(\ref{eq:fd-2})-(\ref{eq:fd-3}) simplify to those given in \cite{Brivio:2017bnu}.
%

It is relevant in this context to comment on the number of flat directions obtained under variations of the fitted datasets, in particular when considering subsets of the data listed in Table~\ref{tab:ew-datasets}.
Table~\ref{tab:FD_data_var} indicates the  number of directions in the parameter space constrained by different choices of the input dataset
entering the SMEFT fit in the absence of other experimental information. 
Even though the number of flat directions remains constant regardless of whether only $\alpha_{\mathrm{EW}}$, only Bhabha, or both are added on top of the $Z$-pole and the $W$ branching ratios, we decide to include all four datasets since we have no a priori reason to prefer one over the other.

\begin{table}[t]
\renewcommand{\arraystretch}{1.3}
    \begin{center}
\begin{tabular}{ l|c }
\toprule
Input dataset & Constrained directions \\
\midrule
$Z-$pole EWPOs & 12/18  \\
$Z-$pole EWPOs + $\alpha_{\rm EW}$ & 13/18 \\
$Z-$pole EWPOs + Bhabha & 14/18 \\
$Z-$pole EWPOs + Bhabha + $\alpha_{\rm EW}$ & 14/18 \\
$Z-$pole EWPOs + Br($W$) & 14/18 \\
$Z-$pole EWPOs + Br($W$) + $\alpha_{\rm EW}$ & 15/18 \\
$Z-$pole EWPOs + Br($W$) + Bhabha & 15/18 \\
$Z-$pole EWPOs + Br($W$) + Bhabha + $\alpha_{\rm EW}$  & 15/18 \\
\bottomrule
\end{tabular}
\end{center}
    \caption{
    The number of directions
in the parameter space constrained by different choices of the input dataset entering the SMEFT fit, 
and consisting of subsets of the full dataset of Table~\ref{tab:ew-datasets},
in the absence of other experimental information.
Under the {\sc\small SMEFiT} flavour assumptions, electroweak precision
observables and related measurements
are sensitive to up to 18
independent directions in the space
of Wilson coefficients.
The last three rows result in the same number of constrained directions, since $\alpha_{\rm EW}$ and Bhabha scattering provide equivalent information.
    }
      \label{tab:FD_data_var}
\end{table}

To demonstrate the consistency of our implementation with previous results, we present in App.~\ref{app-benchmarking} the results of a comparison  of our implementation in the {\sc\small SMEFiT} framework with~\cite{Brivio:2017bnu}, finding good agreement.

\section{The SMEFiT3.0 global analysis and projections for the HL-LHC}
\label{sec:updated_global_fit}

Here we present {\sc\small SMEFiT3.0},
an updated version of the global SMEFT analysis from~\cite{Ethier:2021bye,Giani:2023gfq}.
The major differences as compared with these previous analyses are two-fold.
First, the improved treatment of EWPOs as described in Sect.~\ref{sec:ewpos}.
Second, the inclusion of recent measurements of Higgs, diboson, and top quark production data from the LHC Run II, several of them based on its full integrated luminosity of $\mathcal{L}=139$ fb$^{-1}$.
In this section we start by describing the main features of the new LHC Run II datasets added to the global fit (Sect.~\ref{subsec:new_datasets}); then we quantify the impact of the new LHC data and of the  updated implementation of the EWPOs at the level of EFT coefficients (Sect.~\ref{subsec:impact_ewpos}); and finally we extend the SMEFT analysis with  dedicated projections for HL-LHC measurements
(Sect.~\ref{sec:impact_hl_lhc_data}), see also App.~\ref{app:hl_lhc_projections}.

\subsection{Experimental dataset}
\label{subsec:new_datasets}

Firstly, we describe the new LHC datasets from Run II which enter the updated global SMEFT analysis and which complement those already included in~\cite{Ethier:2021bye,Giani:2023gfq}.
For consistency with previous studies, and to ensure that QCD calculations at the highest available accuracy can be deployed, for top quark and Higgs boson production we restrict ourselves to parton-level measurements.
For diboson production, we consider instead particle-level distributions, for which NNLO QCD predictions are available for the SM~\cite{Grazzini:2019jkl}.

In the case of top quark production observables, we include the same datasets as in the recent EFT and PDF analysis of the top quark sector from the  {\sc\small PBSP} collaboration~\cite{Kassabov:2023hbm}.
These top quark measurements are extended
with additional datasets that have become available since the release of that study.
Theoretical higher-order QCD calculations and EFT cross-sections for these top quark production datasets are also taken from~\cite{Kassabov:2023hbm}, extended when required to the wider operator basis considered here. 

\begin{table}[t]
  \centering
  \small
   \renewcommand{\arraystretch}{1.30}
  \begin{tabular}{|C{3.9cm}|C{6.6cm}|C{2.5cm}|C{2.5cm}|}
  \toprule
 \multirow{2}{*}{Category}   &  \multirow{2}{*}{Processes}    &  \multicolumn{2}{|c|}{$n_{\rm dat}$}     \\
    &     &  {\sc\small SMEFiT2.0} &  {\sc\small SMEFiT3.0}     \\
    \toprule
    \multirow{7}{*}{Top quark production}   &  $t\bar{t}+X$    &  94  &115\\
    &  $t\bar{t}Z$, $t\bar{t}W$    & 14& 21\\
    &  $t\bar{t}\gamma$    & -& 2 \\
    &   single top (inclusive)   & 27 &28\\
    &  $tZ, tW$   &  9&13\\
    &  $t\bar{t}t\bar{t}$, $t\bar{t}b\bar{b}$    & 6 &12\\
    &  {\bf Total}    & {\bf 150 } &{\bf 191 }  \\
    \midrule
    \multirow{3.3}{*}{Higgs production} & Run I signal strengths  &22 & 22 \\
    \multirow{3.1}{*}{and decay} & Run II  signal strengths  & 40 & 36 (*)\\
    & Run II, differential distributions \& STXS  & 35 & 71\\
    &  {\bf Total}    & {\bf 97} & {\bf 129}\\
    \midrule
    \multirow{3}{*}{Diboson production} & LEP-2 &40 & 40 \\
     & LHC & 30& 41 \\
    &  {\bf Total}    & {\bf 70} & {\bf 81}  \\
      \midrule
    \multirow{1}{*}{EWPOs} & LEP-2 & - & 44  \\
    \bottomrule
   Baseline dataset     & {\bf Total}      & {\bf 317} & {\bf 445}\\
\bottomrule
  \end{tabular}
  \caption{\small The number of data points $n_{\rm dat}$ in the baseline dataset
    for each of the categories of processes considered in this work.
    We compare the values in the current analysis ({\sc\small SMEFiT3.0}) with those with its predecessor {\sc\small SMEFiT2.0}~\cite{Ethier:2021bye,Giani:2023gfq}.
    Recall that, in  {\sc\small SMEFiT2.0}, the EWPOs were accounted for in an approximate manner. 
    (*) 4 data points from the CMS Run II Higgs dataset were removed because they cannot be described by a multi-Gaussian distribution.
 \label{eq:table_dataset_overview}
}
\end{table}


Table~\ref{eq:table_dataset_overview} indicates the number of data points $n_{\rm dat}$ in the baseline dataset for each of the categories of processes considered here.
We compare these values in the current analysis ({\sc\small SMEFiT3.0}) with those with its predecessor {\sc\small SMEFiT2.0}~\cite{Ethier:2021bye,Giani:2023gfq}.
From  this overview, one observes that the current analysis has $n_{\rm dat}=445$, up from $n_{\rm dat}=317$ in the previous fit.
The processes that dominate this increase in input cross-sections are top quark production ($n_{\rm dat}$ increasing by 41 points), Higgs production (by 32) and the EWPOs, which in SMEFiT2.0 were accounted for in an approximate manner.

We briefly describe the main features of the new Higgs boson, diboson and top quark datasets included here and the settings of the associated theory calculations.
These are summarised in Table~\ref{eq:input_datasets_higgs}, where we indicate the naming convention, the center-of-mass energy and integrated luminosity, details on the production and decay channels involved, the fitted observables, the number of data points and the corresponding publication reference. 


\begin{table}[t]
  \centering
  \footnotesize
   \renewcommand{\arraystretch}{1.55}
  \begin{tabular}{c|c|c|c|c|c|c}
   \toprule
 Dataset   &  $\sqrt{s}$ (TeV) & $\mathcal{L}$ $\lp\rm{fb}^{-1}\rp$  & Info  &  Observables  & $n_{\rm dat}$ & ref.    \\
\midrule
  \multirow{3}{*}{ {\tt ATLAS\_STXS\_RunII\_13TeV\_2022} }  &\multirow{3}{*}{13} &\multirow{3}{*}{139}  &\multirow{3}{*}{
  $gg$F, VBF, $Vh$, $t\bar{t}h$, $th$} & $d\sigma/dp_T^h$    &   \multirow{3}{*}{36}    &  \multirow{3}{*}{ \cite{ATLAS:2022vkf} } \\
     &  &  &  & $d\sigma/dm_{jj}$&     &    \\
     &  &  &  & $d\sigma/dp_T^V$&     &    \\
     \midrule
     \midrule
   {\tt CMS\_WZ\_pTZ\_13TeV\_2022}  & 13 & 137  &
  $WZ$, fully leptonic & $1/\sigma d\sigma/dp_T^Z$    &    11   &   \cite{CMS:2021icx}  \\
     \midrule
     \midrule
     {\tt     CMS\_tt\_13TeV\_ljets\_inc}  & 13 & 137&
   $\ell+\rm{jets}$ & $\sigma(t\bar{t})$    &    1   &   \cite{CMS:2021vhb}  \\
   {\tt     CMS\_tt\_13TeV\_Mtt}  & 13 & 137&
   $\ell+\rm{jets}$ & $1/\sigma d\sigma/dm_{t\bar{t}}$    &    15   &   \cite{CMS:2021vhb}  \\
    \midrule
    \midrule
   {\tt     CMS\_tt\_13TeV\_asy}  & 13 & 138 &
  $\ell+\rm{jets}$ & $A_C$    &    3   &   \cite{CMS:2022ged}  \\
   {\tt     ATLAS\_tt\_13TeV\_asy\_2022}  & 13 & 139 &
  $\ell+\rm{jets}$ & $A_C$    &    5   &   \cite{ATLAS:2022waa}  \\
       \midrule
     \midrule
   {\tt     ATLAS\_Whel\_13TeV}  & 13 & 139 &
  $W$-helicity fraction & $F_0, F_L$    &    2   &   \cite{ATLAS:2022rms}  \\
       \midrule
    \midrule
   {\tt    ATLAS\_ttZ\_13TeV\_pTZ}  & 13 & 139 & $t\bar{t} Z$
   & $d\sigma/dp_T^Z$    &    7   &   \cite{ATLAS:2021fzm}  \\
      \midrule
    \midrule
   {\tt    ATLAS\_tta\_8TeV}  & 8 & 20.2 &
  Inclusive & $\sigma\lp t\bar{t}\gamma\rp$    &    1   &   \cite{ATLAS:2017yax}  \\
   {\tt    CMS\_tta\_8TeV}  & 8 &  19.7&
  Inclusive & $\sigma\lp t\bar{t}\gamma\rp$    &    1   &   \cite{CMS:2017tzb}  \\
        \midrule
    \midrule
   {\tt    ATLAS\_tttt\_13TeV\_slep\_inc}  & 13 & 139 &
  single-lepton & $\sigma_{\rm tot}\lp t\bar{t}t\bar{t}\rp$    &    1   &   \cite{ATLAS:2021kqb}  \\
   {\tt   CMS\_tttt\_13TeV\_slep\_inc}  & 13 & 35.8 &
 single-lepton & $\sigma_{\rm tot}\lp t\bar{t}t\bar{t}\rp$    &    1   &   \cite{CMS:2019jsc}  \\
 {\tt ATLAS\_tttt\_13TeV\_2023}&13&$139$&multi-lepton&$\sigma_{\rm tot}(t\bar{t}t\bar{t})$&1&\cite{ATLAS:2023ajo}\\
{\tt CMS\_tttt\_13TeV\_2023}&13&$139$&same-sign or multi-lepton&$\sigma_{\rm tot}(t\bar{t}t\bar{t})$&1&\cite{CMS:2023ftu}\\
   {\tt   CMS\_ttbb\_13TeV\_dilepton\_inc}  & 13 & 35.9 &
  dilepton & $\sigma_{\rm tot}\lp t\bar{t}b\bar{b}\rp$    &    1   &   \cite{CMS:2020grm}  \\
   {\tt     CMS\_ttbb\_13TeV\_ljets\_inc}  & 13 & 35.9 &
 $\ell+\rm{jets}$ & $\sigma_{\rm tot}\lp t\bar{t}b\bar{b}\rp$    &    1   &   \cite{CMS:2020grm}  \\
     \midrule
    \midrule
   {\tt     ATLAS\_t\_sch\_13TeV\_inc}  & 13 & 139 &
  $s$-channel & $\sigma_{\rm tot}\lp t + \bar{t}\rp$    &    1   &   \cite{ATLAS:2022wfk}  \\
      \midrule
    \midrule
   {\tt     CMS\_tZ\_13TeV\_pTt}  & 13 & 138 &
  dilepton& $d\sigma_{\rm fid}(tZj)/dp_T^t$    &    3   &   \cite{CMS:2021ugv}  \\
   {\tt     CMS\_tW\_13TeV\_slep\_inc}  & 13 & 36 &
 single-lepton & $\sigma_{\rm tot}(tW)$    &    1   &   \cite{CMS:2021vqm}  \\
    \bottomrule
    \end{tabular}
  \caption{\small Description of the new Higgs boson production and decay, diboson and top-quark datasets added to the global SMEFT fit presented in this work.
For each dataset, we indicate the naming convention, its center-of-mass energy and integrated luminosity, details on the production and decay channels involved, the fitted observables, the number of data points $n_{\rm dat}$, and the publication reference.
  See App.~\ref{app:hl_lhc_projections} for a discussion of which Run II datasets enter the projections for the HL-LHC.
     \label{eq:input_datasets_higgs}
  }
\end{table}


\paragraph{Higgs production and decay.}
We include the recent Simplified Template Cross Section (STXS) measurements from ATLAS~\cite{ATLAS:2022vkf}, based on the full Run II luminosity.
All relevant production modes accessible at the LHC (Run II) are considered: $gg$F, VBF, $Vh$, $t\bar{t}h$, and $th$, each of them in all available decay modes.
This Higgs production and decay dataset, which comes with the detailed breakdown of correlated systematic errors (both experimental and theoretical), adds $n_{\rm dat}=36$ data points to the global fit dataset.
The SM cross-sections are taken from the same ATLAS publication~\cite{ATLAS:2022vkf} while we evaluate the linear and quadratic EFT cross-sections using {\sc\small mg5\_aMC@NLO}~\cite{Alwall:2014hca} interfaced to {\sc\small SMEFT@NLO}~\cite{Degrande:2020evl}, with consistent settings with the rest of the observables considered in the fit.
This Higgs dataset is one of the inputs for the most extensive EFT interpretation of their data carried out by ATLAS to date~\cite{ATL-PHYS-PUB-2022-037,ATLAS:2024lyh}. 

\paragraph{Diboson production.}
We include the CMS measurement of the $p_T^Z$ differential distribution in $WZ$ production at $\sqrt{s}=13$ TeV presented in~\cite{CMS:2021icx} and based
on the full Run II luminosity of $\mathcal{L}=137$ fb$^{-1}$.
The measurement is carried out in the 
fully leptonic final state ($\ell^+\ell^- \ell' \bar{\nu}_{\ell'}$) and consists of $n_{\rm dat}=11$ data points. 
The SM theory calculations include NNLO QCD and NLO electroweak corrections and are taken from~\cite{CMS:2021icx}, while the same settings as for Higgs production are employed for the EFT cross-sections. 

\paragraph{Top quark production.}
As mentioned above, here
 we consider the same top quark production datasets entering the analysis of~\cite{Kassabov:2023hbm}, in most cases corresponding to the full Run II integrated luminosity and extended when required to measurements that have become available after the release of that analysis.
As listed in the dataset overview of Table~\ref{eq:input_datasets_higgs}, 
we include the normalised differential $m_{t\bar{t}}$ distribution from CMS in the lepton+jets final state~\cite{CMS:2021vhb}; the charge asymmetries $A_C$ from ATLAS and CMS in the $\ell$+jets final state~\cite{CMS:2022ged,ATLAS:2022waa}; 
the $W$ helicity fractions from ATLAS~\cite{ATLAS:2022rms}; 
the $p_T^Z$ distribution in $t\bar{t}Z$
associated production from ATLAS~\cite{ATLAS:2021fzm}; the inclusive
$t\bar{t}\gamma$ cross-sections from ATLAS and CMS~\cite{ATLAS:2017yax,CMS:2017tzb}; 
the four-top cross-sections from ATLAS and CMS in the single-lepton and multi-lepton final states~\cite{ATLAS:2021kqb,CMS:2019jsc,ATLAS:2023ajo,CMS:2023ftu}; the $t\bar{t}b\bar{b}$ cross-sections from CMS in the dilepton and $\ell$+jets channels~\cite{CMS:2020grm,CMS:2020grm};
the $s$-channel single-top cross-section from ATLAS~\cite{ATLAS:2022wfk}; and finally the single-top associated production cross-sections for $tZ$ and $tW$ from CMS~\cite{CMS:2021ugv,CMS:2021vqm}.
In all cases, state-of-the-art SM and EFT cross-sections from~\cite{Kassabov:2023hbm} are used,  extended whenever required to the additional directions in the EFT parameter space considered here. 

\subsection{The {\sc\small SMEFiT3.0} global analysis}
\label{subsec:impact_ewpos}
\label{sec:impact_newdata}

We now present the results of {\sc\small SMEFiT3.0}, which provides the baseline for the subsequent studies with HL-LHC and FCC-ee projections. 
We study its main properties, including the data versus theory agreement and its consistency with the SM expectations.
We assess the stability of the results with respect to the order in the EFT expansion adopted (linear versus quadratic), compare individual (one parameter) versus global (marginalised) bounds on the EFT coefficients, map the correlation patterns, quantify the impact of the new data added in comparison with {\sc\small SMEFiT2.0}, and investigate the fit stability with respect to the details of the EWPO implementation. 
All the presented results are based on the Bayesian inference module of {\sc\small SMEFiT} implemented via the Nested Sampling algorithm, which provides our default fitting strategy~\cite{Ethier:2021bye,Giani:2023gfq}.
Table~\ref{table:fit_settings} provides an overview of the input settings adopted for each of the fits discussed in this section and the next one.

\begin{figure}[htbp]
    \center
    \includegraphics[width=.94\linewidth]{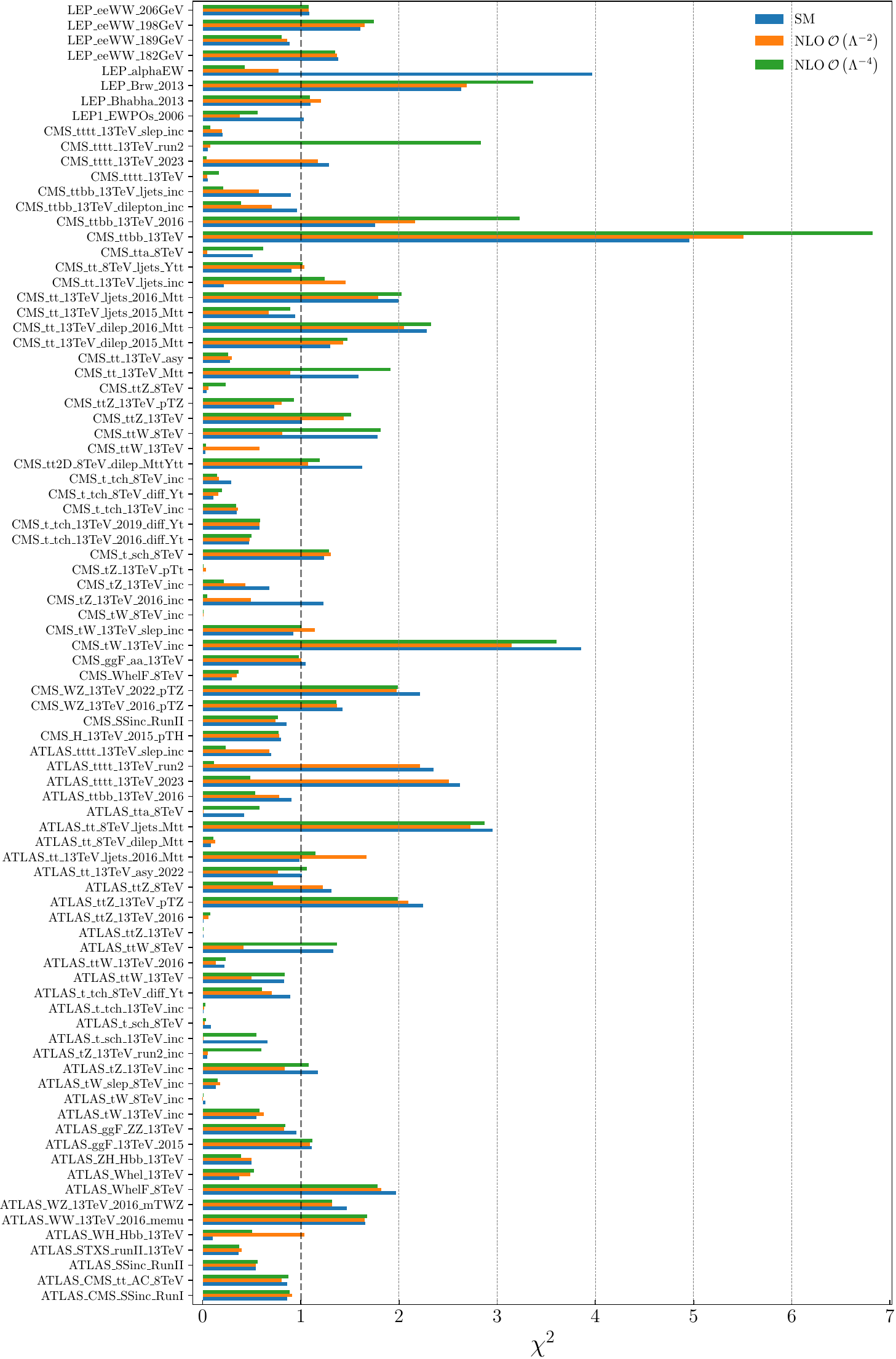}
    \caption{The values of the $\chi^2/n_{\rm dat}$ for the datasets entering the  {\sc\small SMEFiT3.0} analysis. 
    We compare the results based on the SM prediction with the outcome of the SMEFT fits, both at linear and quadratic order in the EFT expansion. 
  See also Table~\ref{eq:chi2-baseline-grouped} for the corresponding results grouped in terms of physical processes.
    }
\label{fig:chi2_lin_vs_quad}
\end{figure}

\paragraph{Fit quality.}
Fig.~\ref{fig:chi2_lin_vs_quad} indicates the values of the $\chi^2/n_{\rm dat}$ for all datasets entering {\sc\small SMEFiT3.0}. 
We compare the values based on the SM predictions with the outcome of the EFT fits, both at linear and quadratic order.
Whenever available, theoretical uncertainties are also included.
The dashed vertical line corresponds to the $\chi^2/n_{\rm dat}=1$ reference.
Note that most of the datasets included in Fig.~\ref{fig:chi2_lin_vs_quad} are composed by just one or a few cross-sections, explaining some of the large fluctuations shown.
The results of Fig.~\ref{fig:chi2_lin_vs_quad} are then tabulated in Table~\ref{eq:chi2-baseline-grouped} at the level of the groups of processes entering the fit. 

\begin{table}[htbp]
  \centering
  \small
   \renewcommand{\arraystretch}{1.70}
   \begin{tabular}
   {l|C{1.0cm}|C{1.8cm}|C{2.7cm}|C{2.7cm}}
     \toprule
        \multirow{2}{*}{ Dataset}   & \multirow{2}{*}{$ n_{\rm dat}$} & \multirow{2}{*}{ $\chi^2_{\rm SM}/n_{\rm dat}$} &  $\chi^2_{\rm EFT}/n_{\rm dat}$   & $\chi^2_{\rm EFT}/n_{\rm dat}$     \\
      &   &   & $\mathcal{O}\lp \Lambda^{-2}\rp$ &  $\mathcal{O}\lp \Lambda^{-4}\rp$  \\
        \toprule
 $t\bar{t}$ inclusive  & 115& 1.365 & 1.193   & 1.386  \\
  $t\bar{t}+\gamma$  & 2  & 0.465  & 0.027   & 0.598  \\
 $t\bar{t}+V$  & 21  & 1.200  & 1.100   & 1.165   \\
 single-top inclusive & 28  & 0.439   & 0.393   & 0.407   \\
 single-top $+V$ &  13 & 0.663  & 0.540   & 0.562  \\
 $t\bar{t}b\bar{b}$ \& $t\bar{t}t\bar{t}$        &  12 & 1.396  & 1.386   & 1.261  \\
 \midrule
 Higgs production \& decay & 129  & 0.687   & 0.692   & 0.676  \\
 \midrule
 Diboson (LEP+LHC)  & 81  & 1.481  & 1.429   & 1.436  \\
 LEP + SLD & 44  & 1.237  &  0.942  & 1.002  \\
 \midrule
 {\bf Total}  & {\bf 445}  &  {\bf 1.087  }   & {\bf 0.992 }  & {\bf 1.048 } \\
\bottomrule
\end{tabular}
\caption{\small Summary of the $\chi^2/n
_{\rm dat}$ results displayed in Fig.~\ref{fig:chi2_lin_vs_quad} in terms of the groups of processes entering the fit. 
\label{eq:chi2-baseline-grouped}
}
\end{table}

The $\chi^2$ values collected in Fig.~\ref{fig:chi2_lin_vs_quad} and Table~\ref{eq:chi2-baseline-grouped} indicate that, for most of the datasets considered here, the SM predictions are in good agreement with the experimental data.
This agreement remains the same, or it is further improved, at the level of the (linear or quadratic) EFT fits.
However, for some datasets, the SM $\chi^2$ turns out to be poor.
In most cases, this happens for datasets containing one or a few cross-section points.
Datasets with a poor $\chi^2$ to the SM include 
{\tt CMS\_ttbb\_13TeV}, {\tt CMS\_tW\_13TeV inc}, {\tt ATLAS\_tttt 13TeV\_2023}, {\tt ATLAS\_tt\_8TeV\_ljets\_Mtt}, and {\tt ATLAS\_ttZ\_13TeV\_pTZ.}
For these datasets, a counterpart from the complementary experiment is also part of the fit and agreement with the SM is found there, suggesting some tension between the ATLAS and CMS measurements.
See also the discussions in~\cite{Kassabov:2023hbm} for the top quark datasets at the light of the covariance matrix decorrelation method~\cite{Kassabov:2022pps}.
This interpretation is supported by the fact that, for these datasets with a poor $\chi^2$ to the SM prediction, accounting for EFT effects does not improve the agreement with the data.
In such cases, the poor $\chi^2$ values may be explained by either internal inconsistencies~\cite{Kassabov:2022pps} or originates from tensions between different measurements of the same process.

Concerning the LEP measurements, good agreement with the SM is observed with the only exception of the electroweak coupling constant $\alpha_{\rm EW}$ and the $W$ branching ratios.
While the $\chi^2$ to the former observable improves markedly once EFT corrections are accounted for, the opposite appears to be true for the LEP $W$-boson branching ratios.
We recall here that, for the $W$ branching fractions in Eq.~(\ref{eq:Wmass_BR}), possible invisible decay channels are not accounted for. 

In Table~\ref{eq:chi2-baseline-grouped}, we present the $\chi^2$ values grouped by physical process, comparing the SM with the best fit parameters found in the SMEFT fits. Notably, the SM demonstrates a commendable per data point $\chi^2=1.087$, a value that further refines to 0.992 and 1.048 for the linear and quadratic EFT fits, respectively, with $n_{\text{eft}}=45$ and 50 parameters in each case. 
While the $\chi^2$ values of the Higgs production and decay dataset are similar in the SM and in the linear and quadratic EFT fits, and likewise for diboson production, more variation is found for the top-quark production datasets, specially for inclusive $t\bar{t}$ production.
In the case of the EWPOs, there is a clear reduction of the $\chi^2$ per data point in the EFT fit as compared to the baseline SM predictions.
It is worth emphasizing that these $\chi^2$ values do not  imply that the EFT model offers a superior description to the data. 
Indeed, a thorough hypothesis test mandates normalizing the $\chi^2$ by the number of degrees of freedom $n_{\text{dof}} = n_{\text{dat}} - n_{\text{eft}}$. 
In this sense, the SM, having a good $\chi^2$ with less parametric freedom, remains the preferred model to describe the available data.

\begin{figure}[t]
    \center
    \includegraphics[width=0.49\textwidth]{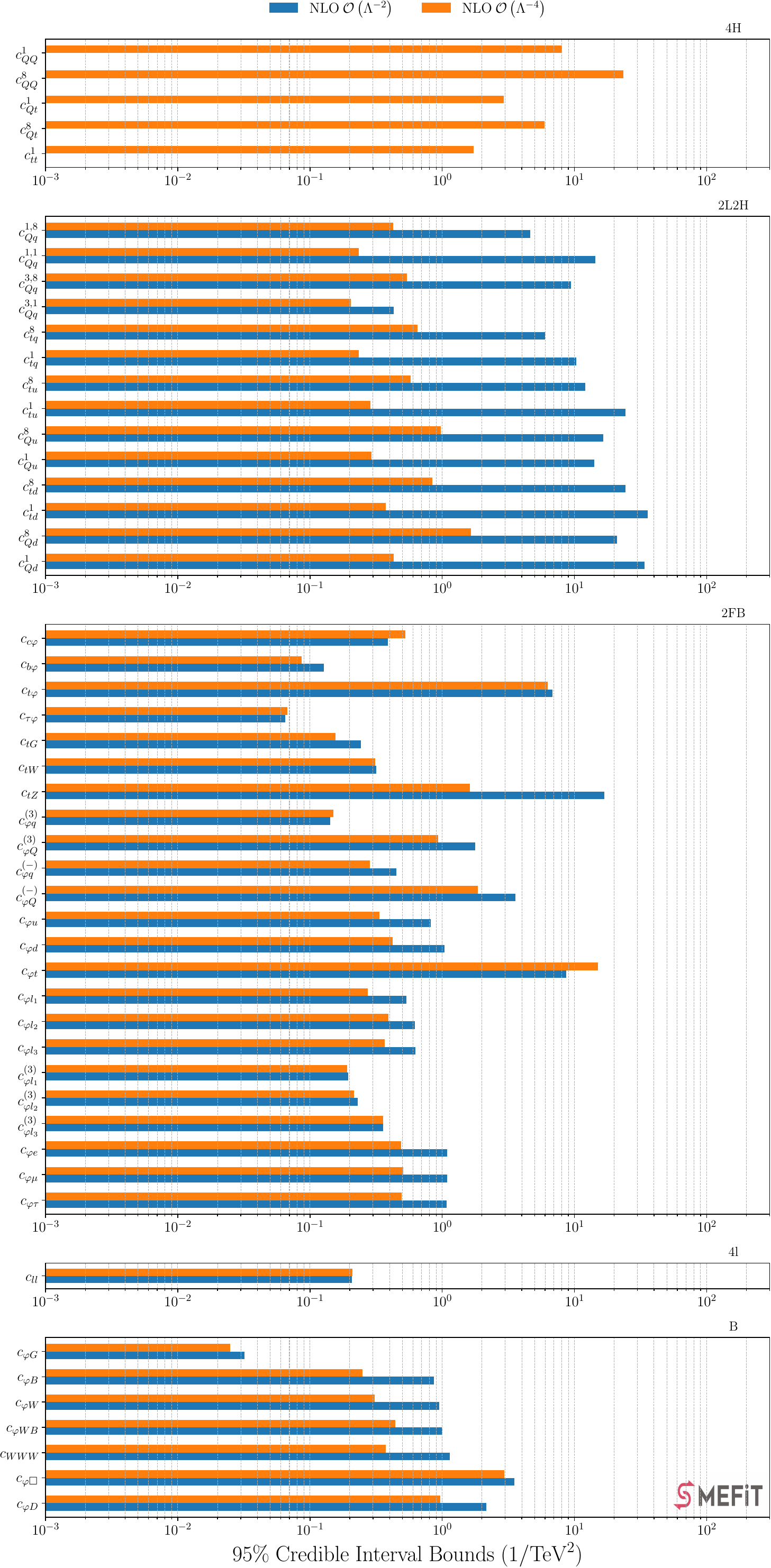}
    \includegraphics[width=0.49\textwidth]{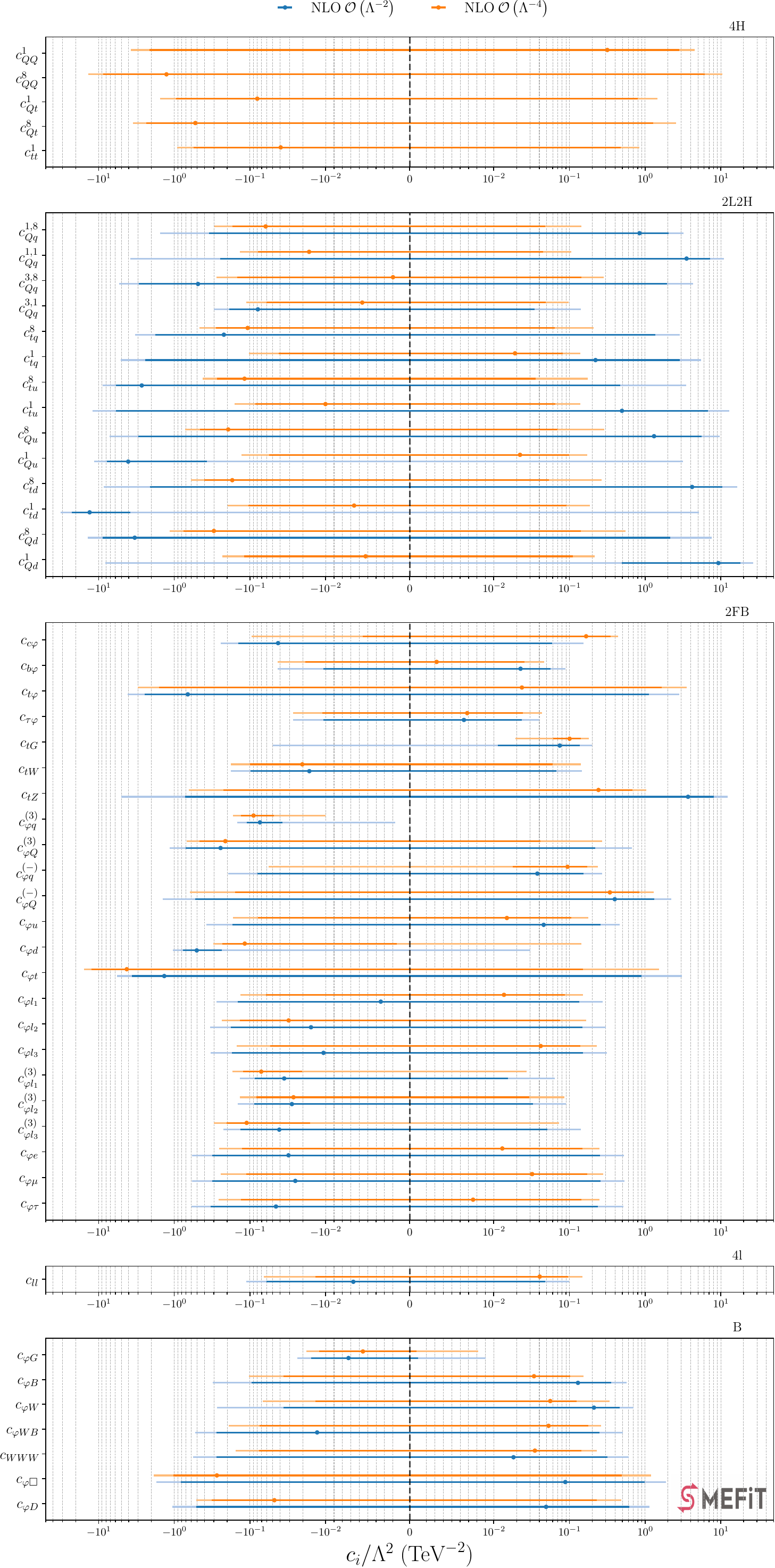}
    \caption{Left: The length of the 95\% credible intervals, expressed in units of 1/TeV$^{2}$, for the $n_{\rm eft}=50$ coefficients entering the fit, both for linear and for quadratic (marginalised) analyses. 
    From top to bottom we display the two-fermion, two-light-two-heavy quark, four-heavy quark (except for the linear fit), and the purely bosonic coefficients.
    Right: the coefficients $c_i/\Lambda^{2}$ for the same fits, where the thicker (thinner) line indicates the 68\% CL (95\% CL) uncertainties.
    The vertical dashed line indicates the SM expectation for the values of these coefficients.
    }
\label{fig:smefit30_marginalised_bounds}
\end{figure}

\paragraph{Constraints on the EFT operators.}
Fig.~\ref{fig:smefit30_marginalised_bounds} displays the results of {\sc\small SMEFiT3.0} at the level of the $n_{\rm eft}=45~(50)$ operators entering the analysis at the linear (quadratic) EFT level.
The right panel shows the best-fit values and the  68\% and 95\% credible intervals (CI), both for the linear and for the quadratic baseline fits.
The reported bounds are extracted from a global fit with all coefficients being varied simultaneously, and then the resultant posterior distributions are marginalised down to individual coefficients.
From top to bottom, we display
the four-heavy quark, two-light-two-heavy quark, two-fermion, four-lepton, and purely bosonic
coefficients.
The corresponding information on the magnitude of the 95\% CI is provided in the left panel of Fig.~\ref{fig:smefit30_marginalised_bounds}.

The bounds displayed in Fig.~\ref{fig:smefit30_marginalised_bounds} are also collected in Table~\ref{tab:smefit30_all_bounds}, where  for completeness we also include the individual bounds obtained from one parameter fits to the data (with all other coefficients set to zero).
It is worth noting that for some operators at the quadratic EFT level the 95\% CI bounds are disjoint, indicating the presence of degenerate solutions. 
For the four-heavy operators, in the linear fit one can only display the individual bounds, since in this sector the SMEFT displays flat directions (for the available data) unless quadratic corrections are included. 

\begin{table}[htbp]
    \centering
    \scriptsize
    \renewcommand{\arraystretch}{1.24}
    \begin{tabular}{l|C{0.8cm}|C{2.3cm}|C{2.3cm}|C{4.1cm}|C{4.1cm}}
        \multirow{2}{*}{Class} & \multirow{2}{*}{DoF}
        & \multicolumn{2}{c|}{ 95\% CI bounds, $\mathcal{O}\lp \Lambda^{-2}\rp$} &
        \multicolumn{2}{c}{95\% CI bounds, $\mathcal{O}\lp \Lambda^{-4}\rp$,} \\
        &                         & Individual          & Marginalised      & Individual                            & Marginalised      \\ \toprule
\multirow{5}{*}{4H}
 & $c_{QQ}^{1}$& [1.648, 24.513] & \textemdash & [-2.403, 2.153]& [-3.765, 4.487] \\ \cline{2-6}
 & $c_{QQ}^{8}$& [3.343, 63.182] & \textemdash & [-7.196, 6.533]& [-13.586, 10.491] \\ \cline{2-6}
 & $c_{Qt}^{1}$& [-509.511, 211.968] & \textemdash & [-1.945, 1.958]& [-1.546, 1.455] \\ \cline{2-6}
 & $c_{Qt}^{8}$& [1.632, 21.393] & \textemdash & [-4.415, 3.607]& [-3.500, 2.549] \\ \cline{2-6}
 & $c_{tt}^{1}$& [0.768, 12.075] & \textemdash & [-1.201, 1.070]& [-0.919, 0.836] \\ \cline{2-6}
\hline
\multirow{14}{*}{2L2H}
 & $c_{Qq}^{1,8}$& [-0.363, 0.201]& [-1.547, 3.207]& [-0.292, 0.141]& [-0.296, 0.144] \\ \cline{2-6}
 & $c_{Qq}^{1,1}$& [-1.154, 0.096]& [-3.820, 11.011]& [-0.150, 0.096]& [-0.136, 0.105] \\ \cline{2-6}
 & $c_{Qq}^{3,8}$& [-1.285, 0.417]& [-5.313, 4.288]& [-0.355, 0.229]& [-0.278, 0.282] \\ \cline{2-6}
 & $c_{Qq}^{3,1}$& [-0.128, 0.106]& [-0.301, 0.141]& [-0.092, 0.080]& [-0.112, 0.097] \\ \cline{2-6}
 & $c_{tq}^{8}$& [-0.639, 0.236]& [-3.270, 2.885]& [-0.459, 0.180]& [-0.467, 0.208] \\ \cline{2-6}
 & $c_{tq}^{1}$& [0.176, 1.188]& [-5.092, 5.481]& [-0.073, 0.160]& [-0.104, 0.139] \\ \cline{2-6}
 & $c_{tu}^{8}$& [-0.675, 0.247]& [-8.866, 3.490]& [-0.439, 0.179]& [-0.422, 0.175] \\ \cline{2-6}
 & $c_{tu}^{1}$& [-1.622, 0.214]& [-12.084, 12.836]& [-0.178, 0.126]& [-0.159, 0.139] \\ \cline{2-6}
 & $c_{Qu}^{8}$& [-1.567, 0.076]& [-7.200, 9.684]& [-0.702, 0.211]& [-0.715, 0.289] \\ \cline{2-6}
 & $c_{Qu}^{1}$& [0.210, 1.596]& [-11.379, 3.183]& [-0.101, 0.193]& [-0.129, 0.171] \\ \cline{2-6}
 & $c_{td}^{8}$& [-1.677, 0.206]& [-8.511, 16.583]& [-0.685, 0.244]& [-0.603, 0.266] \\ \cline{2-6}
 & $c_{td}^{1}$& [-3.955, -0.251]& [-31.597, 5.147]& [-0.234, 0.172]& [-0.198, 0.186] \\ \cline{2-6}
 & $c_{Qd}^{8}$& [-3.147, -0.091]& [-13.997, 7.530]& [-1.108, 0.326]& [-1.158, 0.549] \\ \cline{2-6}
 & $c_{Qd}^{1}$& [0.840, 3.755]& [-8.140, 26.827]& [-0.149, 0.242]& [-0.230, 0.216] \\ \cline{2-6}
\hline
\multirow{23}{*}{2FB}
 & $c_{c \varphi}$& [-0.022, 0.120]& [-0.243, 0.154]& [-0.000, 0.373]& [-0.094, 0.442] \\ \cline{2-6}
 & $c_{b \varphi}$& [-0.007, 0.040]& [-0.043, 0.088]& [-0.008, 0.036]   & [-0.043, 0.046] \\ \cline{2-6}
 & $c_{t \varphi}$& [-1.199, 0.327]& [-4.142, 2.831]& [-1.168, 0.333]& [-3.035, 3.527] \\ \cline{2-6}
 & $c_{\tau \varphi}$& [-0.027, 0.036]& [-0.027, 0.040]&[-0.024, 0.041]& [-0.027, 0.043] \\ \cline{2-6}
 & $c_{tG}$& [0.004, 0.084]& [-0.050, 0.199]& [0.003, 0.080]& [0.019, 0.180] \\ \cline{2-6}
 & $c_{tW}$& [-0.087, 0.029]& [-0.180, 0.147]& [-0.082, 0.029]& [-0.177, 0.141] \\ \cline{2-6}
 & $c_{tZ}$& [-0.034, 0.102]& [-4.999, 12.276]& [-0.038, 0.094]& [-0.645, 1.027] \\ \cline{2-6}
 & $c_{\varphi q}^{(3)}$& [-0.015, 0.012]& [-0.147, -0.002]& [-0.015, 0.012]& [-0.166, -0.010] \\ \cline{2-6}
 & $c_{\varphi Q}^{(3)}$& [-0.016, 0.023]& [-1.155, 0.665]& [-0.016, 0.023]& [-0.685, 0.271] \\ \cline{2-6}
 & $c_{\varphi q}^{(-)}$& [-0.121, 0.119]& [-0.193, 0.269]& [-0.118, 0.119]& [-0.056, 0.239] \\ \cline{2-6}
 & $c_{\varphi Q}^{(-)}$& [-0.031, 0.046]& [-1.427, 2.224]& [-0.031, 0.047]& [-0.620, 1.292] \\ \cline{2-6}
 & $c_{\varphi u}$& [-0.071, 0.081]& [-0.375, 0.461]& [-0.077, 0.079]& [-0.168, 0.177] \\ \cline{2-6}
 & $c_{\varphi d}$& [-0.140, 0.071]& [-1.038, 0.030]& [-0.137, 0.072]& [-0.303, 0.143] \\ \cline{2-6}
 & $c_{\varphi t}$& [-2.855, 1.036]& [-5.750, 3.084]& [-4.000, 0.872]& [-15.638, 1.532] \\ \cline{2-6}
 & $c_{\varphi l_1}$& [-0.009, 0.012]& [-0.276, 0.273]& [-0.008, 0.012]& [-0.133, 0.150] \\ \cline{2-6}
 & $c_{\varphi l_2}$& [-0.031, 0.017]& [-0.334, 0.302]& [-0.030, 0.017]& [-0.237, 0.166] \\ \cline{2-6}
 & $c_{\varphi l_3}$& [-0.035, 0.025]& [-0.329, 0.311]& [-0.034, 0.025]& [-0.150, 0.231] \\ \cline{2-6}
 & $c_{\varphi l_1}^{(3)}$& [-0.015, 0.009]& [-0.136, 0.064]& [-0.015, 0.009]& [-0.170, 0.027] \\ \cline{2-6}
 & $c_{\varphi l_2}^{(3)}$& [-0.031, 0.002]& [-0.146, 0.089]& [-0.031, 0.002]& [-0.137, 0.085] \\ \cline{2-6}
 & $c_{\varphi l_3}^{(3)}$& [-0.039, 0.017]& [-0.225, 0.141]& [-0.039, 0.017]& [-0.298, 0.073] \\ \cline{2-6}
 & $c_{\varphi e}$& [-0.025, 0.001]& [-0.583, 0.527]& [-0.025, 0.001]& [-0.254, 0.248] \\ \cline{2-6}
 & $c_{\varphi \mu}$& [-0.021, 0.039]& [-0.582, 0.533]& [-0.021, 0.038]& [-0.245, 0.277] \\ \cline{2-6}
 & $c_{\varphi \tau}$& [-0.045, 0.024]& [-0.597, 0.512]& [-0.045, 0.024]& [-0.261, 0.248] \\ \cline{2-6}
\hline
\multirow{1}{*}{4l}
 & $c_{ll}$& [-0.008, 0.037]& [-0.112, 0.100]& [-0.008, 0.037]& [-0.066, 0.149] \\ \cline{2-6}
\hline
\multirow{7}{*}{B}
 & $c_{\varphi G}$& [-0.001, 0.005]& [-0.024, 0.009]& [-0.002, 0.005]& [-0.018, 0.008] \\ \cline{2-6}
 & $c_{\varphi B}$& [-0.005, 0.002]& [-0.310, 0.573]& [-0.005, 0.002] $\cup$ [0.085, 0.092]& [-0.103, 0.152]  \\ \cline{2-6}
 & $c_{\varphi W}$& [-0.018, 0.006]& [-0.273, 0.707]& [-0.017, 0.006] $\cup$ [0.282, 0.305] & [-0.067, 0.338] \\ \cline{2-6}
 & $c_{\varphi WB}$& [-0.007, 0.003]& [-0.525, 0.504]& [-0.007, 0.003]& [-0.190, 0.263] \\ \cline{2-6}
 & $c_{WWW}$& [-0.479, 0.607]& [-0.565, 0.609]& [-0.155, 0.197]& [-0.156, 0.230] \\ \cline{2-6}
 & $c_{\varphi \Box}$& [-0.416, 1.193]& [-1.715, 1.879]& [-0.429, 1.141]& [-1.856, 1.199] \\ \cline{2-6}
 & $c_{\varphi D}$& [-0.027, -0.003]& [-1.063, 1.149]& [-0.027, -0.003]& [-0.513, 0.483] \\ \cline{2-6}
\hline
    \end{tabular}
    \caption{\small The 95\% CI bounds on the
    EFT coefficients
    considered in this analysis.
    The reported bounds correspond to $\Lambda=1$ TeV and can thus be rescaled for any other value of $\Lambda$.
    We present results both for linear EFT fits, $\mathcal{O}\lp \Lambda^{-2}\rp$, and for quadratic EFT fits, $\mathcal{O}\lp \Lambda^{-4}\rp$.
    In each case, we indicate both individual bounds (from one-parameter fits) and marginalised bounds (from a simultaneous determination of the full set of $n_{\rm eft}$ coefficients). 
    }
    \label{tab:smefit30_all_bounds}
\end{table}

\begin{figure}[htbp]
    \center
    \includegraphics[width=.77\linewidth]{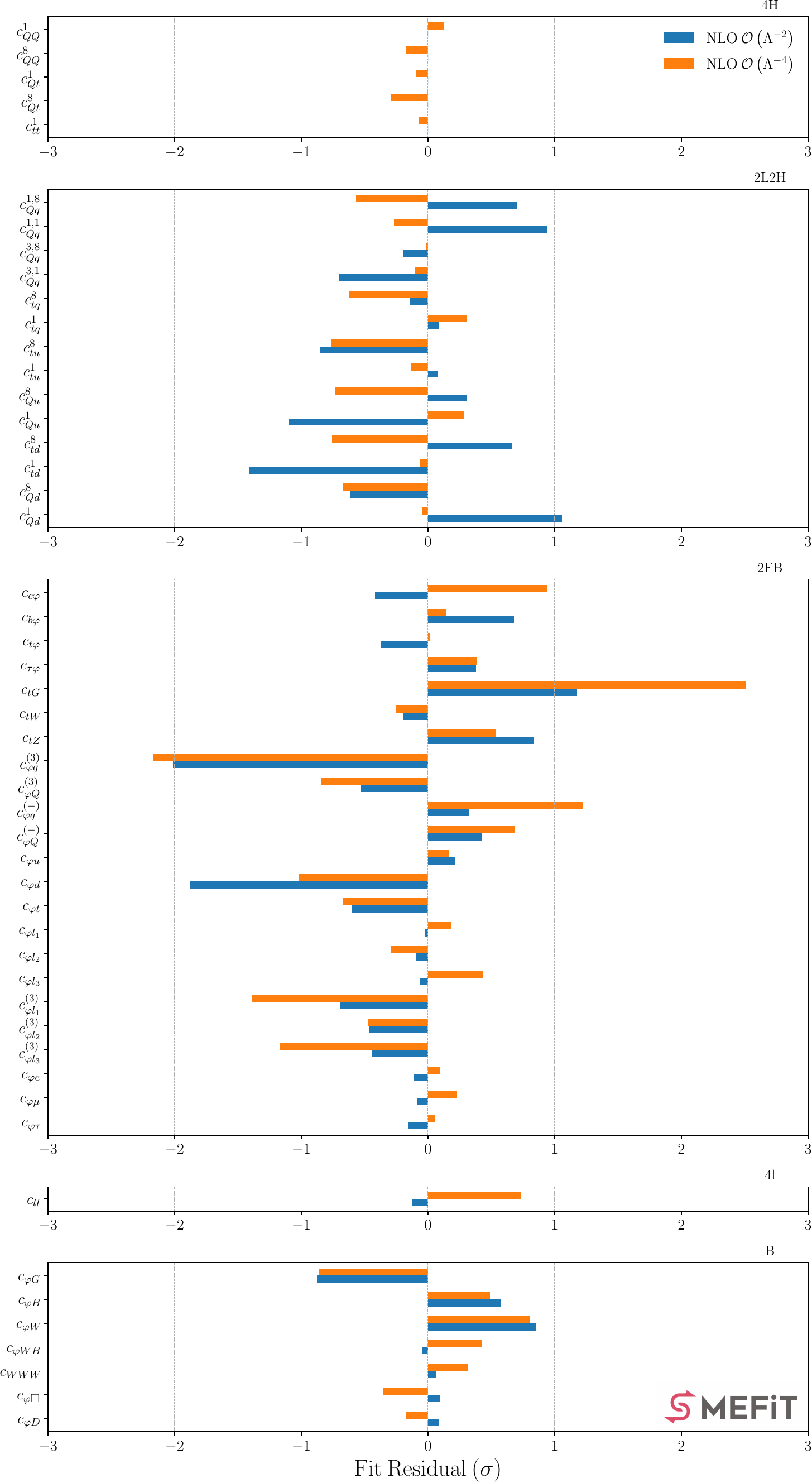}
    \caption{The residuals between the fit results and the SM expectations defined as in Eq.~(\ref{eq:fit_residuals}), for the operators entering Fig.~\ref{fig:smefit30_marginalised_bounds} and for both linear and quadratic fits.
    These residuals are normalised to
    the 68\% CI intervals, hence a residual larger than 2 (in absolute value) indicates a coefficient that disagrees with the SM at the 95\% CI.
    }
\label{fig:pull_lin_vs_quad}
\end{figure}

To further quantify the agreement between the SMEFT fit results and the corresponding SM expectations, Fig.~\ref{fig:pull_lin_vs_quad} displays the fit residuals defined as
\be
\label{eq:fit_residuals}
P_i \equiv 2 \lp \frac{  \left\langle c_i\right\rangle  - c_i^{(\rm SM)}}{
 \lc c_i^{\rm min}, c_i^{\rm max} \rc^{68\%~{\rm CI}}
}\rp \, , \qquad i=1,\ldots, n_{\rm eft}  \, ,
\ee
in the same format as that of Fig.~\ref{fig:smefit30_marginalised_bounds} for both linear and quadratic fits.
Given that $c_i^{(\rm SM)}=0$ and that Eq.~(\ref{eq:fit_residuals}) is normalised to
    the 68\% CI (which in linear fits correspond to the standard deviation $\sigma$), 
a residual larger than 2 (in absolute value) indicates a coefficient that does not agree with the SM at the 95\% CI. An alternative measure to quantify the agreement between the SMEFT and the SM that may be explored in future work is provided by the Bayes factor \cite{Trotta:2008qt}, which compares two models based on the ratio of their Bayesian evidence. 

Several observations can be derived from the inspection of Figs.~\ref{fig:smefit30_marginalised_bounds} and~\ref{fig:pull_lin_vs_quad} as well as Table~\ref{tab:smefit30_all_bounds}.
First, the fit residuals evaluated in Fig.~\ref{fig:pull_lin_vs_quad}, consistently with Fig.~\ref{fig:smefit30_marginalised_bounds}, confirm that in general there is a good agreement between the EFT fit results and the Standard Model.
For the purely bosonic, four-lepton, four-heavy, and two-light-two-heavy operators, the fit residuals satisfy $|P_i|\lsim 1$,  the only exception being $c_{td}^1$ in the linear fit for which $P_i\sim 1.5$.
Somewhat larger residuals are found for a subset of the two-fermion operators, in particular for the chromomagnetic operator coefficient $c_{tG}$ (only in the quadratic fit), for $c_{\varphi q}^{(3)}$, and for $c_{\varphi d}$ (only in the linear fit).
For these coefficients, the values of $|P_i|$ range between 2.0 and 2.5. 
Below we investigate the origin of these large fit residuals. 

One also finds that quadratic EFT corrections improve the bounds on most operators entering the fit, with a particularly marked impact for the two-light-two-heavy operators.
The only exceptions of this trend are $c_{\varphi t}$, which is poorly constrained to begin with, and the charm Yukawa $c_{c\varphi}$.
For both coefficients, the worse bounds arising in the quadratic fit are explained by the appearance of a second, degenerate solution, as demonstrated by the corresponding posterior distributions displayed in Fig.~\ref{fig:new-vs-old-quad}.
Such degenerate solutions may arise~\cite{Ethier:2021bye} when quadratic corrections become comparable in magnitude with opposite sign to the linear EFT cross-section, a configuration formally equivalent to setting $c_i=0$ and hence reproducing the SM.

As is well known~\cite{Hartland:2019bjb}, quadratic EFT corrections also allow one to bound the four-heavy operator coefficients  $c_{tt}^{1}$, $c_{Qt}^{1}$, $c_{Qt}^{8}$, $c_{QQ}^{1}$, and $c_{QQ}^{8}$.
Within a $\mathcal{O}\lp \Lambda^{-2}\rp$ fit, only two linear combinations of these four-heavy operators can be instead constrained, leaving three flat directions.
These considerations do not hold for one-parameter fits, where the four-heavy operators can be separately constrained. 
From Fig.~\ref{fig:smefit30_marginalised_bounds}, one can also see that for some operators the effects of the quadratic EFT corrections are essentially negligible, indicating that the linear (interference) cross-section dominates the sensitivity.
Specifically, operators for which quadratic corrections are small are the four-lepton operator $c_{\ell\ell}$, the two-fermion operators  $c_{\varphi q}^{3}$, $c_{\varphi \ell_i}^{3}$ (with $i=1,\,2,\,3$), $c_{tW}$, and the tau and top Yukawa couplings, $c_{\tau\varphi}$ and $c_{t\varphi}$ respectively. This is caused by small relative quadratic corrections with respect to the linear ones. For instance, in case of the STXS1.2 measurements from ATLAS \cite{ATLAS:2022vkf}, the ratio of quadratic to linear corrections in $c_{t\varphi}$ range from $\mathcal{O}\left(10^{-1}\right)$ up to $\mathcal{O}\left(10^{-2}\right)$ for $c_{t\varphi}=1$.

The comparison between global (marginalised) and individual (one-parameter) fit results reported in Table~\ref{tab:smefit30_all_bounds} indicates that, for the linear EFT fits, one-parameter bounds are always tighter than the marginalised ones.
The differences between individual and marginalised bounds span a wide range of variation, from $c_{WWW}$, which essentially shows no difference, to $c_{\varphi D}$, with individual bounds tighter by two orders of magnitude as compared to the marginalised counterparts.
Concerning the quadratic EFT fits, for the purely bosonic and two-fermion operators the situation is similar as in the linear case, with individual bounds either (much) tighter than the marginalised ones or essentially unchanged (as is the case for $c_{WWW}$
and $c_{\tau \varphi}$, for example).
The situation is somewhat different for the four-heavy and two-light-two-heavy operators. 
For the latter, the marginalised and individual bounds are now similar to each other, as opposed to the linear fit case.
For the four-heavy operators, the marginalised bounds are either similar or a bit broader than the individual ones, except in the case of $c_{Qt}^{(8)}$, whose 95\% CI bounds $[-4.4,3.6]$ (individual) improve to $[-3.5,2.5]$ (marginalised), hence by a factor of approximately 30\%.
In such cases the correlations with other parameters entering the global fit improve the overall sensitivity compared to the one-parameter fits.

\begin{figure}[t]
    \center
    \includegraphics[width=.9\linewidth]{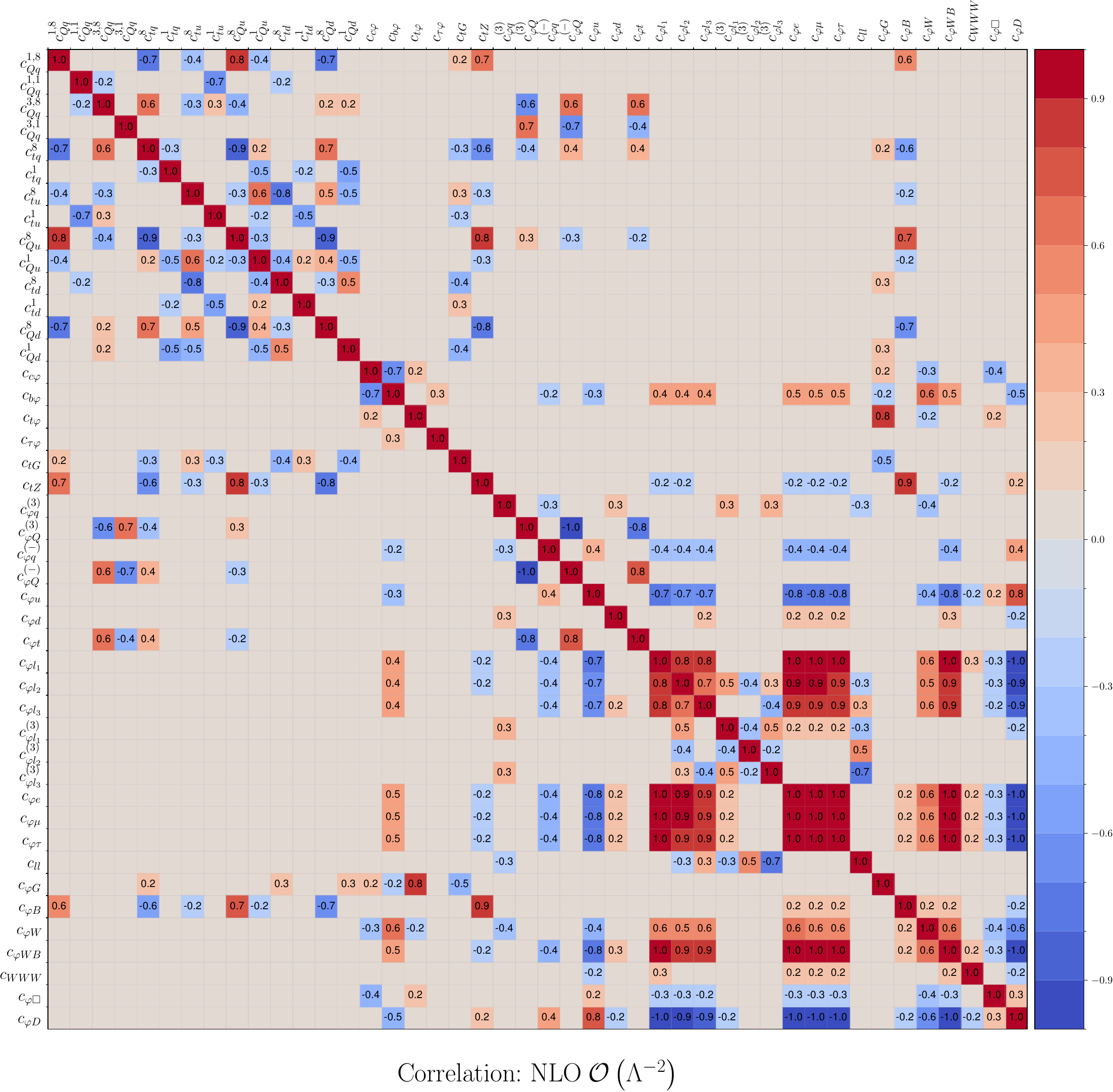}
    \caption{The correlation matrix for the $n_{\rm eff}=45$ coefficients associated to the linear {\sc\small SMEFiT3.0} baseline analysis.
    To facilitate visualisation, the EFT coefficients whose correlation with all other coefficients is $<0.2$ are removed from the plot. 
    }
\label{fig:corr_smefit3_lin}
\end{figure}

\begin{figure}[t]
    \center
    \includegraphics[width=.9\linewidth]{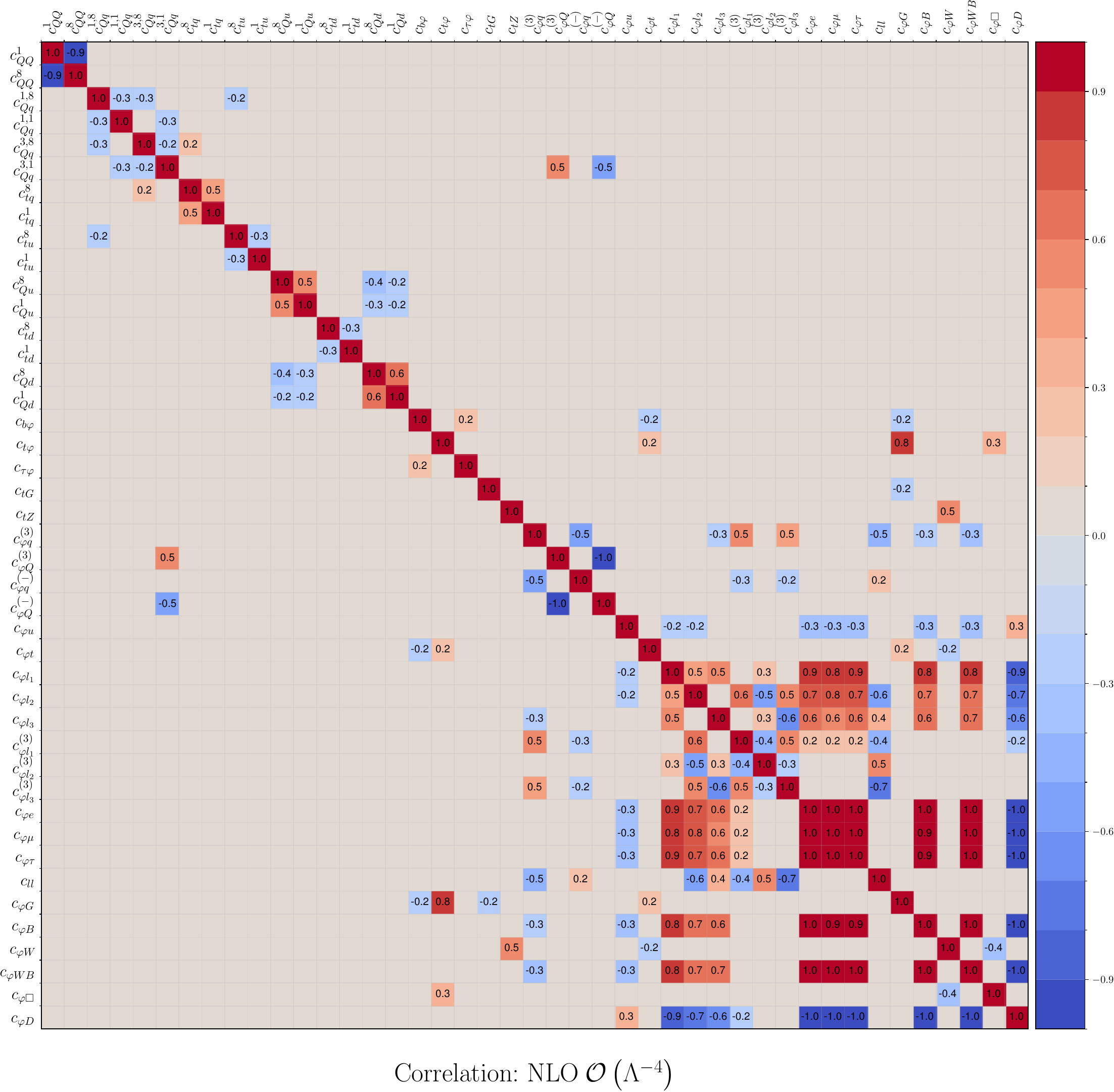}
    \caption{Same as Fig.~\ref{fig:corr_smefit3_lin}, now
    for the {\sc\small SMEFiT3.0} baseline analysis carried out with $\mathcal{O}\lp \Lambda^{-4}\rp$ corrections.
    }
\label{fig:corr_smefit3_quad}
\end{figure}

When interpreting the bounds on the EFT coefficients and the associated residuals displayed in Figs.~\ref{fig:smefit30_marginalised_bounds} and~\ref{fig:pull_lin_vs_quad}, one should recall that in general there are potentially large correlations between them.
To illustrate these, Fig.~\ref{fig:corr_smefit3_lin}~(\ref{fig:corr_smefit3_quad}) displays the entries of the correlation matrix, $\rho_{ij}$, for the $n_{\rm eft}=45$ (50) coefficients associated to the linear (quadratic) {\sc\small SMEFiT3.0} baseline analysis.
To facilitate visualisation, entries with $|\rho_{ij}|<0.2$ (negligible correlations) are not shown in the plot. 

The most noticeable feature comparing Figs.~\ref{fig:corr_smefit3_lin} and~\ref{fig:corr_smefit3_quad} is the fact that correlations become significantly weaker in the quadratic fit, especially for the two-light-two-heavy top quark operators as already noticed in~\cite{Hartland:2019bjb}, but also for some purely bosonic and two-fermion operators.
Nevertheless, some large correlations remain also in the quadratic fit, and for instance $c_{\varphi D}$ is strongly anti-correlated with most of the operators constrained by the EWPOs.
We recall that the correlation patterns in Figs.~\ref{fig:corr_smefit3_lin} and~\ref{fig:corr_smefit3_quad}  depend on the specific fitted dataset, and in particular these patterns change qualitatively once we include the FCC-ee projections in Sect.~\ref{sec:results}.

\paragraph{Coefficients with large residuals.}
As mentioned above, the fit residual analysis of Fig.~\ref{fig:pull_lin_vs_quad} indicates that three Wilson coefficients, namely $c_{tG}$ (in the quadratic fit), $c_{\varphi d}$ (in the linear fit), and $c_{\varphi q}^{3}$ (in both cases) do not agree with the SM expectation at the 95\% CL, with pulls of $P_i \simeq +2.5,-1.9$, and $-2.1$ respectively. 
The corresponding individual (one-parameter) analysis of Table~\ref{tab:smefit30_all_bounds} indicates that for these coefficients the pulls are $P_i \simeq +2.0,-0.3$, and $-0.2$ respectively, when fitted setting all other operators to zero.
Therefore, the pull on $c_{tG}$ in the quadratic case is somewhat reduced in individual fits but does not go away, while the large pulls on $c_{\varphi q}^{3}$ and $c_{\varphi d}$ completely disappear in the one-parameter fits.
The latter result indicates that the pulls of $c_{\varphi q}^{3}$ and $c_{\varphi d}$ found in the global fit arise as a consequence of the correlations with other fit parameters.

In the case of the chromomagnetic operator coefficient $c_{tG}$, the tension with the SM which arises in the quadratic fit was already present in previous versions of our analysis~\cite{Ethier:2021bye,Giani:2023gfq} and is known to be driven by the CMS top-quark double-differential distributions in $(y_{t\bar{t}},m_{t\bar{t}})$ at $8$~TeV from~\cite{CMS:2017iqf}.
In the context of a linear EFT fit, the obtained residual is consistent with the SM, a finding also in agreement with the independent analysis carried out in~\cite{Kassabov:2023hbm}.
Indeed, if this CMS double-differential $t\bar{t}$ 8 TeV measurement is excluded from the quadratic fit, $c_{tG}$ becomes fully consistent with the SM expectation.
We also note that this dataset, with $\chi^2_{\rm SM}/n_{\rm dat}\simeq 1.7$ for $n_{\rm dat}=16$ points, improves down to $\chi^2_{\rm EFT}/n_{\rm dat}\simeq 1$ after the fit.
Given that $c_{tG}$ modifies the overall normalisation of top-quark pair production, rather than the shape of the distributions, this result may imply that the normalisation of this 2D CMS measurement is in tension with that of other $t\bar{t}$ measurements included in the fit.
All in all, it appears unlikely that this large pull on $c_{tG}$ obtained in the quadratic fit is related to a genuine BSM signal.

Regarding the $c_{\varphi d}$ and $c_{\varphi q}^{(3)}$ Wilson coefficients, we note pulls of approximately $-1.9$ and $-2.1$, respectively, in the linear fit. However, in the quadratic fit, the pull for $c_{\varphi d}$ decreases to around $-1.0$. Notably, the individual constraints are instead entirely consistent with the SM.
This pattern arises from the predominance of LEP data in individual fits, where no deviations from the SM are apparent. However, in a comprehensive global fit, the LEP data exhibit strong inter-coefficient correlations, leading to a notable reduction in their constraining effectiveness.
As is well-established, the complementary nature of LHC diboson measurements to EWPO is crucial to break several of these correlations. For this reason, the LHC diboson data, despite being less precise, can have a surprising impact on the bounds of the EFT coefficients affecting LEP observables. We have confirmed that these measurements are indeed responsible for the observed pulls in the global fit.

\begin{figure}[t]
    \centering
    \includegraphics[width=\linewidth]{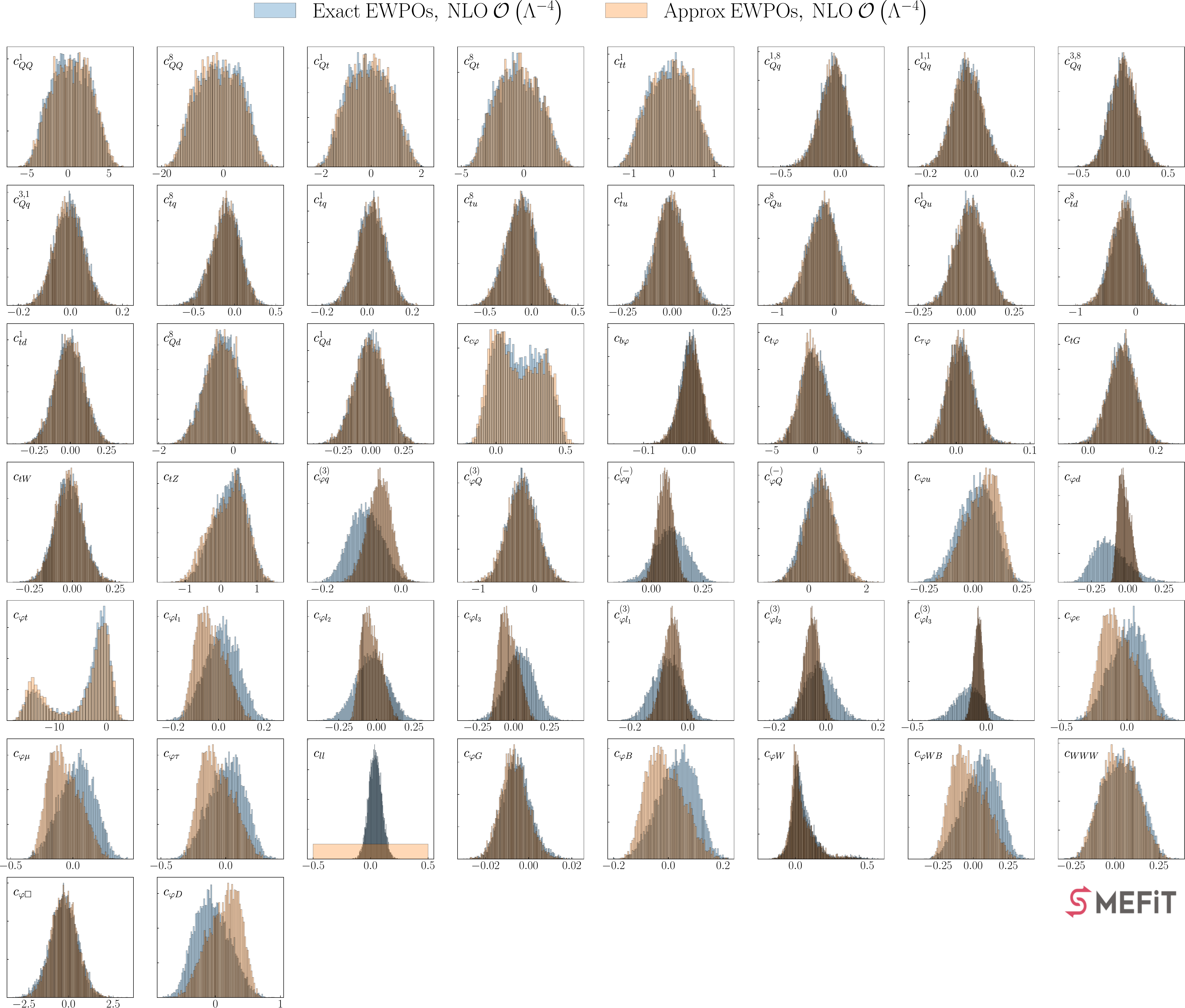}
    \caption{Posterior distributions associated to the $n_{\rm eft}=50$ Wilson coefficients constrained in the 
 {\sc\small SMEFiT3.0} global analysis, carried out at $\mathcal{O}\lp \Lambda^{-4}\rp$ in the EFT expansion.
 A simultaneous determination of all coefficients is performed and then one marginalises for individual degrees of freedom. 
 The baseline results, performed with the exact implementation of the EWPOs, are compared with the approximated implementation used in~\cite{Ethier:2021bye}.
    }
    \label{fig:updated-global-fit-quad}
\end{figure}

\begin{figure}[t]
    \centering
    \includegraphics[width=\linewidth]{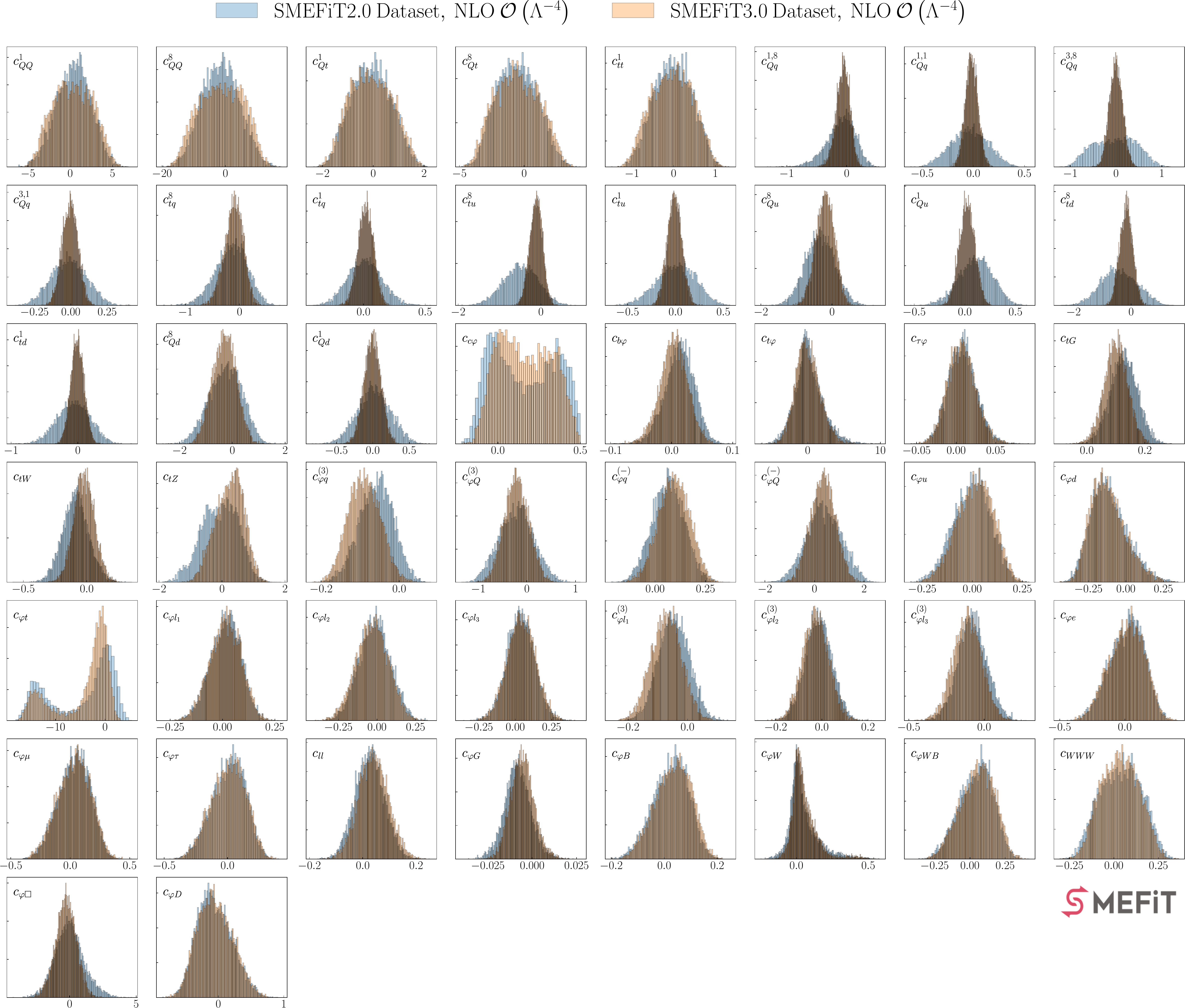}
    \caption{Same as Fig.~\ref{fig:updated-global-fit-quad} now comparing the results of the global analysis based
    on the {\sc\small SMEFiT2.0} and {\sc\small SMEFiT3.0} datasets, all other settings and theory calculations kept the same.
    In particular, in both cases the baseline settings includes
   the exact implementation of the EWPOs.
    }
    \label{fig:new-vs-old-quad}
\end{figure}

\paragraph{Exact versus approximate implementation of the EWPOs.}
Fig.~\ref{fig:updated-global-fit-quad} displays a comparison at the level of the posterior distributions on the Wilson coefficients between the new implementation of the EWPOs presented in Sect.~\ref{sec:ewpos} and used in {\sc\small SMEFiT3.0} and the previous, approximate implementation entering  {\sc\small SMEFiT2.0} and based on imposing the restrictions in Eq.~\eqref{eq:2independents}.
In both cases, these posteriors correspond to global marginalised fits carried out at $\mathcal{O}\lp \Lambda^{-4}\rp$ in the EFT expansion, see also Table~\ref{table:fit_settings}.

From this comparison one observes that the exact implementation of the EWPOs does not lead to major qualitative differences in the posterior distributions. 
Nevertheless, the approximate implementation of the EWPOs was in some cases too aggressive, and when replaced by the exact implementation one observes how the associated posterior distributions may display a broadening, as is the case for instance for the $c_{\varphi \ell_3}^{(3)}$ and $c_{\varphi d}$ coefficients.
Other EFT degrees of freedom for which the posterior distributions are modified following the exact implementation of the EWPOs are $c_{\varphi \ell_3},\,c_{\varphi \ell_2}^{(3)},\,c_{\varphi q}^{(3)},\, c_{\varphi q}^{(-)},\,$ and $c_{\ell\ell}$.  
In particular, we note that the four-lepton coefficient $c_{\ell\ell}$ was set to zero in the approximate implementation, while now it enters as an independent degree of freedom.
 
Two-light-two-heavy and four-heavy operators are constrained mostly by a set of processes not sensitive to the operators entering the EWPOs, i.e. top pair production and four-heavy quark production. 
There is limited cross-talk between the four-heavy operators and those entering the EWPOs, and hence the posteriors of the former remain unchanged comparing the two fits.
Furthermore, for other operators which are not directly sensitive to the EWPOs, we have verified that the residual observed differences arise from their correlations within the global fit with coefficients modifying the electroweak sector of the SMEFT (and they are  hence absent in one-parameter individual fits), see also Figs.~\ref{fig:corr_smefit3_lin} and~\ref{fig:corr_smefit3_quad}.

We conclude from this analysis that, at the level of sensitivity that global SMEFT fits such as the one presented in this work are achieving, it is crucial to properly account for the constraints provided by the precise EWPOs from electron-positron colliders. 

\paragraph{Impact of new LHC Run II data.}
Next we quantify the impact of the new LHC Run II measurements included in the analysis, in comparison with {\sc\small SMEFiT2.0}, and listed in Sect.~\ref{subsec:new_datasets}.
To this end, we compare the baseline global SMEFT fit with a variant in which the input dataset is reduced to match that used in our previous analyses~\cite{Ethier:2021bye,Giani:2023gfq}.
In both cases, methodological settings and theory calculations are kept identical, and in particular both fits include the exact implementation of the EWPOs, quadratic EFT effects, and NLO QCD corrections to the EFT
cross-sections, see also Table~\ref{table:fit_settings}.
Hence the only difference between the two results concerns the LHC Run II data being fitted.

Fig.~\ref{fig:new-vs-old-quad} displays the same comparison of the results of the global analysis based on the {\sc\small SMEFiT2.0} and {\sc\small SMEFiT3.0} datasets.
The most marked impact of the new data is observed for the two-light-two-heavy four-fermion operators, where the narrower posterior distributions reflect improved bounds by a factor between 2 and 3 compared to the fit to the {\sc\small SMEFiT2.0} dataset, depending on the specific operator.
In all cases, the posterior distributions for the two-light-two-heavy operators remain consistent with the SM expectation at the 68\% CI interval, see also Fig.~\ref{fig:pull_lin_vs_quad}.

Other coefficients for which the new data brings in moderate improvements include the charm Yukawa $c_{c\varphi}$ (thanks to the latest Run II measurements which constrain the Higgs branching ratios and hence the total Higgs width), $c_{tZ}$  and $c_{\varphi t}$ (from the new $t\bar{t}Z$ dataset).
For the other coefficients, the impact of the new datasets is minor.
In particular, the latest measurements on $t\bar{t}t\bar{t}$ and $t\bar{t}b\bar{b}$  leave the posterior distributions of the four-heavy-fermion operators essentially unchanged.\footnote{This is explained by tensions between the individual measurements. 
If the same fit is carried out with Level-0 pseudo-data, see App.~\ref{app:hl_lhc_projections}, one observes a clear improvement induced by the latest $t\bar{t}t\bar{t}$ and $t\bar{t}b\bar{b}$ measurements. }

Furthermore, one notes that the bound on the triple-gauge coupling operator $\mathcal{O}_{WWW}$ does not improve upon the inclusion of diboson production measurements based on the full Run II luminosity.
However, we should emphasise that including diboson production in proton-proton collisions in the fit plays an important role in breaking flat directions from EWPOs, for example in the $\lp c_{\varphi q}^{(3)}, c_{\varphi q}^{(-)}\rp$ plane. 
As we will also see in Sect.~\ref{sec:impact_hl_lhc_data}, (HL)-LHC diboson measurements are crucial in improving the bounds on various two-light-fermion coefficients.  

\subsection{Projections for the HL-LHC}
\label{sec:impact_hl_lhc_data}

We now assess the impact of projected HL-LHC measurements when added on top of the {\sc\small SMEFiT3.0} baseline fit.
These projections are constructed following the procedure described in App.~\ref{app:hl_lhc_projections}, where we also list the processes considered.
In a nutshell, we take existing Run II measurements for a given process, focusing on datasets obtained from the highest luminosity, and extrapolate their statistical and systematic uncertainties to the HL-LHC data-taking period. 
Specifically, the statistical uncertainties in the projected pseudo-data are reduced by a factor depending on the ratio of luminosities, while systematic uncertainties are reduced by a fixed factor (taken to be 1/2 in our case) based on the expected performance improvement of the detectors. 

Within the adopted procedure, we maintain the settings and binning of the original Run II analysis unchanged, and assume the SM as the underlying theory.
We note that our projections are not optimised, and in particular with a higher luminosity one could also extend the kinematic coverage of the high-$p_T$ regions~\cite{Durieux:2022cvf}, adopt a finer binning, or attempt multi-differential measurements.
Nevertheless, our approach benefits from being exhaustive and systematic, and is also readily extendable once new Run II and III measurements become available. 

Since the considered HL-LHC projections assume the SM as the underlying theoretical description, and to avoid introducing possible inconsistencies, we generate Level-1 SM pseudo-data for the full {\sc\small SMEFiT3.0} dataset and use it to produce a baseline fit for the subsequent inclusions of the HL-LHC pseudo-data, see also Table~\ref{table:fit_settings}.
In Level-1, the pseudo-data is fluctuated randomly within uncertainties around the central SM theoretical prediction (see App.~\ref{app:hl_lhc_projections}).
This is the same strategy adopted in the closure tests entering the NNPDF proton structure analyses~\cite{NNPDF:2014otw, NNPDF:2021njg}.
A dataset consistent with the SM as underlying theory throughout enables to cleanly separate the sensitivity of the projected data to the SMEFT parameter space from other possible factors, such as dataset inconsistency, eventual BSM signals, or the interplay with QCD uncertainties such as those associated to the PDFs.
We have verified that, in the SMEFT analyses based on pseudo-data generated this way, the fit quality satisfies $\chi^2/n_{\rm dat}\sim 1$ as expected (see Fig.~\ref{fig:closure_test_dist}), both for the baseline fit and once the HL-LHC (and later the FCC-ee and CEPC) projections are included. 
For this reason, in the following we only present results for the relative reduction of the uncertainties associated to the Wilson coefficients, since the central values are by construction consistent with the SM expectations. 

In the rest of this section and in the following one, we present results in terms of $R_{\delta c_i}$, defined as the ratio between the magnitude of the 95\% CL interval for a given EFT coefficient $c_i$, to that of the same quantity in the baseline fit:
\be
\label{eq:RatioC}
R_{\delta c_i} = \frac{ \lc c_i^{\rm min}, c_i^{\rm max} \rc^{95\%~{\rm CI}}~({\rm baseline+\text{HL-LHC}})}{
 \lc c_i^{\rm min}, c_i^{\rm max} \rc^{95\%~{\rm CI}}~({\rm baseline})
} \, , \qquad i=1,\ldots, n_{\rm eft} \, .
\ee
In the case of a disjoint 95\% CI interval, we add up the magnitudes of the separate regions.
From the definition of Eq.~(\ref{eq:RatioC}) it is clear that, for a given  coefficient $c_i$, the smaller the value of $R_{\delta c_i}$ the more significant the impact of the new data.

As mentioned above, central values are by construction in agreement with the SM expectation ($c_i^{\rm (SM)} =0$) within uncertainties, and hence it is not necessary to display them in these comparisons.
The ratio Eq.~(\ref{eq:RatioC}) can be evaluated both in one-parameter fits as well as in the global fit followed by marginalisation. 
In contrast with Level-1 pseudo-data, Level-0 pseudo-data has central values identical to the SM predictions, and do not account for statistical fluctuations in the experimental measurements;  for completeness,
we  also display Level-0 projection results in App.~\ref{app:hl_lhc_projections}.

\begin{figure}[t]
    \centering
    \includegraphics[width=0.85\linewidth]{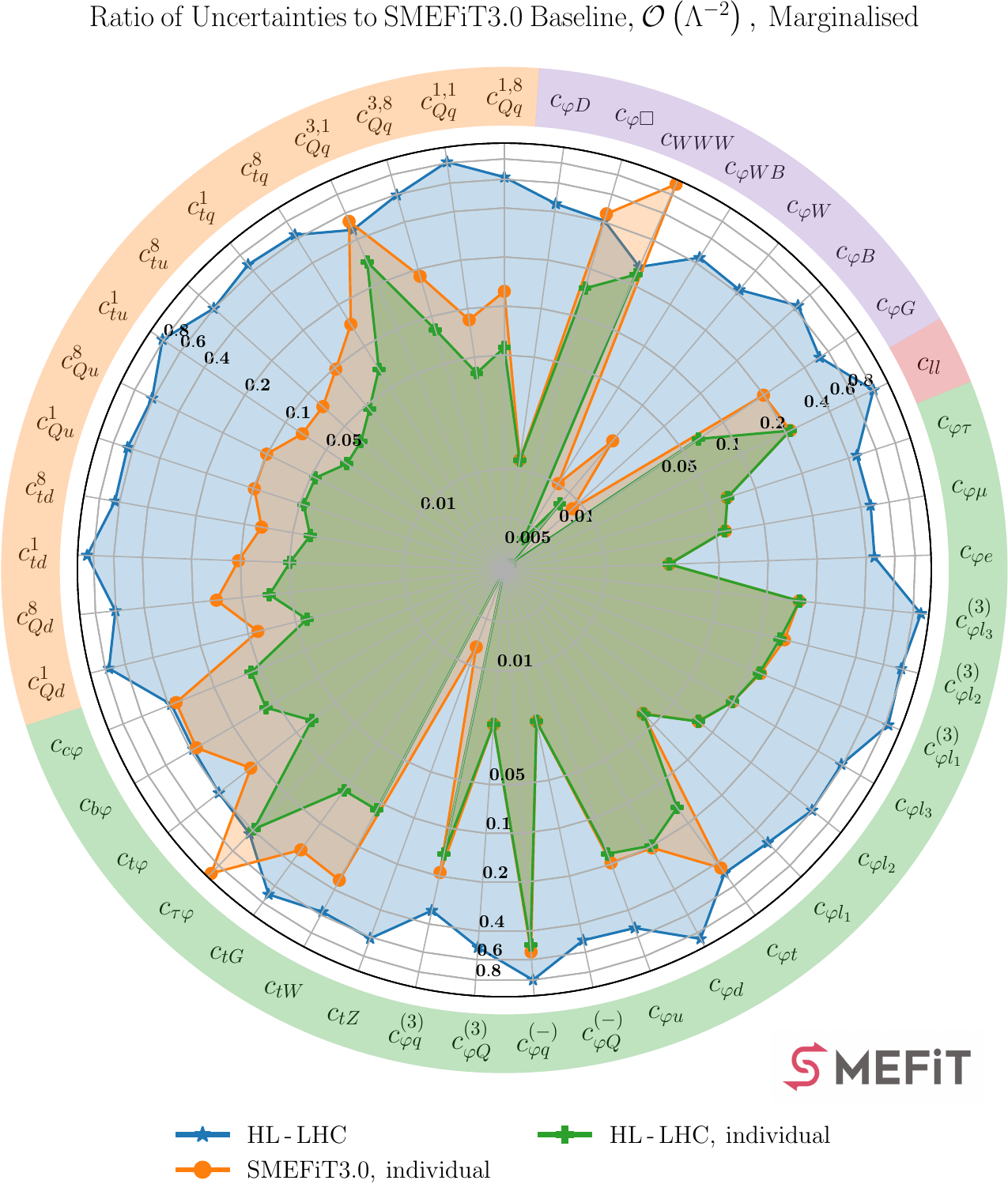}
    \caption{The ratio of uncertainties $R_{\delta c_i}$, defined in Eq.~(\ref{eq:RatioC}), for the $n_{\rm eft}=45$ coefficients entering the linear EFT fit, quantifying the impact of the HL-LHC projections when added on top of the {\sc\small SMEFiT3.0}
    baseline. 
    We display both the results of one-parameter fits and those of the marginalised analysis.
    The different color codes indicate the relevant groups of SMEFT operators: two-light-two-heavy operators (orange), two-fermion operators (green), purely
    bosonic operators (purple), and the four-lepton operator $c_{\ell \ell}$ (red).
    Note that here  the baseline is a fit to pseudo-data for the {\sc\small SMEFiT3.0} dataset generated assuming the SM, rather than the fit to real data, see also Table~\ref{table:fit_settings}.
   }
    \label{fig:spider_nlo_lin_glob}
\end{figure}

\begin{figure}[t]
    \centering
    \includegraphics[width=0.85\linewidth]{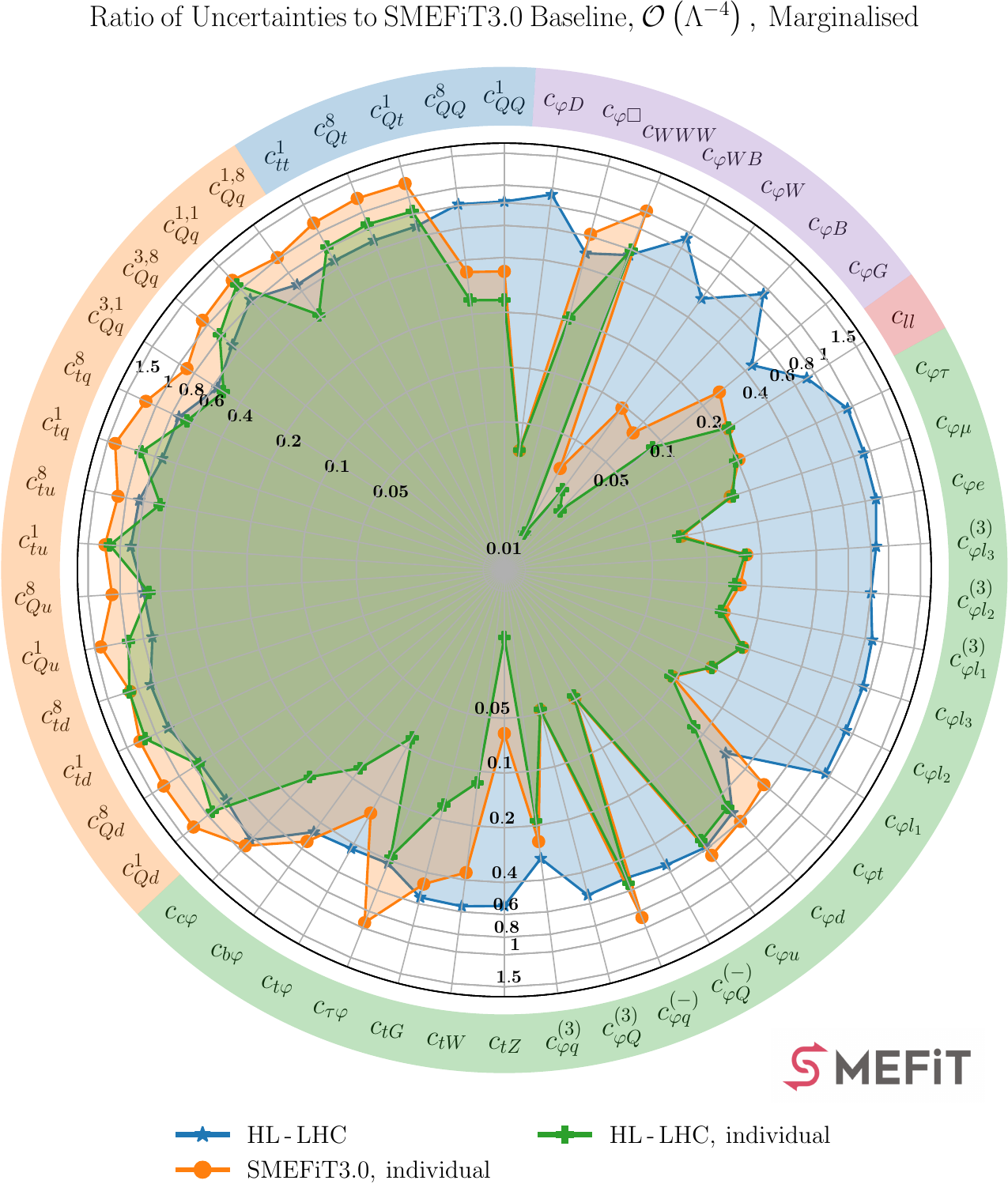}
    \caption{Same as Fig.~\ref{fig:spider_nlo_lin_glob}, now in the case of the analysis with quadratic EFT corrections included. 
    Note that in comparison with the linear fits,
    we now have an extra group of operators (indicated in blue), namely the four-heavy four-fermion operators
    that are constrained by $t\bar{t}t\bar{t}$
    and $t\bar{t}b\bar{b}$ production data.   
 }
    \label{fig:spider_nlo_quad_glob}
\end{figure}

Fig.~\ref{fig:spider_nlo_lin_glob} displays the ratio of uncertainties $R_{\delta c_i}$,  Eq.~(\ref{eq:RatioC}), for the $n_{\rm eft}=45$ Wilson coefficients entering the linear EFT fit, quantifying the impact of the HL-LHC projections when added on top of the {\sc\small SMEFiT3.0}
    baseline. 
We display both the results of the global fit, as well as those of one-parameter fits where all other coefficients are set to zero. 
Whenever available, as in the rest of this work, NLO QCD corrections for the EFT cross-sections are accounted for.
Then in Fig.~\ref{fig:spider_nlo_quad_glob} we
show the same comparison now in the case of the analysis with quadratic EFT corrections included in the theory calculations.\footnote{The counterpart of Fig.~\ref{fig:spider_nlo_quad_glob} based on Level-0 pseudo-data is provided in App.~\ref{app:hl_lhc_projections}.} 

To facilitate visualisation, in Figs.~\ref{fig:spider_nlo_lin_glob}  and~\ref{fig:spider_nlo_quad_glob} results are presented with a ``spider plot'' format, with the different colours on the perimeter indicating the relevant groups of SMEFT operators: two-light-two-heavy four-fermion operators, two-fermion operators, purely bosonic operators, four-heavy four-fermion operators, and the four-lepton operator $c_{\ell \ell}$.
Recall that the four-heavy operators, constrained by $t\bar{t}t\bar{t}$  and $t\bar{t}b\bar{b}$ production data, are excluded from the linear fit due to its lack of sensitivity.
In this plotting format, coefficients whose values for $R_{\delta c_i}$ are closer to the center of the plot correspond to the operators which are the most constrained by the HL-LHC projections, in the sense of the largest reduction of the corresponding uncertainties.
These plots adopt a logarithmic scale for the radial coordinate, to better highlight the large variations between the $R_{\delta c_i}$  values obtained for the different coefficients.

Several observations are worth drawing from the results of Figs.~\ref{fig:spider_nlo_lin_glob} and~\ref{fig:spider_nlo_quad_glob}.
Considering first the linear EFT fits, 
one observes that the projected HL-LHC observables are expected to improve the precision in the determination of the considered Wilson coefficients by an amount which ranges between around 20\% and a factor 3, depending on the specific operator, in the global marginalised fits. 
For instance, we find values of $R_{\delta c_i}\simeq 0.3$ for the triple gauge coupling $c_{WWW}$ and of $R_{\delta c_i}\simeq 0.4$ for the charm, bottom, and tau Yukawa couplings $c_{b\varphi}$, $c_{c\varphi}$, and $c_{\tau \varphi}$.\footnote{We note that our HL-LHC projections do not include measurements directly sensitive to the $h\to c\bar{c}$ decay.} 
For the two-light-two-heavy operators bounds, driven by $t\bar{t}$ distributions, $R_{\delta c_i}$ ranges between 0.8 and 0.5 hence representing up to a factor two of improvement. 
For the operators driven by top quark production, our estimate of the impact of the HL-LHC data can be compared to the results of~\cite{deBlas:2022ofj,Durieux:2022cvf}. 
Their analysis includes dedicated HL-LHC observables which especially help for the two-light-two-heavy operators, resulting in tighter bounds as compared to our fit by around a factor two for some of these coefficients.
%
\begin{figure}[t]
    \centering
    \includegraphics[width=0.6\linewidth]{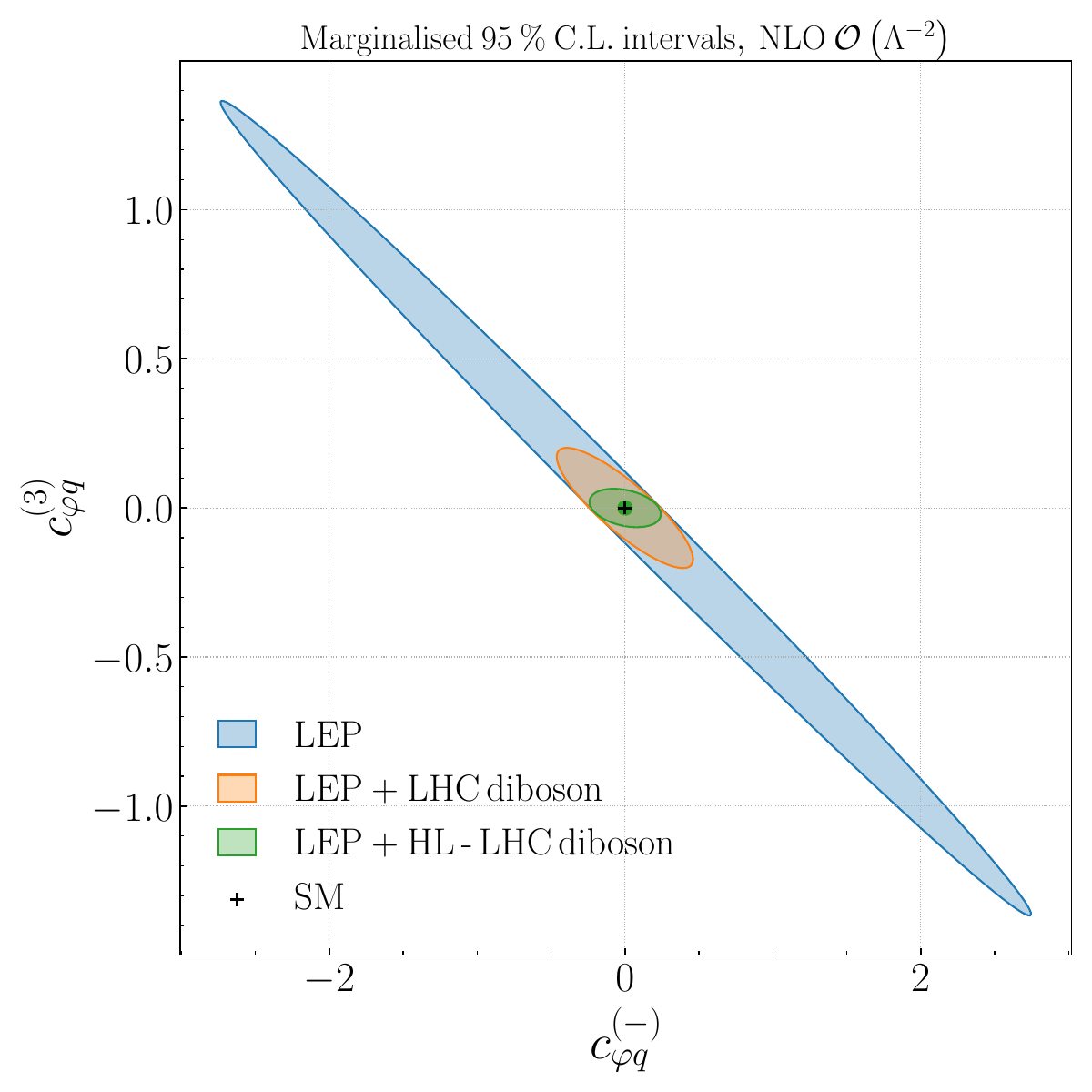}
    \caption{The $95\%$ CI marginalised intervals in the $\lp c_{\varphi q}^{(3)}, c_{\varphi q}^{(-)}\rp$ plane from linear EFT fits to different datasets. 
    We compare the result of a LEP-only fit (blue) with those adding either the LHC Run II diboson data (orange) or the HL-LHC diboson projections (green).
    The three fits are carried out with Level-0 pseudo-data, see App.~\ref{app:hl_lhc_projections}.}
    \label{fig:diboson_effect}
\end{figure}

Additionally, HL-LHC measurements can improve the bounds imposed by EWPOs along directions that are a linear combination of individual coefficients, as illustrated by the $\lp c_{\varphi q}^{(3)}, c_{\varphi q}^{(-)}\rp$ analysis of Fig.~\ref{fig:diboson_effect} where we show the impact of diboson production on  the $95\%$ CI intervals in the  $\lp c_{\varphi q}^{(3)}, c_{\varphi q}^{(-)}\rp$ plane. 
We compare the marginalized bounds from a global linear LEP-only fit with those resulting from combining LEP with either LHC Run II or the HL-LHC diboson data. 
For consistency, the three fits are carried out with Level-0 pseudo-data.
LEP data results in a quasi-flat direction in this plane, which is then well constrained by diboson data at LHC Run II (and subsequently at the HL-LHC), confirming the long-predicted complementarity between LEP and LHC diboson measurements~\cite{Falkowski:2015jaa, Butter:2016cvz, Alioli:2017nzr,Franceschini:2017xkh,Banerjee:2018bio,Grojean:2018dqj}.

The broad reach of the HL-LHC program is illustrated by the fact that essentially all operators considered have associated tighter bounds even in the conservative analysis we perform.  
The comparison between the linear EFT marginalised and individual bounds displayed in Fig.~\ref{fig:spider_nlo_lin_glob} indicates that in the one-parameter fits the sensitivity is typically much better than in the global fit, in some cases by more than an order of magnitude, for example for the $c_{\varphi W B}$, $c_{\varphi B}$,
 $c_{\varphi W}$, and $c_{tZ}$ coefficients.
The exception of this trend are coefficients which are determined by specific subsets of measurements which do not affect other degrees of freedom, such as $c_{WWW}$, constrained from diboson data, and $c_{\tau\varphi}$, constrained only from $h \to \tau\tau$ decays.
This comparison between individual fits to the {\sc\small SMEFiT3.0} and HL-LHC datasets also highlights which operators are constrained mostly by the EWPOs, namely the two-light-fermion operators, the four-lepton operator $c_{\ell\ell}$, and the purely bosonic operator $c_{\varphi \square}$.
For these coefficients, the improvements found in the global marginalised HL-LHC fit arise from indirect improvements in correlated coefficients.

Fig.~\ref{fig:spider_nlo_quad_glob} presents the same comparison as that in Fig.~\ref{fig:spider_nlo_lin_glob} now with the quadratic EFT corrections accounted for, and including also the results for the four-heavy four-fermion operators. 
Quadratic fits break degeneracies and correlations present in the linear fit. 
Hence, some operators not well probed at (HL-)LHC, such as the two-lepton ones, show an $R_{\delta c_i}$ closer to $1$ than in the linear case.
Likewise, for these operators $R_{\delta c_i}$ is unchanged in the individual fits before and after the inclusion of the HL-LHC projections. 

\begin{figure}[t]
    \centering
    \includegraphics[width=.99\linewidth]{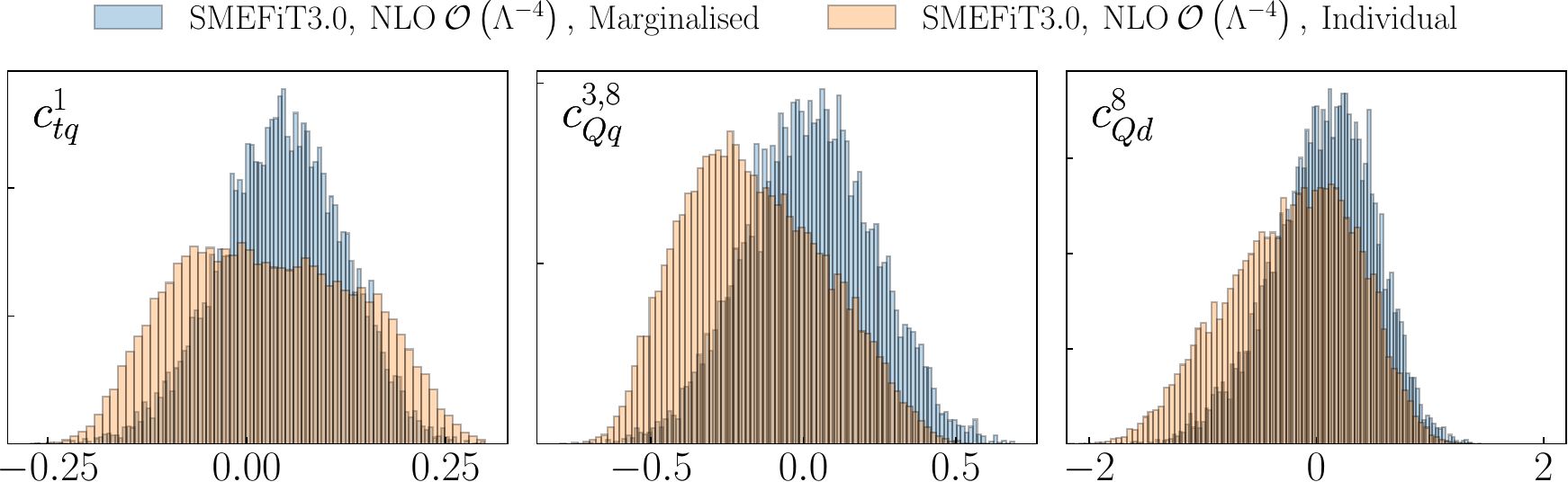}
    \caption{Posterior distributions for three operators which in the quadratic EFT analysis display looser individual (one-parameter) bounds in comparison with corresponding marginalised bounds.
    We show the two-light-two-heavy operators $c_{tq}^1$, $c_{Qq}^{3,8}$, and $c_{Qd}^8$.
    The two fits considered are based on the same global dataset.
    }
    \label{fig:post-ind-glob}
\end{figure}

%
A feature of Fig.~\ref{fig:spider_nlo_quad_glob} is that for some operators the individual bounds are looser than the marginalised ones, albeit by a moderate amount (up to 30\%).
This is the case for most of the two-light-two-heavy operators, and visible both with the {\sc\small SMEFiT3.0} dataset and for the fits with HL-LHC pseudo-data. 
To investigate the origin of this feature,  Fig.~\ref{fig:post-ind-glob} shows posterior distributions for three operators (the two-light-two-heavy operators $c_{tq}^1$, $c_{Qq}^{3,8}$, and $c_{Qd}^8$) which in the quadratic EFT analysis display looser individual bounds in comparison with corresponding marginalised bounds.
For these operators, the marginalised fits lead to narrower posterior distributions, explaining the observed more stringent constraints. This may happen whenever marginalising over the Wilson coefficients leads to additional probability close to the central value of the individual fit, making the posterior more localised.
Finally, we note that in scenarios relevant to the matching to UV models, which involve a subset of EFT operators, the relevant constraints would be in between the global and the individual bounds shown in Fig.~\ref{fig:spider_nlo_quad_glob}.

Overall, our analysis indicates that, in the context of a global SMEFT fit, the extrapolation of Run II measurements to the HL-LHC results into broadly improved bounds, ranging between 20\% and a factor 3 better depending on the specific coefficient.
Qualitatively similar improvements arising from HL-LHC constraints  will be observed once matching to UV models in Sect.~\ref{subsec:uvmodels}.
We note again that the HL-LHC constraints derived here are conservative, as they may be significantly improved through optimised analyses, exploiting features not accessible with the Run II dataset.

\section{The impact  of future $e^+e^-$ colliders on the SMEFT}
\label{sec:results}

We now present the quantitative assessment of the constraints on the SMEFT coefficients provided by measurements to be carried out at the two proposed high-energy $e^+e^-$ circular colliders, FCC-ee and CEPC.
The baseline for these projections is the global SMEFT analysis augmented with the dedicated HL-LHC projections from Sect.~\ref{sec:impact_hl_lhc_data}.
Here first of all we describe the FCC-ee and CEPC observables and running scenarios considered, for which we follow the recent Snowmass study~\cite{deBlas:2022ofj} with minor modifications.
Then we present results at the level of SMEFT coefficients, highlighting the correlation between LHC- and $e^+e^-$-driven constraints.

\subsection{Observables and running scenarios}
\label{sec:future_projections}

Several recent studies~\cite{deBlas:2019rxi,DeBlas:2019qco,deBlas:2022ofj,Durieux:2022cvf} have assessed the 
physics potential of the various
proposed leptonic colliders, including FCC-ee, ILC, CLIC, CEPC, and a muon collider, 
in terms of global fits to SMEFT coefficients and in some cases also matched to UV-complete models.
Here we describe the projections for FCC-ee and CEPC measurements that will be used to constrain the SMEFT parameter space.
We focus on these two colliders as 
representative examples of 
possible new leptonic colliders, though the same strategy can be straightforwardly applied to any other future facility.

The Future Circular Collider~\cite{Benedikt:2020ejr,FCC:2018byv} in its electron-positron mode (FCC-ee)~\cite{FCC:2018evy,Bernardi:2022hny}, originally known as TLEP~\cite{TLEPDesignStudyWorkingGroup:2013myl},
is a proposed electron-positron collider operating in a tunnel of approximately 90 km of circumference in the CERN site and based on well-established accelerator technologies similar to those of LEP. 
Running at several center-of-mass energies  is envisaged, starting from the $Z$-pole all the way up to $\sqrt{s}=365$~GeV, above the top-quark pair production threshold. 
Possible additional runs at $\sqrt{s}=125$ GeV (Higgs pole) and for $\sqrt{s}< m_Z$  (for QCD studies) are under consideration for the FCC-ee.
This circular $e^+e^-$ collider would represent the first stage of a decades-long scientific exploitation of the same tunnel, eventually followed by a $\sqrt{s}\sim 100$ TeV proton-proton collider (FCC-hh).
Here we adopt the same scenarios for the FCC-ee running as in the Snowmass study of~\cite{deBlas:2022ofj} but updated to consider 4 interaction points (IPs), along the lines of the recent midterm feasibility report~\cite{FCCfeasibility}.\footnote{Different running scenarios for the FCC-ee are being discussed, including the integrated luminosity at each $\sqrt{s}$, and therefore the contents of Table~\ref{tab:FCCee_runs} may change as the project matures.} 
We summarise the running scenarios in Table~\ref{tab:FCCee_runs}.
As done in previous studies~\cite{deBlas:2019rxi}, we combine the information associated to the data taken at the runs with $\sqrt{s}=350$~GeV and $365$~GeV and denote this combination as ``$\sqrt{s}=365$~GeV'' in the following.

\begin{table}[t]
    \centering
      \renewcommand{\arraystretch}{1.50}
    \begin{tabular}{|l|l|l|c|}
    \toprule
         \multirow{2}*{Energy $(\sqrt{s})$ } & \multicolumn{2}{c|}{$\mathcal{L}_{\rm int}$ (Run time)} & \multirow{2}*{$\mathcal{L}_{\rm FCC-ee}/\mathcal{L}_{\rm CEPC}$ } 
         \\
        & FCC-ee (4 IPs) & CEPC (2 IPs) & \\
    \hline
  $91$~GeV ($Z$-pole) & $300$~ab$^{-1}$ (4 years)  & $100$~ab$^{-1}$ (2 years) & 3\\
          $161$~GeV ($2\,m_W$) & $20$~ab$^{-1}$ (2 years) & $6$~ab$^{-1}$ (1 year) & 3.3\\
          $240$~GeV & $10$~ab$^{-1}$ (3 years) & $20$~ab$^{-1}$ (10 years) & 0.5\\
          $350$~GeV & $0.4$~ab$^{-1}$ (1 year) & $0.2$~ab$^{-1}$  & 2 \\
          $365$~GeV ($2\,m_t$) & $3$~ab$^{-1}$ (4 years) & $1$~ab$^{-1}$ (5 years) & 3\\
         \bottomrule
    \end{tabular}
    \caption{The running scenarios considered in our analysis for the FCC-ee and the CEPC, following~\cite{deBlas:2022ofj,Bernardi:2022hny,CEPCPhysicsStudyGroup:2022uwl} and the mid-term FCC feasibility report~\cite{FCCfeasibility}.
    Our projections assume 4 interaction points (IPs) for the FCC-ee and 2 for the CEPC.
    For each center of mass energy $\sqrt{s}$,
    we indicate the expected  luminosity as well as the number
    of years in which this luminosity will be collected.
    The last column displays the ratio between the expected integrated luminosities at the FCC-ee and the CEPC.
    When presenting our results, we combine the information associated to the data taken at the runs with $\sqrt{s}=350$~GeV and $365$~GeV and denote this combination as ``$\sqrt{s}=365$~GeV''.
    \label{tab:FCCee_runs}}
\end{table}

The Circular Electron Positron Collider (CEPC)~\cite{An:2018dwb} is a proposed electron-positron collider to be built and operated in China.
The current plan envisages a collider tunnel of around 100 km, and it would operate in stages at different center of mass energies, with a maximum of $\sqrt{s}=365$ GeV above  the top-quark pair production threshold.
The current baseline design assumes two interaction points for the CEPC.
In the same manner as for the FCC-ee,  Table~\ref{tab:FCCee_runs} indicates the expected integrated luminosity (and number of years required to achieve it) in the current running scenarios for each value of the center-of-mass energy $\sqrt{s}$.
The main differences between the projected statistical uncertainties for the FCC-ee and CEPC arise from the different data-taking plans as well as the different number of IPs. 
For instance, CEPC plans a longer running period at $\sqrt{s}=240$ GeV, which would lead to a reduction of statistical errors as compared to the FCC-ee observables corresponding to the same center-of-mass energy.

In the last column of Table~\ref{tab:FCCee_runs} we display the
ratio between the integrated luminosities at the FCC-ee and the CEPC, $\mathcal{L}_{\rm FCC-ee}/\mathcal{L}_{\rm CEPC}$, for each of the data-taking periods at a common center of mass energy.
The FCC-ee is expected to accumulate a luminosity 3 times larger than the CEPC for the runs at $\sqrt{s}=91$ GeV, 161 GeV, and 365 GeV, while for $\sqrt{s}=240$ GeV it would accumulate half of the CEPC luminosity, given that the latter is planned to run for 10 years as opposed to the 3 years of the FCC-ee.

In our analysis, we consider five different classes of observables that are accessible at high-energy circular electron-positron colliders such as the FCC-ee and the CEPC.
These are the EWPOs at the $Z$-pole; light fermion (up to $b$ quarks and $\tau$ leptons) pair production; Higgs boson production in both the $hZ$ and $h\nu\nu$ channels; gauge boson pair production;  and top quark pair production.
Diboson  ($W^+W^-$) production becomes available at $\sqrt{s}=161$ GeV ($WW$ threshold), Higgs production opens up at $\sqrt{s}=240$ GeV, and top quark pair production is accessible starting from $\sqrt{s}=350$ GeV, above the $t\bar{t}$ threshold.

Among these processes, the $Z$-pole EWPOs, light fermion-pair, $W^+W^-$, and Higgs production data are included at the level of inclusive cross-sections, accounting also for the corresponding branching fractions.
The complete list of observables considered, together with the projected experimental uncertainties entering the fit, are collected in App.~\ref{app:observables}.
For diboson and top quark pair production, we consider also unbinned normalised measurements within the optimal observables approach, described in App.~\ref{app:optimal}.
 We  briefly review below these groups of processes.

\paragraph{EWPOs at the $Z$-pole.}
The $Z$-pole electroweak precision observables that would be measured at the FCC-ee and CEPC coincide with those already measured by LEP and SLD and summarised in Table~\ref{tab:ew-datasets}.
The main difference is the greatly improved precision that will be achieved at future electron-positron colliders, due to the increased luminosity and the expected reduction of systematic uncertainties.
Specifically, here we include projections for the QED coupling constant at the $Z$-pole, $\alpha(m_Z)$; the decay widths of the
$W$ and $Z$ bosons, $\Gamma_W$
and $\Gamma_Z$; the asymmetry between vector
and axial couplings $A_f$
for $f=e,\mu,\tau,c,b$;
the total cross-section
for $e^+e^-\to {\rm hadrons}$,
$\sigma_{\rm had}$; 
and the partial decay widths ratio to the total hadronic width $R_f = \Gamma_f/\Gamma_{\mathrm{had}}$
for $f=b,c$ and $R_\ell=\Gamma_\ell/\Gamma_{\mathrm{had}}$ for $\ell=e,\mu,\tau$.
The projected experimental sensitivities to each of these EWPOs at the FCC-ee and CEPC  are collected in Table~\ref{tab:FCCee_EWPOs}.

\paragraph{Light fermion pair production above the $Z$-pole.}
The light fermion pair production measurements considered here, $e^+e^- \to f\bar{f}$, consist of both the total cross sections, $\sigma_{\rm tot}(f\bar{f})$, and the corresponding forward-backward asymmetries, $A^f_{\rm FB}$, with $f=e,\mu,\tau,c,b$, defined in Eq.~(\ref{eq:Rbc}).
The absolute statistical uncertainties for these observables for measurements at FCC-ee and CEPC  at $\sqrt{s}=240$ and 365 GeV are listed in Table~\ref{tab:FCCee_2fprod}.
As for the rest of projections, the corresponding central values are taken from the SM predictions.
The production of top quark pairs, available at $\sqrt{s}=350$ GeV and $365$ GeV, is discussed separately below. 

\paragraph{Higgs production.}
We consider here Higgs production in the two dominant mechanisms relevant for electron-positron colliders, namely associated production with a $Z$ boson, also known as Higgsstrahlung,
\be
e^+e^- \to Zh \, ,
\ee
and in the vector-fusion mode via $W^+W^-$ fusion,
\be
e^+e^- \to \bar{\nu}_eW^+\nu_e W^- \to 
\bar{\nu}_e \nu_e h \, .
\ee
For $\sqrt{s}=240$ GeV, the total Higgs production cross-section is fully dominated by $Zh$ production, while for $\sqrt{s}=365$ GeV the VBF contribution reaches up to 25\% of the total cross-section.
Higgs production via the $ZZ$ fusion channel, $e^+e^- \to e^+e^- ZZ \to e^+e^-h$, is suppressed by a factor 10 in comparison with $WW$ fusion (for both values of $\sqrt{s}$) and is therefore neglected in this analysis.

These two production modes are included in the fit for all the decay modes that become accessible at electron-positron colliders: $c\bar{c}$, $b\bar{b}$, $gg$, $\mu^+\mu^-$, $\tau^+\tau^-$, $ZZ$, $WW$,  $\gamma\gamma$, and $\gamma Z$.
For $\sqrt{s}=240$ GeV, only the dominant decay channel to $b\bar{b}$ is accessible in the vector-boson fusion mode.
We note that possible Higgs decays into invisible final states are also constrained at $e^+e^-$ colliders, a unique feature possible due to the fact that the initial-state energy of the collision $\sqrt{s}$ is precisely known, by means of the direct measurement of the $\sigma_{Zh}$ cross-section via the $Z$-tagged recoil method. 
However, in the present analysis, given that we assume no invisible BSM decays of the Higgs boson, such a direct measurement of the $\sigma_{Zh}$ cross-section does not provide any additional constraints in the EFT parameter space.

Projections for Higgs production measurements are included at the level of inclusive cross-sections times branching ratio, $\sigma\times \text{BR}_X$ (signal strengths), separately for 
$e^+e^- \to Zh$ and $e^+e^- \to 
\bar{\nu}_e \nu_e h$.
Projections for the total inclusive $\sigma_{Zh}$ cross-section are also considered.
The information from differential distributions of the Higgsstrahlung process could in principle also be included, however, its impact on the SMEFT fit is limited~\cite{Durieux:2017rsg,deBlas:2019rxi} and hence we neglect them. 
The expected relative experimental precision for these  Higgs production and decay signal strengths measurements at the FCC-ee and CEPC for $\sqrt{s}=240$ and 365 GeV is summarised in Table~\ref{tab:FCCee_Higgs}. 

\paragraph{Gauge boson pair production.}
We consider weak boson pair production, specifically in the $e^+e^-\to W^+W^-$ final state.
Diboson production in $e^+e^-$ collisions, already measured at LEP, enables the study of the electroweak gauge structure via the search for anomalous triple gauge couplings (aTGCs). 
We include the information from this process by means of a two-fold procedure, considering separately the overall signal strengths, namely the fiducial cross-section for $W^+W^-$ production times the $W$ leptonic and hadronic branching ratios listed in  Table~\ref{tab:FCCee_WW}, and the information contained in the shape of the differential distributions.
The latter is accounted for via the optimal observables strategy described in App.~\ref{app:optimal}.
Projections corresponding to the three center-of-mass energies relevant for diboson production, $\sqrt{s}=161$ GeV, 240 GeV, and 365 GeV, are included.

In our analysis, we assume that the $W$ boson does not have any exotic (invisible) decay modes and thus we impose that the separate leptonic and hadronic branching fractions add up to unity, 
\begin{equation}
    \text{BR}_{W\to e\nu} + \text{BR}_{W\to \mu\nu} + \text{BR}_{W\to \tau\nu} + \text{BR}_{W\to q\bar{q}}=1 \, .
\label{eq:no_exotic_Wdecay}
\end{equation}
This requirement allows one to determine the expected precision for the measurement of the fiducial cross-section and all the branching ratios from the measurement of the relevant $WW$ decay channels, see also Table~6 in~\cite{deBlas:2019rxi}.

\paragraph{Top quark pair production.}
The top quark pair production process, $e^+e^- \to t\bar{t}$, becomes accessible above $\sqrt{s}=2m_t\simeq 350$ GeV, the kinematic production threshold.
Measurements of this process provide information on the electroweak couplings of the top quark, complementing existing LHC measurements, and also enable a determination of the top quark mass with excellent ($\lsim 20$ MeV) precision. 
Similarly to the case of gauge boson pair production, here we include this process in the fit in terms of (now absolute) unbinned measurements within the optimal observables framework for the dominant $e^+e^- \to t\bar{t}\to W^+bW^-\bar{b}$ final state, see App.~\ref{app:optimal} for details.
We neglect systematic uncertainties and adopt the same settings on the acceptance, identification, and reconstruction efficiencies as in the Snowmass EFT study of~\cite{Durieux:2022cvf}. 
Since here we do not consider normalised observables, it is not necessary to include separately the signal strengths for $e^+e^- \to t\bar{t}$ production and decay as done for the $W^+W^-$ case.

\paragraph{Implementation.}
A new module has been added to {\sc\small SMEFiT} which enables the integration of external, user-provided likelihoods into the figure of merit entering the global fit.
We have verified, in the specific case of $W^+W^-$ production, that using such an external likelihood is equivalent to the baseline implementation of datasets in {\sc\small SMEFiT} and based on separate data tables for the theory calculations and for the experimental measurements.
This new functionality is used here to include the constraints from the optimal observables in $t\bar{t}$ production by reusing the results derived in~\cite{Durieux:2018tev,Durieux:2022cvf,deBlas:2022ofj} with adjustments whenever required.

\subsection{Impact on the SMEFT coefficients}
\label{subsec:coeffs}

\begin{figure}[t]
    \centering
    \includegraphics[width=0.85\linewidth]{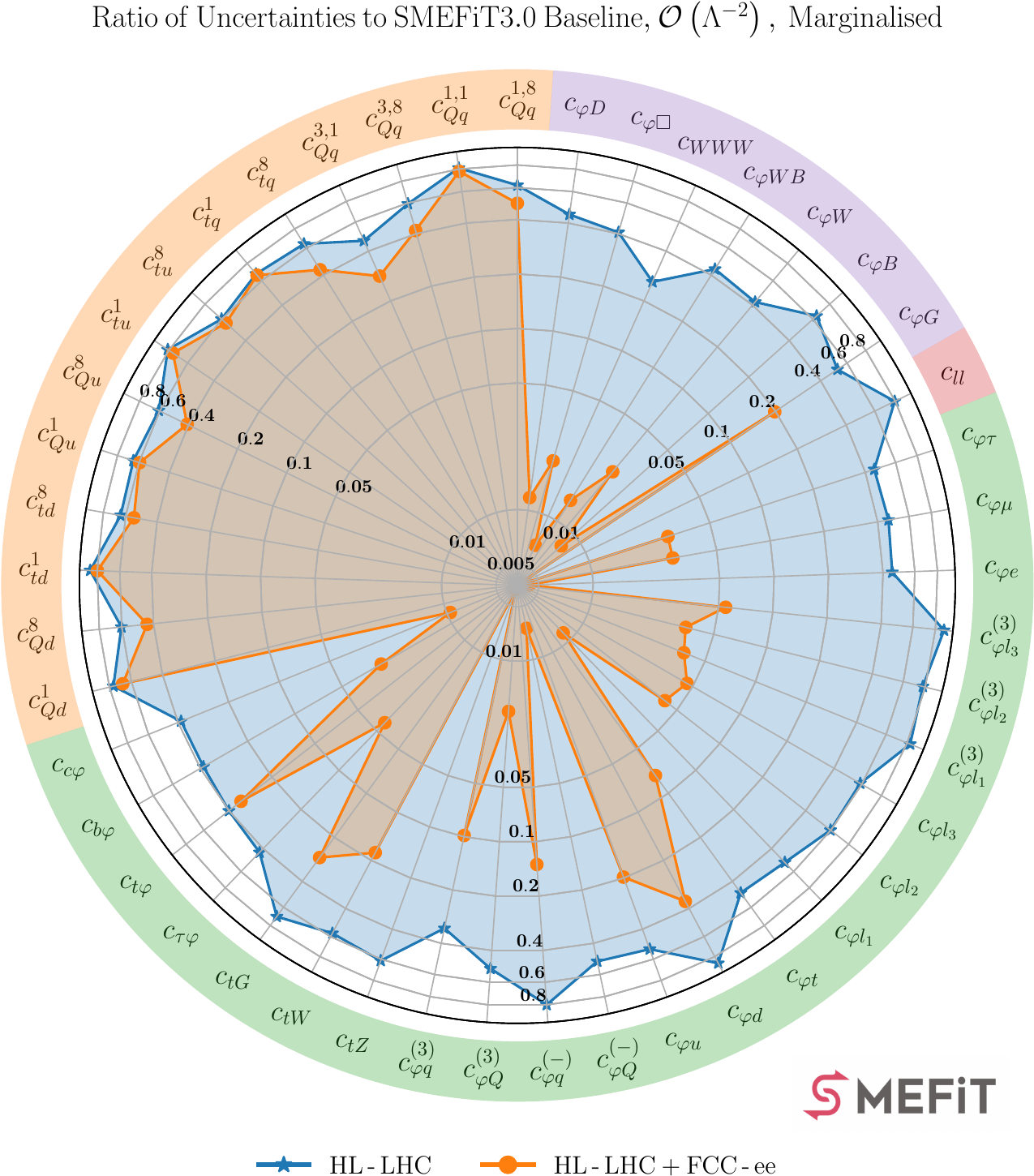}
    \caption{Same as Fig.~\ref{fig:spider_nlo_lin_glob} for the marginalised bounds on the SMEFT operators from a global $\mathcal{O}\lp \Lambda^{-2}\rp$  analysis, displaying the ratio of uncertainties to the 
    {\sc\small SMEFiT3.0} baseline of fits which include first the HL-LHC projections and subsequently both the HL-LHC and the FCC-ee observables. 
    All fits shown here are carried out based on Level-1 pseudo-data.
    }
    \label{fig:spider_fcc_nlo_lin_glob}
\end{figure}

Next we quantify the impact on the EFT coefficients of measurements from the FCC-ee and CEPC when added on top of the baseline fit extended with the HL-LHC projections discussed in Sect.~\ref{sec:impact_hl_lhc_data}.
Since our projections reproduce the assumed SM theory for the fitted observables by construction, one is interested only in the reduction of the length of the 95\% CI interval for the relevant operators, as quantified by the ratio $R_{\delta c_i}$ in Eq.~(\ref{eq:RatioC}), and hence we do not display the fit central values.
We focus first on the FCC-ee, and then compare with the CEPC results.

\begin{figure}[t]
    \centering
    \includegraphics[width=0.85\linewidth]{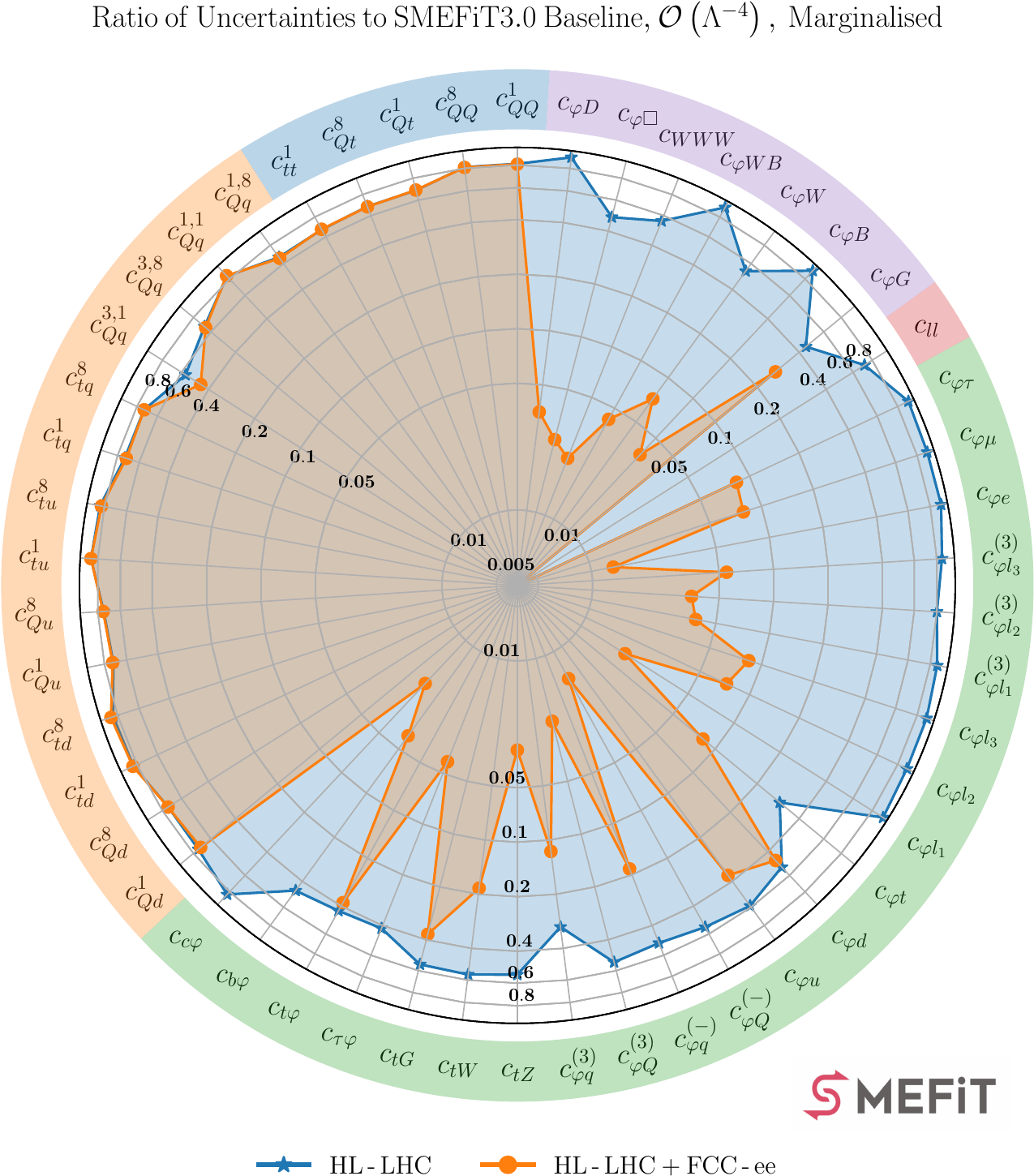}
    \caption{Same as Fig.~\ref{fig:spider_fcc_nlo_lin_glob} in the 
    case of the fits with EFT cross-sections including both the linear, $\mathcal{O}\lp \Lambda^{-2}\rp$, 
    and the quadratic, $\mathcal{O}\lp \Lambda^{-4}\rp$, corrections associated to the dimension-six operators considered in the analysis.}
    \label{fig:spider_fcc_nlo_quad_glob}
\end{figure}

Fig.~\ref{fig:spider_fcc_nlo_lin_glob} (\ref{fig:spider_fcc_nlo_quad_glob}) displays, in the same format as that of Fig.~\ref{fig:spider_nlo_lin_glob}, the marginalised bounds on the SMEFT operators from a global linear (quadratic) analysis, displaying the ratio of uncertainties to the  {\sc\small SMEFiT3.0} baseline of fits which include, first, the HL-LHC projections, and subsequently, both the HL-LHC and the FCC-ee observables, see Table~\ref{table:fit_settings}.
Inspection of Fig.~\ref{fig:spider_fcc_nlo_lin_glob} demonstrates the substantial impact that FCC-ee measurements would provide on the SMEFT parameter space in comparison with a post-HL-LHC baseline.
Indeed, the FCC-ee would constrain a wide range of directions in the EFT coefficients, except for the four-quark operators.
For the latter only moderate indirect improvements are expected, which go away in the quadratic fits shown in Fig.~\ref{fig:spider_fcc_nlo_quad_glob}. This is due to the fact that no FCC-ee observables included in the fit are sensitive to four-quark operators, and thus any impact of FCC-ee on their constraints arises through marginalisation. 

From Fig.~\ref{fig:spider_fcc_nlo_lin_glob} one also observes how the bounds on some of the purely bosonic, four-lepton, and two-fermion operators achieved at the FCC-ee could improve on the HL-LHC ones by almost two orders of magnitude. 
For instance, the 95\% CI interval for the coefficient $c_{WWW}$, which modifies the triple gauge boson interactions, is reduced by a factor $R_{\delta c_i} \simeq 0.3$ at the end of HL-LHC and then down to $R_{\delta c_i}=0.008$ at the FCC-ee, corresponding to a relative improvement on the bound by a factor around $40$. 
Likewise, our analysis finds values of $R_{\delta c_i} \simeq 0.4,\,0.4$ and $0.6$ at the end of the HL-LHC for the coefficients $c_{\varphi e}$, $c_{c\varphi}$,
 and  $c_{\varphi B}$
respectively, which subsequently go down to 
$R_{\delta c_i} \simeq 0.005,\, 0.01$ and $0.008$ upon the inclusion of the FCC-ee pseudo-data.
This translates into relative improvements by factors of around $80$, $40$, and $70$ for each EFT coefficient, respectively. 
While these are only representative examples, they highlight how precision measurements at FCC-ee will provide stringent constraints on the SMEFT parameter space, markedly improving on the limits achievable at the HL-LHC.

In Figs.~\ref{fig:spider_fcc_nlo_lin_glob} and~\ref{fig:spider_fcc_nlo_quad_glob}, the impact of the FCC-ee measurements on the EFT coefficients is presented in terms of the relative improvement with respect the {\sc\small SMEFiT3.0} baseline.
In order to compare with previous related studies, it is useful to also provide the absolute magnitude of the resulting bounds, in addition to their relative improvement as compared to the baseline.
With this motivation, Table~\ref{tab:fcc_all_bounds} indicates the 95\% CI upper bounds on the EFT coefficients obtained from the fits including both the HL-LHC and the FCC-ee projections, assuming $\Lambda=1$ TeV.
In analogy with Table~\ref{tab:smefit30_all_bounds}, we
provide these bounds for linear and quadratic EFT fits and both at the individual and marginalised level.
The impact of the FCC-ee is clearly visible specially for the purely bosonic and
two-fermion operators, with most of them constrained to be $|c|\lsim 0.1$ (for $\Lambda=1$ TeV) in the global marginalised fit, and several of them at the $|c|\lsim 10^{-2}$ level or better.
We demonstrate in Sect.~\ref{subsec:uvmodels} how these constraints on the Wilson coefficients translate into the mass reach for UV-complete models.
     
\begin{figure}[htbp]
    \centering
    \includegraphics[width=0.90\linewidth]{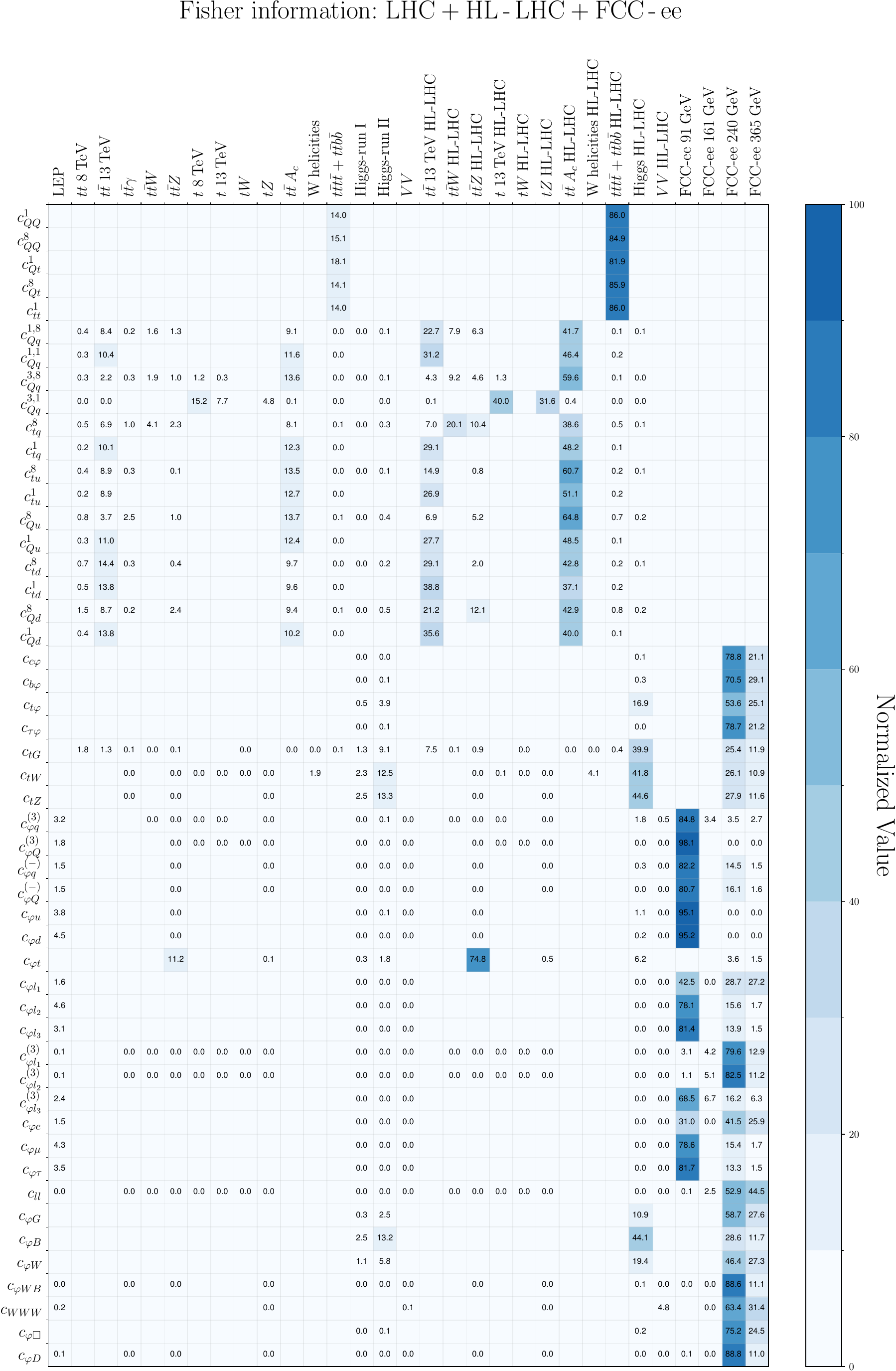}
    \caption{Diagonal entries of the Fisher information matrix evaluated at $\mathcal{O}\lp \Lambda^{-2}\rp$ in the EFT expansion, Eq.~(\ref{eq:fisherinformation2}), for the complete dataset (LEP EWPOs + LHC$_{\rm Run1}$ + LHC$_{\rm Run2}$ + HL-LHC + FCC-ee) considered in this work.
    Each row is normalised to 100. 
    The LHC$_{\rm Run2}$ and HL-LHC datasets are displayed separately.
    For the FCC-ee observables, we evaluate the Fisher matrix individually for each of the relevant $\sqrt{s}$ values. 
    Empty entries indicate lack of direct sensitivity, while ``0.0'' entries indicate a non-zero sensitivity which is less than 0.05\% in magnitude.
    }
    \label{fig:fisher_fcc_lin}
\end{figure}

\begin{table}[htbp]
 \centering
  \scriptsize
   \renewcommand{\arraystretch}{1.24}
   \begin{tabular}{l|C{0.8cm}|C{2.3cm}|C{2.3cm}|C{4.0cm}|C{4.0cm}}
     \multirow{2}{*}{Class}   &  \multirow{2}{*}{DoF}
     &  \multicolumn{2}{c|}{ 95\% CL bounds, $\mathcal{O}\lp \Lambda^{-2}\rp$} &
     \multicolumn{2}{c}{95\% CL bounds, $\mathcal{O}\lp \Lambda^{-4}\rp$,} \\ 
 &  & Individual & Marginalised &  Individual & Marginalised  \\ \toprule
 \multirow{23}{*}{2FB}
 & $c_{c \varphi}$& 2.82$\cdot 10^{-3}$& 3.69$\cdot 10^{-3}$& 2.79$\cdot 10^{-3}$& 3.68$\cdot 10^{-3}$ \\ \cline{2-6}
 & $c_{b \varphi}$& 1.9$\cdot 10^{-3}$& 3.61$\cdot 10^{-3}$& 1.87$\cdot 10^{-3}$& 3.4$\cdot 10^{-3}$ \\ \cline{2-6}
 & $c_{t \varphi}$& 2.96$\cdot 10^{-1}$& 2.18& 2.88$\cdot 10^{-1}$& 2.19 \\ \cline{2-6}
 & $c_{\tau \varphi}$& 1.92$\cdot 10^{-3}$& 2.85$\cdot 10^{-3}$& 1.93$\cdot 10^{-3}$& 2.79$\cdot 10^{-3}$ \\ \cline{2-6}
 & $c_{tG}$& 2.38$\cdot 10^{-2}$& 6.89$\cdot 10^{-2}$& 2.45$\cdot 10^{-2}$& 6.52$\cdot 10^{-2}$ \\ \cline{2-6}
 & $c_{tW}$& 3.83$\cdot 10^{-3}$& 5.82$\cdot 10^{-2}$& 3.74$\cdot 10^{-3}$& 6.02$\cdot 10^{-2}$ \\ \cline{2-6}
 & $c_{tZ}$& 4.79$\cdot 10^{-3}$& 6.92$\cdot 10^{-2}$& 4.7$\cdot 10^{-3}$& 7.19$\cdot 10^{-2}$ \\ \cline{2-6}
 & $c_{\varphi q}^{(3)}$& 4.9$\cdot 10^{-3}$& 1.39$\cdot 10^{-2}$& 4.84$\cdot 10^{-3}$& 1.29$\cdot 10^{-2}$ \\ \cline{2-6}
 & $c_{\varphi Q}^{(3)}$& 5.28$\cdot 10^{-3}$& 3.5$\cdot 10^{-2}$& 5.32$\cdot 10^{-3}$& 1.92$\cdot 10^{-2}$ \\ \cline{2-6}
 & $c_{\varphi q}^{(-)}$& 2.89$\cdot 10^{-2}$& 6.03$\cdot 10^{-2}$& 2.88$\cdot 10^{-2}$& 4.06$\cdot 10^{-2}$ \\ \cline{2-6}
 & $c_{\varphi Q}^{(-)}$& 9.37$\cdot 10^{-3}$& 2.46$\cdot 10^{-2}$& 9.69$\cdot 10^{-3}$& 2.44$\cdot 10^{-2}$ \\ \cline{2-6}
 & $c_{\varphi u}$& 2.99$\cdot 10^{-2}$& 1.54$\cdot 10^{-1}$& 2.99$\cdot 10^{-2}$& 8.18$\cdot 10^{-2}$ \\ \cline{2-6}
 & $c_{\varphi d}$& 4.62$\cdot 10^{-2}$& 3.74$\cdot 10^{-1}$& 4.47$\cdot 10^{-2}$& 1.65$\cdot 10^{-1}$ \\ \cline{2-6}
 & $c_{\varphi t}$& 3.72$\cdot 10^{-1}$& 6.65$\cdot 10^{-1}$& 3.76$\cdot 10^{-1}$& 6.55$\cdot 10^{-1}$ \\ \cline{2-6}
 & $c_{\varphi l_1}$& 2.38$\cdot 10^{-3}$& 4.59$\cdot 10^{-3}$& 2.39$\cdot 10^{-3}$& 4.38$\cdot 10^{-3}$ \\ \cline{2-6}
 & $c_{\varphi l_2}$& 1.03$\cdot 10^{-2}$& 2.55$\cdot 10^{-2}$& 1.01$\cdot 10^{-2}$& 2.56$\cdot 10^{-2}$ \\ \cline{2-6}
 & $c_{\varphi l_3}$& 1.05$\cdot 10^{-2}$& 2.79$\cdot 10^{-2}$& 1.06$\cdot 10^{-2}$& 2.8$\cdot 10^{-2}$ \\ \cline{2-6}
 & $c_{\varphi l_1}^{(3)}$& 8.97$\cdot 10^{-4}$& 7.45$\cdot 10^{-3}$& 8.93$\cdot 10^{-4}$& 7.06$\cdot 10^{-3}$ \\ \cline{2-6}
 & $c_{\varphi l_2}^{(3)}$& 9.79$\cdot 10^{-4}$& 8.1$\cdot 10^{-3}$& 9.73$\cdot 10^{-4}$& 7.8$\cdot 10^{-3}$ \\ \cline{2-6}
 & $c_{\varphi l_3}^{(3)}$& 8.59$\cdot 10^{-3}$& 1.97$\cdot 10^{-2}$& 8.72$\cdot 10^{-3}$& 1.86$\cdot 10^{-2}$ \\ \cline{2-6}
 & $c_{\varphi e}$& 2.52$\cdot 10^{-3}$& 4.62$\cdot 10^{-3}$& 2.54$\cdot 10^{-3}$& 4.62$\cdot 10^{-3}$ \\ \cline{2-6}
 & $c_{\varphi \mu}$& 1.18$\cdot 10^{-2}$& 3.0$\cdot 10^{-2}$& 1.24$\cdot 10^{-2}$& 3.02$\cdot 10^{-2}$ \\ \cline{2-6}
 & $c_{\varphi \tau}$& 1.28$\cdot 10^{-2}$& 3.01$\cdot 10^{-2}$& 1.29$\cdot 10^{-2}$& 2.95$\cdot 10^{-2}$ \\ \cline{2-6}
\hline
\multirow{14}{*}{2L2H}
 & $c_{Qq}^{1,8}$& 2.59$\cdot 10^{-1}$& 2.34& 3.06$\cdot 10^{-1}$& 4.69$\cdot 10^{-1}$ \\ \cline{2-6}
 & $c_{Qq}^{1,1}$& 5.81$\cdot 10^{-1}$& 1.16$\cdot 10^{1}$& 3.64$\cdot 10^{-1}$& 2.82$\cdot 10^{-1}$ \\ \cline{2-6}
 & $c_{Qq}^{3,8}$& 7.95$\cdot 10^{-1}$& 3.96& 6.56$\cdot 10^{-1}$& 5.34$\cdot 10^{-1}$ \\ \cline{2-6}
 & $c_{Qq}^{3,1}$& 1.27$\cdot 10^{-1}$& 1.27$\cdot 10^{-1}$& 1.3$\cdot 10^{-1}$& 1.14$\cdot 10^{-1}$ \\ \cline{2-6}
 & $c_{tq}^{8}$& 4.12$\cdot 10^{-1}$& 2.65& 4.55$\cdot 10^{-1}$& 5.14$\cdot 10^{-1}$ \\ \cline{2-6}
 & $c_{tq}^{1}$& 4.83$\cdot 10^{-1}$& 6.87& 2.83$\cdot 10^{-1}$& 2.19$\cdot 10^{-1}$ \\ \cline{2-6}
 & $c_{tu}^{8}$& 4.38$\cdot 10^{-1}$& 6.97& 5.83$\cdot 10^{-1}$& 7.19$\cdot 10^{-1}$ \\ \cline{2-6}
 & $c_{tu}^{1}$& 8.81$\cdot 10^{-1}$& 1.89$\cdot 10^{1}$& 4.57$\cdot 10^{-1}$& 3.57$\cdot 10^{-1}$ \\ \cline{2-6}
 & $c_{Qu}^{8}$& 7.51$\cdot 10^{-1}$& 6.81& 7.44$\cdot 10^{-1}$& 8.21$\cdot 10^{-1}$ \\ \cline{2-6}
 & $c_{Qu}^{1}$& 6.8$\cdot 10^{-1}$& 8.65& 3.5$\cdot 10^{-1}$& 2.68$\cdot 10^{-1}$ \\ \cline{2-6}
 & $c_{td}^{8}$& 9.38$\cdot 10^{-1}$& 1.32$\cdot 10^{1}$& 1.35& 1.03 \\ \cline{2-6}
 & $c_{td}^{1}$& 1.79& 2.94$\cdot 10^{1}$& 6.2$\cdot 10^{-1}$& 4.61$\cdot 10^{-1}$ \\ \cline{2-6}
 & $c_{Qd}^{8}$& 1.43& 9.3& 1.32& 1.27 \\ \cline{2-6}
 & $c_{Qd}^{1}$& 1.49& 2.32$\cdot 10^{1}$& 5.0$\cdot 10^{-1}$& 3.96$\cdot 10^{-1}$ \\ \cline{2-6}
\hline
\multirow{5}{*}{4H}
 & $c_{QQ}^{1}$& 7.71 & \textemdash & 1.94& 6.75 \\ \cline{2-6}
 & $c_{QQ}^{8}$& 2.15$\cdot 10^{1}$ & \textemdash & 5.82& 2.02$\cdot 10^{1}$ \\ \cline{2-6}
 & $c_{Qt}^{1}$& 2.95$\cdot 10^{2}$ & \textemdash & 1.65& 1.36 \\ \cline{2-6}
 & $c_{Qt}^{8}$& 6.74 & \textemdash & 3.44& 2.75 \\ \cline{2-6}
 & $c_{tt}^{1}$& 3.77 & \textemdash & 9.62$\cdot 10^{-1}$& 8.01$\cdot 10^{-1}$ \\ \cline{2-6}
\hline
\multirow{1}{*}{4L}
 & $c_{ll}$& 6.98$\cdot 10^{-4}$& 8.05$\cdot 10^{-4}$& 6.99$\cdot 10^{-4}$& 8.06$\cdot 10^{-4}$ \\ \cline{2-6}
\hline
\multirow{7}{*}{B}
 & $c_{\varphi G}$& 9.83$\cdot 10^{-4}$& 6.22$\cdot 10^{-3}$& 9.63$\cdot 10^{-4}$& 6.34$\cdot 10^{-3}$ \\ \cline{2-6}
 & $c_{\varphi B}$& 1.97$\cdot 10^{-3}$& 6.65$\cdot 10^{-3}$& 1.97$\cdot 10^{-3}$& 6.67$\cdot 10^{-3}$ \\ \cline{2-6}
 & $c_{\varphi W}$& 5.13$\cdot 10^{-3}$& 2.27$\cdot 10^{-2}$& 5.1$\cdot 10^{-3}$& 1.98$\cdot 10^{-2}$ \\ \cline{2-6}
 & $c_{\varphi WB}$& 2.77$\cdot 10^{-4}$& 1.33$\cdot 10^{-2}$& 2.73$\cdot 10^{-4}$& 1.32$\cdot 10^{-2}$ \\ \cline{2-6}
 & $c_{WWW}$& 7.17$\cdot 10^{-3}$& 7.83$\cdot 10^{-3}$& 7.03$\cdot 10^{-3}$& 7.88$\cdot 10^{-3}$ \\ \cline{2-6}
 & $c_{\varphi \Box}$& 3.91$\cdot 10^{-2}$& 7.0$\cdot 10^{-2}$& 4.0$\cdot 10^{-2}$& 6.72$\cdot 10^{-2}$ \\ \cline{2-6}
 & $c_{\varphi D}$& 6.1$\cdot 10^{-4}$& 2.45$\cdot 10^{-2}$& 6.05$\cdot 10^{-4}$& 2.46$\cdot 10^{-2}$ \\ \cline{2-6}
\hline
\end{tabular}
\caption{\small The 95\% CI upper bounds on the
     EFT coefficients obtained from
     the fits including both the HL-LHC and the FCC-ee (Level-1) pseudo-data,
     see also Figs.~\ref{fig:spider_fcc_nlo_lin_glob} and~\ref{fig:spider_fcc_nlo_quad_glob} for the corresponding graphical representation (relative to the baseline bounds), for  $\Lambda=1$ TeV.
     We present results both for linear and quadratic EFT fits and at the individual and marginalised level,
     in analogy with Table~\ref{tab:smefit30_all_bounds} for the baseline {\sc\small SMEFiT3.0} analysis.
     }
\label{tab:fcc_all_bounds}
\end{table}


\paragraph{Fisher information analysis.}
In order to better scrutinise the interplay between the constraints provided by the (HL-)LHC measurements, on the one hand, and by the FCC-ee ones, on the other hand, it is illustrative to evaluate the Fisher information matrix~\cite{Ethier:2021bye,Ellis:2020unq}.
The entries of the Fisher information matrix at $\mathcal{O}\lp \Lambda^{-2}\rp$ in the EFT expansion are given by
 \be
\label{eq:fisherinformation2}
I_{ij} = \sum_{m=1}^{n_{\rm dat}} \frac{\sigma^{\rm (eft)}_{m,i}\sigma^{\rm (eft)}_{m,j}}{\delta_{{\rm exp},m}^2} \, , \qquad i,j=1,\ldots,n_{\rm eft},
\ee
with $\sigma^{\rm (eft)}_{m,i}$ being the linear EFT correction to the SM cross-section for the $m$-th data point associated to the $i$-th Wilson coefficient, and $\delta_{{\rm exp},m}$ indicates the total experimental uncertainty. 
Eq.~(\ref{eq:fisherinformation2}) should be evaluated separately for each dataset or group of processes entering the analysis; the larger the value of the diagonal entry $I_{ii}$, the more impactful this dataset will be for the $i$-th EFT coefficient. 

Fig.~\ref{fig:fisher_fcc_lin} displays the diagonal entries of the linear Fisher
information matrix, Eq.~(\ref{eq:fisherinformation2}), computed over the  complete dataset considered in this work: LEP EWPOs, LHC$_{\rm Run1}$, LHC$_{\rm Run2}$, HL-LHC, and the FCC-ee.
In this plot each row is normalised to 100. 
Note that the LHC Run II datasets and the HL-LHC projections are displayed separately.
For the FCC-ee projections, we evaluate the Fisher matrix individually for each of the relevant $\sqrt{s}$ values.

From the entries of the Fisher information matrix it can be observed how for all LHC processes being considered, the HL-LHC projections display the largest Fisher information, reflecting the expected reduction of statistical and systematic uncertainties. 
The table also confirms that the LEP EWPOs carry a small amount of information once the FCC-ee projections are included in the fit.
Within the global EFT fit, the FCC-ee observables would provide the dominant constraints for all the two-fermion and purely bosonic operators, except for $c_{tG}$, $c_{tW}$, $c_{tZ}$, $c_{\varphi B}$, and $c_{\varphi t}$, where the constraints from HL-LHC processes are still expected to dominate. For instance, in the case of $c_{tZ}$ and $c_{tW}$, the largest sensitivity comes in through the $H\rightarrow \gamma\gamma$ and $H\rightarrow \gamma Z$ decay channels probed by the Higgs HL-LHC projections.
The two-light-two-heavy operators are entirely dominated by HL-LHC, mostly from inclusive $t\bar{t}$ production and by the charge asymmetry measurements $A_C$.

Concerning the impact of the FCC-ee datasets for different center-of-mass energies, Fig.~\ref{fig:fisher_fcc_lin} reveals that the bulk of the constraints on the SMEFT parameter space should be provided by the runs at $\sqrt{s}=91$ GeV and at $\sqrt{s}=240$ GeV, with relevant information arising also from the  $\sqrt{s}=365$ GeV run.
Fig.~\ref{fig:fisher_fcc_lin} highlights the interplay between runs at different values of $\sqrt{s}$ to constrain new physics through the SMEFT interpretation: when their information is combined, the resulting picture is sharper than that of any individual run. 
On the other hand, only moderate information would be provided by the run at the $WW$ threshold ($\sqrt{s}=161$ GeV), and even for the triple gauge operator coefficient $c_{WWW}$ it would be the $Zh$ run that dominates the sensitivity.
Hence, from the viewpoint of EFT analyses, the $\sqrt{s}=161$ GeV run appears to be the less impactful.

The Fisher information matrix shown in Fig.~\ref{fig:fisher_fcc_lin} quantifies the relative sensitivity of various processes to specific operators based on their linear EFT contributions, and corresponds to what one would expect to find in linear individual fits.
However, it cannot capture the complete picture encompassed by the global fit at the marginalised level, and in particular it does not account for the information on the correlations between operators.
Indeed, in the marginalised fits,  these correlations will often modify the picture as compared with expectations based on the Fisher information matrix. 
We also note that Eq.~(\ref{eq:fisherinformation2}) receives additional contributions when evaluated for quadratic EFT fits, which are not necessarily subdominant in comparison with the linear contributions.

\paragraph{Impact of quadratic EFT corrections.}
The analog of Fig.~\ref{fig:spider_fcc_nlo_lin_glob} in the case of EFT fits with quadratic corrections is displayed in Fig.~\ref{fig:spider_fcc_nlo_quad_glob}.
Also here one observes a significant reduction of the bounds on the EFT coefficients upon inclusion of the FCC-ee observables.
The improvements observed for some four-quark operators in the linear fit of Fig.~\ref{fig:spider_fcc_nlo_lin_glob} mostly disappear when including quadratic corrections demonstrating that the improvement of the linear fit constraints arise from indirect  correlations with other degrees of freedom.
Indeed, the quadratic fit result illustrates how the FCC-ee observables do not have any direct sensitivity on the four-quark operators, both for the two-light-two-heavy and for the four-heavy ones.

From Table~\ref{tab:fcc_all_bounds} one notices that the differences between linear and quadratic bounds are in general reduced as compared to the case of the {\sc\small SMEFiT3.0} results collected in Table~\ref{tab:smefit30_all_bounds}. 
This feature is explained by the improved precision of the FCC-ee measurements: since we assume the SM in the pseudo-data,  the best-fit values of the Wilson coefficients move closer to zero with smaller uncertainties, and hence the quadratic terms become less significant. 
We note however that for a subset of operators, such as the two-light-two-heavy ones, which are not constrained by the FCC-ee measurements the discrepancy between linear and quadratic remains large. 

\paragraph{Disentangling the impact of datasets with fixed $\sqrt{s}$.}
As indicated in Table~\ref{tab:FCCee_runs}, the FCC-ee plans to operate sequentially, collecting data at different center-of-mass energies, $\sqrt{s}$, starting at the $Z$-pole and then increasing the energy up to the $t\bar{t}$ threshold.
Plans to define different running scenarios are also being considered, for example directly starting as a Higgs factory with the $\sqrt{s}=240$ GeV run and only later running at the $Z$-pole energy.
It is therefore relevant to disentangle, at the level of the global SMEFT fit, the separate impact of datasets with a given $\sqrt{s}$ value to evaluate the advantages and disadvantages of the proposed running scenarios, see also the Fisher information matrix in Fig.~\ref{fig:fisher_fcc_lin} that applies in the case of individual fits.

\begin{figure}[t]
    \centering
    \includegraphics[width=0.99\linewidth]{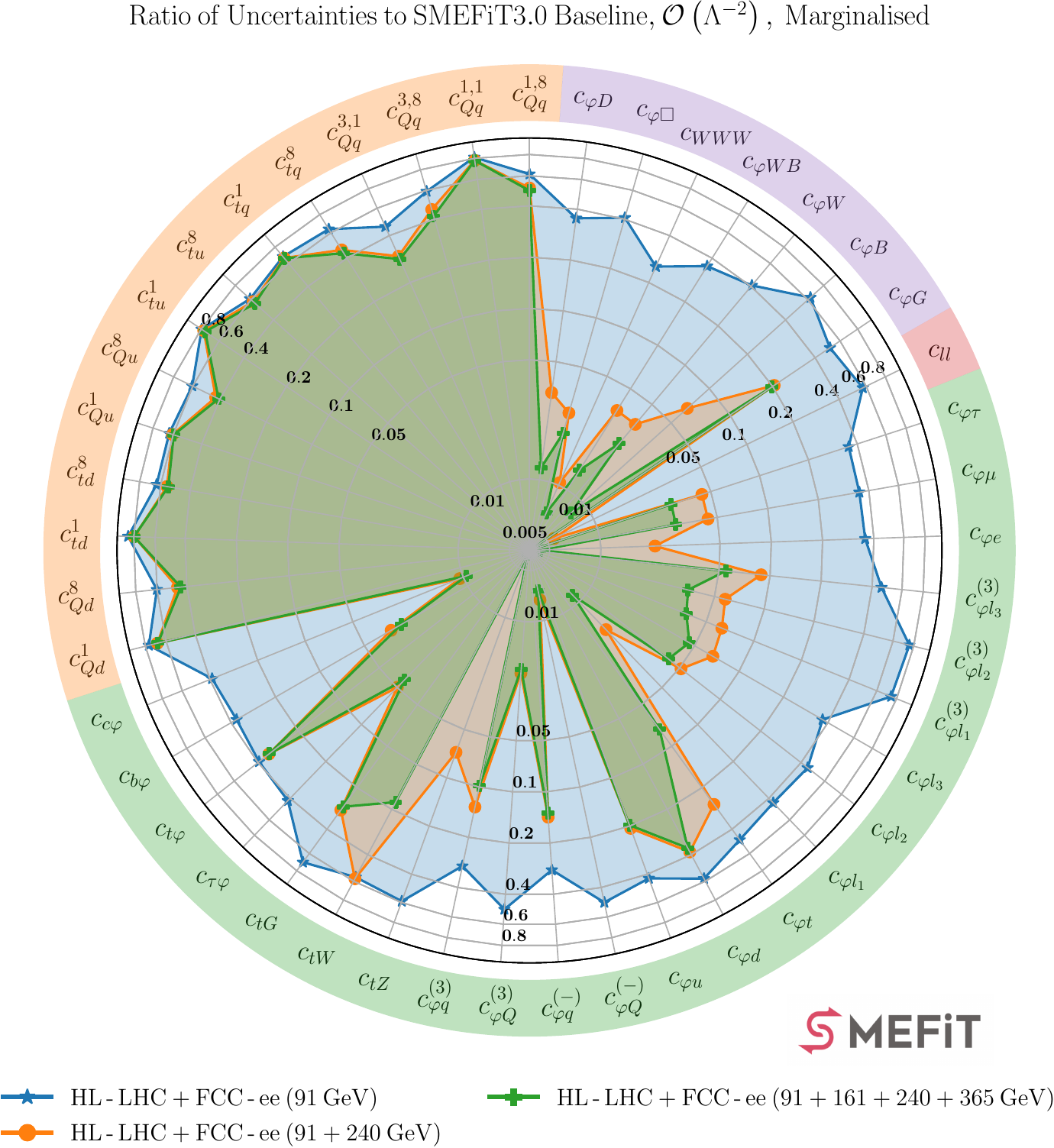}
    \caption{Same as Fig.~\ref{fig:spider_fcc_nlo_lin_glob}, now comparing the sequential impact of the separate $\sqrt{s}$ runs at the FCC-ee with respect to the baseline fit.
    We display the effects of adding the projected FCC-ee dataset at, first, $\sqrt{s}=91$ GeV (blue), followed by adding $\sqrt{s}=240$ GeV (orange) and finally adding both $\sqrt{s}=161$ and 365 GeV (green), which completes the FCC-ee dataset listed in Table~\ref{tab:FCCee_runs}.
    }
\label{fig:spider_fcc_energy_variations}
\end{figure}

Fig.~\ref{fig:spider_fcc_energy_variations} illustrates the sequential impact of the datasets collected at different values of $\sqrt{s}$ at the FCC-ee.
First we show the values of the ratio $R_{\delta c_i}$ when only the $Z$-pole EWPOs at $\sqrt{s}=91$ GeV are included in the fit, then when also the Higgs factory dataset from the $\sqrt{s}=240$ GeV is accounted for, and finally for the full FCC-ee dataset, which includes also the $WW$ run at 161 GeV and the $t\bar{t}$ run at  365 GeV. 
Fig.~\ref{fig:spider_fcc_energy_variations} indicates that the largest impact is obtained when the Higgs, diboson, and fermion-pair production data collected at 240 GeV are included in the fit together with the $Z$-pole run.
We also observe how the measurements from the $\sqrt{s}=161$ GeV and 365 GeV runs are necessary to achieve the ultimate constraining potential of the FCC-ee in the EFT parameter space, with several operators experiencing a marked improvement of the associated bounds. 
This breakdown demonstrates the interplay between the information provided by the FCC-ee runs at the various proposed center-of-mass energies in a global SMEFT fit. We have verified that the equivalent of Fig.~\ref{fig:spider_fcc_energy_variations} in the case of individual fits is consistent with the Fisher information matrix displayed in Fig.~\ref{fig:fisher_fcc_lin}.

\begin{figure}[t]
    \centering
    \includegraphics[width=0.85\linewidth]{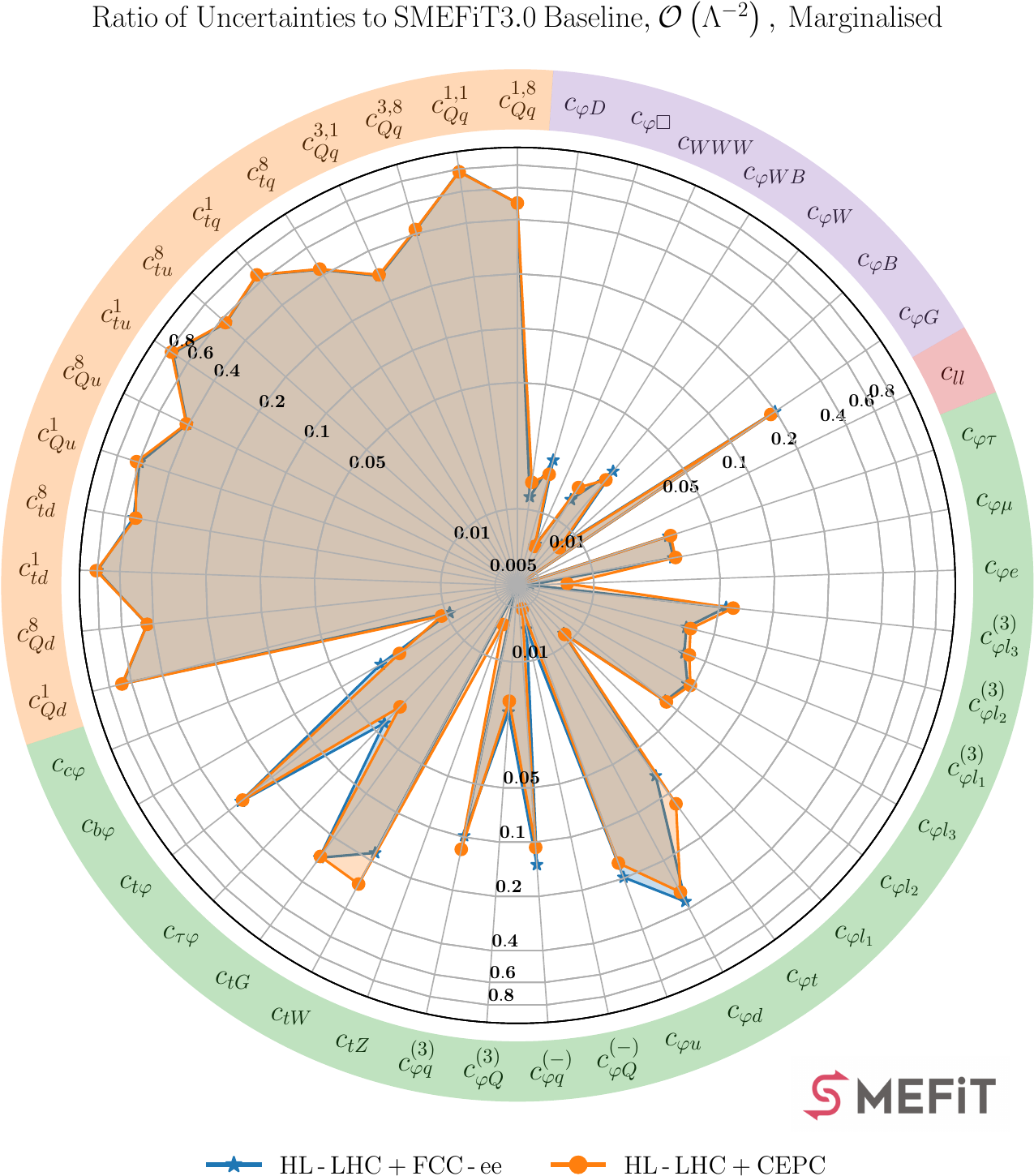}
    \caption{Same as Fig.~\ref{fig:spider_fcc_nlo_lin_glob} now
    comparing the impact of the FCC-ee and CEPC datasets.
    }
    \label{fig:spider_cepc_nlo_lin_glob}
\end{figure}

\paragraph{FCC-ee impact compared to the CEPC.}
We next study the impact of the CEPC measurements on the SMEFT coefficients, relative to that obtained in the FCC-ee  case and shown in Fig.~\ref{fig:spider_fcc_nlo_lin_glob}.
Fig.~\ref{fig:spider_cepc_nlo_lin_glob} displays the constraints provided by both colliders when added on top of the same HL-LHC baseline dataset, always relative to the {\sc\small SMEFiT3.0} baseline fit.
In general, a similar constraining power is obtained, consistent with the lack of major differences between their running plans (see Table~\ref{tab:FCCee_runs}), though the FCC-ee bounds tend to be better than those from CEPC.
The same qualitative conclusions are obtained in the case in which the analysis of Fig.~\ref{fig:spider_cepc_nlo_lin_glob}
is carried with quadratic EFT corrections.

When performing this comparison, we noticed that the total experimental uncertainties provided by the FCC-ee and CEPC collaborations for the Snowmass study, which we adopt in our analysis, differ by more than the scaling of integrated luminosities, as
would be expected in the case of purely statistical uncertainties without further correction factors.
This is illustrated by the Fisher information matrix, defined in Eq.~(\ref{eq:fisherinformation2}) at the linear EFT level, when evaluated in terms of ratios between the two experiments.
If one takes the ratio of the diagonal entries of the  Fisher information matrix between the FCC-ee and the CEPC, given that both colliders share the same theory predictions, one 
obtains
 \be
\label{eq:fisherinformation2_ratio_2}
I_{ii}^{\rm (FCC-ee)}/I_{ii}^{\rm (CEPC)} =  \sum_{m=1}^{n_{\rm dat}} \lp \delta_{{\rm exp},m}^2\rp_{\rm CEPC} \Bigg/\sum_{m'=1}^{n_{\rm dat}} \lp \delta_{{\rm exp},m'}^2\rp_{\rm FCC-ee}  \, , \qquad i=1,\ldots,n_{\rm eft},
\ee
namely the ratio of total experimental uncertainties squared.
For observables with only  statistical uncertainties, and assuming that eventual acceptance corrections cancel out, this ratio should reproduce  the corresponding ratios of integrated luminosities reported in Table~\ref{tab:FCCee_runs}, that is,
 \be
\label{eq:fisherinformation2_ratio_statistical}
I_{ii}^{\rm (FCC-ee)}/I_{ii}^{\rm (CEPC)}\simeq \lp \mathcal{L}_{\rm FCC-ee}/ \mathcal{L}_{\rm CEPC} \rp \, .
 \ee
 
\begin{figure}[t]
    \centering
    \includegraphics[width=\linewidth]{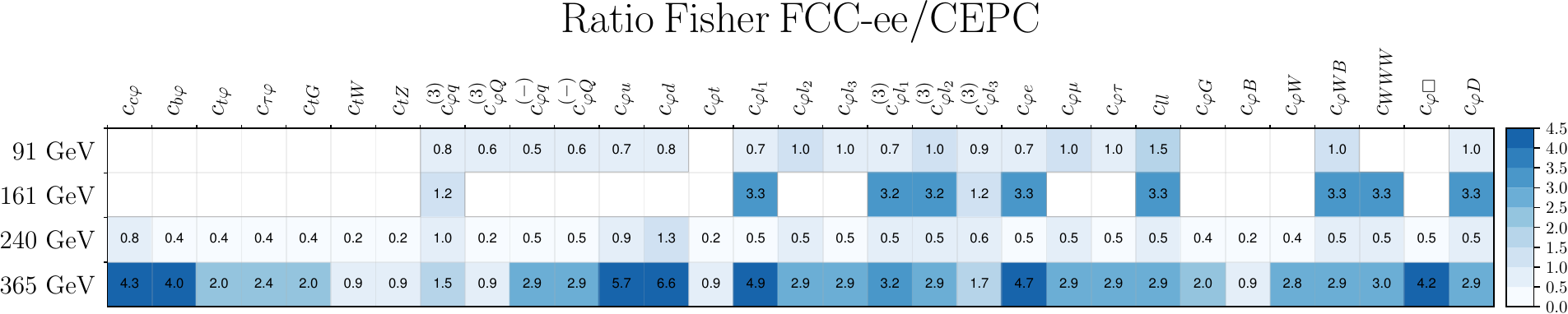}
    \caption{The ratio of the entries of the Fisher information matrix between the FCC-ee and the CEPC, Eq.~(\ref{eq:fisherinformation2_ratio_2}),
    evaluated separately for the observables entering the four center of mass energies $\sqrt{s}$ considered.
    Since projections for both colliders share theory predictions, this ratio is equivalent to the ratio of total experimental uncertainties squared.
    In turn, if the latter contains only the statistical uncertainties, the entries of the table should match the corresponding ratios of integrated luminosities from Table~\ref{tab:FCCee_runs}.
    }
    \label{fig:fisher_cepc_fccee_energies}
\end{figure}

Fig.~\ref{fig:fisher_cepc_fccee_energies} displays the ratio defined in Eq.~(\ref{eq:fisherinformation2_ratio_2}),  evaluated separately for the observables entering the four data-taking periods considered.
Values of Eq.~(\ref{eq:fisherinformation2_ratio_2}) below unity indicate EFT coefficients for which the CEPC observables should be more constraining than the FCC-ee ones. 
A deviation from the luminosity scaling for the $\sqrt{s}=91$ GeV run is expected, and indeed observed, since each collaboration makes different assumptions for their systematic uncertainties. 
On the other hand, since at $\sqrt{s}= 161, 240$ and 365 GeV the uncertainties considered are purely statistical, for these observables the ratios in Fig.~\ref{fig:fisher_cepc_fccee_energies} are expected to follow Eq.~(\ref{eq:fisherinformation2_ratio_statistical}). 
While in some cases this in indeed true, in particular for the runs at $\sqrt{s}=161$ GeV and $240$ GeV, in other cases there are larger differences. 
Particularly noticeable are those arising in the data-taking period at $\sqrt{s}=365$ GeV. There one expects a ratio of around $3$ purely on the luminosity scaling, but actually one obtains a range of values between 0.9 and 6.6.

It is beyond the scope of this work to scrutinise the origin of these differences: they could be explained by different assumptions on the experimental selection procedure and acceptance cuts, for example.
Nevertheless, the analysis of Fig.~\ref{fig:fisher_cepc_fccee_energies} highlights that in general the relative impact in the SMEFT parameter space of the projected FCC-ee and CEPC pseudo-data differs from the expectations based on a pure luminosity scaling.

\section{Constraints on UV-complete models through the SMEFT}
\label{subsec:uvmodels}

We now quantify the constraints that LHC Run II measurements and future collider projections impose on the parameter space of representative UV-complete scenarios.
To this aim, we benefit from the integration of {\sc\small SMEFiT} with {\sc\small matchmakereft} \cite{Carmona:2021xtq} via the {\sc\small Match2Fit} interface presented in~\cite{terHoeve:2023pvs}.
We consider results for the (indirect) mass reach for new heavy particles at the HL-LHC and FCC-ee obtained from the tree-level matching of a wide range of one-particle extensions of the SM.
We also present results for the reach in the UV couplings for the one-loop matching of a subset of the same one-particle extensions and for the tree-level matching of a multi-particle extension of the SM.
The corresponding results for the CEPC are qualitatively similar to the FCC-ee ones, consistently with Fig.~\ref{fig:spider_cepc_nlo_lin_glob}, and hence are not shown here.

\begin{table}[htbp]
\renewcommand{\arraystretch}{1.35}
\begin{center}
\begin{tabular}{|c|c|c|}
\toprule
{\bf Model Label} & $\qquad${\bf SM irreps}$\qquad$ & {\bf UV couplings} \\
\hline
\multicolumn{3}{|c|}{\bf Heavy Scalar Models} \\
\hline
$\mathcal{S}$ & $(1,1)_{0\phantom{/-3}}$ & $\kappa_{\mathcal{S}}$\\
$\phi$ & $(1,2)_{1/2\phantom{-}}$ & $\lambda_{\phi},\,\, (y_{\phi}^u)_{33}$    \\
$\Xi$  & $(1,3)_{0\phantom{/-3}}$ & $\kappa_{\Xi}$ \\
$\Xi$  & $(1,3)_{1\phantom{/-3}}$ & $\kappa_{\Xi_1}$ \\
$\omega_1$  & $(3,1)_{-1/3}$ & $\left(y_{\omega_1}^{qq}\right)_{33}$ \\
$\omega_4$  & $(3,1)_{-4/3}$ & $\left(y_{\omega_4}^{uu}\right)_{33}$ \\
$\zeta$  & $(3,3)_{-1/3}$ & $\left(y_{\zeta}^{qq}\right)_{33}$ \\
$\Omega_1$  & $(6,1)_{1/3\phantom{-}}$ & $\left(y_{\Omega_1}^{qq}\right)_{33}$ \\
$\Omega_4$  & $(6,1)_{4/3\phantom{-}}$ & $\left(y_{\omega_4}\right)_{33}$ \\
$\Upsilon$  & $(6,3)_{1/3\phantom{-}}$ & $\left(y_{\Upsilon} \right)_{33}$\\
$\Phi$  & $(8,2)_{1/2\phantom{-}}$ &  $\left(y_{\Phi}^{qu}\right)_{33}$ \\[1mm]
\hline
\multicolumn{3}{|c|}{\bf Heavy Vector Models} \\
\hline
$\mathcal{B}_1$ & $(1,1)_{1\phantom{/-3}}$ & $g_{B_1}^{\varphi}$ \\
$\mathcal{W}$ & $(1,3)_{0\phantom{/-3}}$ & $\left(g_{\mathcal{W}}^\ell\right)_{11},\,\left(g_{\mathcal{W}}^\ell\right)_{22}$,  $g_{\mathcal{W}}^\varphi$, $\left(g_{\mathcal{W}}^q\right)_{33}$ \\
$\mathcal{W}_1$ & $(1,3)_{1\phantom{/-3}}$ & $g_{\mathcal{W}_1}^{\varphi}$ \\
$\mathcal{H}$ & $(8,3)_{0\phantom{/-3}}$ & $\left(g_{\mathcal{H}}\right)_{33}$ \\
$\mathcal{Q}_5$ & $(3,2)_{-5/6}$ & $\left( g_{\mathcal{Q}_5}^{uq} \right)_{33}$ \\
$\mathcal{Y}_5$ & $(\bar{6},2)_{-5/6}$ & $\left( g_{\mathcal{Y}_5} \right)_{33}$ \\
\hline
\multicolumn{3}{|c|}{\bf Heavy Fermion Models} \\
\hline
 $N$ & $(1,1)_{0\phantom{/-3}}$ & $\left(\lambda_N^e\right)_3$ \\
$E$ & $(1,1)_{-1\phantom{/3}}$ & $\left(\lambda_E\right)_3$ \\
$\Delta_1$ & $(1,2)_{-1/2}$ & $\left(\lambda_{\Delta_1}\right)_3$\\
$\Delta_3$ & $(1,2)_{-3/2}$ & $\left(\lambda_{\Delta_3}\right)_3$ \\
$\Sigma$ & $(1,3)_{0\phantom{/-3}}$ & $\left(\lambda_{\Sigma}\right)_3$ \\
$\Sigma_1$ & $(1,3)_{-1\phantom{/3}}$ & $\left(\lambda_{\Sigma_1}\right)_3$ \\
$U$ & $(3,1)_{2/3\phantom{-}}$ & $\left(\lambda_{U}\right)_3$ \\
$D$ & $(3,1)_{-1/3}$ & $\left(\lambda_{D}\right)_3$\\
$Q_1$ & $(3,2)_{1/6\phantom{-}}$ & $\left(\lambda_{Q_1}^{u} \right)_3$ \\
$Q_7$ & $(3,2)_{7/6\phantom{-}}$ & $\left(\lambda_{Q_7}\right)_3$ \\
$T_1$ & $(3,3)_{-1/3}$ & $\left(\lambda_{T_1}\right)_3$  \\
 $T_2$ & $(3,3)_{2/3\phantom{-}}$ & $\left(\lambda_{T_2}\right)_3$  \\
\bottomrule
\end{tabular}
\caption{\label{tab:uv_couplings_SMEFiT} 
The one-particle UV-complete models that are considered in this work and matched to the SMEFT at tree level.
For the $\phi$ (heavy scalar) and $T_1$ and $T_2$ (heavy fermion) models, we also provide results based on one-loop matching.
For each model, following~\cite{terHoeve:2023pvs,deBlas:2017xtg} we indicate its label, the gauge group representation of the new heavy particle in the notation $(\textrm{SU}(3),\textrm{SU}(2))_{\textrm{U}(1)}$, and the relevant UV couplings entering the associated Lagrangian.
The model couplings are restricted to respect the {\sc\small SMEFiT} flavour assumption after tree-level matching.
Multi-particle models can be constructed from combining subsets of the one-particle models.
\label{tab:UVmodels_list}
}
\end{center}
\end{table}

\paragraph{Tree-level matching of one-particle extensions.}
First, we provide results for the UV-complete one-particle models considered in~\cite{terHoeve:2023pvs}, each of them associated to a different gauge group representation of the new heavy particle, matched at tree-level to the SMEFT.
Table~\ref{tab:UVmodels_list} displays the one-particle models considered in this work, indicating their label, the SM gauge group representation of the new heavy particle, and the UV couplings entering the associated Lagrangian (following the convention in~\cite{deBlas:2017xtg}). 

We restrict the possible UV couplings to ensure consistency with the {\sc\small SMEFiT} flavour assumption after tree-level matching, and consider only UV models that generate (at tree-level) operators  sensitive to our LHC observables. This excludes operators composed of two leptons and two quarks, as well as those compose of four light quarks. At the level of UV models, this restriction excludes leptoquarks, except for $\omega_{1,4}$, and some charged heavy vector bosons such as $\mathcal{U}_{2,5}$ and $\mathcal{X}$. Moreover, the absence of $\mathcal{O}_{\varphi}$ in our current dataset prevents us from studying e.g. the EW scalar quadruplets $\Theta_{1,3}\sim(1,4)_{\frac{1}{2},\frac{3}{2}}$ with tree-level matching results. This operator can be constrained from di-Higgs production, which we leave for future work.

\begin{figure}[htbp]
    \centering
    \includegraphics[width=.98\linewidth]{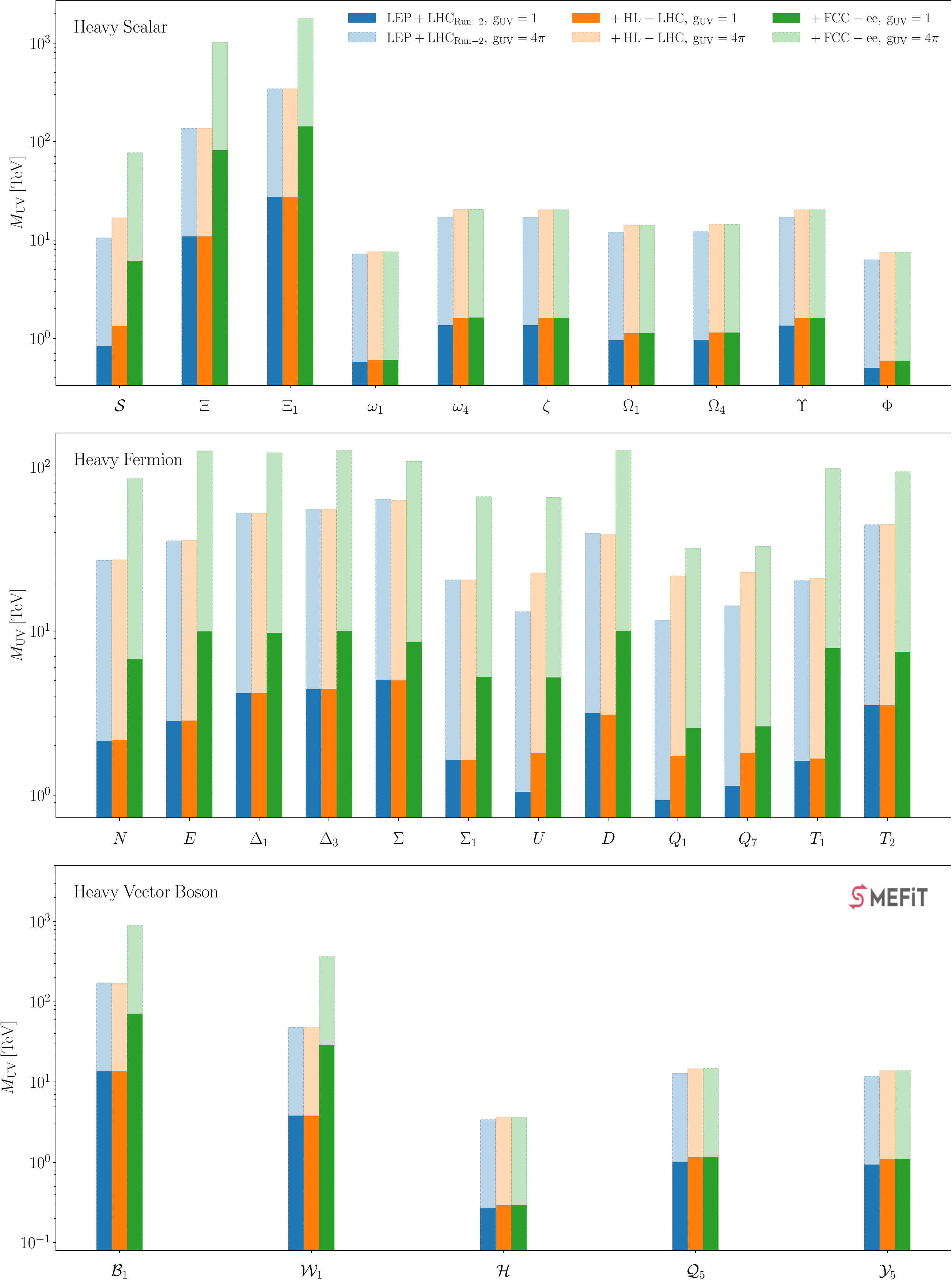}
    \caption{The 95\% CI lower bounds on the heavy particle mass $M_{\rm UV}$ for the one-particle UV-completions of the SM considered in this work, matched to the SMEFT using tree-level relations. 
    In all cases we include corrections up to quadratic order in the EFT expansion.
    From top to bottom, we display results for models with heavy scalars, heavy fermions, and heavy vector bosons, 
    see Table~\ref{tab:UVmodels_list} for
    the definition of each model.
    We present results for SMEFT analyses based on the current dataset (LEP+LHC$_{\rm RunII}$), then for its extension first with HL-LHC projections, and subsequently with the full set of FCC-ee observables.
    We consider two scenarios for the UV coupling constants, $g_{\rm UV}=1$ (darker) and $g_{\rm UV} = 4\pi$ (lighter). 
    Note the logarithmic scale of the $y$-axis.
    }
    \label{fig:uv-bounds}
\end{figure}

To illustrate the reach of the FCC-ee measurements on the parameter space of these UV models, we derive lower bounds on the mass of the heavy particle $M_{\rm UV}$ for  each of the one-particle extensions listed in Table~\ref{tab:UVmodels_list} by assuming a given value for the corresponding UV couplings $g_{\rm UV}$.
For simplicity, we consider only models with a single UV coupling.
The projected $95\%$ CI lower bounds on $M_{\rm UV}$ at FCC-ee are shown in Fig~\ref{fig:uv-bounds} for two limiting assumptions on the value of UV couplings: $g_{\rm UV}=1$ and $g_{\rm UV} = 4\pi$.
The chosen values are on the upper edge of what can be considered as weakly and strongly coupled, and in fact, $g_{\rm UV} > 4\pi$ would violate the perturbative limit.
For each model, we present results for a SMEFT analysis based on Level-1 pseudo-data of the current dataset, LEP+LHC$_{\rm Run\,II}$, and then for its extension first with HL-LHC projections, and subsequently with the complete set of FCC-ee observables.

Several observations can be drawn from perusing the results of Fig.~\ref{fig:uv-bounds}.
On the one hand, we find that the HL-LHC projections improve the mass reach for several models, in particular those that include heavy quark partners such as $U$, $Q_1$, and $Q_7$.
On the other hand, the models that are not affected by HL-LHC fall into two distinct categories.
One is composed of models such as $N$, $E$, $\Delta_{1,3}$, $\Sigma$, $\Sigma_1$, and $D$, that generate a subset of operators which display no improved sensitivity in individual fits to HL-LHC pseudo-data, namely purely leptonic and bosonic operators probed by EWPOs as well as $c_{\varphi Q}^{(-)}$ and $c_{\varphi Q}^{(3)}$.
The other class contains the models $\Xi$, $\Xi_1$, $T_2$, $\mathcal{B}_1$, and $\mathcal{W}_1$ that do generate operators that have associated improved bounds at the HL-LHC, but  where the sensitivity to the UV parameters is driven by operators that are instead insensitive to HL-LHC data, such as $c_{\varphi D}$, $c_{\varphi Q}^{(-)}$ or $c_{\varphi Q}^{(3)}$.

The mass reach enabled by FCC-ee measurements increases markedly for several models as compared to the post-HL-LHC limits, in some cases by up to an order of magnitude.
The largest effects are observed for the UV scenarios that modify the interactions of Higgs and electroweak bosons, which are tightly constrained by the FCC-ee data.
These include the $\mathcal{S}$, $\Xi$, and $\Xi_1$ heavy scalar models; the $N$, $E$, $\Delta_1$, $U$, $D$, $T_1$, and $T_2$ vector-like heavy fermion models; and the $\mathcal{B}_1$ and $\mathcal{W}_1$ heavy vector boson models.
For other scenarios, such as those primarily modifying the four-quark interaction vertices, there is only a marginal information gain provided by FCC-ee measurements, consistently with the findings at the Wilson coefficient level in Fig.~\ref{fig:spider_fcc_nlo_lin_glob}.

In terms of the heavy particle mass reach, we observe that FCC-ee measurements will be sensitive to BSM scales of up to around 100 TeV, 10 TeV, and 70 TeV for some of the studied heavy scalar, fermion, and vector boson UV-completions respectively, in the case of $g_{\rm UV}=1$.
This sensitivity increases to around $10^3$ TeV, 200 TeV, and 800 TeV in the case of strongly coupled one-particle extensions of the SM in the upper limit of the perturbative regime, $g_{\rm UV}=4\pi$. 
One also observes how, at least for the one-particle extensions considered here, at the HL-LHC there is not a large difference between the mass reach associated to direct production (with $m_X \sim 7$ TeV at most) and that associated to the indirect bounds obtained in the EFT framework.
On the other hand, at the FCC-ee the production of new heavy (TeV-scale) particles consistent with the LHC exclusion bounds is kinematically impossible due to the limited $\sqrt{s}$ values available, while the EFT bounds instead reach much higher energies, as illustrated by the examples of Fig.~\ref{fig:uv-bounds}.
This result further confirms the powerful sensitivity to heavy new physics enabled by the high-precision electroweak, Higgs, and top quark measurements to be performed at the FCC-ee highlighted by previous studies.

\paragraph{One-loop matched and multi-particle models.}
Following the discussion on single-particle extensions of the SM matched at tree level, we now evaluate the impact of the FCC-ee data on more general UV completions.
We consider in particular the heavy scalar $\phi$ and the heavy fermion $T_1$ and $T_2$ models, already analysed in Fig.~\ref{fig:uv-bounds}, now matched onto the SMEFT at the one-loop level.
This one-loop matching yields several additional contributions, generally flavour-independent, to bosonic and two-fermion operators as compared to tree-level matching.
One-loop matching contributions can lead to better constraints and, very importantly, allow to constrain otherwise blind directions in the UV parameter space~\cite{terHoeve:2023pvs}.
In addition, we also provide results for a 3-particle model, matched at tree-level, composed by the heavy vector-like fermions $Q_1$, $Q_7$ and the heavy vector boson $\mathcal{W}$ (see Table~\ref{tab:uv_couplings_SMEFiT} for their quantum numbers).

\begin{figure}[t]
    \centering
    \includegraphics[width=0.95\linewidth] {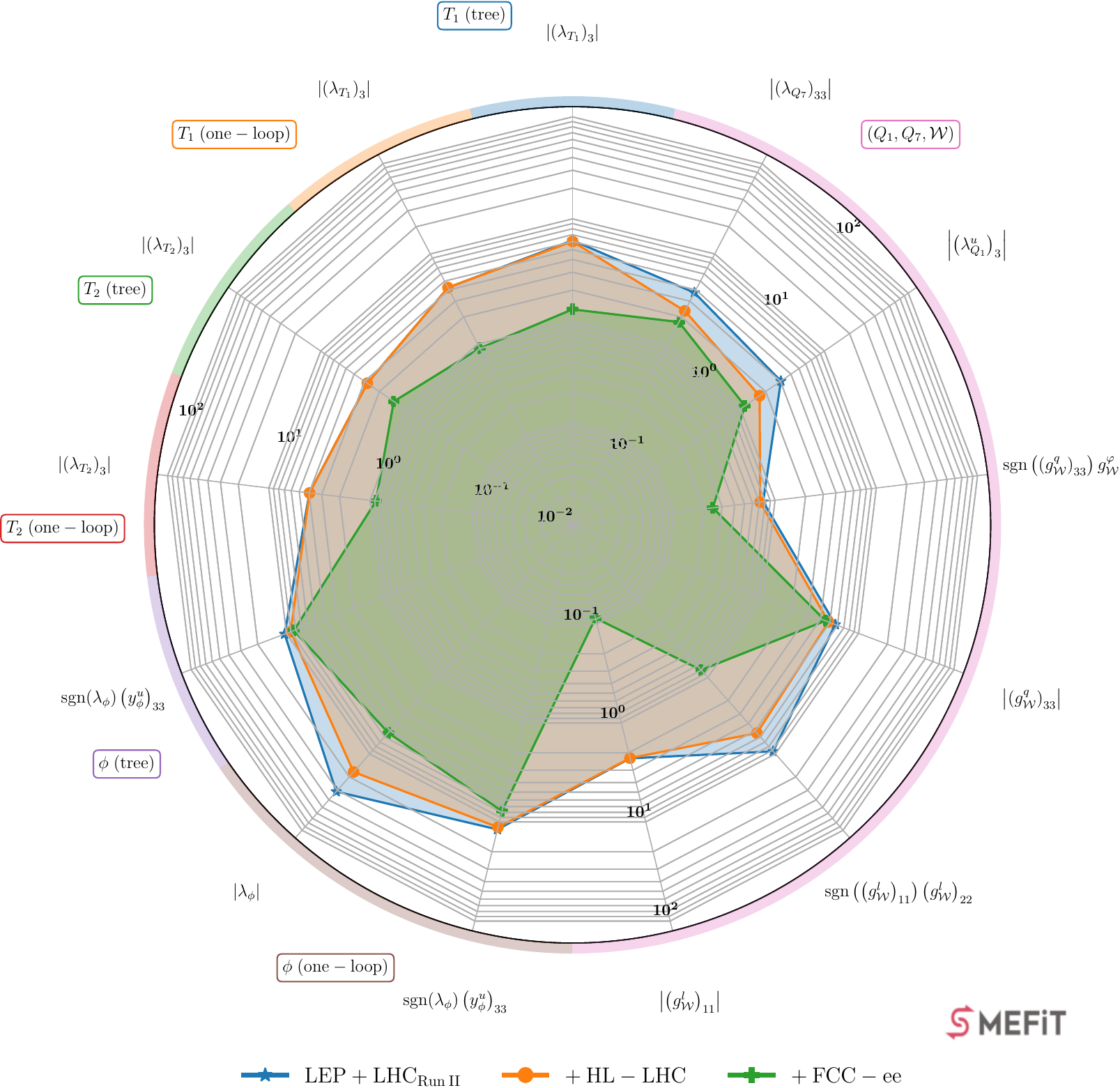}
    \caption{The 95\% CI upper bounds on the UV-invariant couplings of representative models obtained from the {\sc\small SMEFiT3.0} dataset (blue) and from its extension with the HL-LHC (orange) and with both the HL-LHC and FCC-ee (green) projections.
    We consider a 3-particle model, $\lp Q_1,Q_7,\mathcal{W}\rp$, matched at tree level, and three one-particle models, $T_1$, $T_2$, and $\phi$, matched at one loop. 
    For the multi-particle model, we set the masses $m_{Q_1}= 3$~TeV, $m_{Q_7}= 4.5$~TeV, and $m_{\mathcal{W}}= 2.5$~TeV.
    The one-particle extensions matched at one-loop assume a heavy particle mass of $M_{T_1} = M_{T_2} = 10$~TeV and $M_{\phi} = 5$~TeV, representing the indirect mass reach to be probed at the FCC-ee,
    and for completeness, the associated results from tree-level matching are also displayed.
    }
\label{fig:spider_UVbounds_oneloop_matching}
\end{figure}

Fig.~\ref{fig:spider_UVbounds_oneloop_matching} displays the 95\% CI upper bounds on the UV-invariant combination of couplings~\cite{terHoeve:2023pvs} of the considered models obtained from the {\sc\small SMEFiT3.0} dataset and from its extension with first the HL-LHC, and then the HL-LHC+FCC-ee projections.
For the multi-particle model matched at tree level we assume masses of  $m_{Q_1}= 3$~TeV, $m_{Q_7}= 4.5$~TeV, and $m_{\mathcal{W}}= 2.5$~TeV, see also~\cite{terHoeve:2023pvs}.
For the one-particle extensions matched at one-loop we assume $M_{T_1}= M_{T_2} = 10$~TeV and $M_{\phi} = 5$~TeV, which represent the typical mass reach being probed at the FCC-ee for those kinds of heavy particles, see Fig.~{\ref{fig:uv-bounds}}. 
For reference, the corresponding tree-level results with the same $M_{\rm UV}$ are also provided. 

Consistently with Fig.~{\ref{fig:uv-bounds}}, the sensitivity to the heavy fermions $T_1$ and $T_2$ is not improved at the HL-LHC due to being driven by the constraints from LEP data. Instead, the bounds on these models are significantly tightened after adding the FCC-ee projections. 
The inclusion of one-loop matching results does not alter this picture and has a small and generally positive impact on the bounds.
One-loop matching effects are more marked for the $\phi$ heavy scalar model, where it allows one to constrain the additional UV parameter $|\lambda_\phi|$ in all scenarios and with remarkable improvements both at HL-LHC and FCC-ee.
The bounds on $|\lambda_\phi|$ at LHC Run II and HL-LHC are, however, of limited use since they are beyond the perturbative limit, $|\lambda_\phi|<4 \pi$.
The power of precision measurements at FCC-ee is enough to bring down this bound to a strongly-coupled but perturbative regime, thus providing meaningful information on the UV model.
The bound on sgn$(\lambda_\phi)\left(y_{\phi}^{u}\right)_{33}$ is improved at future colliders for tree-level matching by around $15\%$ at each stage, though the logarithmic scale in the plot hides this improvement. 
The difference diminishes when considering one-loop matching results and in all cases the bound is within the perturbative limit.

Finally, the results of the multiparticle model display several interesting features. 
The UV couplings of the heavy fermions, $|(\lambda_{Q_7})_{33}|$ and $|(\lambda_{Q_1}^{u})_{3}| $, receive improved bounds at both HL-LHC and FCC-ee, as expected since they are related to the coefficients $c_{t\varphi}$ and $c_{\varphi t}$ that show a similar behavior. 
The UV invariants from the couplings of the heavy spin-1 vector boson with leptons, $|(g_{\mathcal{W}}^{\ell})_{11}|$ and $\text{sgn}((g_{\mathcal{W}}^{\ell})_{11})(g_{\mathcal{W}}^{\ell})_{22}$, see little to no improvement on their bounds from adding HL-LHC projections, but the bounds become at least one order of magnitude tighter once FCC-ee data is considered. 
A similar situation is found for $\text{sgn}((g_{\mathcal{W}}^{q})_{33})g_{\mathcal{W}} ^ {\varphi}$, whose bound is driven mostly by its contribution to $c_{\varphi \square}$ and $c_{\varphi \ell_1}^{(3)}$ but also affected by its appearance in $c_{\varphi Q}^{(-)}$, $c_{\varphi Q}^{(3)}$ and $c_{b \varphi}$. 
The UV coupling $g_{\mathcal{W}}^{\varphi}$ also contributes at tree level to $c_{t\varphi}$, which connects the fermionic and bosonic sectors in this model.
Such interplay worsens the bound on $\text{sgn} (( g_{\mathcal{W}}^{q} )_{33} )g_{\mathcal{W}} ^ {\varphi}$ by $\sim 30\%$, which highlights the importance of considering models more complex than the one-particle extensions.
The bound on $|(g_{\mathcal{W}}^{q})_{33}|$ is obtained via the Wilson coefficients of four-heavy quark operators such as $c_{QQ}^{1}$ and $c_{QQ}^{8}$. 
Hence, it improves by around $20\%$ when going from LHC to HL-LHC but is not improved further at future lepton colliders.

\section{Summary and outlook}
\label{sec:summary}

In this work we have quantified the impact that measurements at future colliders, first the HL-LHC and then the FCC-ee and the CEPC, would have on the parameter space of the SMEFT in the framework of a global interpretation of particle physics data.
The baseline for these projection studies has been an updated global SMEFT analyses, {\sc\small SMEFiT3.0}, combining the constraints from the EWPOs from LEP and SLD with those provided by the recent LHC Run II measurements on Higgs boson, top quark, and diboson production, in many cases based on the full integrated luminosity.

Our analysis provides bounds on $n_{\rm eft}=50$ independent Wilson coefficients (45 in the linear fits) associated to dimension-six operators, with EFT cross-sections being evaluated either at $\mathcal{O}\lp \Lambda^{-2}\rp$ or $\mathcal{O}\lp \Lambda^{-4}\rp$ and including NLO QCD corrections whenever possible.
These results are enabled by exploiting newly developed functionalities of the {\sc\small SMEFiT} framework, in particular the availability of a projection module which can extrapolate from existing measurements and project them to the settings relevant for other (future) experiments.
This module also makes it possible to carry out global SMEFT analyses based on pseudo-data generated with an arbitrary underlying law.

 We find that the marginalised bounds on the EFT coefficients within the global fit are projected to improve between around 20\% and a factor 3 by the end of the HL-LHC, depending on the operator, with possible further improvements being enabled by optimised analyses not considered here.
 Subsequently, the constraints from the FCC-ee or the CEPC would markedly improve the bounds on most of the purely bosonic and two-fermion operators entering the fit, by up to two orders of magnitude in some cases.
 We have then determined, using the UV matching procedure, how this impact at the EFT coefficient level translates into the parameter space of representative one-particle and multi-particle extensions of the SM matched to the SMEFT.
We find that the stringent constraints on the interactions of the Higgs, $W$ and $Z$ bosons, and top quarks made possible by future $e^+e^-$ circular colliders lead to an indirect mass reach on heavy new particles ranging between a few TeV up to around 100 TeV (for UV couplings $g_{\rm UV}\simeq 1$), depending on the specific UV scenario.
It is hence clear that the precision FCC-ee/CEPC measurements would make possible an extensive indirect exploration of the landscape of UV models containing new heavy particles beyond the direct reach of HL-LHC.

This work could be extended in several directions. 
To begin with, we could consider projections for other proposed future colliders, from the ILC and CLIC to the muon collider at different center-of-mass energies, to assess what are their strengths and weaknesses as compared to the FCC-ee and the CEPC in the context of a global SMEFT analysis and its matching to UV models.
Second, it would be interesting to further explore whether the EFT impact of Higgs and fermion-pair production at leptonic colliders can be enhanced by including differential information, on the same footing as how it is done for the optimal observables for diboson and top-quark pair production.
Third, one may want to include optimised HL-LHC projections, fully exploiting the increase in statistics in ways which cannot be extrapolated from available Run II measurements, such as extended range in differential distributions or finer binning.
We expect these to improve the constraints set by HL-LHC on both the SMEFT parameter space and that of the UV-complete models. 

Fourth, by extending the analysis of the UV-complete scenarios studied here to other models, both at tree level and via one-loop matching, and in particular considering more general multi-particle models, one could further scrutinise the indirect constraints that quantum corrections on FCC-ee observables impose on the masses and couplings of new heavy particles beyond the reach of the HL-LHC.
Fifth, albeit on a longer timescale, one may want to generalise the flavour assumptions on the EFT operators entering the global fit baseline and include data from other processes such as Drell-Yan and $B$-meson decays, and subsequently verify the robustness of the obtained projections for future colliders.
In this context, we note that the FCC-ee will also function as a flavour factory, with huge statistics thanks to the large $B$-mesons sample produced in the $Z$-pole run. 
Finally, the impact of renormalisation group equation running and mixing of the operators would be needed to ensure a more robust bound setting in global fits such as the one we perform here, where a variety of observables with different energy scales are combined.

As the global particle physics community moves closer to identifying the next large scientific infrastructure projects that will inform the field for the coming decades, the availability of methodologies quantifying the physics reach of different colliders represents an essential tool to make informed decisions.
The results of this work demonstrate that the {\sc\small SMEFiT} open-source framework is suitable to contribute to this endeavor.
Within this framework, progress in the global EFT interpretation of the most updated measurements can be kept synchronised with projections for future colliders, in a way that the baseline always represents the state-of-the-art in terms of experimental constraints. 
For these reasons, we expect that projections based on {\sc\small SMEFiT} will provide a valuable contribution to the upcoming ESPPU and related community studies to take place in the coming few years.

\subsection*{Acknowledgments}
We are grateful to M.~Bakker for collaboration in initial stages of this work.
We acknowledge useful discussions with  R. Balasubramanian, J. de Blas, I. Brivio, G. Durieux, C. Grojean,  F. Maltoni, K. Mimasu,  V. Miralles, C. Severi, and M. Vos.
The work of E. C., A. R., M. T. and E. V. is supported by the European Research Council (ERC) under the European
Union’s Horizon 2020 research and innovation programme (Grant agreement No. 949451) and by a Royal
Society University Research Fellowship through grant URF/R1/201553.
The work of L. M. is supported by the European Research Council under the European Union’s
Horizon 2020 research and innovation Programme (grant agreement n.950246).
The work of J. t. H. and T. G. is
supported by the Dutch Research Council (NWO). 
The work of J. R. is supported by the Dutch Research
Council (NWO) and by the Netherlands eScience Center.

\appendix 
\section{Fit input settings}
\label{app-list-fits}

Table~\ref{table:fit_settings} collects the overview of the input settings adopted for the SMEFT fits presented in this work.
For each fit, we indicate its dataset, whether the EFT coefficients are extracted from a global fit and then marginalised or instead from one parameter individual fits, whether we use actual data or SM pseudo-data, and the treatment of EWPOs (exact versus approximate).
In the last column, we list the figures and tables in Sects.~\ref{sec:updated_global_fit} and~\ref{sec:results} of the paper where the corresponding results are displayed. 
For all fits listed in this overview, variants at both $\mathcal{O}\lp \Lambda^{-2}\rp$ and
$\mathcal{O}\lp \Lambda^{-4}\rp$ in the EFT expansion have been produced. 

\begin{table}[htbp]
  \centering
  \footnotesize
   \renewcommand{\arraystretch}{1.70}
   \begin{tabular}
   {l|C{3.4cm}|C{2.1cm}|C{1.4cm}|C{3.4cm}}
     \toprule
  Dataset   & EFT coefficients &  Pseudodata?  & EWPOs  & Figs. \& Tables      \\
       \midrule
 {\sc\small  SMEFiT3.0}  & Global (Marginalised)   &  No  &  Exact   & Figs.~\ref{fig:chi2_lin_vs_quad},~\ref{fig:smefit30_marginalised_bounds},~\ref{fig:pull_lin_vs_quad}, \ref{fig:updated-global-fit-quad},~\ref{fig:new-vs-old-quad}, Tables~\ref{eq:chi2-baseline-grouped}, \ref{tab:smefit30_all_bounds}   \\
 {\sc\small  SMEFiT3.0}  & Individual (one-param)   &  No  &  Exact   &  Table~\ref{tab:smefit30_all_bounds}  \\
 {\sc\small  SMEFiT2.0}  & Global (Marginalised)   &  No  &  Exact   & Fig.~\ref{fig:new-vs-old-quad}  \\
  {\sc\small  SMEFiT3.0}  & Global (Marginalised)   &  No  &  Approx. & 
Fig.~\ref{fig:updated-global-fit-quad}\\
  \midrule
  {\sc\small  SMEFiT3.0}  & Global (Marginalised)   &  Yes  &  Exact   &  Figs.~\ref{fig:spider_nlo_lin_glob},~\ref{fig:spider_nlo_quad_glob} \\
  {\sc\small  SMEFiT3.0}\,+\,HL-LHC  & Global (Marginalised)   &  Yes  &  Exact   & Figs.~\ref{fig:spider_nlo_lin_glob},~\ref{fig:spider_nlo_quad_glob} \\
  {\sc\small  SMEFiT3.0}\,+\,HL-LHC  & Individual (one-param)   &  Yes  &  Exact   &  Figs.~\ref{fig:spider_nlo_lin_glob},~\ref{fig:spider_nlo_quad_glob}\\
  \midrule
   {\sc\small  SMEFiT3.0}\,+\,HL-LHC\,+\,FCC-ee  & Global (Marginalised)   &  Yes  &  Exact  & Figs.~\ref{fig:spider_fcc_nlo_lin_glob},~\ref{fig:fisher_fcc_lin},~\ref{fig:spider_fcc_nlo_quad_glob},~\ref{fig:spider_fcc_energy_variations},~\ref{fig:spider_cepc_nlo_lin_glob} \\ 
    {\sc\small  SMEFiT3.0}\,+\,HL-LHC\,+\,CEPC  & Global (Marginalised)   &  Yes  &  Exact & Fig.~\ref{fig:spider_cepc_nlo_lin_glob}\\ 
\bottomrule
\end{tabular}
\caption{\small Overview of the input settings used in the fits presented in this work.
For each fit, we indicate its dataset, whether the EFT coefficients are extracted from a global fit and then marginalised or instead from one parameter individual fits, whether we use actual data or instead SM pseudo-data is generated, and the treatment of EWPOs (exact versus approximate).
In the last column we list the figures and tables in Sects.~\ref{sec:updated_global_fit} and~\ref{sec:results} of the paper where the corresponding results are displayed. 
For all fits, variants at both $\mathcal{O}\lp \Lambda^{-2}\rp$ and
$\mathcal{O}\lp \Lambda^{-4}\rp$ in the EFT expansion have been produced. 
The HL-LHC dataset is constructed following the projection described in App.~\ref{app:hl_lhc_projections}.
\label{table:fit_settings}
}
\end{table}
\section{SMEFT operator basis}
\label{app-operators}

For completeness, we summarise here the SMEFT operator basis and the corresponding flavour assumptions entering this analysis.
This basis was first defined in~\cite{Hartland:2019bjb} and then extended in~\cite{Ethier:2021bye}, see also~\cite{Giani:2023gfq,terHoeve:2023pvs}.
It is based on the Warsaw basis of dimension-six operators~\cite{Grzadkowski:2010es} along with a U$(2)_q\times$U$(3)_d\times$U$(2)_u\times(\text{U}(1)_\ell\times\text{U}(1)_e)^3$ flavour assumption.

The purely bosonic operators are listed in Table~\ref{tab:oper_bos}, and the two-fermion and four-lepton operators are defined in Table~\ref{tab:oper_ferm_bos} and~\ref{tab:oper_ferm_bos2}. 
The four-fermion operators involving bottom and top quarks that we consider are defined in terms of the following Warsaw basis operators:
\begin{align}
	\qq{1}{qq}{ijkl}
	&= (\bar q_i \gamma^\mu q_j)(\bar q_k\gamma_\mu q_l)
	 \nonumber
	,\\
	\qq{3}{qq}{ijkl}
	&= (\bar q_i \gamma^\mu \tau^I q_j)(\bar q_k\gamma_\mu \tau^I q_l)
 \nonumber
	,\\
	\qq{1}{qu}{ijkl}
	&= (\bar q_i \gamma^\mu q_j)(\bar u_k\gamma_\mu u_l)
         \nonumber
	,\\
	\qq{8}{qu}{ijkl}
	&= (\bar q_i \gamma^\mu T^A q_j)(\bar u_k\gamma_\mu T^A u_l)
         \nonumber
	,\\
	\qq{1}{qd}{ijkl}
	&= (\bar q_i \gamma^\mu q_j)(\bar d_k\gamma_\mu d_l)
         \nonumber
	,\\
	\qq{8}{qd}{ijkl}
	&= (\bar q_i \gamma^\mu T^A q_j)(\bar d_k\gamma_\mu T^A d_l)
        \label{eq:FourQuarkOp} 
	,\\
	\qq{}{uu}{ijkl}
	&=(\bar u_i\gamma^\mu u_j)(\bar u_k\gamma_\mu u_l)
         \nonumber
	,\\
	\qq{1}{ud}{ijkl}
	&=(\bar u_i\gamma^\mu u_j)(\bar d_k\gamma_\mu d_l)
         \nonumber
	,\\
	\qq{8}{ud}{ijkl}
	&=(\bar u_i\gamma^\mu T^A u_j)(\bar d_k\gamma_\mu T^A d_l)
         \nonumber \, ,
\end{align}
Then the  four-fermion coefficients that enter the fit are linear combinations of the Wilson coefficients of the above operators and are defined in Table~\ref{tab:oper_fourtop}. 
In Eq.~(\ref{eq:FourQuarkOp}), the left-handed quark doublets are denoted by $q_i$ for the first two generations and by $Q$ for the third generation, while $t$ denotes a right-handed top-quark field and $u_i, d_i$ the right-handed $u, c$ and $d, s, b$ fields respectively.
The letters $i, j, k, l$ represent the fermion generation indices.
All the quarks are considered to be massless except for the top quark.
Regarding the leptons, $e, \mu, \tau$ stand for the right-handed charged leptons and $\ell_i$ refers to the left-handed leptonic doublets with $i=1,2,3$.

In the conventions that this analysis adopts, the Higgs doublet is indicated by $\varphi$ with vacuum expectation value $v/\sqrt{2}$. 
We define
\begin{equation}
\varphi^\dagger\lra{D}_\mu\,\varphi \equiv \varphi^\dagger D_\mu \varphi - (D_\mu \varphi)^\dagger \varphi\ , \qquad \varphi^\dagger\lra{D}_\mu\,\tau_I \,\varphi \equiv \varphi^\dagger\, \tau_I D_\mu \varphi - (D_\mu \varphi)^\dagger \tau_I \varphi \, ,
\end{equation}
where the covariant derivative is given by:
\begin{equation}
    D_\mu \varphi = \Big(\partial_\mu -i \frac{g}{2} \tau_I W_\mu^I -i \frac{g'}{2}B_\mu \Big) \varphi \, .
\end{equation}
Here $W_\mu,\, B_\mu$ are the electroweak gauge bosons fields, $g$ and $ g'$ are the $SU(2)_L$ and $U(1)_Y$ couplings respectively and $\tau_{I}$ are the Pauli sigma matrices.
Furthermore $G^A_{\mu\nu}$, $W^{\mu\nu}$ and $B^{\mu\nu}$ stand for the $SU(3)_C$, $SU(2)_L$ and $U(1)_Y$ field strength tensors.
$T^A = \frac{1}{2} \lambda^A$ are the $SU(3)$ generators where $\lambda^A$ are the Gell-Mann matrices. 
The strong coupling constant is denoted by $g_s$, $\theta$ represents the weak mixing angle, and $\tau^{\mu\nu} = \frac{1}{2} [\gamma^\mu,\gamma^\nu]$.

\begin{table}[t] 
  \begin{center}
    \renewcommand{\arraystretch}{1.6}
        \begin{tabular}{lll|lll}
          \toprule
          Operator $\quad$ & Coefficient $\quad$ & Definition& Operator $\quad$ & Coefficient $\quad$ & Definition \\
        \midrule
        \midrule
        $\Op{\varphi G}$ & $c_{\varphi G}$  & $\left(\pdp\right)G^{\mu\nu}_{\sss A}\,
        G_{\mu\nu}^{\sss A}$ 
        & 
        $\Op{\varphi \square}$ & $c_{\varphi \square}$ & $\partial_\mu(\pdp)\partial^\mu(\pdp)$ \\
        $\Op{\varphi B}$ & $c_{\varphi B}$ & $\left(\pdp\right)B^{\mu\nu}\,B_{\mu\nu}$
        &
        $\Op{\varphi D}$ & $c_{\varphi D}$ & $(\varphi^\dagger D^\mu\varphi)^\dagger(\varphi^\dagger D_\mu\varphi)$ \\ 
        $\Op{\varphi W}$ &$c_{\varphi W}$ & $\left(\pdp\right)W^{\mu\nu}_{\sss I}\,
        W_{\mu\nu}^{\sss I}$ 
        &
        $\mathcal{O}_{W}$&   $c_{WWW}$ & $\epsilon_{IJK}W_{\mu\nu}^I W^{J,\nu\rho} W^{K,\mu}_\rho$ \\ 
        $\Op{\varphi W B}$ &$c_{\varphi W B}$ & $(\varphi^\dagger \tau_{\sss I}\varphi)\,B^{\mu\nu}W_{\mu\nu}^{\sss I}\,$ \\ 
       \bottomrule
        \end{tabular}
        \caption{Purely bosonic dimension-six operators that
          modify the production and decay of Higgs bosons and
          the interactions of the electroweak gauge bosons.
          For each operator, we indicate its definition in terms of the SM
          fields,
          and the conventions that are used
          both for the operator and for the coefficient. 
          See~\cite{Ethier:2021bye} for more details.
          \label{tab:oper_bos}}
\end{center}
\end{table}

\begin{table}[htbp]
  \begin{center}
    \renewcommand{\arraystretch}{1.45}
    \begin{tabular}{p{1.5cm} p{1.4cm} p{4.4cm} | p{1.5cm} p{1.4cm} p{4.5cm}}
      \toprule
      Operator & Coefficient & $\qquad$ Definition & Operator  & Coefficient & $\qquad$ Definition \\
                \midrule \midrule
      \multicolumn{6}{c}{3rd generation quarks} \\
                \midrule \midrule
    $\Op{\varphi Q}^{(1)}$ & $c_{\varphi Q}^{(1)}$~(*) & $i\big(\varphi^\dagger\lra{D}_\mu\,\varphi\big)
 \big(\bar{Q}\,\gamma^\mu\,Q\big)$ 
 &
 $\Op{tW}$ & $c_{tW}$ & $i\big(\bar{Q}\tau^{\mu\nu}\,\tau_{\sss I}\,t\big)\,
 \tilde{\varphi}\,W^I_{\mu\nu}
 + \text{h.c.}$ \\ 
    $\Op{\varphi Q}^{(3)}$ & $c_{\varphi Q}^{(3)}$  & $i\big(\varphi^\dagger\lra{D}_\mu\,\tau_{\sss I}\varphi\big)
 \big(\bar{Q}\,\gamma^\mu\,\tau^{\sss I}Q\big)$ 
 &
 $\Op{tB}$ & $c_{tB}$~(*) &
 $i\big(\bar{Q}\tau^{\mu\nu}\,t\big)
 \,\tilde{\varphi}\,B_{\mu\nu}
 + \text{h.c.}$\\ 
    $\Op{\varphi t}$ & $c_{\varphi t}$& $i\big(\varphi^\dagger\,\lra{D}_\mu\,\,\varphi\big)
 \big(\bar{t}\,\gamma^\mu\,t\big)$
 &
  $\Op{t G}$ & $c_{tG}$ & $ig{\sss S}\,\big(\bar{Q}\tau^{\mu\nu}\,T_{\sss A}\,t\big)\,
 \tilde{\varphi}\,G^A_{\mu\nu}
 + \text{h.c.}$ \\ 
     $\Op{t \varphi}$ & $c_{t\varphi}$ & $\left(\pdp\right)
 \bar{Q}\,t\,\tilde{\varphi} + \text{h.c.}$ 
 &
  $\Op{b \varphi}$ & $c_{b\varphi}$ & $\left(\pdp\right)
 \bar{Q}\,b\,\varphi + \text{h.c.}$ \\  
                \midrule \midrule
                \multicolumn{6}{c}{1st, 2nd generation quarks} \\
                \midrule \midrule
    $\Op{\varphi q}^{(1)}$ & $c_{\varphi q}^{(1)}$~(*) & $\sum\limits_{\sss i=1,2} i\big(\varphi^\dagger\lra{D}_\mu\,\varphi\big)
 \big(\bar{q}_i\,\gamma^\mu\,q_i\big)$ 
 &
 ${\Op{\varphi d }}$ &
      ${{c_{\varphi d}}}$ & $\sum\limits_{\sss i=1,2,3} i\big(\varphi^\dagger\,\lra{D}_\mu\,\,\varphi\big)
 \big(\bar{d}_i\,\gamma^\mu\,d_i\big)$\\ 
    $\Op{\varphi q}^{(3)}$ & $c_{\varphi q}^{(3)}$ & $\sum\limits_{\sss i=1,2} i\big(\varphi^\dagger\lra{D}_\mu\,\tau_{\sss I}\varphi\big)
 \big(\bar{q}_i\,\gamma^\mu\,\tau^{\sss I}q_i\big)$
 &$\Op{c \varphi}$ & $c_{c \varphi}$ & $\left(\pdp\right)
 \bar{q}_2\,c\,\tilde\varphi + \text{h.c.}$ \\ 
  ${\Op{\varphi u }}$ &
      ${{c_{\varphi u}}}$ & $\sum\limits_{\sss i=1,2} i\big(\varphi^\dagger\,\lra{D}_\mu\,\,\varphi\big)
 \big(\bar{u}_i\,\gamma^\mu\,u_i\big)$\\ 
    
                \midrule \midrule
		      \multicolumn{6}{c}{two-leptons} \\
                \midrule \midrule
    $\Op{\varphi \ell_i}$ & $c_{\varphi \ell_i}$ & $ i\big(\varphi^\dagger\lra{D}_\mu\,\varphi\big)
   \big(\bar{\ell}_i\,\gamma^\mu\,\ell_i\big)$ 
   &
    $\Op{\varphi \mu}$ & $c_{\varphi \mu}$ & $ i\big(\varphi^\dagger\lra{D}_\mu\,\varphi\big)
 \big(\bar{\mu}\,\gamma^\mu\,\mu\big)$ \\  
    $\Op{\varphi \ell_i}^{(3)}$ & $c_{\varphi \ell_i}^{(3)}$ & $ i\big(\varphi^\dagger\lra{D}_\mu\,\tau_{\sss I}\varphi\big)
 \big(\bar{\ell}_i\,\gamma^\mu\,\tau^{\sss I}\ell_i\big)$ 
 &
  $\Op{\varphi \tau}$ & $c_{\varphi \tau}$ & $i\big(\varphi^\dagger\lra{D}_\mu\,\varphi\big)
 \big(\bar{\tau}\,\gamma^\mu\,\tau\big)$ \\  
    $\Op{\varphi e}$ & $c_{\varphi e}$ & $ i\big(\varphi^\dagger\lra{D}_\mu\,\varphi\big)
 \big(\bar{e}\,\gamma^\mu\,e\big)$ 
 &
  $\Op{\tau \varphi}$ & $c_{\tau \varphi}$ & $\left(\pdp\right)
 \bar{\ell_3}\,\tau\,{\varphi} + \text{h.c.}$ \\
                \midrule \midrule
		      \multicolumn{6}{c}{four-leptons} \\
                \midrule \midrule
 $\Op{\ell\ell}$ & $c_{\ell\ell}$ & $\left(\bar \ell_1\gamma_\mu \ell_2\right) \left(\bar \ell_2\gamma^\mu \ell_1\right)$ \\
 
  \bottomrule
\end{tabular}
\caption{Same as Table~\ref{tab:oper_bos}
  for the operators containing two fermion fields, either
  quarks or leptons, as well as the four-lepton operator $\Op{\ell\ell}$.
  The flavour index $i$ runs from 1 to 3.
  The coefficients indicated with (*) in the second column do not correspond to physical degrees of freedom
  in the fit, but are rather replaced by  $c_{\varphi q}^{(-)}$, $c_{\varphi Q}^{(-)}$, and
  $c_{tZ}$ defined in Table~\ref{tab:oper_ferm_bos2}.
\label{tab:oper_ferm_bos}}
\end{center}
\end{table}

\begin{table}[htbp]
  \begin{center}
    \renewcommand{\arraystretch}{1.45}
    \begin{tabular}{l l}
    \toprule
     Coefficient $\qquad$ & Definition\\ \hline
     \midrule
    $c_{\varphi Q}^{(-)}$ & $c_{\varphi Q}^{(1)}-c_{\varphi Q}^{(3)}$\\ 
    $c_{tZ}$ & $-s_\theta \, c_{tB}+ c_\theta \,c_{tW}$ \\ 
    $c_{\varphi q}^{(-)}$ & $c_{\varphi q}^{(1)}-c_{\varphi q}^{(3)}$ \\ 
    \bottomrule
\end{tabular}
\caption{Linear combinations of two-fermion operators listed in Table~\ref{tab:oper_ferm_bos} and that enter at the fit level instead of those marked with (*).
\label{tab:oper_ferm_bos2}}
\end{center}
\end{table}

\begin{table}[htbp] 
  \begin{center}
    \renewcommand{\arraystretch}{1.53}
        \begin{tabular}{ll| ll}
          \toprule
          DoF $\qquad$ &  Definition (in  Warsaw basis notation) & DoF $\qquad$ &  Definition (in  Warsaw basis notation) \\
          \midrule
          \midrule
      $c_{QQ}^1$    &   $2\ccc{1}{qq}{3333}-\frac{2}{3}\ccc{3}{qq}{3333}$ 
      &
      $c_{QQ}^8$       &         $8\ccc{3}{qq}{3333}$\\  
     $c_{Qt}^1$         &         $\ccc{1}{qu}{3333}$
     &
     $c_{Qt}^8$         &         $\ccc{8}{qu}{3333}$\\   
            \midrule      
  $c_{Qq}^{1,8}$       &  	 $\ccc{1}{qq}{i33i}+3\ccc{3}{qq}{i33i}$  
  &
  $c_{Qq}^{1,1}$         &   $\ccc{1}{qq}{ii33}+\frac{1}{6}\ccc{1}{qq}{i33i}+\frac{1}{2}\ccc{3}{qq}{i33i} $   \\    
   $c_{Qq}^{3,8}$         &   $\ccc{1}{qq}{i33i}-\ccc{3}{qq}{i33i} $  
   &
$c_{Qq}^{3,1}$          & 	$\ccc{3}{qq}{ii33}+\frac{1}{6}(\ccc{1}{qq}{i33i}-\ccc{3}{qq}{i33i}) $   \\     
$c_{tq}^{8}$         &  $ \ccc{8}{qu}{ii33}   $ 
  &
$c_{tq}^{1}$       &   $  \ccc{1}{qu}{ii33} $\\   
$c_{tu}^{8}$      &   $2\ccc{}{uu}{i33i}$ 
 &
$c_{tu}^{1}$        &   $ \ccc{}{uu}{ii33} +\frac{1}{3} \ccc{}{uu}{i33i} $ \\   
$c_{Qu}^{8}$         &  $  \ccc{8}{qu}{33ii}$
 &
 $c_{Qu}^{1}$     &  $  \ccc{1}{qu}{33ii}$  \\    
 $c_{td}^{8}$        &   $\ccc{8}{ud}{33jj}$ 
  &
 $c_{td}^{1}$          &  $ \ccc{1}{ud}{33jj}$ \\    
 $c_{Qd}^{8}$        &   $ \ccc{8}{qd}{33jj}$ 
 &
 $c_{Qd}^{1}$         &   $ \ccc{1}{qd}{33jj}$\\
         \bottomrule
  \end{tabular}
  \caption{\small Definition of the four-fermion coefficients that enter in
    the fit in terms of the coefficients of Warsaw basis operators of Eq.~(\ref{eq:FourQuarkOp}).
    These coefficients are classified into four-heavy (upper part) and two-light-two-heavy
    (bottom part) operators. The flavour index $i$ is either 1 or 2, 
    and $j$ is either 1, 2 or 3: with our flavour assumptions,  these coefficients will be the same
    regardless of the specific values that $i$ and $j$ take.
\label{tab:oper_fourtop}}
  \end{center}
\end{table}

\section{EWPOs implementation: benchmarking and validation}
\label{app-benchmarking}

\begin{figure}[t]
\centering\includegraphics[width=0.90\linewidth]{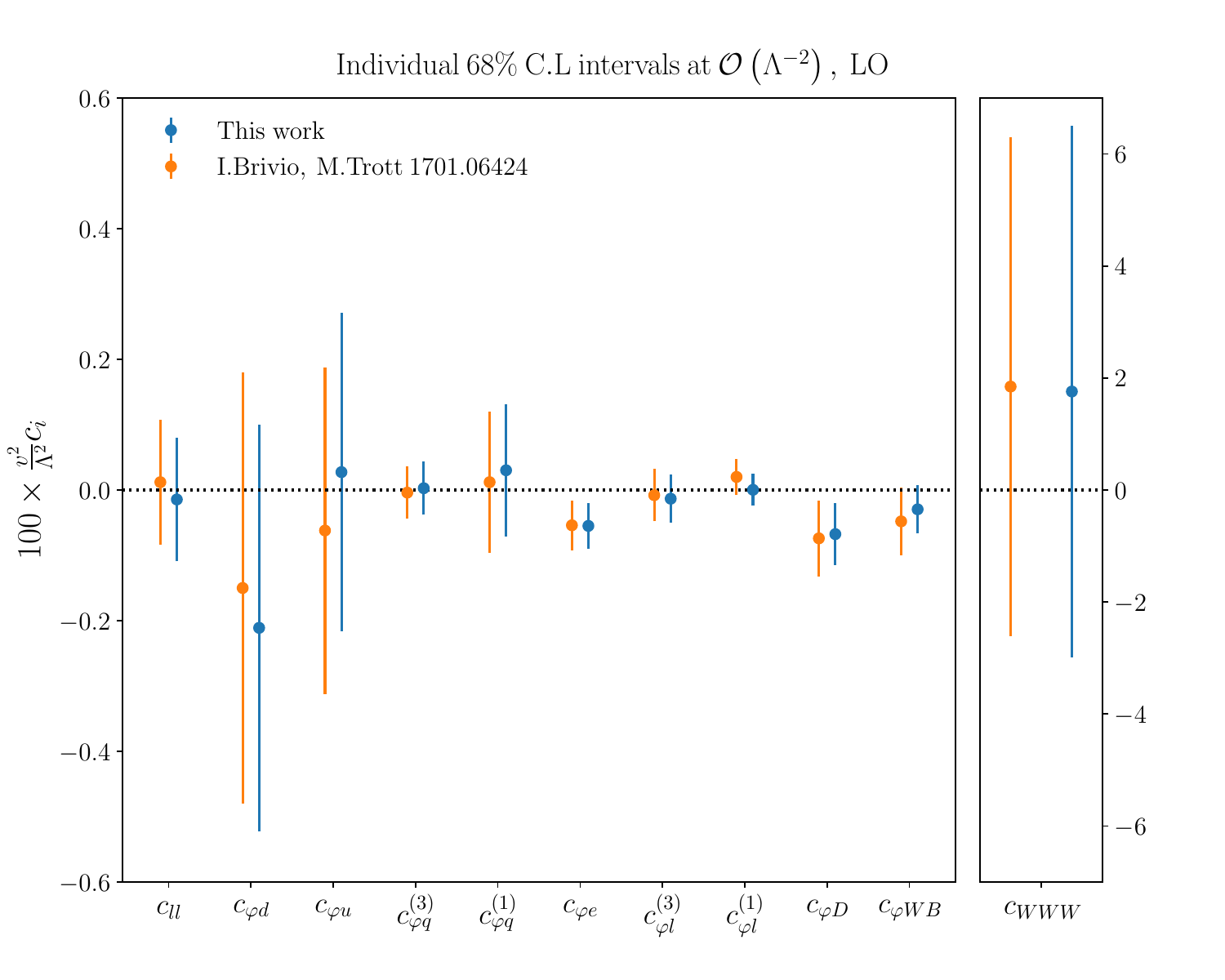}
    \caption{Comparison of a {\sc\small SMEFiT}-based analysis
of the EWPOs from LEP against the corresponding analysis from~\cite{Brivio:2017bnu}.
We display bounds at the $68\%$ CL for one-parameter fits using linear EFT and LO calculations.
One should note that we adopt the convention of~\cite{Brivio:2017bnu} for the
triple gauge operator $O_{WWW}$, which differs from the SMEFT@NLO one by a negative sign. 
For consistency with the settings
adopted in~\cite{Brivio:2017bnu}, the triple
gauge operator coefficient $c_{WWW}$ has been fitted to a subset
of the available LEP data on diboson production composed by only 4 bins.}
    \label{fig:ew-benchmark-comp}
\end{figure}

Our implementation of EWPOs in {\sc\small SMEFiT}, described in Sect.~\ref{sec:ewpos}, has been cross-checked and validated with previous studies, in particular with the analysis of~\cite{Brivio:2017bnu}. 
To highlight the consistency of our implementation with previous results, we provide in Fig.~\ref{fig:ew-benchmark-comp} a comparison of single parameter fits at linear order in the EFT  carried out with {\sc\small SMEFiT} with the EWPO data in Table~\ref{tab:ew-datasets}
against the corresponding analysis from~\cite{Brivio:2017bnu}
based on the same dataset and assuming flavour universality for comparison purposes.
The triple gauge operator $O_{WWW}$ differs from the SMEFT@NLO convention by a minus sign to adopt the same conventions as~\cite{Brivio:2017bnu}, and has been fitted to the same subset of the available LEP data on $WW$ production composed of only 4 bins.
Excellent agreement is found for all operators, further validating the implementation of EWPOs presented in this work.

\section{Projections for the HL-LHC}
\label{app:hl_lhc_projections}

Incorporating projections from future $e^+e^-$ collider measurements into a global SMEFT analysis based on currently available data (LHC Run II with full luminosity) may provide an overly optimistic picture of their impact in the EFT parameter space, given
that HL-LHC program will have already yielded constraints for these operators.
Any impact study for future
colliders therefore needs to build upon a realistic baseline quantifying the constraints on the SMEFT coefficients which will have been achieved by the end of the ``guaranteed'' HL-LHC running operations.

Here we describe the procedure that we have adopted in order to extrapolate existing LHC Run II measurements to the HL-LHC conditions.
These projections are used to quantify  the impact of Higgs, top quark, and diboson production measurements at the HL-LHC on the SMEFT parameter space, and the corresponding results have been presented in Sect.~\ref{sec:impact_hl_lhc_data}.
These projections assume the SM as the underlying theory.
We note that dedicated projections for HL-LHC pseudo-data are available {\it e.g.}~\cite{deBlas:2019rxi,deBlas:2022ofj,DeBlas:2019qco,Durieux:2022cvf} and references therein.
Here we adopt instead a different strategy to ensure consistency with the {\sc\small SMEFiT} analysis settings, based on generating pseudo-data for future LHC runs by means of extrapolating from available Run II datasets.

To this end, we use a new {\sc\small SMEFiT} module generating projections of pseudo-data for future experiments.
This projection is based on extrapolating existing datasets, acquired at the same center-of-mass energy but lower luminosities, for the same underlying process.
The same strategy was adopted  in~\cite{Ethier:2021ydt,Greljo:2021kvv} for the SMEFT impact projections of vector-boson scattering and high-mass Drell-Yan data at the HL-LHC,  as well as in the PDF projections at the  HL-LHC \cite{AbdulKhalek:2018rok} 
and at the Forward Physics Facility of~\cite{Cruz-Martinez:2023sdv,Feng:2022inv}.

This module starts by considering a given available measurement from the LHC Run  II, composed by $n_{\rm bin}$ data points, and with the corresponding SM predictions given by $\mathcal{O}_i^{{\rm (th)}}$.
 The central values for the pseudo-data, denoted by $\mathcal{O}_i^{{\rm (exp)}} $, are obtained
 by fluctuating these theory predictions
 by the fractional statistical ($\delta_i^{\rm (stat)}$) and systematic ($\delta_{k,i}^{\rm (sys)}$)
 uncertainties,
 \begin{equation}
  \label{eq:pseudo_data_v2}
  \mathcal{O}_i^{{\rm (exp)}}
  = \mathcal{O}_i^{{\rm (th)}}
    \left( 1+ r_i \delta_i^{\rm (stat)}
    + \sum_{k=1}^{n_{\rm sys}}
    r_{k,i} \delta_{k,i}^{\rm (sys)}
    \right) \,
    , \qquad i=1,\ldots,n_{\rm bin} \, ,
 \end{equation}
where $r_i$ and $r_{k,i}$ are univariate random Gaussian numbers, whose distribution is such as to reproduce the experimental covariance matrix of the data, and the index $k$ runs over the individual sources of correlated systematic errors. 
We note that theory uncertainties are not included in the pseudo-data generation, and enter only the calculation of the $\chi^2$.

Since one is extrapolating from an existing measurement, whose associated statistical and systematic errors are denoted by $\tilde{\delta}_i^{\rm (stat)}$ and 
$\tilde{\delta}_{k,i}^{\rm (sys)}$, one needs to account for the increased statistics and the expected reduction of the systematic uncertainties for the HL-LHC data-taking period.
The former follows from the increase in integrated luminosity,
\be
\delta_i^{\rm (stat)} = \tilde{\delta}_i^{\rm (stat)} \sqrt{\frac{\mathcal{L}_{\rm Run2}}{\mathcal{L}_{\rm HLLHC}}} \,, \qquad i=1,\ldots, n_{\rm bin} \, ,
\ee
while the reduction of systematic errors is estimated by means of an overall rescaling factor
\be
\delta_{k,i}^{\rm (sys)} = \tilde{\delta}_{k,i}^{\rm (sys)}\times f_{\rm red}^{(k)} \,, \qquad i=1,\ldots, n_{\rm bin} \, ,\quad k=1,\ldots, n_{\rm sys} \, . 
\ee
with $f_{\rm red}^{(k)}$ indicating a correction estimating improvements in the experimental performance, in many cases possible thanks to the larger available event sample.
Here for simplicity we adopt the optimistic scenario considered in the HL-LHC projection studies~\cite{Cepeda:2019klc}, namely $f_{\rm red}^{(k)}=1/2$ for all the datasets.
For datasets without the breakdown of statistical and systematic errors, 
Eq.~(\ref{eq:pseudo_data_v2}) is replaced by 
\begin{equation}
  \label{eq:pseudo_data_v3}
  \mathcal{O}_i^{{\rm (exp)}}
  = \mathcal{O}_i^{{\rm (th)}}
    \left( 1+ r_i \delta_i^{\rm (tot)}
    \right) \,
    , \qquad i=1,\ldots,n_{\rm bin} \, ,
 \end{equation}
 with the total error being reduced
 by a factor $\delta_i^{\rm (tot)}=f_{\rm red}^{{\rm tot}} \times \tilde{\delta}_i^{\rm (tot)}$ with $f_{\rm red}^{{\rm tot}}\sim 1/3$, namely the average of the expected reduction of statistical and systematic uncertainties as compared to the baseline Run II measurements.
 For such datasets, the correlations are neglected in the projections due to the lack of their  breakdown.

 The main benefit of our approach is the possibility to extrapolate the full set of processes entering the current SMEFT global fit, to make sure that all relevant directions in the parameter space are covered in these projections, as well as bypassing the need to evaluate separately the SM and SMEFT theory predictions. 
 One drawback is that it does not account for the possible increase in kinematic range covered by HL-LHC measurements, for example with additional bins in the high-energy region, which can only be considered on a case-by-case basis. 

Tables~\ref{tab:proj-datasets} and~\ref{tab:proj-datasets-top} provide the overview of the LHC Run II datasets on Higgs, top quark, and diboson production datasets  that are adopted as input for the projections at the HL-LHC. 
For each dataset, we specify the corresponding internal {\sc\small SMEFiT} label, the total integrated luminosity, the final state and physical observable, the number of data points and the publication reference.
We only project datasets that are both part of {\sc\small SMEFiT3.0}, as described in Sect.~\ref{sec:updated_global_fit}, and which are based on the highest integrated luminosity for a given process type.
Note that for each process type and final state one can include at most two different projections, given that both ATLAS and CMS will perform independent measurements in the HL-LHC data-taking period.

When generating pseudo-data based on some known underlying law, the total $\chi^2$ values will depend on the random seed used for the pseudo-data generation.
Statistical consistency of the procedure demands that the empirical distribution of these $\chi^2$ values over a large number of different sets of synthetic pseudo-data follows a $\chi^2$ distribution with the appropriate number of degrees of freedom.
This expectation is for instance explicitly verified in the Level-1 closure tests carried out in the context of the NNPDF proton structure analyses~\cite{NNPDF:2021njg,NNPDF:2014otw,DelDebbio:2021whr}.

\begin{figure}[t]
    \centering
    \includegraphics[width=.6\linewidth]{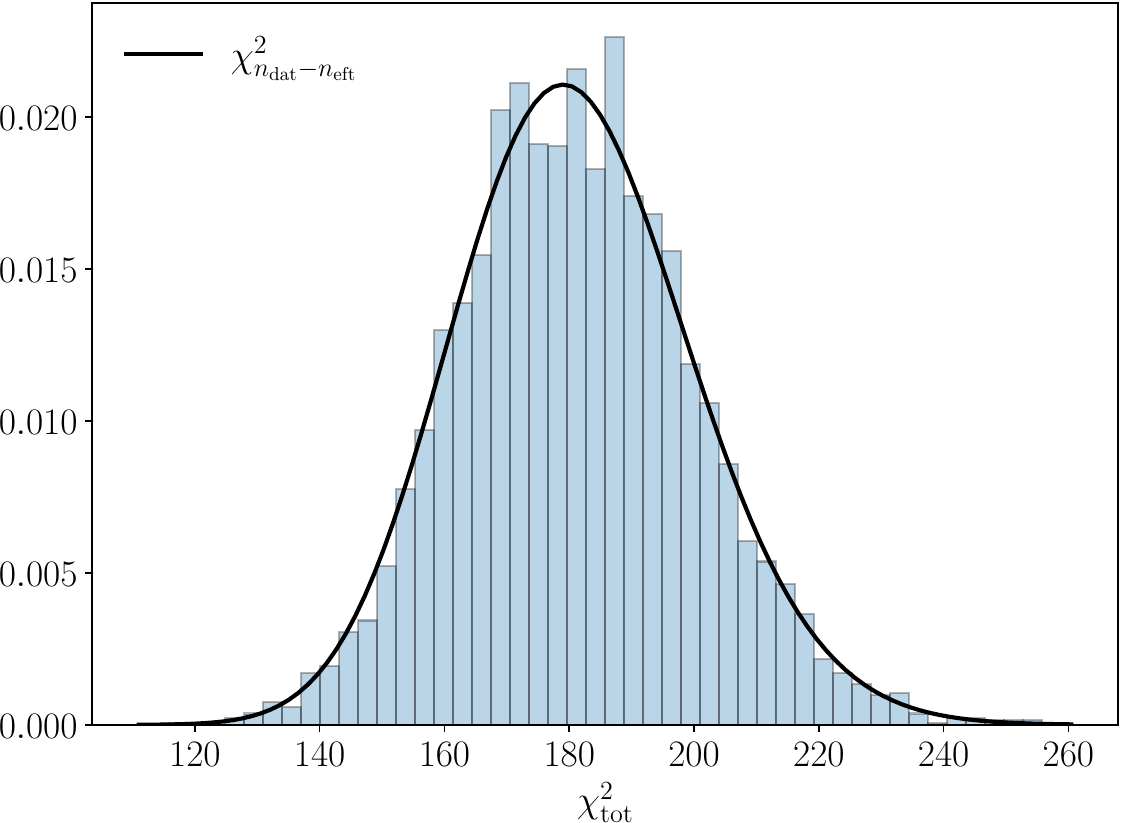}
    \caption{Histogram displaying the distribution of the total $\chi^2_{\rm tot}$ values obtained by repeatedly drawing independent HL-LHC pseudo-datasets following the prescription specified in Eq.~\ref{eq:pseudo_data_v2} and then carrying out the corresponding SMEFT analysis at $\mathcal{O}\lp \Lambda^{-2}\rp$ in the EFT expansion.
    The black curve corresponds to the $\chi^2$ distribution associated to fitting $n_{{\rm dat}}$ data points with $n_{{\rm eft}}$ Wilson coefficients.
    Good agreement between the two distributions is found, demonstrating the statistical consistency of the HL-LHC pseudo-data generation.}
    \label{fig:closure_test_dist}
\end{figure}

To verify this property in the present analysis, Fig.~\ref{fig:closure_test_dist}  displays the histogram with the empirical distribution of the total $\chi^2$ values obtained by repeatedly drawing independent HL-LHC pseudo-datasets, following the prescription specified in this appendix and then carrying out the corresponding SMEFT analysis at linear order in the EFT expansion. 
The black curve corresponds to the  theoretical $\chi^2$ expectation associated to fitting $n_{\rm dat}$ data points with $n_{\rm eft}$ Wilson coefficients.
Excellent agreement between the two distributions is found, demonstrating
the statistical consistency of pseudo-data generation procedure in the {\sc\small SMEFiT} projection module.


\begin{table}[t]
  \centering
  \footnotesize
   \renewcommand{\arraystretch}{1.65}
  \begin{tabular}{l|c|c|c|c|c}
   \toprule
 Dataset  & $\mathcal{L}$ $\lp\rm{fb}^{-1}\rp$  & Info  &  Observables  & $n_{\rm dat}$ & Ref.    \\
\midrule
     \midrule
     \multirow{3}{*}{ {\tt ATLAS\_STXS\_RunII\_13TeV\_2022} }  &\multirow{3}{*}{139}  &\multirow{3}{*}{
  $gg$F, VBF, $Vh$, $t\bar{t}h$, $th$} & $d\sigma/dp_T^h$    &   \multirow{3}{*}{36}    &  \multirow{3}{*}{ \cite{ATLAS:2022vkf} } \\
     &    &  & $d\sigma/dm_{jj}$&     &    \\
     &    &  & $d\sigma/dp_T^V$&     &    \\
    \midrule  
{\tt CMS\_ggF\_aa\_13TeV}&$77.4$&$gg$F, $h\rightarrow \gamma\gamma$&$\sigma_{gg\rm{F}}(p_T^h, N_{\rm jets})$&6&\cite{CMS:2019xnv}\\   
{\tt ATLAS\_ggF\_ZZ\_13TeV}&$79.8$&$gg$F, $h\rightarrow ZZ$&$\sigma_{gg\rm{F}}(p_T^h, N_{\rm jets})$&6&\cite{ATLAS:2019nkf}\\ 
{\tt ATLAS\_ggF\_13TeV\_2015}&$36.1$&$gg$F, $h\rightarrow ZZ, h\rightarrow \gamma\gamma$&$d\sigma(gg\mathrm{F})/dp_T^h$&9&\cite{ATLAS:2018pgp}\\ 
{\tt ATLAS\_WH\_Hbb\_13TeV}&$79.8$&$Wh, h\rightarrow b\bar{b}$&$d\sigma^{(\rm fid)}/dp_T^W$ (stage 1 STXS)&2&\cite{ATLAS:2019yhn}\\
{\tt ATLAS\_ZH\_Hbb\_13TeV}&$79.8$&$Zh, h\rightarrow b\bar{b}$&$d\sigma^{(\rm fid)}/dp_T^Z$ (stage 1 STXS)&2&\cite{ATLAS:2019yhn}\\
{\tt CMS\_H\_13TeV\_2015\_pTH}&$35.9$&$h\rightarrow b\bar{b}, h\rightarrow \gamma \gamma, h\rightarrow ZZ$&$d\sigma/dp_T^h$&9&\cite{CMS:2018gwt}\\
\midrule
{\tt ATLAS\_WW\_13TeV\_2016\_memu}&$36.1$&fully leptonic&$d\sigma^{(\mathrm{fid})}/dm_{e\mu}$&13&\cite{ATLAS:2019rob}\\
{\tt ATLAS\_WZ\_13TeV\_2016\_mTWZ}&$36.1$&fully leptonic&$d\sigma^{(\mathrm{fid})}/dm_T^{WZ}$&6&\cite{ATLAS:2019bsc}\\
{\tt CMS\_WZ\_13TeV\_2016\_pTZ}&$35.9$&fully leptonic&$d\sigma^{(\mathrm{fid})}/dp_T^Z$&11&\cite{CMS:2019efc}\\
{\tt CMS\_WZ\_13TeV\_2022\_pTZ}&$137$&fully leptonic&$d\sigma/dp_T^Z$&11&\cite{CMS:2021icx}\\
\bottomrule
    \end{tabular}
  \caption{\small LHC Run II datasets for Higgs and diboson production ($\sqrt{s}=13$ TeV) used as input to the HL-LHC extrapolations.
  For each dataset, we indicate its internal {\sc\small SMEFiT} label together with its integrated luminosity $\mathcal{L}$ (in $\rm{fb}^{-1}$), the final state or the specific production mechanism, the physical observable, the number of data points, and the corresponding publication reference.
  We only project datasets that are part of {\sc\small SMEFiT3.0}, which are selected on the basis of the highest integrated luminosity for each process type.
     \label{tab:proj-datasets}}
\end{table}

\begin{table}[htbp]
  \centering
   \footnotesize
   \renewcommand{\arraystretch}{1.65}
  \begin{tabular}{l|c|c|c|c|c}
   \toprule
 Dataset  & $\mathcal{L}$ $\lp\rm{fb}^{-1}\rp$  & Info  &  Observables  & $n_{\rm dat}$ & Ref.    \\
\midrule
     \midrule
{\tt ATLAS\_tt\_13TeV\_ljets\_2016\_Mtt}&$36.1$&$
\ell$+jets&$d\sigma/dm_{t\bar{t}}$&7&\cite{ATLAS:2019hxz}\\
{\tt CMS\_tt\_13TeV\_dilep\_2016\_Mtt}&$35.9$&dilepton&$d\sigma/dm_{t\bar{t}}$&7&\cite{CMS:2018adi}\\
{\tt CMS\_tt\_13TeV\_Mtt}&$137$&$
\ell$+jets&$1/\sigma d\sigma/dm_{t\bar{t}}$&14&\cite{CMS:2021vhb}\\
{\tt CMS\_tt\_13TeV\_ljets\_inc}&$137$&$
\ell$+jets&$\sigma(t\bar{t})$&1&\cite{CMS:2021vhb}\\
\midrule
{\tt ATLAS\_tt\_13TeV\_asy\_2022}&$139$&$\ell$ + jets&$A_C$&5&\cite{ATLAS:2022waa}\\
{\tt CMS\_tt\_13TeV\_asy}&$138$&$\ell$ + jets&$A_C$&3&\cite{CMS:2022ged}\\
\midrule
{\tt     ATLAS\_Whel\_13TeV} & 139 &
  $W$-helicity fraction & $F_0, F_L$    &    2   &   \cite{ATLAS:2022rms}\\
\midrule
{\tt ATLAS\_ttbb\_13TeV\_2016}&$36.1$&lepton + jets&$\sigma_{\rm tot}(t\bar{t}b\bar{b})$&1&\cite{ATLAS:2018fwl}\\
{\tt CMS\_ttbb\_13TeV\_2016}&$35.9$&all-jets&$\sigma_{\rm tot}(t\bar{t}b\bar{b})$&1&\cite{CMS:2019eih}\\
{\tt CMS\_ttbb\_13TeV\_dilepton\_inc}&$35.9$&dilepton&$\sigma_{\rm tot}(t\bar{t}b\bar{b})$&1&\cite{CMS:2020grm}\\
{\tt CMS\_ttbb\_13TeV\_ljets\_inc}&$35.9$&lepton + jets&$\sigma_{\rm tot}(t\bar{t}b\bar{b})$&1&\cite{CMS:2020grm}\\
\midrule
{\tt ATLAS\_tttt\_13TeV\_run2}&$139$&multi-lepton&$\sigma_{\rm tot}(t\bar{t}t\bar{t})$&1&\cite{ATLAS:2020hpj}\\
{\tt CMS\_tttt\_13TeV\_run2}&$137$&same-sign or multi-lepton&$\sigma_{\rm tot}(t\bar{t}t\bar{t})$&1&\cite{CMS:2019rvj}\\
{\tt ATLAS\_tttt\_13TeV\_slep\_inc}&$139$&single-lepton&$\sigma_{\rm tot}(t\bar{t}t\bar{t})$&1&\cite{ATLAS:2021kqb}\\
{\tt CMS\_tttt\_13TeV\_slep\_inc}&$35.8$&single-lepton&$\sigma_{\rm tot}(t\bar{t}t\bar{t})$&1&\cite{CMS:2019jsc}\\
{\tt ATLAS\_tttt\_13TeV\_2023}&$139$&multi-lepton&$\sigma_{\rm tot}(t\bar{t}t\bar{t})$&1&\cite{ATLAS:2023ajo}\\
{\tt CMS\_tttt\_13TeV\_2023}&$139$&same-sign or multi-lepton&$\sigma_{\rm tot}(t\bar{t}t\bar{t})$&1&\cite{CMS:2023ftu}\\
\midrule
  {\tt CMS\_ttZ\_13TeV\_pTZ}&$77.5$&$t\bar{t}Z$&$d\sigma(t\bar{t}Z)/dp_T^Z$&4&\cite{CMS:2019too}\\
{\tt ATLAS\_ttZ\_13TeV\_pTZ}&$139$&$t\bar{t}Z$&$d\sigma(t\bar{t}Z)/dp_T^Z$&7&\cite{ATLAS:2021fzm}\\
\midrule
{\tt ATLAS\_ttW\_13TeV\_2016}&$36.1$&$t\bar{t}W$&$\sigma_{\mathrm{tot}}(t\bar{t}W)$&1&\cite{ATLAS:2019fwo}\\
{\tt CMS\_ttW\_13TeV}&$35.9$&$t\bar{t}W$&$\sigma_{\mathrm{tot}}(t\bar{t}W)$&1&\cite{CMS:2017ugv}\\
\midrule
{\tt ATLAS\_t\_tch\_13TeV\_inc}&$3.2$&$t$-channel&$\sigma_{\mathrm{tot}}(tq), \sigma_{\mathrm{tot}}(\bar{t}q)$&2&\cite{ATLAS:2016qhd}\\ 
{\tt CMS\_t\_tch\_13TeV\_2019\_diff\_Yt}&$35.9$&$t$-channel&$d\sigma/d|y_t|$&5&\cite{CMS:2019jjp}\\ 
{\tt ATLAS\_t\_sch\_13TeV\_inc}&$139$&$s$-channel&$\sigma(t+\bar{t})$&1&\cite{ATLAS:2022wfk}\\ 
\midrule
{\tt ATLAS\_tW\_13TeV\_inc}&$3.2$&multi-lepton&$\sigma_{\mathrm{tot}}(tW)$&1&\cite{ATLAS:2016ofl}\\
{\tt CMS\_tW\_13TeV\_inc}&$35.9$&multi-lepton&$\sigma_{\mathrm{tot}}(tW)$&1&\cite{CMS:2018amb}\\
{\tt CMS\_tW\_13TeV\_slep\_inc}&$36$&single-lepton&$\sigma_{\mathrm{tot}}(tW)$&1&\cite{CMS:2021vqm}\\
\midrule
{\tt ATLAS\_tZ\_13TeV\_run2\_inc}&$139$&multi-lepton + jets&$\sigma_{\rm fid}(t\ell^+\ell^- q)$&1&\cite{ATLAS:2020bhu}\\
{\tt     CMS\_tZ\_13TeV\_pTt}  & 138 &
  multi-lepton + jets& $d\sigma_{\rm fid}(tZj)/dp_T^t$    &    3   &   \cite{CMS:2021ugv}  \\
\bottomrule
    \end{tabular}
  \caption{\small Same as Table \ref{tab:proj-datasets} for processes involving top quark production.
     \label{tab:proj-datasets-top}}
\end{table}


By analogy with the terminology adopted in the NNPDF closure test analyses, here we denote projections based on fluctuating the pseudo-data around the associated theory predictions, Eqns.~(\ref{eq:pseudo_data_v2}) and~(\ref{eq:pseudo_data_v3}), as `Level-1' projections.
This is the baseline strategy used in the EFT projections presented in this paper, and corresponds to the expected sensitivity obtained from the analysis of the actual data in the case of perfect compatibility with the SM theory predictions.
An alternative strategy, which corresponds to what is done in other projection studies, corresponds to fitting without fluctuating the pseudo-data,  namely  
\begin{equation}
  \label{eq:pseudo_data_v4}
  \mathcal{O}_i^{{\rm (exp)}}
  = \mathcal{O}_i^{{\rm (th)}} \,
    , \qquad i=1,\ldots,n_{\rm bin} \, ,
 \end{equation}
instead of Eqns.~(\ref{eq:pseudo_data_v2}) and~(\ref{eq:pseudo_data_v3}).
We denote projections based on Eq.~(\ref{eq:pseudo_data_v4}) as `Level-0' projections.
These are now independent of the random seed used for the pseudo-data generation and result in a post-fit value of $\chi^2\simeq 0$, but may be overly optimistic since they miss the layer of statistical fluctuations which accompanies any real measurement.

We have assessed the impact  of HL-LHC and FCC-ee pseudo-data based on either Level-1 or Level-0 projections, finding in general excellent agreement.
The only differences arise at the level of the individual fits carried out at the $\mathcal{O}\lp \Lambda^{-4}\rp$ level, where for some coefficients the impact of the pseudo-data becomes somewhat more marked in the Level-0 projections.
%
\begin{figure}[t]
    \centering
    \includegraphics[width=0.85\linewidth]{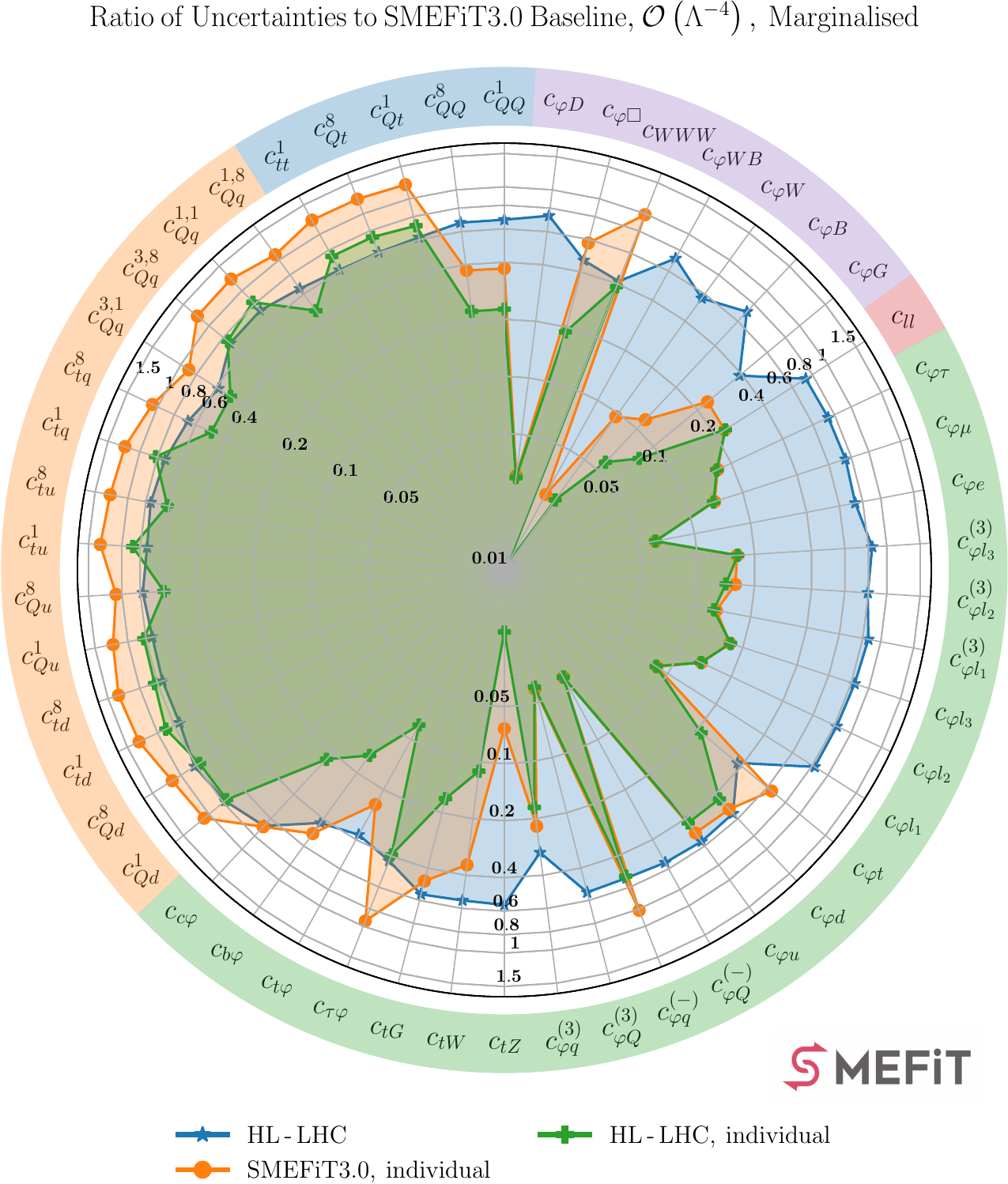}
    \caption{Same as Fig.~\ref{fig:spider_nlo_quad_glob}, but now based on Level-0 pseudo-data projections.
    }
    \label{fig:spider_hllhc_nlo_quad_L0}
\end{figure}
This feature is illustrated by
Fig.~\ref{fig:spider_hllhc_nlo_quad_L0}, which
provides the same information as the HL-LHC projections of Fig.~\ref{fig:spider_nlo_quad_glob} now based on Level-0 pseudo-data. 

Comparing pair-wise these two plots, one finds that the sensitivity estimates for the marginalised (quadratic) fits are unchanged, but that some differences are visible for the individual fits.
For instance, in the Level-0 projections one can see a clear improvement in the bounds associated to the two-light-two-heavy operators thanks to the HL-LHC pseudo-data, which is not necessarily the case in the corresponding Level-1 projections.
Level-0 projections quantifying the impact of the FCC-ee pseudo-data are very similar to the Level-1 ones shown in Fig.~\ref{fig:spider_fcc_nlo_quad_glob}, and hence are not considered here.
All in all, differences between Level-1 and Level-0 projections are very moderate and restricted mostly to the two-light-two-heavy and four-heavy operators in the individual quadratic EFT fits with HL-LHC pseudo-data. 
\section{FCC-ee and CEPC projections: observables and uncertainties}
\label{app:observables}

Here we provide the  list of observables that are being considered in the SMEFT analyses with FCC-ee or CEPC pseudo-data presented in Sect.~\ref{sec:results}.
We describe only the measurements included at the level of total cross-sections or signal strengths, since differential
distributions in $W^+W^-$ and $t\bar{t}$ production are accounted for by means of the unbinned optimal observables formalism and discussed in App.~\ref{app:optimal}.
For each process, we also indicate the experimental uncertainties that have been provided by the corresponding experimental collaborations.

To begin with, Table~\ref{tab:FCCee_EWPOs} displays the $Z$-pole ($\sqrt{s}=91.2$ GeV) electroweak precision observables for the FCC-ee and CEPC considered in this analysis, corresponding to the running scenarios of these two colliders described in Sect.~\ref{sec:future_projections}.
For each observable, we indicate the expected absolute ($\Delta$) or relative ($\delta$) statistical and experimental systematic uncertainties.

\begin{table}[h!]
\centering
\small 
 \renewcommand{\arraystretch}{1.30}
\begin{tabular}{|l|l|l|}
\toprule
\multicolumn{3}{|c|}{$Z$-pole EWPOs ($\sqrt{s}=91.2$ GeV)} \\ 
\midrule
\multirow{2}{*}{$\qquad\quad \mathcal{O}_i$}  & \multicolumn{2}{c|}{$\delta/\Delta~\mathcal{O}_i$}  \\
  & FCC-ee $\qquad\qquad\qquad$  & CEPC $\qquad\qquad\qquad$ \\\hline
\multicolumn{1}{|c|}{$\alpha (m_Z)^{-1}\, (\times10^3)$} &  $\Delta= 2.7$ (1.2)       & $\Delta=17.8$ \\
\multicolumn{1}{|c|}{$\Gamma_W$ (MeV)}                          &  $\Delta= 0.85$ (0.3)       & $\Delta= 1.8$ (0.9)  \\
\multicolumn{1}{|c|}{$\Gamma_Z$ (MeV)}                          &   $\Delta= 0.0028$ (0.025)      &  $\Delta=0.005$ (0.025)  \\
\multicolumn{1}{|c|}{$ A_e\,(\times10^5)$}                          &    $\Delta= 0.5$ (2)     &  $\Delta=  1.5  $   \\
\multicolumn{1}{|c|}{$ A_\mu\,(\times10^5)$}                          &  $\Delta= 1.6$ (2.2)       &   $\Delta= 3.0 ~(1.8) $  \\
\multicolumn{1}{|c|}{$ A_\tau\,(\times10^5)$}                          &   $\Delta= 0.35$ (20)      & $\Delta= 1.2~ (6.9)$ \\
\multicolumn{1}{|c|}{$ A_b\,(\times10^5)$}                         &   $\Delta= 1.7$ (21)      &  $\Delta= 3~ (21)$  \\ 
\multicolumn{1}{|c|}{$A_c\,(\times10^5)$} 
               &   $\Delta= 14$ (15)      &  
               $\Delta=  6 ~(30)$  \\ 
\multicolumn{1}{|c|}{$ \sigma_{\text{had}}^{0}$ (pb)}                & $\Delta= 0.025~(4)$ &  $\Delta= 0.05 ~(2)$ \\
\multicolumn{1}{|c|}{$R_e\,(\times 10^3)$}                        & $\delta=0.0028$ (0.3) &  $\delta=0.003~(0.2)$ \\
\multicolumn{1}{|c|}{$R_\mu\,(\times 10^3)$} & $\delta=0.0021$ (0.05) &  $\delta=0.003 ~(0.1)$ \\
\multicolumn{1}{|c|}{$R_\tau\,(\times 10^3)$} & $\delta=0.0021$ (0.1) & $\delta=0.003 ~(0.1)$ \\
\multicolumn{1}{|c|}{$ R_b\,(\times 10^3)$} & $\delta=0.001$ (0.3)  & $\delta=0.005 ~(0.2)$ \\
\multicolumn{1}{|c|}{$ R_c\,(\times 10^3)$} & $\delta=0.011$ (1.5)  & $\delta=0.02$ (1) \\
\bottomrule
\end{tabular}
\caption{The $Z$-pole ($\sqrt{s}=91.2$ GeV) electroweak precision observables for the FCC-ee and CEPC considered in this analysis, 
 corresponding to the running scenarios described in Sect.~\ref{sec:future_projections}.
For each observable, we indicate the expected statistical (systematic) experimental uncertainty, either in absolute units
($\Delta$) 
or relative to the central value ($\delta$).
See Sect.~\ref{sec:ewpos} for the definitions of these observables.
}
\label{tab:FCCee_EWPOs}
\end{table}

Then Table~\ref{tab:FCCee_2fprod} provides analogous information to Table~\ref{tab:FCCee_EWPOs} for
light fermion pair production
at $\sqrt{s}=240$ and 365 GeV.
For each of the di-fermion final states considered ($e^+e^-$, $\mu^+\mu^-$,
$\tau^+\tau^-$, $c\bar{c}$, and $b\bar{b}$), we indicate the absolute uncertainty both for the total cross-section and for
the forward-backward asymmetry.
For each final state we assume a total selection efficiency of 0.98, 0.98, 0.9, 0.03 and 0.15 respectively, and an acceptance angle around the beam corresponding to  $|\cos(\theta)| = 0.9, 0.95, 0.9, 0.9, 0.9$, for all energies and collider scenarios.   
The central values are assumed to correspond to the SM predictions, as for all the other observables considered in this work.

\begin{table}[h!]
\centering
\small
 \renewcommand{\arraystretch}{1.30}
\begin{tabular}{|l|l|l|l|l|}
\toprule
\multicolumn{5}{|c|}{$e^+e^- \to f\bar{f}$} \\ 
\midrule
& \multicolumn{2}{|c|}{$\sqrt{s}=240$ GeV} 
& \multicolumn{2}{c|}{$\sqrt{s}=365$ GeV }\\ 
\midrule
$\qquad \quad \mathcal{O}_i$  & $\Delta_{\exp}\mathcal{O}_i$ (FCC-ee) & $\Delta_{\exp}\mathcal{O}_i$ (CEPC)
&
$\Delta_{\exp}\mathcal{O}_i$ (FCC-ee) & $\Delta_{\exp}\mathcal{O}_i$ (CEPC)
\\ 
\midrule
$\sigma_{\rm tot}(e^+e^-)$ [fb] & 2.29 & 1.62 & 2.74 & 4.68 \\
$A_{\rm FB}(e^+e^-)$ & $9.79\cdot 10^{-6}$ & $6.92\cdot 10^{-6}$ & $2.83\cdot 10^{-5}$ & $4.83\cdot 10^{-5}$ \\
$\sigma_{\rm tot}(\mu^+\mu^-)$ [fb] & 0.405 & 0.287 & 0.48 & 0.82 \\
$A_{\rm FB}(\mu^+\mu^-)$ & $1.98\cdot 10^{-4}$ & $1.397\cdot 10^{-4}$ & $5.69\cdot 10^{-4}$ & $9.7\cdot 10^{-4}$ \\
$\sigma_{\rm tot}(\tau^+\tau^-)$ [fb] & 0.374 & 0.264 & 0.443 & 0.756 \\
$A_{\rm FB}(\tau^+\tau^-)$ & $2.17\cdot 10^{-4}$ & $1.53\cdot 10^{-4}$ & $6.24\cdot 10^{-4}$ & 0.00106 \\
$\sigma_{\rm tot}(c\bar{c})$ [fb] & 0.088 & 0.062 & 0.102 & 0.175 \\
$A_{\rm FB}(c\bar{c})$ & 0.000813 & $5.74\cdot 10^{-4}$ & 0.00238 & 0.00405 \\
$\sigma_{\rm tot}(b\bar{b})$ [fb]  & 0.151 & 0.107 & 0.171 & 0.29 \\
$A_{\rm FB}(b\bar{b})$ & $4.86\cdot 10^{-4}$ & $3.44\cdot 10^{-4}$ & 0.00142 & 0.00243 \\
\bottomrule
\end{tabular}
\caption{Same as Table~\ref{tab:FCCee_EWPOs} for light-fermion pair production at $\sqrt{s}=240$ and 365 GeV.
For each of the di-fermion final states considered, we indicate the
absolute statistical uncertainty both for
the total cross-section and for
the forward-backward asymmetry $A_{\rm FB}$
defined in Eq.~(\ref{eq:Rbc}).
As for all projections considered in this work, the corresponding central values are set to be equal to the SM predictions.
}
\label{tab:FCCee_2fprod}
\end{table}

We list the projections for Higgs production measurements at the FCC-ee and the CEPC in  Table~\ref{tab:FCCee_Higgs}.
We display the forecast relative experimental uncertainties  for the inclusive production    cross-section in the $Zh$ channel, $\sigma_{Zh}$, and for the corresponding Higgs signal strengths (production times decay)
at $\sqrt{s}=240$ and 365 GeV, for different final states, for both the associated production ($Zh$) and vector-boson fusion ($h\nu\nu$) channels.
The final states considered are $b\bar{b}$,
$c\bar{c}$, $gg$, $ZZ$, $WW$, $\tau^+\tau^-$, $\gamma Z$ and $\gamma\gamma$.
For the $\sqrt{s}=240$ data-taking period, only the  $b\bar{b}$ final state is available in the vector-boson fusion production channel.

\begin{table}[htbp]
\centering
\small
 \renewcommand{\arraystretch}{1.60}
\begin{tabular}{|l|c|c|c|c|}
\toprule
\multicolumn{5}{|c|}{$e^+e^- \to Zh$} \\ 
\midrule
& \multicolumn{2}{|c|}{$\sqrt{s}=240$ GeV} 
& \multicolumn{2}{c|}{$\sqrt{s}=365$ GeV }\\ 
\midrule
$\mathcal{O}_i$  & $\delta_{\exp}\mathcal{O}_i$ (FCC-ee) & $\delta_{\exp}\mathcal{O}_i$ (CEPC)
&
$\delta_{\exp}\mathcal{O}_i$ (FCC-ee) & $\delta_{\exp}\mathcal{O}_i$ (CEPC)
\\ 
\midrule
$\sigma_{Zh}$  & 0.0035 & 0.0026 & 0.0064 & 0.014 \\
$\sigma_{Zh}\times {\rm BR}_{b\bar{b}}$  & 0.0021 & 0.0014 & 0.0035 & 0.009 \\
$\sigma_{Zh}\times {\rm BR}_{c\bar{c}}$  & 0.0156 & 0.0202 & 0.046 & 0.088 \\
$\sigma_{Zh}\times {\rm BR}_{gg}$  & 0.0134 & 0.0081 & 0.0247 & 0.034 \\
$\sigma_{Zh}\times {\rm BR}_{ZZ}$  & 0.0311 & 0.0417 & 0.0849 & 0.2\\
$\sigma_{Zh}\times {\rm BR}_{WW}$  & 0.0085 & 0.0053 & 0.0184 & 0.028 \\
$\sigma_{Zh}\times {\rm BR}_{\tau^+\tau^-}$  & 0.0064 & 0.0042 & 0.0127 & 0.021 \\
$\sigma_{Zh}\times {\rm BR}_{\gamma\gamma}$  & 0.0636 & 0.0302 & 0.127 & 0.11 \\
$\sigma_{Zh}\times {\rm BR}_{\gamma Z}$  & 0.12 & 0.085 & - & -\\
\midrule
\multicolumn{5}{|c|}{$e^+e^- \to h\nu\nu$} \\ 
\midrule
& \multicolumn{2}{|c|}{$\sqrt{s}=240$ GeV} 
& \multicolumn{2}{c|}{$\sqrt{s}=365$ GeV }\\ 
\midrule
$\mathcal{O}_i$  & $\delta_{\exp}\mathcal{O}_i$ (FCC-ee) & $\delta_{\exp}\mathcal{O}_i$ (CEPC)
&
$\delta_{\exp}\mathcal{O}_i$ (FCC-ee) & $\delta_{\exp}\mathcal{O}_i$ (CEPC)
\\ 
\midrule
$\sigma_{h\nu\nu}\times {\rm BR}_{b\bar{b}}$  & 0.0219 & 0.0159 & 0.0064 & 0.011 \\
$\sigma_{h\nu\nu}\times {\rm BR}_{c\bar{c}}$  & - & - & 0.0707 & 0.16 \\
$\sigma_{h\nu\nu}\times {\rm BR}_{gg}$  & - & - & 0.0318 & 0.045 
\\$\sigma_{h\nu\nu}\times {\rm BR}_{ZZ}$  & - & - & 0.0707 & 0.21 \\
$\sigma_{h\nu\nu}\times {\rm BR}_{WW}$  & - & - & 0.0255 & 0.044 \\
$\sigma_{h\nu\nu}\times {\rm BR}_{\tau^+\tau^-}$  & - & - & 0.0566 & 0.042 \\
$\sigma_{h\nu\nu}\times {\rm BR}_{\gamma\gamma}$  & - & - & 0.156 & 0.16 \\
\bottomrule
\end{tabular}
    \caption{Same as Table~\ref{tab:FCCee_EWPOs} for the relative statistical experimental uncertainties projected for the inclusive production
    cross-section in the $Zh$ channel, $\sigma_{Zh}$,
    and for the corresponding
    Higgs signal strengths
    (production times decay)
at $\sqrt{s}=240$ and 365 GeV,
for different final states.
The signal strengths are provided for both the 
associated production ($Zh$)
and the vector-boson fusion ($h\nu\nu$) channels. 
The corresponding central values are taken from the SM predictions.}
    \label{tab:FCCee_Higgs}
\end{table}

Finally, we also consider in our analysis the inclusive cross section of $W^+W^-$ production and the individual leptonic $W$ branching ratios. 
We indicate in Table~\ref{tab:FCCee_WW} the forecast relative uncertainties for these observables at FCC-ee and CEPC for the
three relevant center of mass energies: $\sqrt{s}=161$ GeV, 240 GeV, and 365 GeV.
The leptonic branching ratio uncertainty is assumed to be the same for each of the three lepton generations, while the uncertainty associated to  hadronic $W$ decays is constrained from enforcing the lack of exotic decays, i.e. Eq.~\ref{eq:no_exotic_Wdecay}.
The uncertainties listed in Table~\ref{tab:FCCee_WW} for the FCC-ee were extracted from~\cite{DeBlas:2019qco}, and in the case of CEPC were rescaled by the appropriate luminosity factors to match with its baseline running scenario. 
%
%
We note that the measurements indicated in Table~\ref{tab:FCCee_WW} do not overlap with the optimal observables from $W^+W^-$ production described in App.~\ref{app:optimal}, since the latter are normalised and implemented at the level of undecayed $W$ bosons.

\begin{table}[htbp]
\centering
\small 
 \renewcommand{\arraystretch}{1.40}
\begin{tabular}{|l|c|c|c|c|c|c|}
\toprule
\multicolumn{7}{|c|}{$e^+e^- \to W^+W^-$} \\ 
\midrule
\multirow{2}{*}{ $\mathcal{O}_i$ }  
& \multicolumn{2}{c|}{$\sqrt{s}=161$ GeV} & \multicolumn{2}{c|}{$\sqrt{s}=240$ GeV} 
& \multicolumn{2}{c|}{$\sqrt{s}=365$ GeV }\\ 
\cline{2-7}
 & $\delta_{\exp}$ (FCC-ee) & $\delta_{\exp}$ (CEPC) & $\delta_{\exp}$ (FCC-ee) & $\delta_{\exp}$ (CEPC)
& $\delta_{\exp}$ (FCC-ee) & $\delta_{\exp}$ (CEPC)
\\ 
\midrule
$\sigma_{WW}$  & $1.36\cdot 10^{-4}$ & $2.48\cdot 10^{-4}$ & $1.22\cdot 10^{-4}$ & $8.63\cdot 10^{-5}$ & $2.81\cdot 10^{-4}$ & $4.87\cdot 10^{-4}$ \\
BR$_{W\to\ell_i\nu_i}$ & $2.72\cdot 10^{-4}$ & $4.95\cdot 10^{-4}$ & $2.44\cdot 10^{-4}$ & $1.73\cdot 10^{-4}$ & $5.63\cdot 10^{-4}$ & $9.75\cdot 10^{-4}$ \\
\bottomrule
\end{tabular}
    \caption{Same as Table~\ref{tab:FCCee_EWPOs} for the relative experimental uncertainties associated to the measurements of the inclusive $e^+e^-\to W^+W^-$ cross section and the leptonic $W$ branching ratio at FCC-ee and CEPC.
    The leptonic branching fraction measurements are considered separately for the three
    generations.}
    \label{tab:FCCee_WW}
\end{table}

To summarise, for the projections of both FCC-ee and CEPC, we consider the $Z$-pole EWPOs, fermion pair production at $\sqrt{s}=240$ GeV and $365$ GeV, $W^+W^-$ production (including branching fractions) at $\sqrt{s}=161$ GeV, $240$~GeV and $365$ GeV, and Higgs production and decay in various final states (including signal strengths) at $\sqrt{s}=240$~GeV and $365$ GeV.
Furthermore, we also include the optimal observables for normalised differential distributions in diboson and top-quark pair production, as discussed in App.~\ref{app:optimal}.
\section{Optimal observables at future lepton colliders}
\label{app:optimal}

For processes well described by linear EFT corrections, with negligible systematic uncertainties, and where the kinematic information is available on an event-by-event basis, one can use the technique of ``optimal observables''~\cite{Diehl:1993br}. 
This technique maximizes the sensitivity to the EFT coefficients by exploiting the information contained in the fully differential cross-section. 
In this work, the projected differential measurements of both $W^+W^-$ and $ t\bar t$ production at FCC-ee and CEPC are included in terms of such optimal observables in the same manner as in~\cite{DeBlas:2019qco,deBlas:2022ofj}. 
In this appendix we summarize our treatment of these observables in the global SMEFT analysis, keeping the discussion fully general.

\paragraph{Optimal observable definition.}
For physical observables characterised by a linear dependence on $n_{\rm eft}$ EFT coefficients,  the fully differential distribution can be expressed as
\begin{equation}
    \frac{\rm d \sigma_{\rm th}}{\rm d \Phi} = \left( S_{\rm SM}(\Phi)+\sum_{k=1}^{n_{\rm eft}} c_k S_k(\Phi)\right) \equiv S(\Phi) \, ,
    \label{eq:App_linearEFT}
\end{equation}
where $\Phi$ represents the final-state kinematic variables, $S_{\rm SM}(\Phi)$ is the SM squared matrix element (including phase space factors) and $S_k(\Phi)$  the corresponding linear (interference) EFT matrix element.
Eq.~(\ref{eq:App_linearEFT}) assumes that quadratic EFT corrections and beyond can be neglected, and factors of $\Lambda^{-2}$ have been reabsorbed in the Wilson coefficient $c_k$. 
These observables may be extended to the case of quadratic EFT corrections via Machine Learning techniques~\cite{Chen:2020mev,GomezAmbrosio:2022mpm,Chen:2023ind,Chai:2024zyl}, but we have verified that keeping only the linear piece is a good approximation for the processes we consider here.

We consider now an experiment that measures $n_{\rm ev}$ events, each one characterised by their own set of kinematic final-state variables, ${\Phi_1, ..., \Phi_{n_{\rm ev}}}$, such as rapidities and
transverse momenta. 
Depending on the type of measurement,
$\Phi_i$ could represent parton-level,
particle-level, or detector-level
variables.
The optimal observables are constructed on an event-by-event basis as
\begin{equation}
    \mathcal{O}^{(\rm exp)}_{k,i} =  \frac{S_k(\Phi_i)}{S_{\rm SM}(\Phi_i)} \equiv  \frac{S_{k,i}}{S_{{\rm SM},i}} \, , \quad i=1,\ldots,n_{\rm ev}
    \, , \quad k=1,\ldots,n_{\rm eft}\, ,
\end{equation}
that is, as the ratio between the linear EFT and the SM matrix elements, evaluated at the kinematics of the corresponding event $\Phi_i$.
%
%
The expectation value of these observables 
is then obtained by averaging over
the full set of events, namely
\begin{equation}
    \left\langle \mathcal{O}^{(\rm exp)}_{k}\right\rangle = \frac{1}{n_{\rm ev}} \sum_{i=1}^{n_{\rm ev}} \frac{S_k(\Phi_i)}{S_{\rm SM}(\Phi_i)} \equiv \frac{1}{n_{\rm ev}} \sum_{i=1}^{n_{\rm ev}} \frac{S_{k,i}}{S_{{\rm SM},i}} \, , \qquad k=1,\ldots,n_{\rm eft} \, ,
    \label{app:OO_definition_exp}
\end{equation}
separately for each of the relevant Wilson coefficients.
For a given integrated luminosity 
$\mathcal{L}$ and fiducial (SM) cross-section
 $\sigma_{\rm SM}$, the total number of events $n_{\rm ev}$ follows a Poisson distribution with mean  $\lambda= n_{\rm ev}=\mathcal{L}\,\sigma_{\rm SM}$, where we neglect effects such as acceptances and efficiencies.
In the following, we assume for simplicity that $n_{\rm ev}$ corresponds to the mean value of this distribution.

Theoretical predictions for the expectation value of the optimal observables defined in Eq.~(\ref{app:OO_definition_exp}) can be evaluated either analytically or numerically, 
depending
on the complexity of the calculation.
Here, since we are making projections for future colliders, we assume that the measured experimental values agree with the SM theoretical prediction. 
As explained below, this allows us to rotate both the observables and the inverse covariance matrix in a basis such that the rotated observables
coincide with the Wilson coefficients,
$\widetilde{\mathcal{O}}_k=c_k$.
Expressed in this basis, the inverse experimental covariance matrix takes the simple form~\cite{Diehl:1993br,DeBlas:2019qco},
\be
\label{eq:app_covmat}
\lp {\rm cov}_{\rm exp}\lp c_k, c_{k'} \rp\rp ^{-1} = \mathcal{L}\lp\int \frac{ S_{k} \, S_{k'} }{ S_{\rm SM} }\, {\rm d \Phi} -\frac{1}{\sigma_{\rm SM}} \int S_{k} {\rm d \Phi} \int S_{k'} {\rm d \Phi}  \rp.
\ee
For relatively simple processes, the phase space integrals in Eq.~(\ref{eq:app_covmat}) can be computed analytically. 
The final contribution to the global fit figure of merit arising from
these optimal observables is then
\be
\label{eq:appOO_figmerit}
\Delta \chi^2 \lp {\boldsymbol{c}}\rp = 
\sum_{k,k'=1}^{n_{\rm eft}} c_k\lp {\rm cov}_{\rm exp}\lp c_k, c_{k'} \rp\rp ^{-1}c_{k'} \, ,
\ee
where, since we assume the SM, in the rotated basis the experimental values for the Wilson coefficients vanish.

\paragraph{Observable rotation.}
\label{subsec:app-optimal_general_redef}
The evaluation of the contribution to the total $\chi^2$ from the optimal observables is performed most efficiently in a rotated basis, where these observables are matched to each one of the individual Wilson coefficients entering the theory prediction.
Let one assume that we have a set of $N$ observables that depend linearly on $n_{\rm eft}$ EFT coefficients, that is,
\begin{equation}
    \sigma^{(i)}_{\rm th}=\sigma^{(i)}_{\rm SM} + \sum_{k=1}^{n_{\rm eft}}\sigma^{(i)}_{k} c_k\,,\quad\quad i=1,...,N\,,
    \label{eq:linearobs}
\end{equation}
where $\sigma^{(i)}_{k}$ is the linear EFT cross-section, and again factors of $\Lambda^{-2}$ are reabsorbed in the Wilson coefficient.
For simplicity, we assume that the measured values of these $N$ observables coincide with the corresponding SM prediction, $\sigma^{(i)}_{\text{exp}} = \sigma^{(i)}_{\rm SM}$.
Furthermore, the $N$ observables are distributed according to a multi-Gaussian likelihood.
In this case, we can express the contribution of these $N$ observables to the figure of merit of the global EFT fit as follows
\begin{align}
    \Delta\chi^2 &=
    \sum_{i,j=1}^N \lp \sigma^{(i)}_{\rm exp} - \sigma^{(i)}_{\rm th} \rp
    \lp \text{cov}_{\rm exp}\rp^{-1}_{ij}
    \lp \sigma^{(j)}_{\rm exp} - \sigma^{(j)}_{\rm th} \rp
    \nonumber \\
    &= \sum_{i,j=1}^N\left( \sum_{k=1}^{n_{\rm eft}}\sigma^{(i)}_{k} c_k\right)
     \lp \text{cov}_{\rm exp}\rp^{-1}_{ij}
    \left( \sum_{k'=1}^{n_{\rm eft}}\sigma^{(j)}_{k'} c_{k'}\right) \, ,
            \label{eq:trafochi2}
\end{align}
where $\lp \text{cov}_{\rm exp}\rp_{ij}$
is the experimental covariance matrix 
between the observables $\sigma^{(i)}_{\text{exp}}$
and 
$\sigma^{(j)}_{\text{exp}}$.
If the measured quantities did not agree with their SM predictions, the following steps could still be performed after adding and subtracting the mean of the Wilson coefficients and with an additional constant term in $\Delta \chi^2$.

One can rewrite Eq.~(\ref{eq:trafochi2}) in a more compact manner as follows:
\be
\Delta\chi^2= \sum_{k,k'=1}^{n_{\rm eft}} c_k c_{k'}\sum_{i,j=1}^N \sigma^{(i)}_{k} 
     \lp \text{cov}_{\rm exp}\rp^{-1}_{ij}
     \sigma^{(j)}_{k'} \equiv 
     \sum_{k,k'=1}^{n_{\rm eft}} c_k A^{\sigma}_{kk'} c_{k'} \, ,
\ee
in terms of a new covariance matrix, now in the space of Wilson coefficients, defined as
\be
 A^{\sigma}_{kk'}  = \sum_{i,j=1}^N \sigma^{(i)}_{k} 
     \lp \text{cov}_{\rm exp}\rp^{-1}_{ij}
     \sigma^{(j)}_{k'} \, ,\qquad k,k'=1,\ldots,n_{\rm eft} \, ,
     \label{eq:trafochi22}
\ee
and obtained by multiplying the linear  EFT cross-sections with the inverse of the experimental covariance matrix.
The result of Eq.~(\ref{eq:trafochi22})
indicates that one can define new observables in terms of individual Wilson coefficients,
\begin{equation}
    \widetilde{\sigma}^{(k)} = c_k \,,\qquad k=1,\ldots,n_{\rm eft} \, ,
\end{equation}
with $A^{\sigma}_{kk'}$ as inverse covariance matrix, that is
\be
{\rm cov}_{\rm exp}\lp \widetilde{\sigma}^{(k)},\widetilde{\sigma}^{(k')} \rp^{-1} = A^{\sigma}_{kk'} \, .
\ee
Note that the original 
 covariance matrix
 $\lp \text{cov}_{\rm exp}\rp_{ij}$ lives
in the space $N\times N$ spanned
by the experimental data, while
the covariance matrix for
the redefined observables, 
${\rm cov}_{\rm exp}\lp \widetilde{\sigma}^{(k)},\widetilde{\sigma}^{(k')}\rp$, instead lives in the $n_{\rm eft}\times n_{\rm eft}$
space spanned by the Wilson coefficients.
This transformation is only possible
when EFT corrections beyond the linear correction are neglected.

The main advantage of this redefinition is that one can easily evaluate the covariance matrix $A^\sigma_{kk'}$, in the scenario where the observables are statistically limited. Then, Eq.~(\ref{eq:trafochi22}) accounts for them in the global fit, as has been done in Eq.~(\ref{eq:appOO_figmerit}) above.

\paragraph{Implementation.}
The optimal observable framework reviewed in this appendix is used to include in the fit the FCC-ee and CEPC projections for differential distributions in $W^+W^-$ and $t\bar t$ production. 

In the case of $W^+W^-$ production, we consider measurements taken from the $\sqrt{s}=161$~GeV ($W$ pair-production threshold), $240$~GeV and $365$~GeV data-taking periods, each one for the integrated luminosities indicated in Table~\ref{tab:FCCee_runs}. 
We include the fully leptonic and semi-leptonic $W$ decay channels and treat on the same footing the three lepton generations.
We assume that the experiment will be able to measure the momenta of all the final state fermions by assuming on-shell $W$s, i.e. a perfect neutrino reconstruction.
However, we account for the inability to distinguish the quarks produced by a hadronic $W$ decay by
implementing the corresponding angle-folding effect.
Furthermore, we assume an overall signal selection efficiency of $45\%$~and a detector acceptance cut on the polar angle of $|\cos\theta |<0.9$ and $0.95$ for jets and leptons respectively, in agreement with previous studies~\cite{DeBlas:2019qco,deBlas:2022ofj}. The selection efficiency is implemented by rescaling the luminosity while the acceptance cuts determine the phase-space region over which we integrate.
We computed the inverse covariance matrix by performing the required phase space integrals by suitably modifying the phase-space integrator included in {\sc\small MadGraph5}. 
We validated our results against a method based on the reweighting of LHE events like the one used in~\cite{Durieux:2018tev}.

$W^+W^-$ production at $e^+e^-$ colliders is in principle sensitive to eight of the dimension-six operators considered in the SMEFiT global analysis, namely
\begin{equation}
    \mcO_{\varphi D},\,\mcO_{\varphi  WB},\,\mcO_{WWW},\,\mcO_{\varphi\ell_1},\,\,\mcO_{\varphi e},\,\mcO_{\varphi\ell_1}^{(3)},\,\mcO_{\varphi\ell_2}^{(3)},\,\mcO_{\ell\ell} \, .
\end{equation}
However, the linear combination of coefficients $c_{\varphi\ell_1}^{(3)}+\,c_{\varphi\ell_2}^{(3)}-2c_{\ell\ell}$ only modifies the inclusive cross-section via input parameters shifts, to which the optimal observables defined with normalized differential distributions are insensitive. 
Moreover, the coefficients $c_{\varphi\ell_2}^{(3)}$ and $c_{\ell\ell}$ enter only through the previous linear combination and $c_{\varphi\ell_1}^{(3)}$ modifies on its own the $eeZ$ and $e\nu_e W$ interaction vertices. 
These considerations indicate that the optimal observables considered here have associated two flat directions, which we remove from the inverse covariance before the fit.
All in all, the (normalized) optimal observables in $W^+W^-$ production at FCC-ee/CEPC constrain six EFT coefficients:
\begin{equation}
    c_{\varphi D},\,c_{\varphi{ WB}},\,c_{ WWW},\,c_{\varphi\ell_1},\,\,c_{\varphi e},\,c_{\varphi\ell_1}^{(3)},
\end{equation}
in agreement with previous studies~\cite{DeBlas:2019qco,deBlas:2022ofj}.

In the case of top-quark pair production, we took the inverse covariance matrix defining the optimal observables from the Snowmass study~\cite{deBlas:2022ofj}, used there for the {\sc\small HEPfit}~\cite{DeBlas:2019ehy} interpretation of future collider projections.
These optimal observables are non-zero only for $\sqrt{s}=365$~GeV, just above the top pair production threshold.
While for $W^+W^-$ we used the definition for normalized differential distributions, Eq.~\eqref{eq:app_covmat}, for $t\bar{t}$ we use optimal observables that include the information in the inclusive cross-section, which equates to considering only the first term on the right-hand side of Eq.~\eqref{eq:app_covmat}.
They offer sensitivity to eight CP-even dimension-six operators, four of which are semi-leptonic four-fermion operators. Since the latter are not included in the SMEFiT global fit, we neglect them. Hence, the projections for $e^+e^-\to t\bar{t}$ measurements at the FCC-ee and the CEPC constrain the following coefficients
\begin{equation}
    c_{\varphi Q}^{(-)},\,c_{\varphi t},\,c_{tW},\,c_{tZ} \, ,
\end{equation}
hence providing complementary sensitivity to the LHC measurements of top quark production. 
\section{SMEFT effects in the $W$-boson decay widths}
\label{subsec:app-widths}

Here whenever we evaluate theoretical
predictions for processes involving
the production and decay of $W$ bosons
with {\sc\small mg5\_aMC@NLO}, we assume
a constant value of the $W$ boson decay width.
However, this fixed-width approximation
is theoretically inconsistent, since 
it neglects the corrections that SMEFT
dimension-six operators introduce
at the $W$ width level.
In this appendix, we describe our approach to account {\it a posteriori} for SMEFT effects 
entering the $W$-boson decay widths,
which follows the procedure developed in~\cite{Brivio:2019myy}.
We spell out the procedure in the case of $WW$ production in the fully-leptonic channel, i.e. $W\rightarrow \ell \nu$, but
the same approach applies to any
other process involving $W$ boson decays.

The starting point is to assume that one can factorise production and decay effects
in the $WW$ process by writing
\be
\sigma_{WW\rightarrow 4\ell} = X^{(\rm th)} f(\Gamma_W),
\label{eq:factorised_decay}
\ee
with production effects encoded in
\be
X^{(\rm th)} = X^{(\rm sm)} + \sum_{j=1}^{n_{\rm eft}}c_j X_j^{(\rm eft)} + \sum_{j=1}^{n_{\rm eft}}\sum_{k\ge j}^{n_{\rm eft}}c_j c_k X_{jk}^{(\rm eft)} \, ,
\label{eq:eft_theory}
\ee
with $X^{(\rm sm)}$ indicating the SM prediction and $X_j^{(\rm eft)}$ and  $X_{jk}^{(\rm eft)}$ the linear and quadratic SMEFT corrections respectively for
the $n_{\rm eft}$ dimension-six operators relevant for the analysis.
Effects related to the $W$-boson
widths are then encapsulated in the
$f(\Gamma_W)$ term.
In the narrow-width approximation, one has $f(\Gamma_W) = 1/\Gamma_W^2$ in the case of two decaying $W$s, but the actual dependence on the width is in general more convoluted. 

Next, one assumes that SMEFT effects distort
the SM value of the $W$-boson width,
$\Gamma_W^{(\rm sm)}$, by a moderate
amount only.
This allows us to carry out the following
Taylor expansion:
\be
f(\Gamma_W) = f(\Gamma_W^{(\rm sm)} + \delta \Gamma_W)
\approx f\lp \Gamma_W^{(\rm sm)}\rp\lp 1+ \delta_1 \frac{\delta\Gamma_W}{\Gamma_W^{(\rm sm)}} +  \delta_2 \lp \frac{\delta\Gamma_W}{\Gamma_W^{(\rm sm)}}\rp^2 + \mathcal{O}\lp \frac{\delta\Gamma_W}{\Gamma_W^{(\rm sm)}}\rp^3\rp,
\label{eq:width_func_form}
\ee
where $\delta \Gamma_W$ accounts
for SMEFT effects to the decay width, and where terms cubic on $\delta \Gamma_W$ and beyond are neglected. 
Substituting Eqns.~(\ref{eq:eft_theory}) and~(\ref{eq:width_func_form}) into Eq.~(\ref{eq:factorised_decay}) results into
\begin{align}
\nonumber
\sigma_{WW\rightarrow 4\ell} &= \lp X^{(\rm sm)} + \sum_{j=1}^{n_{\rm eft}}c_j X_j^{(\rm eft)} + \sum_{j=1}^{n_{\rm eft}}\sum_{k\ge j}^{n_{\rm eft}}c_j c_k X_{jk}^{(\rm eft)}\rp f\lp \Gamma_W^{(\rm sm)}\rp \lp 1+ \delta_1 \frac{\delta\Gamma_W}{\Gamma_W^{(\rm sm)}} +  \delta_2 \lp \frac{\delta\Gamma_W}{\Gamma_W^{(\rm sm)}}\rp^2\rp\\
&\equiv \lp \sigma_{WW\rightarrow 4\ell}^{\rm(sm)} + \sum_{j=1}^{n_{\rm eft}}c_j Y_j^{(\rm eft)} + \sum_{j=1}^{n_{\rm eft}}\sum_{k\ge j}^{n_{\rm eft}}c_j c_k Y_{jk}^{(\rm eft)}\rp \lp 1+ \delta_1 \frac{\delta\Gamma_W}{\Gamma_W^{(\rm sm)}} +  \delta_2 \lp \frac{\delta\Gamma_W}{\Gamma_W^{(\rm sm)}}\rp^2\rp,
\label{eq:delta_12}
\end{align}
where we have defined
\be
\sigma_{WW\rightarrow 4\ell}^{\rm(sm)} \equiv X^{\rm(sm)}f\lp \Gamma_W^{\rm(sm)}\rp \qquad \text{and} \qquad Y_i \equiv X_i f\lp \Gamma_W^{\rm(sm)}\rp \, .
\ee
Given that we compute the SMEFT corrections
to the $WW$ process in the fixed-width
approximation (that is, assuming $\delta\Gamma_W = 0$),
we already know the coefficients
$Y_{j}^{(\rm eft)}$ and $Y_{jk}^{(\rm eft)}$.
What remains to be done is to determine
the expansion coefficients $\delta_1$ and $\delta_2$ in Eq.~(\ref{eq:delta_12}). 
To do so, we
evaluate the differential cross-section, $\sigma_{WW\rightarrow 4\ell}$, 
setting to zero all EFT effects
while varying the width $\Gamma_W$ along a fixed set of values $\delta\Gamma_W^{(i)} = \{\delta\Gamma_W^{(1)}, \dots, \delta\Gamma_W^{(n)} \}$ around the SM corresponding to $\delta\Gamma_W=0$.
One can then fit a quartic polynomial to the resulting values
\be
\sigma_{WW\rightarrow 4\ell}^{(i)} = \sigma_{WW\rightarrow 4\ell}^{\rm(sm)} \lp 1+ \delta_1 \frac{\delta\Gamma_W^{(i)}}{\Gamma_W^{(\rm sm)}} +  \delta_2 \lp \frac{\delta\Gamma_W^{(i)}}{\Gamma_W^{(\rm sm)}}\rp^2 +  \delta_3 \lp \frac{\delta\Gamma_W^{(i)}}{\Gamma_W^{(\rm sm)}}\rp^3 +  \delta_4 \lp \frac{\delta\Gamma_W^{(i)}}{\Gamma_W^{(\rm sm)}}\rp^4\rp \, ,\qquad i=1,\dots, n \, ,
\label{eq:fitting_relation}
\ee
to extract the expansion coefficients $\delta_1$ and $\delta_2$. At the level of the polynomial fit, we also include the expansion coefficients $\delta_3$ and $\delta_4$, which are relevant in case the fitting range extends beyond the quadratic regime where cubic and quartic effects introduce sizeable corrections that would otherwise get absorbed into incorrect values of $\delta_1$ and $\delta_2$. In the following, we will not write $\delta_3$ and $\delta_4$ explicitly with the understanding that $\delta_1$ and $\delta_2$ are always determined in a quartic fit.

The last ingredient required is to compute the shifts $\delta\Gamma_W$ in Eq.~(\ref{eq:width_func_form}) as a function of the SMEFT coefficients in our fitting basis.
To this end, we parameterise these shifts as
\be
\delta\Gamma_W = \sum_{j=1}^{n_{\rm eft}}c_j Z_j^{(\rm eft)} + \sum_{j=1}^{n_{\rm eft}}\sum_{k\ge j}^{n_{\rm eft}}c_j c_k Z_{jk}^{(\rm eft)},
\label{eq:width_eft}
\ee
in terms of the linear and quadratic corrections,
and where the coefficients $Z_j$ and $Z_{jk}$ can be computed analytically. 
One then substitutes Eq.~(\ref{eq:width_eft}) into Eq.~(\ref{eq:delta_12}) and up to quadratic corrections this gives:
\begin{align}
\nonumber &\sigma_{WW\rightarrow 4 \ell} = \lp \sigma_{WW\rightarrow 4\ell}^{\rm(sm)} + \sum_{j=1}^{n_{\rm eft}}c_j Y_j^{(\rm eft)} + \sum_{j=1}^{n_{\rm eft}}\sum_{k\ge j}^{n_{\rm eft}}c_j c_k Y_{jk}^{(\rm eft)}\rp \\
&\qquad \times \lp 1+ \frac{\delta_1}{\Gamma_W^{\rm (sm)}}\lp\sum_{j=1}^{n_{\rm eft}}c_j Z_j^{(\rm eft)} + \sum_{j=1}^{n_{\rm eft}}\sum_{k\ge j}^{n_{\rm eft}}c_j c_k Z_{jk}^{(\rm eft)}\rp + \frac{\delta_2}{\lp\Gamma_W^{\rm(sm)}\rp^2}\lp\sum_{j=1}^{n_{\rm eft}}c_j Z_j^{(\rm eft)} \rp^2\rp\, ,
\label{eq:modified_xsec}
\end{align}
where higher-order terms in the expansion beyond quadratic corrections are neglected.
We can now equate terms proportional to the same power of the Wilson coefficients  to determine how the SMEFT corrections to production, $Y_{i}^{(\rm eft)}$ and $Y_{ij}^{(\rm eft)}$ in Eq.~(\ref{eq:delta_12}), are modified in the presence of corrections to the $W$-boson decay width:
\be
\sigma_{WW\rightarrow 4\ell} = \lp \sigma_{WW\rightarrow 4\ell}^{\rm(sm)} + \sum_{j=1}^{n_{\rm eft}}c_j \widetilde{Y}_j^{(\rm eft)} + \sum_{j=1}^{n_{\rm eft}}\sum_{k\ge j}^{n_{\rm eft}}c_j c_k \widetilde{Y}_{jk}^{(\rm eft)}\rp  \, ,
\ee
with
\begin{align}
\label{eq:app_widthscorr_1}
\widetilde{Y}_i &= Y_i^{\rm (eft)} + \frac{\delta_1}{\Gamma_W^{\rm(sm)}}Z_i^{\rm(eft)}\sigma_{WW\rightarrow 4\ell}^{\rm(sm)} \, ,\\
\widetilde{Y}_{ii}^{\rm (eft)} &= Y_{ii}^{\rm (eft)} + \sigma_{WW\rightarrow 4\ell}^{\rm(sm)}\lp\frac{\delta_1}{\Gamma_W^{\rm(sm)}}Z_{ii}^{\rm(eft)}+\frac{\delta_2}{\lp\Gamma_W^{\rm(sm)}\rp^2}\lp Z_{i}^{\rm(eft)}\rp^2\rp + \frac{\delta_1}{\Gamma_W^{\rm(sm)}}Y_i^{\rm(eft)} Z_i^{\rm(eft)} \, ,\\
\nonumber \widetilde{Y}_{ij}^{\rm (eft)} &= Y_{ij}^{\rm (eft)} + \sigma_{WW\rightarrow 4\ell}^{\rm(sm)}\lp\frac{\delta_1}{\Gamma_W^{\rm(sm)}}Z_{ij}^{\rm(eft)}+2\frac{\delta_2}{\lp\Gamma_W^{\rm(sm)}\rp^2}Z_{i}^{\rm(eft)}Z_{j}^{\rm(eft)}\rp \\
&\hspace{3cm} + \frac{\delta_1}{\Gamma_W^{\rm(sm)}}\lp Y_i^{\rm(eft)} Z_j^{\rm(eft)} + Y_j^{\rm(eft)}Z_i^{\rm(eft)} \rp \, ,\qquad i\neq j. \label{eq:app_widthscorr_2}
\end{align}
Eqns.~(\ref{eq:app_widthscorr_1})--(\ref{eq:app_widthscorr_2}) indicate
how to account for SMEFT effects
in the decay width in the case 
of theory calculations carried out
in the fixed-width approximation.

\paragraph{Multiple widths.}
This approach can be generalised to 
processes depending on  multiple decay widths, like $\Gamma_Z$ and $\Gamma_W$ in $WZ$ production. One should modify Eq.~(\ref{eq:fitting_relation}) to

\begin{align}
\sigma_{WZ\rightarrow 4\ell} &=
\sigma_{WZ\rightarrow 4\ell}^{\rm(sm)}  \lp 1+ \delta_1^W \frac{\delta\Gamma_W}{\Gamma_W^{(\rm sm)}} +  \delta_2^W \lp \frac{\delta\Gamma_W}{\Gamma_W^{(\rm sm)}}\rp^2 + \delta_1^Z \frac{\delta\Gamma_Z}{\Gamma_Z^{(\rm sm)}} +  \delta_2^Z \lp \frac{\delta\Gamma_Z}{\Gamma_Z^{(\rm sm)}}\rp^2 + \delta^{ZW}\frac{\delta\Gamma_Z\delta\Gamma_W}{\Gamma_W^{(\rm sm)}\Gamma_Z^{(\rm sm)}}\rp,
\label{eq:fitting_relation_2d}
\end{align}
which introduces an additional fit parameter $\delta^{ZW}$ compared to the single width case that is sensitive to simultaneous variations of the individual decay widths, $\delta\Gamma_Z$ and $\delta\Gamma_W$.
One may determine this parameter in one of two ways.
The first option is to fit a paraboloid directly to Eq.~(\ref{eq:fitting_relation_2d}), which gives simultaneous access to all fit parameters. 

The second option, which we adopt here, fits along three 1D slices, one of which is rotated to cancel the
orthogonal component.
Next, one proceeds like in the single width case to end up with a generalised version of Eq.~(\ref{eq:modified_xsec}):
\begin{align}
\nonumber\sigma_{WZ\rightarrow 4 \ell} &= \lp \sigma_{WZ\rightarrow 4\ell}^{\rm(sm)} + \sum_{j=1}^{n_{\rm eft}}c_j Y_j^{(\rm eft)} + \sum_{j=1}^{n_{\rm eft}}\sum_{k\ge j}^{n_{\rm eft}}c_j c_k Y_{jk}^{(\rm eft)}\rp \\
\nonumber &\qquad \times \left\{ 1+ \frac{\delta_1^W}{\Gamma_W^{\rm (sm)}}\lp\sum_{j=1}^{n_{\rm eft}}c_j Z_j^{(\mathrm{eft}, W)} + \sum_{j=1}^{n_{\rm eft}}\sum_{k\ge j}^{n_{\rm eft}}c_j c_k Z_{jk}^{(\mathrm{ eft},W)}\rp + \frac{\delta_2^W}{\lp\Gamma_W^{\rm(sm)}\rp^2}\lp\sum_{j=1}^{n_{\rm eft}}c_j Z_j^{(\mathrm{eft}, W)} \rp^2\right.\\
\nonumber &\qquad\qquad + \frac{\delta_1^Z}{\Gamma_Z^{\rm (sm)}}\lp\sum_{j=1}^{n_{\rm eft}}c_j Z_j^{(\mathrm{eft}, Z)} + \sum_{j=1}^{n_{\rm eft}}\sum_{k\ge j}^{n_{\rm eft}}c_j c_k Z_{jk}^{(\mathrm{ eft},Z)}\rp + \frac{\delta_2^Z}{\lp\Gamma_Z^{\rm(sm)}\rp^2}\lp\sum_{j=1}^{n_{\rm eft}}c_j Z_j^{(\mathrm{eft}, Z)} \rp^2\\
&\qquad\qquad + \left.\frac{\delta^{ZW}}{\Gamma_Z^{\rm (sm)}\Gamma_W^{\rm (sm)}}\lp\sum_{j=1}^{n_{\rm eft}}c_j Z_j^{(\mathrm{eft}, W)} \rp\lp\sum_{k=1}^{n_{\rm eft}}c_k Z_k^{(\mathrm{eft}, Z)} \rp\right\},
\end{align}
where we have added superscripts $W, Z$ to the width predictions to distinguish the contributions associated from $W$ and $Z$ decays, respectively. 
Collecting powers of the Wilson coefficients leads to the following modified theory predictions:

\begin{align}
\widetilde{Y}_i^{\rm (eft)} &= Y_i^{\rm (eft)} + \sigma_{WW\rightarrow 4\ell}^{\rm(sm)}\sum_{V\in\{W,Z\}}\frac{\delta_1^V}{\Gamma_V^{\rm(sm)}}Z_i^{(\mathrm{eft}, V)} \, ,\\
\nonumber \widetilde{Y}_{ii}^{\rm (eft)} &= _{ii}^{\rm (eft)} + \frac{\delta^{ZW}}{\Gamma_Z^{\rm(sm)}\Gamma_W^{\rm(sm)}}Z_{i}^{(\mathrm{eft}, W)}Z_{i}^{(\mathrm{eft}, Z)}\sigma_{WZ\rightarrow 4\ell}^{\rm(sm)} \\
&+\sum_{V\in\{W,Z\}}\sigma_{WZ\rightarrow 4\ell}^{\rm(sm)}\lp\frac{\delta_1^V}{\Gamma_V^{\rm(sm)}}Z_{ii}^{(\mathrm{eft}, V)}+\frac{\delta_2^V}{\lp\Gamma_V^{\rm(sm)}\rp^2}\lp Z_{i}^{(\mathrm{eft}, V)}\rp^2\rp + \frac{\delta_1^V}{\Gamma_V^{\rm(sm)}}Y_i^{\rm(eft)} Z_i^{(\mathrm{eft}, V)} \, , \\
\nonumber \widetilde{Y}_{ij}^{\rm (eft)} &= Y_{ij}^{\rm (eft)}+ \frac{\delta^{ZW}}{\Gamma_Z^{\rm(sm)}\Gamma_W^{\rm(sm)}}\lp Z_{i}^{(\mathrm{eft}, W)}Z_{j}^{(\mathrm{eft}, Z)} + Z_{j}^{(\mathrm{eft}, W)}Z_{i}^{(\mathrm{eft}, Z)}\rp \sigma_{WZ\rightarrow 4\ell}^{\rm(sm)} \\
\nonumber&+\sum_{V\in\{W,Z\}}\sigma_{WZ\rightarrow 4\ell}^{\rm(sm)}\lp\frac{\delta_1^V}{\Gamma_V^{\rm(sm)}}Z_{ij}^{(\mathrm{eft}, V)}+2\frac{\delta_2^V}{\lp\Gamma_V^{\rm(sm)}\rp^2}Z_{i}^{(\mathrm{eft}, V)}Z_{j}^{(\mathrm{eft}, V)}\rp \\
&\hspace{5cm}+ \frac{\delta_1^V}{\Gamma_V^{\rm(sm)}}\lp Y_i^{\rm(eft)} Z_j^{(\mathrm{eft}, V)} + Y_j^{\rm(eft)}Z_i^{(\mathrm{eft}, V)} \, ,\rp\qquad i\neq j \, .
\end{align}
These expressions make possible to fully account for SMEFT effects in the decay widths of electroweak gauge bosons, also in the case of $WZ$ cross-sections.

\paragraph{EFT branching ratios.}
%
Finally, we provide the expressions to combine SMEFT effects in production and decay within the narrow width approximation, as was assumed in LEP $WW$ production.
In this case, the production of a final state $Y$ from the decay of an intermediate state $X$
can be separated into the product of producing $X$ alone and its branching ratio into the state $Y$,

\be
\sigma(e^+ e^- \rightarrow X \rightarrow Y) = \sigma(e^+ e^- \rightarrow X)\mathrm{BR}(X\rightarrow Y) \, ,
\label{eq:NWA}
\ee
where $\sigma(e^+ e^- \rightarrow X)$ and $\mathrm{BR}(X\rightarrow Y)$ receive EFT corrections given by

\be
\sigma(e^+ e^- \rightarrow X) = \sigma^{(\rm sm)}+\sum_{j=1}c_j\sigma_j^{(\rm eft)} + \sum_{j=1}\sum_{k\ge j}c_jc_k\sigma_{jk}^{\rm(eft)},
\label{eq:eft_xsec}
\ee
and

\be
\mathrm{BR}(X \rightarrow Y) = \mathrm{BR}^{(\rm sm)}+\sum_{j=1}c_j\mathrm{BR}_j^{(\rm eft)} + \sum_{j=1}\sum_{k\ge j}c_jc_k\mathrm{BR}_{jk}^{\rm(eft)} \, ,
\label{eq:eft_br}
\ee
respectively. Substituting Eq.~(\ref{eq:eft_xsec}) and (\ref{eq:eft_br}) into Eq.~(\ref{eq:NWA}) gives
\begin{align}
    \nonumber \sigma(e^+ e^- \rightarrow X \rightarrow Y) &= \sigma^{(\rm sm)} \mathrm{BR}^{(\rm sm)} + \sum_{j=1}c_j\lp\sigma_j^{(\rm eft)}\mathrm{BR}^{(\rm sm)} + \sigma^{(\rm sm)} \mathrm{BR}_j^{(\rm eft)}\rp +\\
    &\hspace{-1cm}\sum_{j=1}\sum_{k\ge j}c_jc_k\lp \sigma_{jk}^{\rm(eft)} \mathrm{BR}^{(\rm sm)} + \sigma^{\rm(sm)}\mathrm{BR}_{jk}^{\rm(eft)}\rp + \sum_{j=1}\sum_{k=1}c_{j}c_k\sigma_j^{\rm(eft)}\mathrm{BR}_k^{\rm(eft)}+\mathcal{O} \lp c^3 \rp.
\end{align}
After collecting equal powers of $c$, we find the following mappings
\begin{align}
    \sigma^{(\rm sm)} &\rightarrow \sigma^{(\rm sm)} \mathrm{BR}^{(\rm sm)}\\
    \sigma_i^{(\rm eft)} &\rightarrow \sigma_i^{(\rm eft)}\mathrm{BR}^{(\rm sm)} + \sigma^{(\rm sm)}\mathrm{BR}_i^{(\rm eft)}\\
    \sigma_{ij}^{\rm (eft)} &\rightarrow
    \begin{cases}
        \sigma_{ij}^{\rm(eft)}\mathrm{BR}^{(\rm sm )}+\sigma^{(\rm sm)}\mathrm{BR}_{ij}^{(\rm eft)} + \sigma_i^{(\rm eft)}\mathrm{BR}_j^{\rm (eft)} + \sigma_j^{(\rm eft)}\mathrm{BR}_i^{\rm (eft)}&i\ne j\\
        \sigma_{ij}^{\rm(eft)}\mathrm{BR}^{(\rm sm )}+\sigma^{(\rm sm)}\mathrm{BR}_{ij}^{(\rm eft)} + \sigma_i^{(\rm eft)}\mathrm{BR}_j^{\rm (eft)}&i= j
    \end{cases}
\end{align}


\begin{thebibliography}{100}

\bibitem{Cepeda:2019klc}
M.~Cepeda et~al., \emph{{Report from Working Group 2}: {Higgs Physics at the
  HL-LHC and HE-LHC}},
  \href{https://doi.org/10.23731/CYRM-2019-007.221}{\emph{CERN Yellow Rep.
  Monogr.} {\bfseries 7} (2019) 221}
  [\href{https://arxiv.org/abs/1902.00134}{{\ttfamily 1902.00134}}].

\bibitem{Azzi:2019yne}
P.~Azzi et~al., \emph{{Report from Working Group 1}: {Standard Model Physics at
  the HL-LHC and HE-LHC}},
  \href{https://doi.org/10.23731/CYRM-2019-007.1}{\emph{CERN Yellow Rep.
  Monogr.} {\bfseries 7} (2019) 1}
  [\href{https://arxiv.org/abs/1902.04070}{{\ttfamily 1902.04070}}].

\bibitem{FCC:2018byv}
{\scshape FCC} collaboration, \emph{{FCC Physics Opportunities}: {Future
  Circular Collider Conceptual Design Report Volume 1}},
  \href{https://doi.org/10.1140/epjc/s10052-019-6904-3}{\emph{Eur. Phys. J. C}
  {\bfseries 79} (2019) 474}.

\bibitem{FCC:2018evy}
{\scshape FCC} collaboration, \emph{{FCC-ee: The Lepton Collider}: {Future
  Circular Collider Conceptual Design Report Volume 2}},
  \href{https://doi.org/10.1140/epjst/e2019-900045-4}{\emph{Eur. Phys. J. ST}
  {\bfseries 228} (2019) 261}.

\bibitem{CEPCPhysicsStudyGroup:2022uwl}
{\scshape CEPC Physics Study Group} collaboration, \emph{{The Physics potential
  of the CEPC. Prepared for the US Snowmass Community Planning Exercise
  (Snowmass 2021)}},  in \emph{{Snowmass 2021}}, 5, 2022,
  \href{https://arxiv.org/abs/2205.08553}{{\ttfamily 2205.08553}}.

\bibitem{Behnke:2013xla}
\emph{{The International Linear Collider Technical Design Report - Volume 1:
  Executive Summary}},  \href{https://arxiv.org/abs/1306.6327}{{\ttfamily
  1306.6327}}.

\bibitem{ILC:2013jhg}
{\scshape ILC} collaboration, \emph{{The International Linear Collider
  Technical Design Report - Volume 2: Physics}},
  \href{https://arxiv.org/abs/1306.6352}{{\ttfamily 1306.6352}}.

\bibitem{Vernieri:2022fae}
C.~Vernieri et~al., \emph{{Strategy for Understanding the Higgs Physics: The
  Cool Copper Collider}},
  \href{https://doi.org/10.1088/1748-0221/18/07/P07053}{\emph{JINST} {\bfseries
  18} (2023) P07053} [\href{https://arxiv.org/abs/2203.07646}{{\ttfamily
  2203.07646}}].

\bibitem{Linssen:2012hp}
\emph{{Physics and Detectors at CLIC: CLIC Conceptual Design Report}},
  \href{https://arxiv.org/abs/1202.5940}{{\ttfamily 1202.5940}}.

\bibitem{FCC:2018vvp}
{\scshape FCC} collaboration, \emph{{FCC-hh: The Hadron Collider}: {Future
  Circular Collider Conceptual Design Report Volume 3}},
  \href{https://doi.org/10.1140/epjst/e2019-900087-0}{\emph{Eur. Phys. J. ST}
  {\bfseries 228} (2019) 755}.

\bibitem{Tang:2015qga}
J.~Tang et~al., \emph{{Concept for a Future Super Proton-Proton Collider}},
  \href{https://arxiv.org/abs/1507.03224}{{\ttfamily 1507.03224}}.

\bibitem{Accettura:2023ked}
C.~Accettura et~al., \emph{{Towards a muon collider}},
  \href{https://doi.org/10.1140/epjc/s10052-023-11889-x}{\emph{Eur. Phys. J. C}
  {\bfseries 83} (2023) 864}
  [\href{https://arxiv.org/abs/2303.08533}{{\ttfamily 2303.08533}}].

\bibitem{Aime:2022flm}
C.~Aime et~al., \emph{{Muon Collider Physics Summary}},
  \href{https://arxiv.org/abs/2203.07256}{{\ttfamily 2203.07256}}.

\bibitem{LHeC:2020van}
{\scshape LHeC, FCC-he Study Group} collaboration, \emph{{The Large
  Hadron\textendash{}Electron Collider at the HL-LHC}},
  \href{https://doi.org/10.1088/1361-6471/abf3ba}{\emph{J. Phys. G} {\bfseries
  48} (2021) 110501} [\href{https://arxiv.org/abs/2007.14491}{{\ttfamily
  2007.14491}}].

\bibitem{AbdulKhalek:2021gbh}
R.~Abdul~Khalek et~al., \emph{{Science Requirements and Detector Concepts for
  the Electron-Ion Collider}: {EIC Yellow Report}},
  \href{https://doi.org/10.1016/j.nuclphysa.2022.122447}{\emph{Nucl. Phys. A}
  {\bfseries 1026} (2022) 122447}
  [\href{https://arxiv.org/abs/2103.05419}{{\ttfamily 2103.05419}}].

\bibitem{TLEPDesignStudyWorkingGroup:2013myl}
{\scshape TLEP Design Study Working Group} collaboration, \emph{{First Look at
  the Physics Case of TLEP}},
  \href{https://doi.org/10.1007/JHEP01(2014)164}{\emph{JHEP} {\bfseries 01}
  (2014) 164} [\href{https://arxiv.org/abs/1308.6176}{{\ttfamily 1308.6176}}].

\bibitem{EuropeanStrategyforParticlePhysicsPreparatoryGroup:2019qin}
R.~K. Ellis et~al., \emph{{Physics Briefing Book}: {Input for the European
  Strategy for Particle Physics Update 2020}},
  \href{https://arxiv.org/abs/1910.11775}{{\ttfamily 1910.11775}}.

\bibitem{Narain:2022qud}
M.~Narain et~al., \emph{{The Future of US Particle Physics - The Snowmass 2021
  Energy Frontier Report}},  \href{https://arxiv.org/abs/2211.11084}{{\ttfamily
  2211.11084}}.

\bibitem{Grzadkowski:2010es}
B.~Grzadkowski, M.~Iskrzynski, M.~Misiak and J.~Rosiek, \emph{{Dimension-Six
  Terms in the Standard Model Lagrangian}},
  \href{https://doi.org/10.1007/JHEP10(2010)085}{\emph{JHEP} {\bfseries 10}
  (2010) 085} [\href{https://arxiv.org/abs/1008.4884}{{\ttfamily 1008.4884}}].

\bibitem{Brivio:2017vri}
I.~Brivio and M.~Trott, \emph{{The Standard Model as an Effective Field
  Theory}}, \href{https://doi.org/10.1016/j.physrep.2018.11.002}{\emph{Phys.
  Rept.} {\bfseries 793} (2019) 1}
  [\href{https://arxiv.org/abs/1706.08945}{{\ttfamily 1706.08945}}].

\bibitem{Isidori:2023pyp}
G.~Isidori, F.~Wilsch and D.~Wyler, \emph{{The standard model effective field
  theory at work}},
  \href{https://doi.org/10.1103/RevModPhys.96.015006}{\emph{Rev. Mod. Phys.}
  {\bfseries 96} (2024) 015006}
  [\href{https://arxiv.org/abs/2303.16922}{{\ttfamily 2303.16922}}].

\bibitem{Falkowski:2019tft}
A.~Falkowski and R.~Rattazzi, \emph{{Which EFT}},
  \href{https://doi.org/10.1007/JHEP10(2019)255}{\emph{JHEP} {\bfseries 10}
  (2019) 255} [\href{https://arxiv.org/abs/1902.05936}{{\ttfamily
  1902.05936}}].

\bibitem{Cohen:2020xca}
T.~Cohen, N.~Craig, X.~Lu and D.~Sutherland, \emph{{Is SMEFT Enough?}},
  \href{https://doi.org/10.1007/JHEP03(2021)237}{\emph{JHEP} {\bfseries 03}
  (2021) 237} [\href{https://arxiv.org/abs/2008.08597}{{\ttfamily
  2008.08597}}].

\bibitem{DeBlas:2019ehy}
J.~De~Blas et~al., \emph{{$\texttt{HEPfit}$: a code for the combination of
  indirect and direct constraints on high energy physics models}},
  \href{https://doi.org/10.1140/epjc/s10052-020-7904-z}{\emph{Eur. Phys. J. C}
  {\bfseries 80} (2020) 456}
  [\href{https://arxiv.org/abs/1910.14012}{{\ttfamily 1910.14012}}].

\bibitem{Ellis:2020unq}
J.~Ellis, M.~Madigan, K.~Mimasu, V.~Sanz and T.~You, \emph{{Top, Higgs, Diboson
  and Electroweak Fit to the Standard Model Effective Field Theory}},
  \href{https://doi.org/10.1007/JHEP04(2021)279}{\emph{JHEP} {\bfseries 04}
  (2021) 279} [\href{https://arxiv.org/abs/2012.02779}{{\ttfamily
  2012.02779}}].

\bibitem{Brivio:2021alv}
I.~Brivio, S.~Bruggisser, E.~Geoffray, W.~Kilian, M.~Kr\"amer, M.~Luchmann
  et~al., \emph{{From Models to SMEFT and Back?}},
  \href{https://doi.org/10.21468/SciPostPhys.12.1.036}{\emph{SciPost Phys.}
  {\bfseries 12} (2022) 036}
  [\href{https://arxiv.org/abs/2108.01094}{{\ttfamily 2108.01094}}].

\bibitem{Giani:2023gfq}
T.~Giani, G.~Magni and J.~Rojo, \emph{{SMEFiT: a flexible toolbox for global
  interpretations of particle physics data with effective field theories}},
  \href{https://doi.org/10.1140/epjc/s10052-023-11534-7}{\emph{Eur. Phys. J. C}
  {\bfseries 83} (2023) 393}
  [\href{https://arxiv.org/abs/2302.06660}{{\ttfamily 2302.06660}}].

\bibitem{deBlas:2017xtg}
J.~de~Blas, J.~C. Criado, M.~Perez-Victoria and J.~Santiago, \emph{{Effective
  description of general extensions of the Standard Model: the complete
  tree-level dictionary}},
  \href{https://doi.org/10.1007/JHEP03(2018)109}{\emph{JHEP} {\bfseries 03}
  (2018) 109} [\href{https://arxiv.org/abs/1711.10391}{{\ttfamily
  1711.10391}}].

\bibitem{DasBakshi:2018vni}
S.~Das~Bakshi, J.~Chakrabortty and S.~K. Patra, \emph{{CoDEx: Wilson
  coefficient calculator connecting SMEFT to UV theory}},
  \href{https://doi.org/10.1140/epjc/s10052-018-6444-2}{\emph{Eur. Phys. J. C}
  {\bfseries 79} (2019) 21} [\href{https://arxiv.org/abs/1808.04403}{{\ttfamily
  1808.04403}}].

\bibitem{Carmona:2021xtq}
A.~Carmona, A.~Lazopoulos, P.~Olgoso and J.~Santiago, \emph{{Matchmakereft:
  automated tree-level and one-loop matching}},
  \href{https://doi.org/10.21468/SciPostPhys.12.6.198}{\emph{SciPost Phys.}
  {\bfseries 12} (2022) 198}
  [\href{https://arxiv.org/abs/2112.10787}{{\ttfamily 2112.10787}}].

\bibitem{Fuentes-Martin:2022jrf}
J.~Fuentes-Mart\'\i{}n, M.~K\"onig, J.~Pag\`es, A.~E. Thomsen and F.~Wilsch,
  \emph{{A Proof of Concept for Matchete: An Automated Tool for Matching
  Effective Theories}},  \href{https://arxiv.org/abs/2212.04510}{{\ttfamily
  2212.04510}}.

\bibitem{terHoeve:2023pvs}
J.~ter Hoeve, G.~Magni, J.~Rojo, A.~N. Rossia and E.~Vryonidou, \emph{{The
  automation of SMEFT-assisted constraints on UV-complete models}},
  \href{https://doi.org/10.1007/JHEP01(2024)179}{\emph{JHEP} {\bfseries 01}
  (2024) 179} [\href{https://arxiv.org/abs/2309.04523}{{\ttfamily
  2309.04523}}].

\bibitem{Ellis:2015sca}
J.~Ellis and T.~You, \emph{{Sensitivities of Prospective Future e+e- Colliders
  to Decoupled New Physics}},
  \href{https://doi.org/10.1007/JHEP03(2016)089}{\emph{JHEP} {\bfseries 03}
  (2016) 089} [\href{https://arxiv.org/abs/1510.04561}{{\ttfamily
  1510.04561}}].

\bibitem{deBlas:2016nqo}
J.~de~Blas, M.~Ciuchini, E.~Franco, S.~Mishima, M.~Pierini, L.~Reina et~al.,
  \emph{{Electroweak precision constraints at present and future colliders}},
  \href{https://doi.org/10.22323/1.282.0690}{\emph{PoS} {\bfseries ICHEP2016}
  (2017) 690} [\href{https://arxiv.org/abs/1611.05354}{{\ttfamily
  1611.05354}}].

\bibitem{Ellis:2017kfi}
J.~Ellis, P.~Roloff, V.~Sanz and T.~You, \emph{{Dimension-6 Operator Analysis
  of the CLIC Sensitivity to New Physics}},
  \href{https://doi.org/10.1007/JHEP05(2017)096}{\emph{JHEP} {\bfseries 05}
  (2017) 096} [\href{https://arxiv.org/abs/1701.04804}{{\ttfamily
  1701.04804}}].

\bibitem{Durieux:2017rsg}
G.~Durieux, C.~Grojean, J.~Gu and K.~Wang, \emph{{The leptonic future of the
  Higgs}}, \href{https://doi.org/10.1007/JHEP09(2017)014}{\emph{JHEP}
  {\bfseries 09} (2017) 014}
  [\href{https://arxiv.org/abs/1704.02333}{{\ttfamily 1704.02333}}].

\bibitem{Barklow:2017suo}
T.~Barklow, K.~Fujii, S.~Jung, R.~Karl, J.~List, T.~Ogawa et~al.,
  \emph{{Improved Formalism for Precision Higgs Coupling Fits}},
  \href{https://doi.org/10.1103/PhysRevD.97.053003}{\emph{Phys. Rev. D}
  {\bfseries 97} (2018) 053003}
  [\href{https://arxiv.org/abs/1708.08912}{{\ttfamily 1708.08912}}].

\bibitem{Barklow:2017awn}
T.~Barklow, K.~Fujii, S.~Jung, M.~E. Peskin and J.~Tian,
  \emph{{Model-Independent Determination of the Triple Higgs Coupling at e+e-
  Colliders}}, \href{https://doi.org/10.1103/PhysRevD.97.053004}{\emph{Phys.
  Rev. D} {\bfseries 97} (2018) 053004}
  [\href{https://arxiv.org/abs/1708.09079}{{\ttfamily 1708.09079}}].

\bibitem{DiVita:2017vrr}
S.~Di~Vita, G.~Durieux, C.~Grojean, J.~Gu, Z.~Liu, G.~Panico et~al., \emph{{A
  global view on the Higgs self-coupling at lepton colliders}},
  \href{https://doi.org/10.1007/JHEP02(2018)178}{\emph{JHEP} {\bfseries 02}
  (2018) 178} [\href{https://arxiv.org/abs/1711.03978}{{\ttfamily
  1711.03978}}].

\bibitem{Chiu:2017yrx}
W.~H. Chiu, S.~C. Leung, T.~Liu, K.-F. Lyu and L.-T. Wang, \emph{{Probing 6D
  operators at future e$^{-}$e$^{+}$ colliders}},
  \href{https://doi.org/10.1007/JHEP05(2018)081}{\emph{JHEP} {\bfseries 05}
  (2018) 081} [\href{https://arxiv.org/abs/1711.04046}{{\ttfamily
  1711.04046}}].

\bibitem{Durieux:2018tev}
G.~Durieux, M.~Perell\'o, M.~Vos and C.~Zhang, \emph{{Global and optimal probes
  for the top-quark effective field theory at future lepton colliders}},
  \href{https://doi.org/10.1007/JHEP10(2018)168}{\emph{JHEP} {\bfseries 10}
  (2018) 168} [\href{https://arxiv.org/abs/1807.02121}{{\ttfamily
  1807.02121}}].

\bibitem{deBlas:2019rxi}
J.~de~Blas et~al., \emph{{Higgs Boson Studies at Future Particle Colliders}},
  \href{https://doi.org/10.1007/JHEP01(2020)139}{\emph{JHEP} {\bfseries 01}
  (2020) 139} [\href{https://arxiv.org/abs/1905.03764}{{\ttfamily
  1905.03764}}].

\bibitem{DeBlas:2019qco}
J.~De~Blas, G.~Durieux, C.~Grojean, J.~Gu and A.~Paul, \emph{{On the future of
  Higgs, electroweak and diboson measurements at lepton colliders}},
  \href{https://doi.org/10.1007/JHEP12(2019)117}{\emph{JHEP} {\bfseries 12}
  (2019) 117} [\href{https://arxiv.org/abs/1907.04311}{{\ttfamily
  1907.04311}}].

\bibitem{LCCPhysicsWorkingGroup:2019fvj}
{\scshape LCC Physics Working Group} collaboration, \emph{{Tests of the
  Standard Model at the International Linear Collider}},
  \href{https://arxiv.org/abs/1908.11299}{{\ttfamily 1908.11299}}.

\bibitem{Durieux:2019rbz}
G.~Durieux, A.~Irles, V.~Miralles, A.~Pe\~nuelas, R.~P\"oschl, M.~Perell\'o
  et~al., \emph{{The electro-weak couplings of the top and bottom quarks
  \textemdash{} Global fit and future prospects}},
  \href{https://doi.org/10.1007/JHEP12(2019)098}{\emph{JHEP} {\bfseries 12}
  (2019) 98} [\href{https://arxiv.org/abs/1907.10619}{{\ttfamily 1907.10619}}].

\bibitem{Jung:2020uzh}
S.~Jung, J.~Lee, M.~Perell\'o, J.~Tian and M.~Vos, \emph{{Higgs, top quark, and
  electroweak precision measurements at future e+e- colliders: A combined
  effective field theory analysis with renormalization mixing}},
  \href{https://doi.org/10.1103/PhysRevD.105.016003}{\emph{Phys. Rev. D}
  {\bfseries 105} (2022) 016003}
  [\href{https://arxiv.org/abs/2006.14631}{{\ttfamily 2006.14631}}].

\bibitem{deBlas:2021jlt}
J.~de~Blas, \emph{{New physics at the FCC-ee: indirect discovery potential}},
  \href{https://doi.org/10.1140/epjp/s13360-021-01847-5}{\emph{Eur. Phys. J.
  Plus} {\bfseries 136} (2021) 897}.

\bibitem{MuonCollider:2022xlm}
{\scshape Muon Collider} collaboration, \emph{{The physics case of a 3 TeV muon
  collider stage}},  \href{https://arxiv.org/abs/2203.07261}{{\ttfamily
  2203.07261}}.

\bibitem{deBlas:2022ofj}
J.~de~Blas, Y.~Du, C.~Grojean, J.~Gu, V.~Miralles, M.~E. Peskin et~al.,
  \emph{{Global SMEFT Fits at Future Colliders}},
  \href{https://arxiv.org/abs/2206.08326}{{\ttfamily 2206.08326}}.

\bibitem{Allwicher:2023shc}
L.~Allwicher, C.~Cornella, B.~A. Stefanek and G.~Isidori, \emph{{New Physics in
  the Third Generation: A Comprehensive SMEFT Analysis and Future Prospects}},
  \href{https://arxiv.org/abs/2311.00020}{{\ttfamily 2311.00020}}.

\bibitem{Gu:2020thj}
J.~Gu and L.-T. Wang, \emph{{Sum Rules in the Standard Model Effective Field
  Theory from Helicity Amplitudes}},
  \href{https://doi.org/10.1007/JHEP03(2021)149}{\emph{JHEP} {\bfseries 03}
  (2021) 149} [\href{https://arxiv.org/abs/2008.07551}{{\ttfamily
  2008.07551}}].

\bibitem{Gu:2020ldn}
J.~Gu, L.-T. Wang and C.~Zhang, \emph{{Unambiguously Testing Positivity at
  Lepton Colliders}},
  \href{https://doi.org/10.1103/PhysRevLett.129.011805}{\emph{Phys. Rev. Lett.}
  {\bfseries 129} (2022) 011805}
  [\href{https://arxiv.org/abs/2011.03055}{{\ttfamily 2011.03055}}].

\bibitem{ALEPH:2005ab}
{\scshape ALEPH, DELPHI, L3, OPAL, SLD, LEP Electroweak Working Group, SLD
  Electroweak Group, SLD Heavy Flavour Group} collaboration, \emph{{Precision
  electroweak measurements on the $Z$ resonance}},
  \href{https://doi.org/10.1016/j.physrep.2005.12.006}{\emph{Phys. Rept.}
  {\bfseries 427} (2006) 257}
  [\href{https://arxiv.org/abs/hep-ex/0509008}{{\ttfamily hep-ex/0509008}}].

\bibitem{Altarelli:1991fk}
G.~Altarelli, R.~Barbieri and S.~Jadach, \emph{{Toward a model independent
  analysis of electroweak data}},
  \href{https://doi.org/10.1016/0550-3213(92)90376-M}{\emph{Nucl. Phys. B}
  {\bfseries 369} (1992) 3}.

\bibitem{Baak:2012kk}
M.~Baak, M.~Goebel, J.~Haller, A.~Hoecker, D.~Kennedy, R.~Kogler et~al.,
  \emph{{The Electroweak Fit of the Standard Model after the Discovery of a New
  Boson at the LHC}},
  \href{https://doi.org/10.1140/epjc/s10052-012-2205-9}{\emph{Eur. Phys. J. C}
  {\bfseries 72} (2012) 2205}
  [\href{https://arxiv.org/abs/1209.2716}{{\ttfamily 1209.2716}}].

\bibitem{Brivio:2017bnu}
I.~Brivio and M.~Trott, \emph{{Scheming in the SMEFT... and a
  reparameterization invariance!}},
  \href{https://doi.org/10.1007/JHEP07(2017)148}{\emph{JHEP} {\bfseries 07}
  (2017) 148} [\href{https://arxiv.org/abs/1701.06424}{{\ttfamily
  1701.06424}}].

\bibitem{Dawson:2018dcd}
S.~Dawson, C.~Englert and T.~Plehn, \emph{{Higgs Physics: It ain't over till
  it's over}}, \href{https://doi.org/10.1016/j.physrep.2019.05.001}{\emph{Phys.
  Rept.} {\bfseries 816} (2019) 1}
  [\href{https://arxiv.org/abs/1808.01324}{{\ttfamily 1808.01324}}].

\bibitem{Butter:2016cvz}
A.~Butter, O.~J.~P. \'Eboli, J.~Gonzalez-Fraile, M.~C. Gonzalez-Garcia,
  T.~Plehn and M.~Rauch, \emph{{The Gauge-Higgs Legacy of the LHC Run I}},
  \href{https://doi.org/10.1007/JHEP07(2016)152}{\emph{JHEP} {\bfseries 07}
  (2016) 152} [\href{https://arxiv.org/abs/1604.03105}{{\ttfamily
  1604.03105}}].

\bibitem{Azatov:2019xxn}
A.~Azatov, D.~Barducci and E.~Venturini, \emph{{Precision diboson measurements
  at hadron colliders}},
  \href{https://doi.org/10.1007/JHEP04(2019)075}{\emph{JHEP} {\bfseries 04}
  (2019) 075} [\href{https://arxiv.org/abs/1901.04821}{{\ttfamily
  1901.04821}}].

\bibitem{Cirigliano:2016nyn}
V.~Cirigliano, W.~Dekens, J.~de~Vries and E.~Mereghetti, \emph{{Constraining
  the top-Higgs sector of the Standard Model Effective Field Theory}},
  \href{https://doi.org/10.1103/PhysRevD.94.034031}{\emph{Phys. Rev. D}
  {\bfseries 94} (2016) 034031}
  [\href{https://arxiv.org/abs/1605.04311}{{\ttfamily 1605.04311}}].

\bibitem{Aoude:2020dwv}
R.~Aoude, T.~Hurth, S.~Renner and W.~Shepherd, \emph{{The impact of flavour
  data on global fits of the MFV SMEFT}},
  \href{https://doi.org/10.1007/JHEP12(2020)113}{\emph{JHEP} {\bfseries 12}
  (2020) 113} [\href{https://arxiv.org/abs/2003.05432}{{\ttfamily
  2003.05432}}].

\bibitem{Crivellin:2020ebi}
A.~Crivellin, F.~Kirk, C.~A. Manzari and M.~Montull, \emph{{Global Electroweak
  Fit and Vector-Like Leptons in Light of the Cabibbo Angle Anomaly}},
  \href{https://doi.org/10.1007/JHEP12(2020)166}{\emph{JHEP} {\bfseries 12}
  (2020) 166} [\href{https://arxiv.org/abs/2008.01113}{{\ttfamily
  2008.01113}}].

\bibitem{Bruggisser:2021duo}
S.~Bruggisser, R.~Sch\"afer, D.~van Dyk and S.~Westhoff, \emph{{The Flavor of
  UV Physics}}, \href{https://doi.org/10.1007/JHEP05(2021)257}{\emph{JHEP}
  {\bfseries 05} (2021) 257}
  [\href{https://arxiv.org/abs/2101.07273}{{\ttfamily 2101.07273}}].

\bibitem{Bruggisser:2022rhb}
S.~Bruggisser, D.~van Dyk and S.~Westhoff, \emph{{Resolving the flavor
  structure in the MFV-SMEFT}},
  \href{https://doi.org/10.1007/JHEP02(2023)225}{\emph{JHEP} {\bfseries 02}
  (2023) 225} [\href{https://arxiv.org/abs/2212.02532}{{\ttfamily
  2212.02532}}].

\bibitem{Cirigliano:2022qdm}
V.~Cirigliano, W.~Dekens, J.~de~Vries, E.~Mereghetti and T.~Tong,
  \emph{{Beta-decay implications for the W-boson mass anomaly}},
  \href{https://doi.org/10.1103/PhysRevD.106.075001}{\emph{Phys. Rev. D}
  {\bfseries 106} (2022) 075001}
  [\href{https://arxiv.org/abs/2204.08440}{{\ttfamily 2204.08440}}].

\bibitem{Bartocci:2023nvp}
R.~Bartocci, A.~Biek\"otter and T.~Hurth, \emph{{A global analysis of the SMEFT
  under the minimal MFV assumption}},
  \href{https://arxiv.org/abs/2311.04963}{{\ttfamily 2311.04963}}.

\bibitem{Cirigliano:2023nol}
V.~Cirigliano, W.~Dekens, J.~de~Vries, E.~Mereghetti and T.~Tong,
  \emph{{Anomalies in global SMEFT analyses. A case study of first-row CKM
  unitarity}}, \href{https://doi.org/10.1007/JHEP03(2024)033}{\emph{JHEP}
  {\bfseries 03} (2024) 033}
  [\href{https://arxiv.org/abs/2311.00021}{{\ttfamily 2311.00021}}].

\bibitem{Bellafronte:2023amz}
L.~Bellafronte, S.~Dawson and P.~P. Giardino, \emph{{The importance of flavor
  in SMEFT Electroweak Precision Fits}},
  \href{https://doi.org/10.1007/JHEP05(2023)208}{\emph{JHEP} {\bfseries 05}
  (2023) 208} [\href{https://arxiv.org/abs/2304.00029}{{\ttfamily
  2304.00029}}].

\bibitem{Garosi:2023yxg}
F.~Garosi, D.~Marzocca, A.~R. S\'anchez and A.~Stanzione, \emph{{Indirect
  constraints on top quark operators from a global SMEFT analysis}},
  \href{https://doi.org/10.1007/JHEP12(2023)129}{\emph{JHEP} {\bfseries 12}
  (2023) 129} [\href{https://arxiv.org/abs/2310.00047}{{\ttfamily
  2310.00047}}].

\bibitem{Jenkins:2013zja}
E.~E. Jenkins, A.~V. Manohar and M.~Trott, \emph{{Renormalization Group
  Evolution of the Standard Model Dimension Six Operators I: Formalism and
  lambda Dependence}},
  \href{https://doi.org/10.1007/JHEP10(2013)087}{\emph{JHEP} {\bfseries 10}
  (2013) 087} [\href{https://arxiv.org/abs/1308.2627}{{\ttfamily 1308.2627}}].

\bibitem{Jenkins:2013wua}
E.~E. Jenkins, A.~V. Manohar and M.~Trott, \emph{{Renormalization Group
  Evolution of the Standard Model Dimension Six Operators II: Yukawa
  Dependence}}, \href{https://doi.org/10.1007/JHEP01(2014)035}{\emph{JHEP}
  {\bfseries 01} (2014) 035} [\href{https://arxiv.org/abs/1310.4838}{{\ttfamily
  1310.4838}}].

\bibitem{Alonso:2013hga}
R.~Alonso, E.~E. Jenkins, A.~V. Manohar and M.~Trott, \emph{{Renormalization
  Group Evolution of the Standard Model Dimension Six Operators III: Gauge
  Coupling Dependence and Phenomenology}},
  \href{https://doi.org/10.1007/JHEP04(2014)159}{\emph{JHEP} {\bfseries 04}
  (2014) 159} [\href{https://arxiv.org/abs/1312.2014}{{\ttfamily 1312.2014}}].

\bibitem{Aoude:2022aro}
R.~Aoude, F.~Maltoni, O.~Mattelaer, C.~Severi and E.~Vryonidou,
  \emph{{Renormalisation group effects on SMEFT interpretations of LHC data}},
  \href{https://doi.org/10.1007/JHEP09(2023)191}{\emph{JHEP} {\bfseries 09}
  (2023) 191} [\href{https://arxiv.org/abs/2212.05067}{{\ttfamily
  2212.05067}}].

\bibitem{Hartland:2019bjb}
N.~P. Hartland, F.~Maltoni, E.~R. Nocera, J.~Rojo, E.~Slade, E.~Vryonidou
  et~al., \emph{{A Monte Carlo global analysis of the Standard Model Effective
  Field Theory: the top quark sector}},
  \href{https://doi.org/10.1007/JHEP04(2019)100}{\emph{JHEP} {\bfseries 04}
  (2019) 100} [\href{https://arxiv.org/abs/1901.05965}{{\ttfamily
  1901.05965}}].

\bibitem{Ethier:2021ydt}
J.~J. Ethier, R.~Gomez-Ambrosio, G.~Magni and J.~Rojo, \emph{{SMEFT analysis of
  vector boson scattering and diboson data from the LHC Run II}},
  \href{https://doi.org/10.1140/epjc/s10052-021-09347-7}{\emph{Eur. Phys. J. C}
  {\bfseries 81} (2021) 560}
  [\href{https://arxiv.org/abs/2101.03180}{{\ttfamily 2101.03180}}].

\bibitem{Ethier:2021bye}
{\scshape SMEFiT} collaboration, \emph{{Combined SMEFT interpretation of Higgs,
  diboson, and top quark data from the LHC}},
  \href{https://doi.org/10.1007/JHEP11(2021)089}{\emph{JHEP} {\bfseries 11}
  (2021) 089} [\href{https://arxiv.org/abs/2105.00006}{{\ttfamily
  2105.00006}}].

\bibitem{vanBeek:2019evb}
S.~van Beek, E.~R. Nocera, J.~Rojo and E.~Slade, \emph{{Constraining the SMEFT
  with Bayesian reweighting}},
  \href{https://doi.org/10.21468/SciPostPhys.7.5.070}{\emph{SciPost Phys.}
  {\bfseries 7} (2019) 070} [\href{https://arxiv.org/abs/1906.05296}{{\ttfamily
  1906.05296}}].

\bibitem{Grojean:2018dqj}
C.~Grojean, M.~Montull and M.~Riembau, \emph{{Diboson at the LHC vs LEP}},
  \href{https://doi.org/10.1007/JHEP03(2019)020}{\emph{JHEP} {\bfseries 03}
  (2019) 020} [\href{https://arxiv.org/abs/1810.05149}{{\ttfamily
  1810.05149}}].

\bibitem{Banerjee:2018bio}
S.~Banerjee, C.~Englert, R.~S. Gupta and M.~Spannowsky, \emph{{Probing
  Electroweak Precision Physics via boosted Higgs-strahlung at the LHC}},
  \href{https://doi.org/10.1103/PhysRevD.98.095012}{\emph{Phys. Rev. D}
  {\bfseries 98} (2018) 095012}
  [\href{https://arxiv.org/abs/1807.01796}{{\ttfamily 1807.01796}}].

\bibitem{Berthier:2015gja}
L.~Berthier and M.~Trott, \emph{{Consistent constraints on the Standard Model
  Effective Field Theory}},
  \href{https://doi.org/10.1007/JHEP02(2016)069}{\emph{JHEP} {\bfseries 02}
  (2016) 069} [\href{https://arxiv.org/abs/1508.05060}{{\ttfamily
  1508.05060}}].

\bibitem{Corbett:2021eux}
T.~Corbett, A.~Helset, A.~Martin and M.~Trott, \emph{{EWPD in the SMEFT to
  dimension eight}}, \href{https://doi.org/10.1007/JHEP06(2021)076}{\emph{JHEP}
  {\bfseries 06} (2021) 076}
  [\href{https://arxiv.org/abs/2102.02819}{{\ttfamily 2102.02819}}].

\bibitem{Awramik:2003rn}
M.~Awramik, M.~Czakon, A.~Freitas and G.~Weiglein, \emph{{Precise prediction
  for the W boson mass in the standard model}},
  \href{https://doi.org/10.1103/PhysRevD.69.053006}{\emph{Phys. Rev. D}
  {\bfseries 69} (2004) 053006}
  [\href{https://arxiv.org/abs/hep-ph/0311148}{{\ttfamily hep-ph/0311148}}].

\bibitem{Freitas:2014hra}
A.~Freitas, \emph{{Higher-order electroweak corrections to the partial widths
  and branching ratios of the Z boson}},
  \href{https://doi.org/10.1007/JHEP04(2014)070}{\emph{JHEP} {\bfseries 04}
  (2014) 070} [\href{https://arxiv.org/abs/1401.2447}{{\ttfamily 1401.2447}}].

\bibitem{LEP-2}
{The ALEPH, DELPHI, L3, OPAL Collaborations, the LEP Electroweak Working
  Group}, \emph{{Electroweak Measurements in Electron-Positron Collisions at
  W-Boson-Pair Energies at LEP}}, {\emph{Phys. Rept.} {\bfseries 532} (2013)
  119} [\href{https://arxiv.org/abs/1302.3415}{{\ttfamily 1302.3415}}].

\bibitem{PDG}
{\scshape Particle Data Group} collaboration, \emph{{Review of Particle
  Physics}}, \href{https://doi.org/10.1093/ptep/ptac097}{\emph{PTEP} {\bfseries
  2022} (2022) 083C01}.

\bibitem{Efrati:2015eaa}
A.~Efrati, A.~Falkowski and Y.~Soreq, \emph{{Electroweak constraints on
  flavorful effective theories}},
  \href{https://doi.org/10.1007/JHEP07(2015)018}{\emph{JHEP} {\bfseries 07}
  (2015) 018} [\href{https://arxiv.org/abs/1503.07872}{{\ttfamily
  1503.07872}}].

\bibitem{Hahn:2000kx}
T.~Hahn, \emph{{Generating Feynman diagrams and amplitudes with FeynArts 3}},
  \href{https://doi.org/10.1016/S0010-4655(01)00290-9}{\emph{Comput. Phys.
  Commun.} {\bfseries 140} (2001) 418}
  [\href{https://arxiv.org/abs/hep-ph/0012260}{{\ttfamily hep-ph/0012260}}].

\bibitem{Mertig:1990an}
R.~Mertig, M.~Bohm and A.~Denner, \emph{{FEYN CALC: Computer algebraic
  calculation of Feynman amplitudes}},
  \href{https://doi.org/10.1016/0010-4655(91)90130-D}{\emph{Comput. Phys.
  Commun.} {\bfseries 64} (1991) 345}.

\bibitem{Shtabovenko:2016sxi}
V.~Shtabovenko, R.~Mertig and F.~Orellana, \emph{{New Developments in FeynCalc
  9.0}}, \href{https://doi.org/10.1016/j.cpc.2016.06.008}{\emph{Comput. Phys.
  Commun.} {\bfseries 207} (2016) 432}
  [\href{https://arxiv.org/abs/1601.01167}{{\ttfamily 1601.01167}}].

\bibitem{Shtabovenko:2020gxv}
V.~Shtabovenko, R.~Mertig and F.~Orellana, \emph{{FeynCalc 9.3: New features
  and improvements}},
  \href{https://doi.org/10.1016/j.cpc.2020.107478}{\emph{Comput. Phys. Commun.}
  {\bfseries 256} (2020) 107478}
  [\href{https://arxiv.org/abs/2001.04407}{{\ttfamily 2001.04407}}].

\bibitem{Degrande:2020evl}
C.~Degrande, G.~Durieux, F.~Maltoni, K.~Mimasu, E.~Vryonidou and C.~Zhang,
  \emph{{Automated one-loop computations in the standard model effective field
  theory}}, \href{https://doi.org/10.1103/PhysRevD.103.096024}{\emph{Phys. Rev.
  D} {\bfseries 103} (2021) 096024}
  [\href{https://arxiv.org/abs/2008.11743}{{\ttfamily 2008.11743}}].

\bibitem{Alwall:2014hca}
J.~Alwall, R.~Frederix, S.~Frixione, V.~Hirschi, F.~Maltoni, O.~Mattelaer
  et~al., \emph{{The automated computation of tree-level and next-to-leading
  order differential cross sections, and their matching to parton shower
  simulations}}, \href{https://doi.org/10.1007/JHEP07(2014)079}{\emph{JHEP}
  {\bfseries 07} (2014) 079} [\href{https://arxiv.org/abs/1405.0301}{{\ttfamily
  1405.0301}}].

\bibitem{Carrazza:2019sec}
S.~Carrazza, C.~Degrande, S.~Iranipour, J.~Rojo and M.~Ubiali, \emph{{Can New
  Physics hide inside the proton?}},
  \href{https://doi.org/10.1103/PhysRevLett.123.132001}{\emph{Phys. Rev. Lett.}
  {\bfseries 123} (2019) 132001}
  [\href{https://arxiv.org/abs/1905.05215}{{\ttfamily 1905.05215}}].

\bibitem{Kassabov:2023hbm}
Z.~Kassabov, M.~Madigan, L.~Mantani, J.~Moore, M.~Morales~Alvarado, J.~Rojo
  et~al., \emph{{The top quark legacy of the LHC Run II for PDF and SMEFT
  analyses}}, \href{https://doi.org/10.1007/JHEP05(2023)205}{\emph{JHEP}
  {\bfseries 05} (2023) 205}
  [\href{https://arxiv.org/abs/2303.06159}{{\ttfamily 2303.06159}}].

\bibitem{Greljo:2021kvv}
A.~Greljo, S.~Iranipour, Z.~Kassabov, M.~Madigan, J.~Moore, J.~Rojo et~al.,
  \emph{{Parton distributions in the SMEFT from high-energy Drell-Yan tails}},
  \href{https://doi.org/10.1007/JHEP07(2021)122}{\emph{JHEP} {\bfseries 07}
  (2021) 122} [\href{https://arxiv.org/abs/2104.02723}{{\ttfamily
  2104.02723}}].

\bibitem{NNPDF:2021njg}
{\scshape NNPDF} collaboration, \emph{{The path to proton structure at 1\%
  accuracy}}, \href{https://doi.org/10.1140/epjc/s10052-022-10328-7}{\emph{Eur.
  Phys. J. C} {\bfseries 82} (2022) 428}
  [\href{https://arxiv.org/abs/2109.02653}{{\ttfamily 2109.02653}}].

\bibitem{Altmannshofer:2021qrr}
W.~Altmannshofer and P.~Stangl, \emph{{New physics in rare B decays after
  Moriond 2021}},
  \href{https://doi.org/10.1140/epjc/s10052-021-09725-1}{\emph{Eur. Phys. J. C}
  {\bfseries 81} (2021) 952}
  [\href{https://arxiv.org/abs/2103.13370}{{\ttfamily 2103.13370}}].

\bibitem{Grazzini:2019jkl}
M.~Grazzini, S.~Kallweit, J.~M. Lindert, S.~Pozzorini and M.~Wiesemann,
  \emph{{NNLO QCD + NLO EW with Matrix+OpenLoops: precise predictions for
  vector-boson pair production}},
  \href{https://doi.org/10.1007/JHEP02(2020)087}{\emph{JHEP} {\bfseries 02}
  (2020) 087} [\href{https://arxiv.org/abs/1912.00068}{{\ttfamily
  1912.00068}}].

\bibitem{ATLAS:2022vkf}
{\scshape ATLAS} collaboration, \emph{{A detailed map of Higgs boson
  interactions by the ATLAS experiment ten years after the discovery}},
  \href{https://doi.org/10.1038/s41586-022-04893-w}{\emph{Nature} {\bfseries
  607} (2022) 52} [\href{https://arxiv.org/abs/2207.00092}{{\ttfamily
  2207.00092}}].

\bibitem{CMS:2021icx}
{\scshape CMS} collaboration, \emph{{Measurement of the inclusive and
  differential WZ production cross sections, polarization angles, and triple
  gauge couplings in pp collisions at $ \sqrt{s} $ = 13 TeV}},
  \href{https://doi.org/10.1007/JHEP07(2022)032}{\emph{JHEP} {\bfseries 07}
  (2022) 032} [\href{https://arxiv.org/abs/2110.11231}{{\ttfamily
  2110.11231}}].

\bibitem{CMS:2021vhb}
{\scshape CMS} collaboration, \emph{{Measurement of differential $t \bar t$
  production cross sections in the full kinematic range using lepton+jets
  events from proton-proton collisions at $\sqrt {s}$ = 13\,\,TeV}},
  \href{https://doi.org/10.1103/PhysRevD.104.092013}{\emph{Phys. Rev. D}
  {\bfseries 104} (2021) 092013}
  [\href{https://arxiv.org/abs/2108.02803}{{\ttfamily 2108.02803}}].

\bibitem{CMS:2022ged}
{\scshape CMS} collaboration, \emph{{Measurement of the tt\textasciimacron{}
  charge asymmetry in events with highly Lorentz-boosted top quarks in pp
  collisions at s=13 TeV}},
  \href{https://doi.org/10.1016/j.physletb.2023.137703}{\emph{Phys. Lett. B}
  {\bfseries 846} (2023) 137703}
  [\href{https://arxiv.org/abs/2208.02751}{{\ttfamily 2208.02751}}].

\bibitem{ATLAS:2022waa}
{\scshape ATLAS} collaboration, \emph{{Evidence for the charge asymmetry in pp
  \textrightarrow{} $ t\overline{t} $ production at $ \sqrt{s} $ = 13 TeV with
  the ATLAS detector}},
  \href{https://doi.org/10.1007/JHEP08(2023)077}{\emph{JHEP} {\bfseries 08}
  (2023) 077} [\href{https://arxiv.org/abs/2208.12095}{{\ttfamily
  2208.12095}}].

\bibitem{ATLAS:2022rms}
{\scshape ATLAS} collaboration, \emph{{Measurement of the polarisation of W
  bosons produced in top-quark decays using dilepton events at s=13 TeV with
  the ATLAS experiment}},
  \href{https://doi.org/10.1016/j.physletb.2023.137829}{\emph{Phys. Lett. B}
  {\bfseries 843} (2023) 137829}
  [\href{https://arxiv.org/abs/2209.14903}{{\ttfamily 2209.14903}}].

\bibitem{ATLAS:2021fzm}
{\scshape ATLAS} collaboration, \emph{{Measurements of the inclusive and
  differential production cross sections of a top-quark\textendash{}antiquark
  pair in association with a Z~boson at $\sqrt{s} = 13$~TeV with the ATLAS
  detector}}, \href{https://doi.org/10.1140/epjc/s10052-021-09439-4}{\emph{Eur.
  Phys. J. C} {\bfseries 81} (2021) 737}
  [\href{https://arxiv.org/abs/2103.12603}{{\ttfamily 2103.12603}}].

\bibitem{ATLAS:2017yax}
{\scshape ATLAS} collaboration, \emph{{Measurement of the $ t\overline{t}\gamma
  $ production cross section in proton-proton collisions at $ \sqrt{s}=8 $ TeV
  with the ATLAS detector}},
  \href{https://doi.org/10.1007/JHEP11(2017)086}{\emph{JHEP} {\bfseries 11}
  (2017) 086} [\href{https://arxiv.org/abs/1706.03046}{{\ttfamily
  1706.03046}}].

\bibitem{CMS:2017tzb}
{\scshape CMS} collaboration, \emph{{Measurement of the semileptonic $
  \mathrm{t}\overline{\mathrm{t}} $ + \ensuremath{\gamma} production cross
  section in pp collisions at $ \sqrt{s}=8 $ TeV}},
  \href{https://doi.org/10.1007/JHEP10(2017)006}{\emph{JHEP} {\bfseries 10}
  (2017) 006} [\href{https://arxiv.org/abs/1706.08128}{{\ttfamily
  1706.08128}}].

\bibitem{ATLAS:2021kqb}
{\scshape ATLAS} collaboration, \emph{{Measurement of the t$ \overline{t} $t$
  \overline{t} $ production cross section in $pp$ collisions at $ \sqrt{s} $ =
  13 TeV with the ATLAS detector}},
  \href{https://doi.org/10.1007/JHEP11(2021)118}{\emph{JHEP} {\bfseries 11}
  (2021) 118} [\href{https://arxiv.org/abs/2106.11683}{{\ttfamily
  2106.11683}}].

\bibitem{CMS:2019jsc}
{\scshape CMS} collaboration, \emph{{Search for the production of four top
  quarks in the single-lepton and opposite-sign dilepton final states in
  proton-proton collisions at $ \sqrt{s} $ = 13 TeV}},
  \href{https://doi.org/10.1007/JHEP11(2019)082}{\emph{JHEP} {\bfseries 11}
  (2019) 082} [\href{https://arxiv.org/abs/1906.02805}{{\ttfamily
  1906.02805}}].

\bibitem{ATLAS:2023ajo}
{\scshape ATLAS} collaboration, \emph{{Observation of four-top-quark production
  in the multilepton final state with the ATLAS detector}},
  \href{https://doi.org/10.1140/epjc/s10052-023-11573-0}{\emph{Eur. Phys. J. C}
  {\bfseries 83} (2023) 496}
  [\href{https://arxiv.org/abs/2303.15061}{{\ttfamily 2303.15061}}].

\bibitem{CMS:2023ftu}
{\scshape CMS} collaboration, \emph{{Observation of four top quark production
  in proton-proton collisions at s=13TeV}},
  \href{https://doi.org/10.1016/j.physletb.2023.138290}{\emph{Phys. Lett. B}
  {\bfseries 847} (2023) 138290}
  [\href{https://arxiv.org/abs/2305.13439}{{\ttfamily 2305.13439}}].

\bibitem{CMS:2020grm}
{\scshape CMS} collaboration, \emph{{Measurement of the cross section for
  $\text{t}\bar{\text{t}}$ production with additional jets and b jets in pp
  collisions at $\sqrt{s}=$ 13 TeV}},
  \href{https://doi.org/10.1007/JHEP07(2020)125}{\emph{JHEP} {\bfseries 07}
  (2020) 125} [\href{https://arxiv.org/abs/2003.06467}{{\ttfamily
  2003.06467}}].

\bibitem{ATLAS:2022wfk}
{\scshape ATLAS} collaboration, \emph{{Measurement of single top-quark
  production in the s-channel in proton\textendash{}proton collisions at $
  \sqrt{s} $ = 13 TeV with the ATLAS detector}},
  \href{https://doi.org/10.1007/JHEP06(2023)191}{\emph{JHEP} {\bfseries 06}
  (2023) 191} [\href{https://arxiv.org/abs/2209.08990}{{\ttfamily
  2209.08990}}].

\bibitem{CMS:2021ugv}
{\scshape CMS} collaboration, \emph{{Inclusive and differential cross section
  measurements of single top quark production in association with a Z boson in
  proton-proton collisions at $ \sqrt{s} $ = 13 TeV}},
  \href{https://doi.org/10.1007/JHEP02(2022)107}{\emph{JHEP} {\bfseries 02}
  (2022) 107} [\href{https://arxiv.org/abs/2111.02860}{{\ttfamily
  2111.02860}}].

\bibitem{CMS:2021vqm}
{\scshape CMS} collaboration, \emph{{Observation of tW production in the
  single-lepton channel in pp collisions at $ \sqrt{s} $ = 13 TeV}},
  \href{https://doi.org/10.1007/JHEP11(2021)111}{\emph{JHEP} {\bfseries 11}
  (2021) 111} [\href{https://arxiv.org/abs/2109.01706}{{\ttfamily
  2109.01706}}].

\bibitem{ATL-PHYS-PUB-2022-037}
{\scshape ATLAS} collaboration, \emph{{Combined effective field theory
  interpretation of Higgs boson and weak boson production and decay with ATLAS
  data and electroweak precision observables}},  tech. rep., CERN, Geneva,
  2022.

\bibitem{ATLAS:2024lyh}
{\scshape ATLAS} collaboration, \emph{{Interpretations of the ATLAS
  measurements of Higgs boson production and decay rates and differential
  cross-sections in $pp$ collisions at $\sqrt{s}=13$ TeV}},
  \href{https://arxiv.org/abs/2402.05742}{{\ttfamily 2402.05742}}.

\bibitem{Kassabov:2022pps}
Z.~Kassabov, E.~R. Nocera and M.~Wilson, \emph{{Regularising experimental
  correlations in LHC data: theory and application to a global analysis of
  parton distributions}},
  \href{https://doi.org/10.1140/epjc/s10052-022-10932-7}{\emph{Eur. Phys. J. C}
  {\bfseries 82} (2022) 956}
  [\href{https://arxiv.org/abs/2207.00690}{{\ttfamily 2207.00690}}].

\bibitem{Trotta:2008qt}
R.~Trotta, \emph{{Bayes in the sky: Bayesian inference and model selection in
  cosmology}}, \href{https://doi.org/10.1080/00107510802066753}{\emph{Contemp.
  Phys.} {\bfseries 49} (2008) 71}
  [\href{https://arxiv.org/abs/0803.4089}{{\ttfamily 0803.4089}}].

\bibitem{CMS:2017iqf}
{\scshape CMS} collaboration, \emph{{Measurement of double-differential cross
  sections for top quark pair production in pp collisions at $\sqrt{s} = 8$
  $\,\text {TeV}$ and impact on parton distribution functions}},
  \href{https://doi.org/10.1140/epjc/s10052-017-4984-5}{\emph{Eur. Phys. J. C}
  {\bfseries 77} (2017) 459}
  [\href{https://arxiv.org/abs/1703.01630}{{\ttfamily 1703.01630}}].

\bibitem{Durieux:2022cvf}
G.~Durieux, A.~G. Camacho, L.~Mantani, V.~Miralles, M.~M. L\'opez,
  M.~Ll\'acer~Moreno et~al., \emph{{Snowmass White Paper: prospects for the
  measurement of top-quark couplings}},  in \emph{{Snowmass 2021}}, 5, 2022,
  \href{https://arxiv.org/abs/2205.02140}{{\ttfamily 2205.02140}}.

\bibitem{NNPDF:2014otw}
{\scshape NNPDF} collaboration, \emph{{Parton distributions for the LHC Run
  II}}, \href{https://doi.org/10.1007/JHEP04(2015)040}{\emph{JHEP} {\bfseries
  04} (2015) 040} [\href{https://arxiv.org/abs/1410.8849}{{\ttfamily
  1410.8849}}].

\bibitem{Falkowski:2015jaa}
A.~Falkowski, M.~Gonzalez-Alonso, A.~Greljo and D.~Marzocca, \emph{{Global
  constraints on anomalous triple gauge couplings in effective field theory
  approach}}, \href{https://doi.org/10.1103/PhysRevLett.116.011801}{\emph{Phys.
  Rev. Lett.} {\bfseries 116} (2016) 011801}
  [\href{https://arxiv.org/abs/1508.00581}{{\ttfamily 1508.00581}}].

\bibitem{Alioli:2017nzr}
S.~Alioli, M.~Farina, D.~Pappadopulo and J.~T. Ruderman, \emph{{Catching a New
  Force by the Tail}},
  \href{https://doi.org/10.1103/PhysRevLett.120.101801}{\emph{Phys. Rev. Lett.}
  {\bfseries 120} (2018) 101801}
  [\href{https://arxiv.org/abs/1712.02347}{{\ttfamily 1712.02347}}].

\bibitem{Franceschini:2017xkh}
R.~Franceschini, G.~Panico, A.~Pomarol, F.~Riva and A.~Wulzer,
  \emph{{Electroweak Precision Tests in High-Energy Diboson Processes}},
  \href{https://doi.org/10.1007/JHEP02(2018)111}{\emph{JHEP} {\bfseries 02}
  (2018) 111} [\href{https://arxiv.org/abs/1712.01310}{{\ttfamily
  1712.01310}}].

\bibitem{Benedikt:2020ejr}
M.~Benedikt, A.~Blondel, P.~Janot, M.~Mangano and F.~Zimmermann, \emph{{Future
  Circular Colliders succeeding the LHC}},
  \href{https://doi.org/10.1038/s41567-020-0856-2}{\emph{Nature Phys.}
  {\bfseries 16} (2020) 402}.

\bibitem{Bernardi:2022hny}
G.~Bernardi et~al., \emph{{The Future Circular Collider: a Summary for the US
  2021 Snowmass Process}},  \href{https://arxiv.org/abs/2203.06520}{{\ttfamily
  2203.06520}}.

\bibitem{FCCfeasibility}
\emph{{FCC Feasibility Study Mid-Term Report - Executive Summary. Scientific
  Policy Committee - Three-Hundred-and-Thirty-Sixth Meeting}},  tech. rep.,
  2023.

\bibitem{An:2018dwb}
F.~An et~al., \emph{{Precision Higgs physics at the CEPC}},
  \href{https://doi.org/10.1088/1674-1137/43/4/043002}{\emph{Chin. Phys. C}
  {\bfseries 43} (2019) 043002}
  [\href{https://arxiv.org/abs/1810.09037}{{\ttfamily 1810.09037}}].

\bibitem{AbdulKhalek:2018rok}
R.~Abdul~Khalek, S.~Bailey, J.~Gao, L.~Harland-Lang and J.~Rojo, \emph{{Towards
  Ultimate Parton Distributions at the High-Luminosity LHC}},
  \href{https://doi.org/10.1140/epjc/s10052-018-6448-y}{\emph{Eur. Phys. J. C}
  {\bfseries 78} (2018) 962}
  [\href{https://arxiv.org/abs/1810.03639}{{\ttfamily 1810.03639}}].

\bibitem{Cruz-Martinez:2023sdv}
J.~M. Cruz-Martinez, M.~Fieg, T.~Giani, P.~Krack, T.~M\"akel\"a, T.~R.
  Rabemananjara et~al., \emph{{The LHC as a Neutrino-Ion Collider}},
  \href{https://doi.org/10.1140/epjc/s10052-024-12665-1}{\emph{Eur. Phys. J. C}
  {\bfseries 84} (2024) 369}
  [\href{https://arxiv.org/abs/2309.09581}{{\ttfamily 2309.09581}}].

\bibitem{Feng:2022inv}
J.~L. Feng et~al., \emph{{The Forward Physics Facility at the High-Luminosity
  LHC}}, \href{https://doi.org/10.1088/1361-6471/ac865e}{\emph{J. Phys. G}
  {\bfseries 50} (2023) 030501}
  [\href{https://arxiv.org/abs/2203.05090}{{\ttfamily 2203.05090}}].

\bibitem{DelDebbio:2021whr}
L.~Del~Debbio, T.~Giani and M.~Wilson, \emph{{Bayesian approach to inverse
  problems: an application to NNPDF closure testing}},
  \href{https://doi.org/10.1140/epjc/s10052-022-10297-x}{\emph{Eur. Phys. J. C}
  {\bfseries 82} (2022) 330}
  [\href{https://arxiv.org/abs/2111.05787}{{\ttfamily 2111.05787}}].

\bibitem{CMS:2019xnv}
{\scshape CMS} collaboration, \emph{{Measurements of Higgs boson production via
  gluon fusion and vector boson fusion in the diphoton decay channel at
  $\sqrt{s} = 13$ TeV}}, .

\bibitem{ATLAS:2019nkf}
{\scshape ATLAS} collaboration, \emph{{Combined measurements of Higgs boson
  production and decay using up to $80$ fb$^{-1}$ of proton-proton collision
  data at $\sqrt{s}=$ 13 TeV collected with the ATLAS experiment}},
  \href{https://doi.org/10.1103/PhysRevD.101.012002}{\emph{Phys. Rev. D}
  {\bfseries 101} (2020) 012002}
  [\href{https://arxiv.org/abs/1909.02845}{{\ttfamily 1909.02845}}].

\bibitem{ATLAS:2018pgp}
{\scshape ATLAS} collaboration, \emph{{Combined measurement of differential and
  total cross sections in the $H \rightarrow \gamma \gamma$ and the $H
  \rightarrow ZZ^* \rightarrow 4\ell$ decay channels at $\sqrt{s} = 13$ TeV
  with the ATLAS detector}},
  \href{https://doi.org/10.1016/j.physletb.2018.09.019}{\emph{Phys. Lett. B}
  {\bfseries 786} (2018) 114}
  [\href{https://arxiv.org/abs/1805.10197}{{\ttfamily 1805.10197}}].

\bibitem{ATLAS:2019yhn}
{\scshape ATLAS} collaboration, \emph{{Measurement of VH, $ \mathrm{H}\to
  \mathrm{b}\overline{\mathrm{b}} $ production as a function of the
  vector-boson transverse momentum in 13 TeV pp collisions with the ATLAS
  detector}}, \href{https://doi.org/10.1007/JHEP05(2019)141}{\emph{JHEP}
  {\bfseries 05} (2019) 141}
  [\href{https://arxiv.org/abs/1903.04618}{{\ttfamily 1903.04618}}].

\bibitem{CMS:2018gwt}
{\scshape CMS} collaboration, \emph{{Measurement and interpretation of
  differential cross sections for Higgs boson production at $\sqrt{s} =$ 13
  TeV}}, \href{https://doi.org/10.1016/j.physletb.2019.03.059}{\emph{Phys.
  Lett. B} {\bfseries 792} (2019) 369}
  [\href{https://arxiv.org/abs/1812.06504}{{\ttfamily 1812.06504}}].

\bibitem{ATLAS:2019rob}
{\scshape ATLAS} collaboration, \emph{{Measurement of fiducial and differential
  $W^+W^-$ production cross-sections at $\sqrt{s}=13$ TeV with the ATLAS
  detector}}, \href{https://doi.org/10.1140/epjc/s10052-019-7371-6}{\emph{Eur.
  Phys. J. C} {\bfseries 79} (2019) 884}
  [\href{https://arxiv.org/abs/1905.04242}{{\ttfamily 1905.04242}}].

\bibitem{ATLAS:2019bsc}
{\scshape ATLAS} collaboration, \emph{{Measurement of $W^{\pm}Z$ production
  cross sections and gauge boson polarisation in $pp$ collisions at $\sqrt{s} =
  13$ TeV with the ATLAS detector}},
  \href{https://doi.org/10.1140/epjc/s10052-019-7027-6}{\emph{Eur. Phys. J. C}
  {\bfseries 79} (2019) 535}
  [\href{https://arxiv.org/abs/1902.05759}{{\ttfamily 1902.05759}}].

\bibitem{CMS:2019efc}
{\scshape CMS} collaboration, \emph{{Measurements of the pp $\to$ WZ inclusive
  and differential production cross section and constraints on charged
  anomalous triple gauge couplings at $\sqrt{s} =$ 13 TeV}},
  \href{https://doi.org/10.1007/JHEP04(2019)122}{\emph{JHEP} {\bfseries 04}
  (2019) 122} [\href{https://arxiv.org/abs/1901.03428}{{\ttfamily
  1901.03428}}].

\bibitem{ATLAS:2019hxz}
{\scshape ATLAS} collaboration, \emph{{Measurements of top-quark pair
  differential and double-differential cross-sections in the $\ell$+jets
  channel with $pp$ collisions at $\sqrt{s}=13$ TeV using the ATLAS detector}},
  \href{https://doi.org/10.1140/epjc/s10052-019-7525-6}{\emph{Eur. Phys. J. C}
  {\bfseries 79} (2019) 1028}
  [\href{https://arxiv.org/abs/1908.07305}{{\ttfamily 1908.07305}}].

\bibitem{CMS:2018adi}
{\scshape CMS} collaboration, \emph{{Measurements of $\mathrm{t\overline{t}}$
  differential cross sections in proton-proton collisions at $\sqrt{s}=$ 13 TeV
  using events containing two leptons}},
  \href{https://doi.org/10.1007/JHEP02(2019)149}{\emph{JHEP} {\bfseries 02}
  (2019) 149} [\href{https://arxiv.org/abs/1811.06625}{{\ttfamily
  1811.06625}}].

\bibitem{ATLAS:2018fwl}
{\scshape ATLAS} collaboration, \emph{{Measurements of inclusive and
  differential fiducial cross-sections of $ t\overline{t} $ production with
  additional heavy-flavour jets in proton-proton collisions at $ \sqrt{s} $ =
  13 TeV with the ATLAS detector}},
  \href{https://doi.org/10.1007/JHEP04(2019)046}{\emph{JHEP} {\bfseries 04}
  (2019) 046} [\href{https://arxiv.org/abs/1811.12113}{{\ttfamily
  1811.12113}}].

\bibitem{CMS:2019eih}
{\scshape CMS} collaboration, \emph{{Measurement of the
  $\mathrm{t\bar{t}}\mathrm{b\bar{b}}$ production cross section in the all-jet
  final state in pp collisions at $\sqrt{s} =$ 13 TeV}},
  \href{https://doi.org/10.1016/j.physletb.2020.135285}{\emph{Phys. Lett. B}
  {\bfseries 803} (2020) 135285}
  [\href{https://arxiv.org/abs/1909.05306}{{\ttfamily 1909.05306}}].

\bibitem{ATLAS:2020hpj}
{\scshape ATLAS} collaboration, \emph{{Evidence for $t\bar{t}t\bar{t}$
  production in the multilepton final state in proton\textendash{}proton
  collisions at $\sqrt{s}=13$ $\text {TeV}$ with the ATLAS detector}},
  \href{https://doi.org/10.1140/epjc/s10052-020-08509-3}{\emph{Eur. Phys. J. C}
  {\bfseries 80} (2020) 1085}
  [\href{https://arxiv.org/abs/2007.14858}{{\ttfamily 2007.14858}}].

\bibitem{CMS:2019rvj}
{\scshape CMS} collaboration, \emph{{Search for production of four top quarks
  in final states with same-sign or multiple leptons in proton-proton
  collisions at $\sqrt{s}=$ 13 TeV}},
  \href{https://doi.org/10.1140/epjc/s10052-019-7593-7}{\emph{Eur. Phys. J. C}
  {\bfseries 80} (2020) 75} [\href{https://arxiv.org/abs/1908.06463}{{\ttfamily
  1908.06463}}].

\bibitem{CMS:2019too}
{\scshape CMS} collaboration, \emph{{Measurement of top quark pair production
  in association with a Z boson in proton-proton collisions at $\sqrt{s}=$ 13
  TeV}}, \href{https://doi.org/10.1007/JHEP03(2020)056}{\emph{JHEP} {\bfseries
  03} (2020) 056} [\href{https://arxiv.org/abs/1907.11270}{{\ttfamily
  1907.11270}}].

\bibitem{ATLAS:2019fwo}
{\scshape ATLAS} collaboration, \emph{{Measurement of the $t\bar{t}Z$ and
  $t\bar{t}W$ cross sections in proton-proton collisions at $\sqrt{s}=13$ TeV
  with the ATLAS detector}},
  \href{https://doi.org/10.1103/PhysRevD.99.072009}{\emph{Phys. Rev. D}
  {\bfseries 99} (2019) 072009}
  [\href{https://arxiv.org/abs/1901.03584}{{\ttfamily 1901.03584}}].

\bibitem{CMS:2017ugv}
{\scshape CMS} collaboration, \emph{{Measurement of the cross section for top
  quark pair production in association with a W or Z boson in proton-proton
  collisions at $\sqrt{s} =$ 13 TeV}},
  \href{https://doi.org/10.1007/JHEP08(2018)011}{\emph{JHEP} {\bfseries 08}
  (2018) 011} [\href{https://arxiv.org/abs/1711.02547}{{\ttfamily
  1711.02547}}].

\bibitem{ATLAS:2016qhd}
{\scshape ATLAS} collaboration, \emph{{Measurement of the inclusive
  cross-sections of single top-quark and top-antiquark $t$-channel production
  in $pp$ collisions at $\sqrt{s}$ = 13 TeV with the ATLAS detector}},
  \href{https://doi.org/10.1007/JHEP04(2017)086}{\emph{JHEP} {\bfseries 04}
  (2017) 086} [\href{https://arxiv.org/abs/1609.03920}{{\ttfamily
  1609.03920}}].

\bibitem{CMS:2019jjp}
{\scshape CMS} collaboration, \emph{{Measurement of differential cross sections
  and charge ratios for t-channel single top quark production in
  proton\textendash{}proton collisions at $\sqrt{s}=13\,\text {Te}\text {V}$}},
  \href{https://doi.org/10.1140/epjc/s10052-020-7858-1}{\emph{Eur. Phys. J. C}
  {\bfseries 80} (2020) 370}
  [\href{https://arxiv.org/abs/1907.08330}{{\ttfamily 1907.08330}}].

\bibitem{ATLAS:2016ofl}
{\scshape ATLAS} collaboration, \emph{{Measurement of the cross-section for
  producing a W boson in association with a single top quark in pp collisions
  at $ \sqrt{s}=13 $ TeV with ATLAS}},
  \href{https://doi.org/10.1007/JHEP01(2018)063}{\emph{JHEP} {\bfseries 01}
  (2018) 063} [\href{https://arxiv.org/abs/1612.07231}{{\ttfamily
  1612.07231}}].

\bibitem{CMS:2018amb}
{\scshape CMS} collaboration, \emph{{Measurement of the production cross
  section for single top quarks in association with W bosons in proton-proton
  collisions at $ \sqrt{s}=13 $ TeV}},
  \href{https://doi.org/10.1007/JHEP10(2018)117}{\emph{JHEP} {\bfseries 10}
  (2018) 117} [\href{https://arxiv.org/abs/1805.07399}{{\ttfamily
  1805.07399}}].

\bibitem{ATLAS:2020bhu}
{\scshape ATLAS} collaboration, \emph{{Observation of the associated production
  of a top quark and a $Z$ boson in $pp$ collisions at $\sqrt{s} = 13$ TeV with
  the ATLAS detector}},
  \href{https://doi.org/10.1007/JHEP07(2020)124}{\emph{JHEP} {\bfseries 07}
  (2020) 124} [\href{https://arxiv.org/abs/2002.07546}{{\ttfamily
  2002.07546}}].

\bibitem{Diehl:1993br}
M.~Diehl and O.~Nachtmann, \emph{{Optimal observables for the measurement of
  three gauge boson couplings in e+ e- ---\ensuremath{>} W+ W-}},
  \href{https://doi.org/10.1007/BF01555899}{\emph{Z. Phys. C} {\bfseries 62}
  (1994) 397}.

\bibitem{Chen:2020mev}
S.~Chen, A.~Glioti, G.~Panico and A.~Wulzer, \emph{{Parametrized classifiers
  for optimal EFT sensitivity}},
  \href{https://doi.org/10.1007/JHEP05(2021)247}{\emph{JHEP} {\bfseries 05}
  (2021) 247} [\href{https://arxiv.org/abs/2007.10356}{{\ttfamily
  2007.10356}}].

\bibitem{GomezAmbrosio:2022mpm}
R.~Gomez~Ambrosio, J.~ter Hoeve, M.~Madigan, J.~Rojo and V.~Sanz,
  \emph{{Unbinned multivariate observables for global SMEFT analyses from
  machine learning}},
  \href{https://doi.org/10.1007/JHEP03(2023)033}{\emph{JHEP} {\bfseries 03}
  (2023) 033} [\href{https://arxiv.org/abs/2211.02058}{{\ttfamily
  2211.02058}}].

\bibitem{Chen:2023ind}
S.~Chen, A.~Glioti, G.~Panico and A.~Wulzer, \emph{{Boosting likelihood
  learning with event reweighting}},
  \href{https://doi.org/10.1007/JHEP03(2024)117}{\emph{JHEP} {\bfseries 03}
  (2024) 117} [\href{https://arxiv.org/abs/2308.05704}{{\ttfamily
  2308.05704}}].

\bibitem{Chai:2024zyl}
S.~Chai, J.~Gu and L.~Li, \emph{{From optimal observables to machine learning:
  an effective-field-theory analysis of $e+e^-\rightarrow W^+W^-$ at future
  lepton colliders}},
  \href{https://doi.org/10.1007/JHEP05(2024)292}{\emph{JHEP} {\bfseries 05}
  (2024) 292} [\href{https://arxiv.org/abs/2401.02474}{{\ttfamily
  2401.02474}}].

\bibitem{Brivio:2019myy}
I.~Brivio, T.~Corbett and M.~Trott, \emph{{The Higgs width in the SMEFT}},
  \href{https://doi.org/10.1007/JHEP10(2019)056}{\emph{JHEP} {\bfseries 10}
  (2019) 056} [\href{https://arxiv.org/abs/1906.06949}{{\ttfamily
  1906.06949}}].

\end{thebibliography}

\providecommand{\href}[2]{#2}\begingroup\raggedright\endgroup

\end{document}